\documentclass[12pt]{article}
\usepackage{amsthm,amsmath,latexsym,amssymb,amsfonts,amscd}
\usepackage{graphics,lscape,fancyhdr,array,stmaryrd,euscript,wrapfig}
\usepackage{empheq}
\usepackage{verbatim,slashed}
\numberwithin{equation}{section}
\usepackage{hyperref,setspace}
\usepackage[numbers,sort&compress]{natbib}
\usepackage[nottoc]{tocbibind}
\usepackage{tikz-cd}
\usepackage[all]{xy}
\usepackage{tikz}
\usetikzlibrary{fit}
\usetikzlibrary{patterns}
\usetikzlibrary{arrows.meta}
\usetikzlibrary{decorations.pathmorphing}
\usepackage{multirow}

\usepackage{rotating}% provides a rotate environment

\newcommand{\sign}{\operatorname{sign}}

\numberwithin{equation}{section}
\newcommand{\pl}{\partial}

\newcommand{\Tr}{\mathrm{Tr}}

\newcommand{\fud}[2]{{}^{#1}{}_{#2}\,}
\newcommand{\fdu}[2]{{}_{#1}{}^{#2}\,}
\newcommand{\fudu}[3]{{}^{#1}{}_{#2}{}^{#3}\,}
\newcommand{\fudud}[4]{{}^{#1}{}_{#2}{}^{#3}{}_{#4}\,}
\newcommand{\fdud}[3]{{}_{#1}{}^{#2}{}_{#3}\,}

\newcommand{\brk}{{{\bar{k}}}}

\newcommand{\pll}{\eth}

\newcommand{\besubeqs}{\begin{subequations}}
\newcommand{\esubeqs}{\end{subequations}}

\begin{document}
%%%%%%%%%%%%%%%%%%%%%%%%%%%%%%%%%%%%%%%%%%%%%%%%%%%%%%%%%%%%%
%\pagenumbering{gobble}
\hfill
\vskip 0.01\textheight
\begin{center}
{\large\bfseries 
Dirichlet, Neumann, Mixed and self-dual holography:\\
\vspace{0.2cm}
(self-dual) Yang--Mills theory II}

\vspace{0.4cm}

\vskip 0.03\textheight
\renewcommand{\thefootnote}{\fnsymbol{footnote}}
Evgeny \textsc{Skvortsov}\footnote{Also affiliated with Lebedev Institute of Physics.}  \& Richard van Dongen

\vskip 0.03\textheight

{\em Service de Physique de l'Univers, Champs et Gravitation, \\ Universit\'e de Mons, 20 place du Parc, 7000 Mons, 
Belgium}\\
\vspace*{5pt}

\renewcommand{\thefootnote}{\arabic{footnote}}
\end{center}

\vskip 0.02\textheight

\begin{abstract}
We consider Yang--Mills, Chalmers--Siegel and self-dual Yang--Mills (SDYM) theories within AdS/CFT correspondence. Bulk-to-bulk and boundary-to-bulk propagators are derived in various gauges and for Dirichlet, Neumann, mixed and self-dual boundary conditions. Three- and four-point holographic correlators are computed in the three theories to establish the relation between the observables thereof. This is a companion paper to [arXiv:2602.21658]. 
\end{abstract}

\newpage
\tableofcontents
\newpage

%%%%%%%%%%%%%%%%%%%%%%%%%%%%%%%%%%%%%%%%%%%%%%%%%%%%%%%%%%%%%
\section{Introduction}
\label{sec:intro}
%%%%%%%%%%%%%%%%%%%%%%%%%%%%%%%%%%%%%%%%%%%%%%%%%%%%%%%%%%%%%

In the search for diverse and simple models of the AdS/CFT duality, various extensions and generalizations have been explored. One rich ``degree of freedom'' is the fact that a single bulk theory can be dual to many(fold) of conformal field theories, due to a multitude of boundary conditions one has to impose before computing any AdS/CFT correlators. The relation between double-/multi-trace deformations and Dirichlet/Neumann/mixed boundary conditions has been proposed in \cite{Klebanov:1999tb,Witten:2001ua} for scalar fields. Except for the end points, Dirichlet and Neumann, generic mixed boundary conditions for a scalar field are not conformally-invariant. The $\text{AdS}_4/\text{CFT}_3$ case is special in that one can impose conformally-invariant mixed boundary conditions on gauge fields \cite{Witten:2003ya,Leigh:2003ez,Yee:2004ju,deHaro:2007eg}, which can be extended to gravity and even to higher spins  \cite{Leigh:2003gk,Petkou:2004nu,Compere:2008us,deHaro:2008gp,Giombi:2013yva}. In particular, instanton/self-dual options have been mentioned in \cite{Petkou:2004nu,deHaro:2007eg,Compere:2008us,deHaro:2008gp}. It was also understood that a switch between Dirichlet and Neumann boundary conditions admits a purely CFT interpretation of a Legendre transform \cite{Klebanov:1999tb,Witten:2001ua, Hartman:2006dy, Giombi:2011ya}. A practical importance of this observation is that if one of the dualities is established, the other one follows immediately \cite{Hartman:2006dy,Giombi:2011ya} and does not require any additional bulk computations. 

A new direction within the AdS/CFT duality is to explore the self-duality, see e.g. a handful of the recent works \cite{Skvortsov:2018uru,Sharapov:2022awp,Jain:2024bza,Aharony:2024nqs,Sharma:2025ntb}. Indeed, self-dual theories, such as self-dual Yang--Mills, self-dual gravity and supersymmetric and higher-spin extensions thereof, have many appealing features: the theories are much simpler (even integrable in some sense) than the parent ones; self-dual theories are UV finite, which makes them very good toy-models of the AdS/CFT duality as they can lead to complete (no need for a UV completion) and completely solvable models of the duality; all solutions and amplitudes of self-dual theories belong to the parent ones and, hence, one is always ``in touch'' with the parent theory. In flat space, tree-level amplitudes of self-dual theories are trivial for generic kinematics, but see the recent \cite{Guevara:2026qzd}. However, this is not the case within the AdS/CFT correspondence: the curvature corrections make even tree-level amplitudes nonvanishing. Developing self-dual or chiral holography requires new techniques, some of which have been developed recently, and this paper extends them further. The present paper contains some technical details omitted in \cite{Skvortsov:2026gtq} and many more results.

After the first seminal works that computed AdS/CFT correlators, see e.g. \cite{Liu:1998ty,DHoker:1998bqu,DHoker:1999kzh,DHoker:1999mqo,Arutyunov:2000py}, which were all done in position space, the focus has moved to the hybrid momentum-position space where a Fourier transform is performed with respect to boundary coordinates whereas the Poincare radius $z$ is kept, see e.g. \cite{Raju:2011mp,Raju:2012zs,Raju:2012zr,Albayrak:2023kfk,Albayrak:2023jzl,Armstrong:2020woi}. It was also discovered that the residue of the leading energy pole of an AdS/CFT correlator is just the flat space amplitude of the bulk theory, see e.g. \cite{Maldacena:2011nz,Hijano:2019qmi,Li:2022tby}. Therefore, the flat space limit of bulk theories capture the leading behavior of AdS/CFT correlators. The latter also motivates the development of a variety of spinor-helicity-type techniques \cite{Maldacena:2011nz,Nagaraj:2018nxq,Nagaraj:2019zmk,Nagaraj:2020sji,Skvortsov:2022wzo,Baumann:2024ttn}, which makes the relation to flat space kinematics simpler. Some AdS/CFT amplitudes in the spinor-helicity language have been computed in \cite{Armstrong:2020woi,Albayrak:2023jzl}. 

Since self-duality discriminates between negative and positive helicities, it is important to approach it wisely. A natural framework for self-dual theories has always been the twistor space. However, holographic applications of the twistor space techniques are yet to be developed, but see \cite{Baumann:2024ttn,CarrilloGonzalez:2025qjk}. Self-dual theories are somewhat easier to understand in the light-cone gauge. For example, the self-dual Yang--Mills theory and gravity, when reduced to the physical degrees of freedom by going into the light-cone gauge, are theories of two ``scalar'' fields that represent $\pm1$ or $\pm2$ degrees of freedom. The self-dual truncation keeps, say, the $(++-)$ half of the cubic vertex, omits the other $(--+)$ half and also drops all higher order vertices. It is still nontrivial that the resulting theory is Lorentz-invariant. While very handy in the flat space, light-cone gauge becomes significantly more difficult in $\text{AdS}_4$, see e.g. \cite{Metsaev:2018xip,Lipstein:2023pih,Neiman:2023bkq,Neiman:2024vit,CarrilloGonzalez:2024sto,Chowdhury:2024dcy}, see, however, \cite{Kozaki:2025jrj,Kozaki:2026rdc} for a hybrid approach that makes the light-cone gauge more covariant. 

A covariant formulation of self-dual Yang--Mills theory is provided by the Chalmers--Siegel action \cite{Chalmers:1996rq} and this is the action we use in the paper. For the self-dual gravity it should be advantageous to use Krasnov's formulation \cite{Krasnov:2016emc}, which originates from the Plebansky approach. Both the Krasnov and Chalmers--Siegel formulations have straightforward higher-spin extensions \cite{Krasnov:2021nsq}, originally discovered in the light-cone gauge \cite{Ponomarev:2017nrr}. However, the space of self-dual theories is much bigger than originally thought \cite{Serrani:2025owx}. Also, self-dual theories are smooth in the cosmological constant and some self-dual theories are even conformally invariant, which makes them ideal examples to study the flat space holography. In flat space, self-dual theories can be defined as those with vanishing tree-level amplitudes; as theories featuring only cubic interactions with the total helicity being positive $\sum_i \lambda_i>0$; as those for which the holomorphic celestial OPE is closed \cite{Ren:2022sws,Serrani:2025oaw}.

Self-dual theories raise a number of basic AdS/CFT questions that we are going to answer for the Chalmers--Siegel example: what the possible boundary conditions are and what boundary terms to have a well-defined variational principle are (one subtlety is that the on-shell actions vanish, which is also the case for other first-order theories such as for spin-half and the gravitino); what the Fefferman-Graham expansion in self-dual theories is; what the AdS/CFT dictionary is; what bulk-to-bulk and boundary-to-bulk propagators for the most general mixed boundary conditions are and what happens in the self-dual limit; how AdS/CFT correlators of self-dual theories look like and what the relation between correlators in parent and self-dual theories is. 

One important feature of the hybrid momentum-position space and of the related spinor-helicity language is that it allows one to separate fields into components of definite helicity. The latter is crucial since positive and negative helicity fields have different regularity properties in the interior of anti-de Sitter space. Some of these subtleties have already been observed for chiral fermions $\text{AdS}_4$, see e.g.  \cite{Gripaios:2008rg,Porrati:2009dy,Rattazzi:2009ux,Bolognesi:2011un,Herzog:2018lqz,Foit:2019nsr,Ciccone:2025dqx}, as well as the fact that the on-shell action vanishes and that the two-point function comes from a boundary term \cite{Arutyunov:1998ve,Henneaux:1998ch}.

In the paper we study a family of actions that originate from the Yang--Mills theory $S_{\text{YM}}$, which, up to a boundary term, is equivalent to its chiral part $(F_+)^2$, $F_\pm$ being the (anti)self-dual components of $F$. The latter can be represented in the Chalmers--Siegel form upon introducing an auxiliary (self-dual) field $\Psi$, $S_{\text{Ch.Si.}}$. Upon dropping the last, gauge-invariant, piece, one ends up at SDYM, $S_{\text{SDYM}}$. Schematically, the sequence is
\begin{align}
    S_{\text{YM}}\sim\int (F_+)^2+(F_-)^2 \sim \int (F_+)^2 \rightarrow S_{\text{Ch.Si.}}\sim \int \Psi F_+ -\tfrac{\epsilon}2 \Psi^2 \rightarrow S_{\text{SDYM}}\sim\int \Psi F_+ \,.
\end{align}
In the paper we develop the holographic applications for these two and half theories: the Yang--Mills theory, its Chalmers--Siegel reformulation and SDYM. A short summary of the results (can also be read as the outline) is as follows:
\begin{itemize}

    \item In Section \ref{sec:2.5} we review the family of theories. Even though SDYM can be thought of as a limit of the Chalmers--Siegel theory, the limit is not quite smooth. For any nonvanishing coefficient of the $\Psi^2$-term, the theory is equivalent back to YM. Since an infinite boundary term needs to be dropped as well as taking the limit, SDYM is not a naive limit of YM. It may be better to think of SDYM as of a closed subsector of YM, i.e. a theory that is consistent on its own and captures some of the observables of YM (all solutions of SDYM are solutions of YM and all amplitudes of SDYM are subsets of amplitudes of YM). There exists however, a theory with two complex couplings $g$ and $\bar{g}$ that has YM as a real slice $g=\bar{g}\in \mathbb{R}$, SDYM as the limit $\bar{g}=0$ and the MHV-expansion, which can be interpreted as a deformation of SDYM towards YM, is an expansion in $\bar{g}$. In some sense, this gives an alternative to the Chalmers--Siegel action that is smooth. 

    Another ``theory'' we consider is chiral Yang--Mills theory where interactions are confined to $(F_+)^2$ in the bulk and to a Chern--Simons term on the boundary. This simplifies the AdS/CFT computations (less diagrams) and also isolates the terms that survive in the flat space limit. Note that the leading energy pole of the AdS/CFT correlator should reproduce the flat space amplitude. Genuine bulk exchanges and contact vertices do have the right energy pole, while the boundary vertices and exchanges involving boundary and bulk vertices are softer.

    \item The holographic dictionary can be revealed with the help of the Fefferman-Graham expansion in Section \ref{sec:FG}, i.e. by solving the equations of motion near the boundary. This allows one to identify the boundary data (and to read off conformal properties thereof), i.e. the free functions that can be fixed on the boundary to reconstruct solutions, at least asymptotically close to the boundary. The boundary data gives options for the CFT dual operators. 
    
    For Maxwell theory, the boundary data is well-known: $A_i=z(a_i+z E_i)+...$, which is a gauge field $a_i$ (a more invariant characteristic would be its magnetic field $B_i$) and the electric field $E_i$, $\pl_i E^i=0$ that behaves as a conserved current. The Chalmers--Siegel theory isolates/confines the negative helicity part of both $a_i$ and $E_i$ to $\Psi\sim B+E=F_+$, while the gauge potential $A_\mu$ still carries everything.

    In SDYM, where $\Psi$ is no longer auxiliary, $\Psi$ supports only the negative helicity $F_+=B+E$ component of the current combined with the gauge field ($\Psi$ still satisfies the same equations as $F_+$ but is no longer related to $A_\mu$). $A_\mu$, whose $F_+$ now vanishes, supports the positive helicity. Note that each of $E_i$ and $B_i$ carries two (off-shell) degrees of freedom (the number of transverse and traceless components). SDYM is a first order theory and only Dirichlet conditions are possible. 
    
    The boundary momentum space (with the radial coordinate $z$ kept) allows one to separate helicity eigenstates in a Lorentz-invariant way, which is impossible (non-local, in other words) in the position space. The definite helicity states are important in achieving a well-defined variational problem for SDYM and for extracting the AdS/CFT dictionary.
    
    \item A Lagrangian submanifold of the boundary data needs to be fixed for solutions to be regular in the interior of the bulk. The Dirichlet/Neumann choices are to fix $a_i$ or $E_i$, respectively. Mixed boundary conditions that are conformally-invariant fix 
    \begin{align}
        &B_i \cos(\gamma) +i \sin(\gamma) E_i && \Longleftrightarrow && F_+ e^{i\gamma} +F_- e^{-i\gamma} \,.
    \end{align}
    The self-dual limit corresponds to $\gamma=-i\infty$, where $F_+=B+E$ is not just fixed, but is set to zero. The boundary data should be associated with the components of $A_\mu$ and $\Psi$ that admit regular solutions in the bulk. Such boundary data, in the old YM terms, looks like Dirichlet/Neumann conditions on the positive/negative helicity components of $A_i$. With $a_i$ fixed (Dirichlet) the AdS theory yields $W[a]$ --- a generating functional of correlators of the dual current $J_i=E_i$, which can also be read off as $E_i$ near the boundary. With $E_i$ fixed (Neumann), the AdS theory gives a generating functional of correlators of a gauge field $a_i$ on the boundary. A better object to consider is the associated magnetic field $B_i$, which behaves as a conserved current as well. 

    In SDYM $A_\mu$/$\Psi$ behave as a gauge field/conserved current on the boundary, but only positive/negative helicity components $A_+/\Psi_-$ are allowed to propagate, respectively. Therefore, a natural proposal for self-dual holography is that $A_+/\Psi_-$ are dual to the corresponding components $J_+/a_-$ of the current and a gauge field on the boundary. The magnetic field of $a_-$ gives the missing $J_-$ of the full current on the boundary. This is discussed in Section \ref{sec:AdSCFTdict}.
    
    We also discuss a one-parameter family of conformally-invariant boundary conditions that can be imposed on the scalar field and its shadow and on the spin-half field. This is useful since, in terms of the helicity eigenstates, all these theories --- YM, Ch.Si., spin-half and SDYM --- correspond to theories of several scalar fields and look very close to each other. 

    \item In order to compute holographic correlators one needs bulk-to-bulk and boundary-to-bulk propagators, which are derived in Section \ref{sec:props}. To make sure the results are reliable and to explore options for further applications we have studied a number of gauges in YM, SDYM and Chalmers--Siegel theory. It is very advantageous to use the fact that all these theories are Weyl invariant and, hence, $\text{AdS}_4$ can be modeled as the flat space with a boundary at $z=0$. Some of the usual suspects, e.g. Feynman and Lorenz gauges, can be imposed in the flat space model. This is not the same as to impose them before the Weyl transformation. This way we have got very simple propagators as compared to the ones derived directly in $\text{AdS}_4$, see e.g. \cite{Moga:2025gdy}. Axial gauge is, of course, the one that simplifies the AdS/CFT dictionary, but the propagators are the simplest in the Feynman gauge. The inhomogeneous part of the propagators can be obtained by a Fourier transform of momentum space propagators in flat space with respect to the Euclidean energy that is dual to radial $z$. Further  homogeneous terms are needed to support the required boundary conditions and regularity in the bulk.  

    In order for the most general boundary conditions to be well-defined within the variational principle, one has to supplement the actions with specific boundary terms. The same boundary terms are crucial for reproducing the two-point functions. In practice, one can stick to Feynman--Witten diagrams and never discuss the boundary terms. In general, either with a Chern--Simons term added or with mixed boundary conditions imposed, the two-point functions read
    \begin{align}
        \langle ++ \rangle&=k (a+b) \,, & \langle -- \rangle&=k (a-b) \,,
    \end{align}
    where $a/b$ can be tuned to any value. The parity-odd contribution, $b$, comes either from the theta/Chern--Simons term or from the mixed boundary conditions. The self-dual limit should have $\langle -- \rangle=0$. In general, nonvanishing correlators should have the total helicity positive. 
    
    Despite $\langle -- \rangle=0$ is achievable just with Dirichlet boundary conditions, SDYM should be approached as the self-dual limit of the mixed-boundary condition (having, say, Dirichlet boundary condition plus self-duality over-constrains the system).  

    \item With the preparatory steps above, one can proceed to the computation of Feynman--Witten diagrams in SDYM, Chalmers--Siegel and  Yang--Mills theory in Section \ref{sec:Correlators}. What has a smooth limit to self-duality is the Chalmers--Siegel action, which is equivalent to the Yang--Mills action with a particular value of the theta-term that makes the Lagrangian be just $(F_+)^2$.

    There is a difference between SDYM and YM already at the three-point level unless the theta-term is fine-tuned. We carefully compute the four-point functions in SDYM in axial and Feynman gauge. We also compute the four-point function in YM/Chalmers--Siegel for Dirichlet, Neumann and mixed boundary conditions. In the self-dual limit the result of YM approaches that of SDYM. The gauge (in)dependence is carefully analyzed.
    
\end{itemize}

In order to elaborate on various aspects of the story and to make the exposition self-contained, the paper is supplemented with a number of appendices. There is a number of purely technical ones meant to collect some formulae: Appendix \ref{app:notation} summarizes our notation; Appendix \ref{app:fourier} lists some integrals used in the main part; Appendix \ref{app:Lorenz} discusses Lorenz gauge for YM, which we do not use much in the main text; a discussion of color ordering in SDYM can be found in Appendix \ref{app:color}. Apart from that, there are more conceptual Appendices. General aspects of mixed boundary conditions, boundary terms they require and toy models based on scalar fields can be found in Appendix \ref{app:boundaryterms}. We detail the computation of three- and four-point amplitudes in SDYM, (chiral) YM and Chalmers--Siegel theory in flat space in Appendix \ref{app:flat}. This gives the leading pole of the AdS/CFT results and helps to see how the $(+++-)$-amplitude vanishes. Gauge (in)dependence of the CFT correlators is discussed in Appendix \ref{app:gaugedependendence}. The contribution of composite operators, i.e. $aa$-terms in the dual current $J=*(da+aa)$, is considered in Appendix \ref{app:sdymleftovers}. It is often advantageous to reduce a given theory to the physical degrees of freedom (unitary gauge), which is done in Appendix \ref{app:ham}. Collinear limits of the AdS-correlators are analyzed in Appendix \ref{app:collinear}.

As compared to \cite{Skvortsov:2026gtq}, the present text contains many details of the computations. The propagators are derived here, while in \cite{Skvortsov:2026gtq} some of them are just given. However, it is impossible to completely disentangle the two texts and, also for the sake of completeness, there is an overlap of Sections devoted to the Fefferman--Graham expansions, AdS/CFT dictionary and correlators.

%%%%%%%%%%%%%%%%%%%%%%%%%%%%%%%%%%%%%%%%%%%%%%%%%%%%%%%%%%%%%
\section{Two and half theories}
\label{sec:2.5}
%%%%%%%%%%%%%%%%%%%%%%%%%%%%%%%%%%%%%%%%%%%%%%%%%%%%%%%%%%%%%
All theories we work with are conformally-invariant. Therefore, instead of considering them in the Euclidean anti-de Sitter space in Poincare coordinates $(z,x^i)$ with metric 
\begin{align*}
    ds^2 &= \frac{1}{\lambda z^2}(dz^2+dx_i dx^i)\,, && x^i\in \mathbb{R}^3\,, \quad z>0\,,
\end{align*}
it is advantageous to map them to the half of the flat Euclidean space $\mathbb{R}^4$ with $x^i\in \mathbb{R}^3$, $z>0$. The map trivializes $\sqrt{g}$ and ensures $\nabla_{AA'}=\pl_{AA'}$ for the fields we consider.\footnote{Recall that we only discuss conformally invariant theories in the paper. It is not true that $\nabla_{AA'}=\pl_{AA'}$ for non-conformal fields after the Weyl rescaling. } Above, $\lambda$ is related to the cosmological constant, which will not be used in what follows. Throughout the text we will work in the two-component spinor language, which is especially suitable for making contact with the spinor-helicity formalism and self-duality.  

\paragraph{Yang--Mills theory.} The action of Yang--Mills theory reads
\begin{align}\label{YMaction}
    S_{\text{YM}}&= \frac{a}{g^2}\Tr\int \left(F_{AB}F^{AB} +\bar{F}_{A'B'} \bar{F}^{A'B'}\right)\,,
\end{align}
where the decomposition into (anti)-self-dual components has been made
\begin{equation}
F_{\mu\nu} \equiv F_{AA'BB'} = F_{AB}\epsilon_{A'B'} + \bar{F}_{A'B'}\epsilon_{AB} \,.
\end{equation}
In particular,\footnote{Here, we slowly pass to the two-component spinor language, $A,B,...=1,2$, $A',B',...=1,2$. The indices are raised by $\epsilon_{AB}$ and $\epsilon_{A'B'}$, e.g. $\xi^A=\epsilon^{AB}\xi_B$, $\xi^A\epsilon_{AB}=\xi_B$. Indices on fields separated by a comma, e.g. $X_{A,B}$, are assumed not to possess any symmetry, while symmetrized indices are not separated by a comma, $X_{AB}=X_{BA}$. Out of laziness a group of indices in which a spin-tensor is symmetric or to be symmetrized can be denoted by the same letter, e.g. $\Phi_{AA}\equiv \Phi_{A_1 A_2}$. $3d$ vectors carry indices $a,b,i,j,m,...=1,2,3$ and correspond to symmetric bi-spinors, e.g. $v_i \sim v_{AA}$. } $F_{AB}=\tfrac12(\pl_{AB'} A\fdu{B,}{B'}+\pl_{BB'} A\fdu{A,}{B'}+A_{A,B'}A\fdu{B,}{B'}+A_{B,B'}A\fdu{A,}{B'})$ and $\bar{F}_{A'B'}=\tfrac12(\pl_{BA'} A\fud{B}{,B'}+\pl_{BB'} A\fud{B}{,A'}+A_{B,A'}A\fud{B}{,B'}+A_{B,B'}A\fud{B}{,A'})$. Here, $A_{M,M'}{}\fud{i}{j}$ is understood as taking values in $\mathfrak{gl}(n,\mathbb{C})$ with appropriate reality conditions to be imposed later, if necessary. In terms of the corresponding one-form $dx^{AA'} A_{A,A'}$ the field strength is defined as $F=dA+A\wedge A$ with the matrix product implied. $g$ is the coupling constant, which we will not be using much. $a$ is a numerical normalization factor to produce the canonically normalized kinetic term after replacing $A \rightarrow g A$ and we will also use it to make some formulas look more symmetric. In case of a nontrivial gravitational background it can be more useful to define 
\begin{align}
    F=\tfrac12 H^{AB} F_{AB}+\tfrac12 \bar{H}^{A'B'} \bar{F}_{A'B'}=\tfrac12 F_+ +\tfrac12 F_-
\end{align}
where $H^{AB}=H^{BA}=e\fud{A}{C'}\wedge e^{BC'}$, $\bar{H}^{A'B'}=\bar{H}^{B'A'}=e\fdu{C}{A'}\wedge e^{CB'}$ and $e^{AA'}$ is the vierbein one-form, $\nabla e^{AA'}=0$. In terms of these differential forms the action reads
\begin{align}\label{YMactionB}
    S_{\text{YM}}&= \frac{3a}{vg^2}\Tr\int F_+ \wedge F_+ -F_-\wedge F_- =\frac{3a}{vg^2}\Tr\int (F_++F_-)\wedge (F_+-F_-)\,,
\end{align}
where we can refer to Appendix \ref{app:notation} for our conventions and some identities. In particular, $v$ is the volume conversion factor between the flat measure $d^4x$ and the volume four-form $H_{AB}\wedge H^{AB}$.

For practical calculations it is advantageous to simplify the structure of interactions, which motivates us to introduce another ``theory'' below.

\paragraph{Chiral Yang--Mills theory (cYM).} To begin with, there is a topological invariant
\begin{align}\label{YMtop}
    S_{\text{top}}&= b\,\Tr\int \left(F_{AB}F^{AB} -\bar{F}_{A'B'} \bar{F}^{A'B'}\right)= \frac{3b}{v}\,\Tr\int F_+ \wedge F_+ +F_-\wedge F_- \,.
\end{align}
On the other hand,
\begin{align}
    \Tr \int F\wedge F&=\tfrac14 \Tr \int F_+ \wedge F_+ +F_-\wedge F_-= \Tr\int d\,\left[A \wedge dA +\tfrac23 A^3 \right] \,.
\end{align}
By adding the right amount of the topological invariant
\begin{align}
    \frac{a}{g^2}\Tr \int \left(F_{AB}F^{AB} -\bar{F}_{A'B'} \bar{F}^{A'B'}\right)=\frac{a}{g^2} \frac{12}{v}\Tr \int d\,\left[A \wedge dA +\tfrac23 A^3 \right]\,,
\end{align}
the Yang--Mills theory can be brought into a more chiral form
\begin{align}\label{chiralYM}
    S_{\text{cYM}}&= \frac{2a}{g^2}\,\Tr\int F_{AB}F^{AB} \,.
\end{align}
The topological invariant is perturbatively harmless in flat space, but will lead to detectable deviations from the pure Yang--Mills theory in anti-de Sitter space. The main advantage of chiral Yang--Mills theory is that it serves as a useful intermediate theory between the standard Yang--Mills and self-dual Yang--Mills theories: action \eqref{chiralYM} has simpler interactions, leading to simpler Feynman rules, see below.

\paragraph{Self-dual Yang--Mills theory (SDYM).} Following Chalmers and Siegel \cite{Chalmers:1996rq} one can rewrite the chiral Yang--Mills action by introducing an auxiliary field $\Psi_{AB}=\Psi_{BA}$ as
\begin{align}
    S_{\text{Ch.Si.}}&= \frac{2a}{g^2}2\epsilon\,\Tr\int \left(\Psi_{AB}F^{AB} - \frac{\epsilon}2 \Psi_{AB} \Psi^{AB}\right) \,.
\end{align}
Indeed, $\Psi_{AB}$ can be integrated out since $F_{AB}=\epsilon \Psi_{AB}$ is the equation for $\Psi_{AB}$. Dropping the last term (which is gauge-invariant on its own), we get SDYM\footnote{The following analogy can be helpful. Consider the action of the free particle, $S=\int p_\mu \dot{x}^\mu - \tfrac{e}2(p^2+m^2)$. The massless limit is reminiscent of the Chalmers--Siegel action, $S=\int p_\mu \dot{x}^\mu - \tfrac{e}2p^2 \sim \int \Psi \pl \Phi -\epsilon \Psi^2/2$. The SDYM limit corresponds to the singular gauge $e=0$, where we get a topological theory $S=\int p_\mu \dot{x}^\mu$, except that SDYM is not topological. Nevertheless, $\Psi \sim p$ and $\Phi\sim q$ would be a good analogy. There are also interesting singular gauges in string theory, e.g. the ambitwistor string \cite{Casali:2016atr}.}
\begin{align}
    S_{\text{SDYM}}&= \frac{\tilde{a}}{\tilde{g}}\, \Tr\int \Psi_{AB}F^{AB} \,.
\end{align}
SDYM appears to be the limit of the (chiral) Yang--Mills theory. Indeed, we can send $\epsilon\rightarrow0$ while keeping $4a\epsilon/g^2=\tilde{a}/\tilde{g}$ finite. In fact, $\tilde{a}=1$ seems to be the most natural normalization. However, in this limit the $V^{--+}$ and $V^{++--}$ vertices of Yang--Mills are instantaneously switched off, while the $V^{++-}$ remains. 

Since SDYM is less well-known than Yang--Mills theory and for the purpose of performing the Fefferman-Graham expansion and for deriving the Feynman rules, let us expand the action. To avoid a clash between the gauge potential $A$ and indices by having $A^{A,A'}$ we rename $A$ into $\Phi$. The action reads
\begin{align}\label{SDYMexpanded}
    S_{\text{SDYM}}&= {\tilde{g}^{-1}}\, \Tr\int \Psi^{AA}(\pl_{AC'}\Phi\fdu{A,}{C'} +\Phi_{A,C'}\Phi\fdu{A,}{C'})\,.
\end{align}
In a nontrivial gravitational background the kinetic term is $|e| \Psi^{AA}\nabla_{AC'}\Phi\fdu{A}{C'}$. Upon adding the $\Psi^2$-deformation and rescaling $\Phi \rightarrow \tilde{g}\Phi$ we end up with
\begin{align}\label{SDYMexpandedPert}
    S_{\text{Ch.Si.}}&=\Tr\int \Psi^{AA}(\pl_{AC'}\Phi\fdu{A,}{C'} +\tilde{g}\,\Phi_{A,C'}\Phi\fdu{A,}{C'})-\frac{\epsilon}{2}\Psi^{AB}\Psi_{AB}\,.
\end{align}
Let us consider the latter as a theory with two coupling constants, $\tilde{g}$ and $\epsilon$. Note that $\Psi^2$ looks like an independent gauge-invariant term that can be added to the action. However, $\epsilon$ is not a new coupling constant since its variation can be absorbed by that of $\tilde{g}$ (integrating $\Psi$ out makes it obvious). Nevertheless, $\epsilon=0$ and $\epsilon\neq0$ are two different theories. 

It is very easy to identify the subset of amplitudes of the Yang--Mills theory that is computed by SDYM ($\epsilon=0$): all tree-level amplitudes of type $\mathcal{A}^{++...+-}_{\text{tree}}$ and all-plus one-loop amplitudes $\mathcal{A}^{++...+}_{1-\text{loop}}$. The tree-level amplitudes are known to vanish in flat space. The one-loop amplitudes are nontrivial and are rational.\footnote{Supersymmetry, if introduced, trivializes even these amplitudes and SDYM becomes a 'trivial' theory in flat space.} Let us consider tree-level correlators, which is more suitable for AdS/CFT-applications. At $\epsilon=0$ we find that $\langle \Psi^n \Phi\rangle$ is of order $\tilde{g}^{n-1}$. The $\Psi^2$ vertex allows one to glue two external $\Psi$ legs together (each of which carries $\langle \Psi \Phi \rangle$-propagator of type $\slashed p/p^2$). The net result of this sewing is that the usual Yang--Mills propagator is generated $\slashed p/p^2 \times \slashed p/p^2=p^{-2}$ with the factor of $\epsilon$. Each $\Psi^2$-vertex allows us to increase the number of external $\Phi$s. As a result, $\langle \Psi^n \Phi^m\rangle$ is of order $\tilde{g}^{n+m-2} \epsilon^{m-1}$. In the Yang--Mills theory $\langle \Phi^N \rangle$ is of order $g^{N-2}$. Therefore, the exact match requires $\epsilon=1$ and $\tilde{g}=g$.

\paragraph{Theta term.} It is worth noting that both Yang--Mills theory and SDYM allow for a topological $\theta$-term. The topological term is important for getting the parity-odd structures inside the AdS/CFT correlation functions. While $S_{\text{top}}$ can be added to $S_{\text{YM}}$ as usual, in the case of SDYM there are additional tricks \cite{Losev:2017qrj}. One can start with
\begin{align}
    S_{\text{Ch.Si.}}+ b\,\Tr\int \left(F_{AB}F^{AB} -\bar{F}_{A'B'} \bar{F}^{A'B'}\right)
\end{align}
and absorb $F_{AB}$ via a redefinition $\Psi_{AB}\rightarrow \Psi_{AB}+\alpha F_{AB}$, which is especially easy at $\epsilon=0$. The leftover of the topological term is just $-b\bar{F}_{A'B'} \bar{F}^{A'B'}$. With this in mind, one can perform the Chalmers--Siegel trick while keeping the topological term. We begin with a sum of the Yang--Mills action and the theta-term
\begin{align}\label{YMthetaAction}
    S_{\text{YM},\theta}&=\frac{\alpha}4\,\Tr\int \left(F_{AB}F^{AB} +\bar{F}_{A'B'} \bar{F}^{A'B'}\right)+\frac{\beta}4 \,\Tr\int \left(F_{AB}F^{AB} -\bar{F}_{A'B'} \bar{F}^{A'B'}\right)\,,
\end{align}
where $g$ has already been absorbed into $F=dA+gA^2$. By simple manipulations we get
\begin{align}\label{CSthetaAction}
    S_{\text{Ch.Si.},\theta}&= \frac{(\alpha+\beta)}{4}2\epsilon\,\Tr\int \left(\Psi_{AB}F^{AB} - \frac{\epsilon}2 \Psi_{AB} \Psi^{AB}\right) + \frac{(\alpha-\beta)}4 \,\Tr\int \bar{F}_{A'B'} \bar{F}^{A'B'}\,.
\end{align}
With $\alpha=\beta$ we recover the Chalmers--Siegel action. Another way to proceed is to keep the topological term in its original form, the advantage being in that it generates the Chern--Simons term on the boundary:
\begin{align}\label{CSthetaActionB}
    S_{\text{Ch.Si.},\theta}&= \alpha\epsilon\,\Tr\int \left(\Psi_{AB}F^{AB} - \frac{\epsilon}2 \Psi_{AB} \Psi^{AB}\right) + \tfrac{(\beta-\alpha)}4 \,\Tr\int \left(F_{AB}F^{AB} -\bar{F}_{A'B'} \bar{F}^{A'B'}\right)\,.
\end{align}
Taking the SDYM limit we keep $\alpha \epsilon$ fixed, say new $\alpha$, and $(\beta-\alpha)$ also fixed, say $\varkappa$:
\begin{align}\label{SDYMthetaAction}
    S_{\text{SDYM},\theta}&= \alpha\Tr\int \Psi_{AB}F^{AB} + \frac{\varkappa}4 \,\Tr\int  -\bar{F}_{A'B'} \bar{F}^{A'B'}\,.
\end{align}
In the SDYM limit $F_{AB}=0$ on-shell and the $F_{AB}F^{AB}$-part of the topological term can be dropped. To conclude, YM and SDYM have an additional coupling constant associated with the theta-term.

\paragraph{Light-cone gauge. Extended Yang--Mills theory. } Some of the aspects of the relation between SDYM and Yang--Mills theory are better seen in the light-cone gauge, $A^+=0$. After integrating out $A^-$ since its equations of motions are algebraic (provided, as is necessary in the light-cone gauge, $\pl^+$ is considered invertible), one is left with $A^{1,2}$ that can be repackaged as $A^1\pm i A^2$ into $\phi$ and $\bar \phi$ representing positive and negative helicity fields (they can also be understood as gauge-fixed versions of $\Phi$ and $\Psi$ from above with some components solved for algebraically). The Yang--Mills theory's action in the light-cone gauge reads
\begin{align}
    \mathcal{L}&= \bar\phi \square \phi + g\,V(\bar\phi, \phi,\phi)+ g\, V(\phi, \bar\phi,\bar\phi) +g^2\, V(\phi, \phi, \bar\phi,\bar\phi)\,.
\end{align}
The vertex splits into $++-$ and $--+$ parts and there is a quartic $++--$ vertex. Let us note that in the light-cone gauge one can define a theory with two complex couplings\footnote{The same procedure was discussed in \cite{Skvortsov:2018uru} for a more general example where SDYM was extended to chiral higher-spin gravity, which was a way to see the $3d$ bosonization duality \cite{Giombi:2011kc, Maldacena:2012sf, Aharony:2012nh,Aharony:2015mjs} holographically. In particular, a similar procedure can be applied to gravity with the self-dual gravity as a closed subsector.} 
\begin{align}
    \mathcal{L}&= \bar\phi \square \phi + g\,V(\bar\phi, \phi,\phi)+ \bar g\, V(\phi, \bar\phi,\bar\phi) +g \bar g\, V(\phi, \phi, \bar\phi,\bar\phi)\,.
\end{align}
Whenever $g=\bar{g}=g^*$ is a single real coupling we recover the Yang--Mills theory. However, the action is Lorentz invariant for arbitrary complex $g$ and $\bar g$. Setting $\bar{g}=0$ gives SDYM. This way, YM can be represented as a smooth deformation of SDYM along $\bar{g}$. 

Identifying $\Psi$ with $\bar\phi$ and $\Phi$ with $\phi$, we find that $\langle \Psi^n \Phi\rangle $ is of order $g^{n-1}$. Switching on $\bar{g}$ leads to $\langle \Psi^n \Phi^m\rangle $ being of order $g^{n-1} \bar{g}^{m-1}$. Therefore, to compare to the covariant action, we need to make $\epsilon$ be of order $\bar{g}/g$, which explains the strange nature of $\epsilon$ (it is not a genuine coupling constant unless $\epsilon=0$).

%%%%%%%%%%%%%%%%%%%%%%%%%%%%%%%%%%%%%%%%%%%%%%%%%%%%%%%%%%%%%
\section{Fefferman-Graham analysis}
\label{sec:FG}
%%%%%%%%%%%%%%%%%%%%%%%%%%%%%%%%%%%%%%%%%%%%%%%%%%%%%%%%%%%%%
The Fefferman-Graham analysis reveals the AdS/CFT dictionary between bulk fields and operators in the boundary CFT. For the Maxwell/Yang--Mills theory the results are well-known, see e.g. \cite{Mueck:1998iz,Bianchi:2001kw}, but we rederive them to facilitate comparison with the self-dual case. The main result of the section is the Fefferman-Graham expansion for SDYM. It is also useful to roll back to the Poincare coordinates to correctly read off the boundary asymptotics. 

%%%%%%%%%%%%%%%%%%%%%%%%%%%%%%%%%%%%%%%%%%%%%%%%%%%%%%%%%%%%%
\subsection{Maxwell theory}
\label{sec:}
%%%%%%%%%%%%%%%%%%%%%%%%%%%%%%%%%%%%%%%%%%%%%%%%%%%%%%%%%%%%%
In the two-component spinor language the free equations of motion $\nabla^{\text{AdS}}_\nu F^{\mu\nu}=0$ turn into $\nabla^{\text{AdS}}_{AC'} \bar{F}^{A'C'}=0$ or $\nabla^{\text{AdS}}_{AA'} F^{BA'}=0$ and lead to 
\begin{align} \label{Maxwell}
    \nabla_{\text{AdS}}^{BB'}\nabla^{\text{AdS}}_{BB'}\Phi_{A,A'}-\nabla_{\text{AdS}}^{BB'}\nabla^{\text{AdS}}_{AA'}\Phi_{B,B'}=0 \,.
\end{align}
We now impose the axial gauge $n_\mu \Phi^\mu=0$, where $n_\mu$ is orthogonal to the boundary, i.e. $n_{AA'}=\epsilon_{AA'}$ in our coordinates. This implies $\epsilon^{AA'}\Phi_{A,A'}=0$, i.e. $\Phi_{A,A'}$ is symmetric, $\Phi_{A,B}\equiv \Phi_{AB}$. The equations of motion in the axial gauge read
\begin{subequations}
    \begin{align}
        2(z^2\partial_z^2-2z\partial_z-k^2z^2+2)\Phi_{AA}-z^2k_{AA}k^{BB}\Phi_{BB}=0 \,,\\
        z(1-z\partial_z)k^{BB}\Phi_{BB}=0 \,.
    \end{align}
\end{subequations}
The former arises from projecting \eqref{Maxwell} onto its symmetric part in $AA'$, corresponding to the boundary coordinates, while the latter is the antisymmetric part, which is related to the $z$-direction. We then perform the expansion
\begin{align}
    \Phi_{AA}=z^\Delta\phi_{AA}=z^\Delta (\phi_{AA}^{(0)}+z\phi^{(1)}_{AA}+z^2\phi^{(2)}_{AA}+\dots)
\end{align}
and we find the equations of motion for the field $\phi_{AA}$ to be
\begin{subequations}
    \begin{align}
        2\Big( (\Delta-2)(\Delta-1) + z(2\Delta-2)\partial_z +z^2(\partial_z^2-k^2) \Big)\phi_{AA} - z^2k_{AA}k^{BB}\phi_{BB}&=0 \label{eom1}\,, \\ (1-\Delta-z\partial_z)k^{BB}\phi_{BB}=0\label{eom2}\,.
    \end{align}
\end{subequations}
These equations may be solved order-by-order, i.e.
\begin{subequations}
    \begin{align}
        \big((\Delta-2)(\Delta-1)+n(2\Delta+n-3) \big) \phi^{(n)}_{AA}-k^2\mathcal{P}\fdu{AA}{BB}\phi^{(n-2)}_{BB}=0\,,\label{eomA}\\
        (1-\Delta-n)k^{BB}\phi_{BB}^{(n)}=0\,,\label{eomB}
    \end{align}
\end{subequations}
where
\begin{align}
    \mathcal{P}\fdu{AA}{BB}=\epsilon\fdu{A}{B}\epsilon\fdu{A}{B}+\frac{k_{AA}k^{BB}}{2} \,.
\end{align}
The zeroth order of \eqref{eomA} tells us that there are two independent solutions: the leading order solution $\phi^-_{AA}$ and the subleading $\phi^+_{AA}$, with conformal dimensions
\begin{align}
        \Delta_-&=1 \,, & \Delta_+&=2\,,
\end{align}
respectively. $\Delta=2$ is the conformal dimension of a spin-one primary operator at the unitarity bound in $d=3$ dimensions. Such an operator $J_{a}$ is a conserved spin-1 current, $\pl^m J_{m}=0$. $\Delta=1$ corresponds to the source $\varphi^{m}$, which is also a gauge field on the boundary, $\delta \varphi^{m}=\pl^m \xi$.  
\begin{table}[h!]
\centering
\caption{Expansion coefficients of $\phi^{+}_{AA}$ and $\phi^-_{AA}$.}
\label{tab2}
\vspace{0.2cm}
\begin{tabular}{ | c | c | c | } 
  \hline
            \rule{0pt}{16pt}& $\Delta_-=1$ & $\Delta_+=2$ \\
            \hline
            \noalign{\hrule height 1.5pt} % Thick second horizontal line
            \rule{0pt}{16pt}$n=0$ & - & $k^{BB}\phi^{+(0)}_{BB}=0$ \\
            \hline
            \rule{0pt}{16pt}$n\geq 1$ & $k^{BB}\phi_{BB}^{-(n)}=0$ & $k^{BB}\phi_{BB}^{+(n)}=0$ \\
            \rule{0pt}{16pt}&$\phi_{AA}^{-(n)}=\frac{k^2}{n(n-1)}\mathcal{P}\fdu{AA}{BB}\phi^{-(n-2)}_{BB}$ & $\phi^{+(n)}_{AA}=\frac{k^2}{n(n+1)}\mathcal{P}\fdu{AA}{BB}\phi^{+(n-2)}_{BB}$\\
            \hline
\end{tabular}
\end{table}

The conditions imposed on the solutions by \eqref{eomA} and \eqref{eomB}, are shown in Table \ref{tab2}. For the leading order solution \eqref{eomB} vanishes trivially for $n=0$, leaving the coefficient $\phi^{-(0)}_{AA}$ unconstrained, while all other coefficients satisfy a conservation law. For both solutions, we obtain a relation between expansion coefficients that are separated by two orders. This implies that coefficients of odd order in $z$ in the expansion for $\Phi_{AA}$ are supplied by the coefficients of $\phi^{-}_{AA}$, while the coefficients of even order are the coefficients of $\phi^{+}_{AA}$. In particular, the two solutions never contribute to coefficients of the same order when finitely many terms are taken into account. 

The two branches of solutions are not related to each other near the boundary, but rather by a regularity condition in the bulk. To see this, we first write all coefficients of $\phi^{\pm}_{AA}$ in terms of $\phi^{\pm(0)}_{AA}$.
The conservation laws in Table \ref{tab2} allow us to simplify the recurrence relation for $\phi^{-(n)}_{AA}$:
\begin{align}
    \phi^{-(n)}_{AA}=\frac{k^2}{n(n-1)}\phi^{-(n-2)}_{AA}\,,
\end{align}
when $n>2$ and otherwise
\begin{align}
    \phi^{-(2)}_{AA}=\frac{k^2}{2}\mathcal{P}\fdu{AA}{BB}\phi^{-(0)}_{BB}\,.
\end{align}
We then find the expansion coefficients to be 
\begin{align}
    \phi^{-(2n)}_{AA}=\frac{2k^{2n-2}}{(2n)!}\phi^{-(2)}_{AA}=\frac{k^{2n}}{(2n)!}\mathcal{P}\fdu{AA}{BB}\phi^{-(0)}_{BB}\,.
\end{align}
The function $\phi_{AA}^{-}$ then becomes
\begin{align}\notag
    \begin{aligned}
        z^{\Delta_-}\phi^{-}_{AA}&=z\phi^{-(0)}_{AA}+\sum_{n=1}^\infty\frac{k^{2n}}{(2n)!}\mathcal{P}\fdu{AA}{BB} z^{2n+1}\phi^{-(0)}_{BB}=z\Big(\mathcal{P}\fdu{AA}{BB} \cosh{(kz)}-\frac{k_{AA}k^{BB}}{2k^2} \Big)\phi^{-(0)}_{BB}\,.
    \end{aligned}
\end{align}
The coefficients of the subleading solution are given by
\begin{align}
    \phi^{+(2n)}_{AA}=\frac{k^{2n}}{(2n+1)!}\phi^{+(0)}_{AA}\,,
\end{align}
due to the conservation laws. The closed-form solution then reads
\begin{align}
    z^{\Delta_+}\phi^{+}_{AA} = \sum_{n=0}^\infty \phi^{+(2n)}_{AA}z^{2n+2}=\frac{z\sinh{(kz)}}{k}\phi^{+(0)}_{AA}\,.
\end{align}
It is now straightforward to see that even though the solutions $\phi^{+}_{AA'}$ and $\phi^{-}_{AA'}$ blow up separately in the deep bulk $z\rightarrow\infty$, together they can yield a regular solution. In particular, this happens if $\phi^{+(0)}_{AA}=-k\mathcal{P}\fdu{AA}{BB}\phi^{-(0)}_{BB}$. The full regular solution then reads
\begin{align}
    \Phi_{AA}=\phi^{-}_{AA}z^{\Delta_-}+\phi^{+}_{AA}z^{\Delta_+} = z\Big(\mathcal{P}\fdu{AA}{BB}e^{-kz}-\frac{k_{AA}k^{BB}}{2k^2}\Big)\phi^{-(0)}_{BB}
\end{align}
and is nothing but the boundary-to-bulk propagator. It may be useful to recall that the expansion near the boundary can also be written as $A_i=z(a_i + z E_i)+...$, where $E_i=\pl_z A_i|_{z=0}$ is the electric field (in the axial gauge) and $a_i$ is a $3d$ gauge potential. The regularity condition implies that $E_i=-k \Pi _{ij}a^j$, where $\Pi_{ij}=\delta_{ij}-k_ik_j/k^2$.

%%%%%%%%%%%%%%%%%%%%%%%%%%%%%%%%%%%%%%%%%%%%%%%%%%%%%%%%%%%%%
\subsection{SDYM, negative helicity}
\label{sec:}
%%%%%%%%%%%%%%%%%%%%%%%%%%%%%%%%%%%%%%%%%%%%%%%%%%%%%%%%%%%%%
We start by evaluating the Fefferman-Graham expansion for the negative helicity field $\Psi^{AA}$. Its equation of motion can be read off from \eqref{SDYMexpanded}
\begin{align} \label{eom}
    \nabla^{\text{AdS}}_{BA'}\Psi^{AB}=0
\end{align}
and after applying the $\text{AdS}_4$ covariant derivative, one finds
\begin{align}
    (z k\fdu{B}{B'}+\epsilon\fdu{B}{B'}(2-z\partial_z))\Psi^{AB} = 0 \,.
\end{align}
We expand $\Psi^{AA}$ in powers of $z$ as
\begin{align} \label{expans}
    \Psi^{AA}(k,z)&=z^\Delta \psi^{AA}(k,z) = z^\Delta \Big(\psi^{AA}_{(0)} +\psi^{AA}_{(1)}z+\psi^{AA}_{(2)}z^2+\dots\Big)\,.
\end{align}
In terms of the field $\psi$ the equation of motion, \eqref{eom}, reads 
\begin{align}
    \Big(2-\Delta\Big)\psi^{AB'}(x) + z\Big(k\fdu{B}{B'}-\epsilon\fdu{B}{B'}\partial_z\Big)\psi^{AB}(x) = 0\,.
\end{align}
The first term yields the conformal dimension
$\Delta=2$, which corresponds to a conserved current on the boundary. After imposing this value for the conformal dimension we are left with
\begin{align} \label{nextOrder}
    (k\fdu{B}{B'}-\epsilon\fdu{B}{B'}\partial_z)\psi^{AB}=0\,.
\end{align}
The free indices $A$ and $B'$ can be either symmetrized or anti-symmetrized. The components of $\psi^{AA}$ satisfy in the former case
\begin{align} \label{rec}
    \psi_{(n)}^{AA}=\frac{1}{n}k\fdu{B}{A}\psi_{(n-1)}^{AB} \,,
\end{align}
while in the latter we find the conservation law
\begin{align} \label{cons}
    k_{BB}\psi_{(n)}^{BB}=0 \,.
\end{align}
Together with its conformal dimension, this conservation law confirms that $\Psi^{AA}$ behaves as a conserved spin-one current in the boundary CFT. From the Yang--Mills vantage point we reproduce the $\Delta=2$ asymptotic, but we have lost the $\Delta=1$ one. 

In order to solve equation \eqref{nextOrder}, let us \cite{Maldacena:2011nz} introduce an auxiliary $4d$ light-like momentum $p_{AA'}=k_{AA'}+k\epsilon_{AA'}$.\footnote{Our convention is that $p_\mu p^\mu=-\tfrac12 p_{AA'}p^{AA'}$, $k_i k^i= -\tfrac12 k_{AB}k^{AB}$, $k^2=-\tfrac12 k_{AB}k^{AB}=k_ik^i$. } Since it is light-like it can be factorized as $p_{AA'}=k_A\bar{k}_{A'}$. This allows one to define two $3d$ vectors $k_Ak_B$ and $\bar{k}_{A}\bar{k}_{B}$ that both are $3d$-null and orthogonal to $k^{AB}$. This decomposition is Lorentz-invariant. Such auxiliary light-like momenta are very helpful in defining the spinor-helicity formalism for $3d$ CFTs and $\text{(A)dS}_4$, see e.g. \cite{Maldacena:2011nz,Jain:2021vrv}.\footnote{There is more than one spinor-helicity formalism suitable for $\text{AdS}_4/\text{CFT}_3$ applications, see e.g. \cite{Maldacena:2011nz,Nagaraj:2018nxq,Nagaraj:2019zmk,Nagaraj:2020sji,Skvortsov:2022wzo,Baumann:2024ttn}. } With the ansatz
\begin{align} \label{PsiKappa}
    \psi^{AA} = \psi_{(0)} k^Ak^A e^{-kz} \,,
\end{align}
equation \eqref{nextOrder} gives
$
    (k_{BB'}+k\epsilon_{BB'})k^Ak^B = k_B\bar{k}_{B'}k^Ak^B = 0
$
and similarly for the expansion coefficients of $\psi^{AA}$. Here, $\psi_{(0)}$ is an overall normalization constant, which can be interpreted as the boundary data. The solution can be simplified to
\begin{align}
    \psi^{AA}_{(n)} = -\frac{k}{n} \psi^{AA}_{(n-1)}
\end{align}
and, hence, the expansion \eqref{expans} reads
\begin{align} \label{PsiProp}
    \Psi^{AA} = z^{2} \sum_{n=0}^{\infty}\frac{(-kz)^n}{n!}\psi^{AA}_{(0)} = \psi_{(0)}^-z^{2}e^{-kz}k^{A}k^A\,.
\end{align}
The upper summation bound is justified, as the expansion clearly never terminates, as can be seen from \eqref{nextOrder}. Alternatively, one could try the ansatz
\begin{align}
    \psi^{AA} = \bar{k}^{A}\bar{k}^Ae^{kz} \,,
\end{align}
with the equation of motion reducing to
$
    (k_{BB'}-k\epsilon_{BB'})\bar{k}^A\bar{k}^B= k_{B'}\bar{k}_{B}\bar{k}^A\bar{k}^{B} =0
$. 
Following the same steps as before, one finds the solution
\begin{align}
    \Psi^{AA} = \psi_{(0)}^+z^{2}e^{kz}\bar{k}^{A}\bar{k}^A\,,
\end{align}
which blows up in the bulk. Thus, on imposing regularity in the bulk, the first order formulation prefers the solution \eqref{PsiProp} for the negative helicity field, where the indices are supplied by $k_A$. Note that the third ``vector'' $\psi^{AA}\sim f(z) k^A \bar{k}^{A}$ cannot satisfy the equations of motion since it is not transverse to $k^{AB}$ (in fact, when symmetrized, $k^A \bar{k}^{A}=k^{AA}$). 

To summarize, the complete Fefferman-Graham expansion can be parameterized by a transverse $\psi_0^{AA}$, $k_{AA} \psi^{AA}_0=0$, as\footnote{$\slashed k$ acts as $\slashed k \xi= k\fud{A}{B}\xi^B$, $\slashed k \psi= k\fud{A}{B}\psi^{AB}$. Note that $\slashed k k \equiv k\fud{A}{B} k^B=-k k^A $, $\slashed k \brk \equiv k\fud{A}{B} \brk^B=+k \brk^A $. } 
\begin{align}
    z^{-2}\Psi^{AA}&= \cosh(kz) \psi_0^{AA} +\tfrac{1}{k}\sinh(kz) \slashed k \psi_0^{AA} \,.
\end{align}
This also makes it obvious that $\psi_0^{AA}\sim \brk^A\brk^A$ leads to an irregular solution and $\psi_0^{AA}\sim k^Ak^A$ to a regular one.

%%%%%%%%%%%%%%%%%%%%%%%%%%%%%%%%%%%%%%%%%%%%%%%%%%%%%%%%%%%%%
\subsection{SDYM, positive helicity}
\label{sec:}
%%%%%%%%%%%%%%%%%%%%%%%%%%%%%%%%%%%%%%%%%%%%%%%%%%%%%%%%%%%%%
Next, we derive the Fefferman-Graham expansion for the positive helicity gauge field $\Phi_{A,A'}$ in SDYM. The equations of motion that follow from \eqref{SDYMexpanded} give
\begin{align}
    \nabla^{\text{AdS}}_{AB'}\Phi\fdu{A,}{B'} = \Big( zk_{AB'} + (1-z\partial_z)\epsilon_{AB'} \Big) \Phi\fdu{A,}{B'} = 0\,.
\end{align}
The Fefferman-Graham expansion reads
\begin{align}
    \Phi_{A,A'}(k,z) = z^{\Delta}\phi_{A,A'}(k,z) = z^\Delta \Big( \phi^{(0)}_{A,A'} + \phi^{(1)}_{A,A'} z +\phi^{(2)}_{A,A'} z^2 + \dots \Big)\,.
\end{align}
In terms of the field $\phi_{A,A'}$ the equation of motion reads
\begin{align}
    (1-\Delta)\phi_{A,A} + z (k\fdu{A}{A'}-\epsilon\fdu{A}{A'}\partial_z)\phi_{A,A'} = 0 \,.
\end{align}
Setting $z=0$ yields the conformal dimension $\Delta=1$, 
which agrees with the conformal dimension of a source $\varphi^{m}$ of a spin-1 current $a_{m}$ or of just a gauge field on the boundary. The solution can be obtained from
\begin{align}
    (k\fdu{A}{A'} - \epsilon\fdu{A}{A'}\partial_z)\phi_{A,A'} = 0\,,
\end{align}
which for the components of the expansion reads
\begin{align}
    \phi^{(n)}_{A,A}=\frac{1}{n}k\fdu{A}{A'}\phi^{(n-1)}_{A,A'} \,,
\end{align}
just like \eqref{rec}. However, note that this equation requires the free indices to be symmetrized, as opposed to \eqref{nextOrder}, where we obtained a conservation law from the anti-symmetric part of the equation. Here, such a conservation law cannot be obtained. Indeed, the equation for $\Phi_{A,A'}$ is obtained by varying the action with respect to the symmetric $\Psi^{AA}$ and vice versa. Therefore, there are more equations for $\Psi$ than for $\Phi$. Still, inspired by the solution found for the negative helicity field $\Psi^{AA}$ two solutions can be found, 
\begin{align} \label{PhiProp}
    \begin{aligned}
        \Phi_{A,A'}&=\phi^{(0)}_-ze^{+kz} q_Ak_{A'}\,,\qquad &
        \Phi_{A,A'}&=\phi^{(0)}_+ze^{-kz} q_A\bar{k}_{A'}\,,
    \end{aligned}
\end{align}
where $q_A$ is an arbitrary reference spinor. By requiring regularity in the deep bulk ($z\rightarrow \infty)$, one chooses the latter, which is the boundary-to-bulk propagator. The solutions obey the Lorenz gauge, $\nabla^{\text{AdS}}_{AA'}\Phi^{A,A'}=0$.\footnote{From now on, we always assume $\nabla^{k,z}_{CC'}=k_{CC'}-\pl_z \epsilon_{CC'}$, i.e. we are essentially in the flat space (we omit the labels in the superscript whenever the momentum of the field the derivative acts on is positive and it is clear on which field it acts). It may be important when comparing formulas in different coordinates since the divergence operator is not conformally invariant, but there is always some modified gauge condition, e.g. $\nabla_{CC'}\Phi^{C,C'}+a \epsilon_{CC'} \Phi^{C,C'}=0$ that would be true in $\text{AdS}_4$ had we undone the Weyl transformation.}

\paragraph{Axial gauge.} In the later sections, where we compute the correlation functions, the axial gauge will play an important role. Imposing the axial gauge on a Yang--Mills field, we will require the $z$-component of the gauge field to vanish, which becomes $\epsilon_{AA'}\Phi^{A,A'}=0$ in the spinor language. This forces $\Phi_{A,A'}$ to be symmetric in its indices and we will write it as $\Phi_{AA}$. For the solutions \eqref{PhiProp}, this implies that a specific reference spinor must be chosen. The solutions become
\begin{align}
        \Phi_{AA}&=\phi^{(0)}_-ze^{+kz} k_Ak_{A}\,, &
        \Phi_{AA}&=\phi^{(0)}_+ze^{-kz} \bar{k}_A\bar{k}_{A}\,,
\end{align}
which are now transverse to the three-momentum $k^{AB}$. Again, regularity in the deep bulk forces us to choose the latter solution. The axial gauge condition does not fix a gauge completely and we are left with some residual gauge transformations  $\delta\Phi^{AA}=\nabla_{\text{AdS}}^{AA}\xi=k^{AA}\xi$, which allows us to impose $k \cdot \Phi=0$ on the boundary. The Fefferman-Graham expansion can be summarized by
\begin{align}
    z^{-1}\Phi^{AA}&= \cosh(kz) \phi_0^{AA} -\tfrac{1}{k}\sinh(kz) \slashed k \phi_0^{AA} \,,
\end{align}
where it is assumed that $k_{AA}\phi_0^{AA}=0$ is imposed.

%%%%%%%%%%%%%%%%%%%%%%%%%%%%%%%%%%%%%%%%%%%%%%%%%%%%%%%%%%%%%
\subsection{YM, SDYM, Dirac}
\label{sec:}
%%%%%%%%%%%%%%%%%%%%%%%%%%%%%%%%%%%%%%%%%%%%%%%%%%%%%%%%%%%%%
Since the Fefferman-Graham expansion is the key to building the AdS/CFT dictionary and has not been discussed much for self-dual theories, see, however, \cite{deHaro:2007fg,deHaro:2008gp,Compere:2008us}, it is instructive to confront YM vs. SDYM and compare the latter to Dirac/Weyl massless fermions that share similar features. 

\paragraph{Recap.} For the solutions that behave as $\exp{(-k z)}$ the operator $\nabla_{CC'}$ acts as $k_{CC'}+k\epsilon_{CC'}=k_C\brk_{C'}$, i.e. as would be for plane waves in the flat space. For the blowing up exponent $\exp{(+k z)}$ the operator $\nabla_{CC'}$ acts as $k_{CC'}-k\epsilon_{CC'}=\brk_{C}k_{C'}$. This allows one to quickly see what kind of polarization spin-tensors need to be involved. Thus, the boundary-to-bulk propagator for the negative helicity field is proportional to $k_{A}k_{A}$, while the index structure for the positive helicity field in axial gauge arises from $\bar{k}_{A}\bar{k}_A$:
\begin{align}\label{recapprop}
    \Psi^{AA}&=\psi_{(0)}z^{2}e^{-kz}k^{A}k^A \,, & \Phi_{AA}&=\phi^{(0)}ze^{-kz} \bar{k}_A\bar{k}_{A} \,.
\end{align}
The conformal transformation from the Poincare coordinates on $\text{AdS}_4$ to the half-space model eliminates the explicit $z^2$ and $z$ in \eqref{recapprop} and we will not see them again.

\paragraph{Penrose/Weyl equations.} The equation for $\Psi^{AB}$ is just a particular case of a more general equation \cite{Penrose:1965am} that is valid for all spins
\begin{align}
    \nabla\fdu{B}{A'}\Psi^{A(2s-1)B}&=0\,.
\end{align}
In our convention it describes a helicity-$(-s)$ field. In particular, this applies to $s=\tfrac12$, where it becomes the equation for a single massless Weyl spinor. On the other hand, the Fefferman-Graham expansion of the Dirac equation has been considered in the earlier days of AdS/CFT correspondence \cite{Henningson:1998cd,Mueck:1998iz,Henneaux:1998ch}. A brief summary is that the asymptotic expansion 
\begin{align}
    \chi^\alpha&= z^{\Delta_- } \psi_-^\alpha +... z^{\Delta_+ } \psi_+^\alpha+...
\end{align}
has two falloffs $\Delta_{\pm}=d/2\pm m$. The boundary data $\psi_{\pm}$ are restricted to be chiral spinors (with respect to the radial gamma-matrix, $\gamma_z$), similarly for the conjugate Dirac spinor. The usual quantization is to choose $\psi_-=0$ and interpret $\psi_-$ as a source for an operator of dimension $\Delta_+$, whose expectation value is given by $\psi_+$.

In a bit more detail, which should clarify the situation with the massless limit, let us consider one massive Dirac spinor $\chi^\alpha=(\Phi^{A'},\Psi^A)$ that satisfies
\begin{align}
    \nabla\fdu{B}{B'}\Psi^B +im \Phi^{B'}&=0\,, &
    \nabla\fud{B}{B'}\Phi^{B'}+im \Psi^B&=0 \,.
\end{align}
For spinors the covariant derivative is $\nabla_{AA'}= zk_{AA'}+\epsilon_{AA'}(3/2-z\pl_z)$. It is convenient to replace $\chi^\alpha$ with $z^{3/2}\chi^{\alpha}$, which reduces $\nabla_{AA'}$ to $\nabla_{AA'}= zk_{AA'}- \epsilon_{AA'} z\pl_z$. Looking for where the asymptotic expansion can start we omit $zk_{AA'}$ and balance $z\pl_z \chi$ and $im \chi$, which gives 
\begin{align}\label{chiraldecomposition}
    \chi^\alpha&= z^{-m}\begin{pmatrix}
        iq^{A'} \\ q^A
    \end{pmatrix} +... +z^{+m}\begin{pmatrix}
        -ip^{A'} \\ p^A
    \end{pmatrix}+...\,,
\end{align}
where $q^A$ and $p^A$ are free boundary data. Note that taking $\chi^\alpha=z^w(\phi_0^{A'},  \psi^A_0)$ actually gives
\begin{align}
    -w \psi_0^B+i m \phi_0^B&=0 \,, & +w \phi_0^B+i m \psi_0^B&=0\,,
\end{align}
which implies $\phi_0^A=-i\tfrac{w}{m}\psi_0^A$ and $w^2=m^2$, i.e. $w=\pm m$. Interestingly, the relation between $\phi_0$ and $\psi_0$ has a smooth $m\rightarrow 0$ limit, where the two falloffs become the same, but they still can be separated by the chirality. However, if we start out from the massless theory, we find no trace of this relation and arrive at the Fefferman-Graham-type expansion of the form
\begin{align}\label{FGfermions}
\begin{aligned}
    \Phi^{A'}&=\cosh (kz)\, \phi_0^{A'} -\tfrac1k\sinh(kz) k\fud{A'}{B}\phi_0^B\,, \\
    \Psi^{A}&=\cosh (kz)\, \psi_0^{A} +\tfrac1k\sinh(kz) k\fud{A}{B}\psi_0^B\,.
\end{aligned}
\end{align}
This way, $\phi_0^A$ and $\psi_0^A$ are free boundary data, $\chi^\alpha_0=(\phi_0^{A'},\psi_0^A)$. On the other hand, in the limit $m\rightarrow 0$ we find at $z=0$ for the field  $\chi^\alpha=(i(q-p)^{A'}, (q+p)^A)$ from \eqref{chiraldecomposition}. As long as $k^A$ and $\brk^A$ are linearly independent one can use them to expand the boundary data into the definite helicity components. With the help of $k_{AB}\brk^B=k \bar{k}_A$ and $k_{AB}k^B=-k k_A$, we find
\begin{align}\label{helsplit}
\begin{aligned}
    \Phi^{A'}(\phi_0=k)&= e^{+kz}k^{A'} \,, \qquad \qquad & \Phi^{A'}(\phi_0=\brk)&= e^{-kz}\brk^{A'}\,,\\
    \Psi^{A}(\psi_0=k)&= e^{-kz}k^A \,, & \Psi^{A}(\psi_0=\brk)&= e^{+kz}\brk^A\,.
\end{aligned}
\end{align}
While nothing special happens at $m=0$ to the Dirac spinor, what might look problematic is to consider just a single Weyl spinor.\footnote{In Minkowski signature the Lorentz algebra is $sl(2,\mathbb{C})$ in the bulk and $sl(2,\mathbb{R})$ on the boundary; in Euclidean signature it is  $su(2)\otimes su(2)$ and $su(2)$, respectively, and complex conjugation does not swap the left and right spinors. The problem with Weyl spinors is not the Fefferman-Graham expansion, but the fact that one needs to have a well-defined variational problem, which forces one to introduce boundary terms and boundary conditions that break chiral symmetry.} This case was discussed in \cite{Porrati:2009dy,Foit:2019nsr} and we will come back to this in Section \ref{sec:boundaryterms} when considering the variational problem. 

\paragraph{Helicity decomposition: Fermions vs. SDYM vs. YM.} We need to identify a subspace of the boundary data that is consistent with the regularity in the bulk (it should be a half-dimensional subspace and will be a Lagrangian submanifold once we remember that all the systems we discuss have actions). Such a subspace provides us with a boundary data that can be fixed and corresponds to sources in AdS/CFT. Fixing all of the Fefferman-Graham boundary data is inconsistent with regularity, see e.g. \eqref{FGfermions}. There does not seem to be any way to split Weyl spinors $\Psi^A$, $\Phi^{A'}$ further into two objects locally in the position space. This is where the helicity, which is intrinsically a momentum space notion, comes into the game. Let us introduce 
\begin{align}\label{polarizationspinors}
    \epsilon_+^A(\vec k)&=  \frac{\brk^{A}}{\sqrt{2k}} \,, & \epsilon_-^A(\vec k)&=  \frac{k^{A}}{\sqrt{2k}} \,, & \epsilon_\pm^{AA}(\vec k)&=\epsilon_\pm^A\epsilon_\pm^A \,.
\end{align}
Now, we can decompose Weyl spinors $\Phi^{A'}$, $\Psi^A$,  as
\begin{align}
    \Phi^{A'}(\vec k,z)&= \epsilon_+^A\Phi_+(\vec k,z)+\epsilon_-^A\Phi_-(\vec k,z) \,,&
    \Psi^{A}(\vec k,z)&= \epsilon_+^A\Psi_+(\vec k,z)+\epsilon_-^A\Psi_-(\vec k,z) \,.
\end{align}
The equations of motion simplify to ($D_\pm=\pl_z \pm k$)
\begin{align}\label{eqhel}
    D_+ \Phi_+&=0 \,, & D_- \Psi_+&=0\,, & D_- \Phi_-&=0 \,,& D_+ \Psi_-&=0 \,.
\end{align}
These first order equations have obvious solutions $D_\pm f=0$ $\rightarrow$ $f=e^{\mp kz}$, which reproduces \eqref{helsplit}. At the same time, we can read off the right boundary conditions that are compatible with regularity in the bulk: lucky fields that obey $D_+f=0$ are allowed to fluctuate and satisfy the Dirichlet boundary condition; unlucky fields that obey $D_-f=0$ cannot fluctuate, the only regular solution being zero. 

Let us perform a similar decomposition for the SDYM/Chalmers--Siegel fields $\Phi^{AA}$ (in the unitary gauge)\footnote{We also can call it the complete or physical gauge where the dynamical field obeys $A_z=0$ and $\pl_i A^i=0$, it is discussed in Appendix \ref{app:ham}. Effectively, the Gauss law is solved and non-transverse degrees of freedom are integrated out.} and $\Psi^{AA}$
\begin{align}
\begin{aligned}
           \Phi^{AA}(\vec k,z)&= \epsilon_+^{AA}(\vec k) \Phi_+(\vec k,z)+\epsilon_-^{AA}(\vec k) \Phi_-(\vec k,z)\,, \\
        \Psi^{AA}(\vec k,z)&=\epsilon_+^{AA}(\vec k) \Psi_+(\vec k,z)+\epsilon_-^{AA}(\vec k) \Psi_-(\vec k,z) \,.
\end{aligned}
\end{align}
For the case of SDYM we find exactly the same equations \eqref{eqhel}, the only difference being that for the spin-half case the helicity fields $\Phi_\pm$, $\Psi_\pm$ are Grassmann-odd and for SDYM they are Grassmann-even. The helicity decomposition above gives a simple realization of the Fefferman-Graham expansion, which reduces to $e^{\pm kz}$. Consequently, the only boundary data that is compatible with regularity is given by $\Phi_+$ and $\Psi_-$ at $z=0$. 

The Maxwell theory has the same field $\Phi^{AA}$, but the equations are second order $D_+D_- \Phi_{\pm}=0$, ($D_+D_-=\pl^2_z-k^2$). As a result, the regularity does not get entangled with the helicity structure. At this point, we can summarize all three types of Fefferman-Graham expansion we have encountered:
\besubeqs
\begin{align}
    D_+D_-\phi&=0 \,, && \phi=\cosh(kz) \phi_0 + \tfrac{1}{k}\sinh(kz) \phi_1 \,,\\
    D_+ \phi&=0 \,, && \phi=\exp(-kz) \phi_0 \,,\\
    D_- \psi&=0 \,, && \psi=\exp(+kz)\psi_0 \,.
\end{align}
\esubeqs
For the Maxwell theory $\phi\equiv A^i$, $k_i A^i=0$. Upon introducing $F_+=D_+\phi_+$ one can see that the expansion for $F_+$ collapses onto the expansion for $\psi$ with $\psi_0=k \phi_0+\phi_1$ ($\phi_0\equiv \phi|_{z=0}$ and $\phi_1\equiv \pl_z \phi|_{z=0}$ is the notation for the Dirichlet/Neumann data we will use below) and similarly for $D_-$. Note that $k\phi^0$ is a shadow operator for $\phi^0$, which is what allows us to impose conformally-invariant mixed boundary conditions on the scalar field, see also Appendix \ref{app:boundaryterms}. The full decomposition of the $A_i$ Fefferman-Graham expansion into $F_{AA}$, $\bar{F}_{A'A'}$ and with respect to helicities reads
\[
\begin{tikzcd}[]
0  & \arrow[l, "D_+", swap] \overbrace{ \rule{0pt}{14pt}-e^{-kz}(\Phi_+^1-k\Phi_+^0)}^{\displaystyle\bar{F}_{A'A'}}  & \arrow[l, "D_-", swap] \overbrace{\rule{0pt}{14pt}\Phi_+}^{\displaystyle A_i} \arrow[r, "D_+"] &  \overbrace{\rule{0pt}{14pt}e^{+kz}(\Phi_+^1+k\Phi_+^0)}^{\displaystyle F_{AA}} \arrow[r, leftrightarrow, "\epsilon"] & \overbrace{\rule{0pt}{14pt}\Psi_+}^{\Psi} \arrow[r, "D_-"] & 0\\
0  & \arrow[l, "D_-", swap] -e^{+kz}(\Phi_-^1+k\Phi_-^0)  & \arrow[l, "D_+", swap] \Phi_- \arrow[r, "D_-"] &  e^{-kz}(\Phi_-^1-k\Phi_-^0) \arrow[r, leftrightarrow, "\epsilon"] & \Psi_- \arrow[r, "D_+"] & 0\,.
\end{tikzcd}
\]
At the level of the Maxwell theory, the boundary data corresponds to $A_i^0=A_i(z=0)$ and $A_i^1=\pl_z A_i(z=0)$. Equivalently, in terms of the helicity components it corresponds to $\Phi_\pm^0$ and $\Phi_\pm^1$. $\Psi_\pm^0$ provides a package for half of the boundary data. The self-dual constraint, which appears at $\epsilon=0$, sets this half to zero. However, $\Psi$, which is now detached from $\Phi$, continues to carry the same amount of boundary data. What we would like to stress now, and we will see more evidence later, is that YM theory with self-dual boundary conditions is not a field theory we need. Indeed, half of the degrees of freedom are killed by $F_{AA}|_{z=0}=0$.\footnote{\label{ft:Feq}$F_{AA}|_{z=0}=0$ is not even truly a boundary condition: since $F_{AA}$ obeys the same equations as $\Psi$, one sees that $F_{AA}|_{z=0}=0$ implies $F_{AA}=0$ in the bulk. } The right theory is obtained at $\epsilon=0$ of the Chalmers--Siegel action, which preserves the degrees of freedom. 

As we can see, SDYM is very close to the spin-half example above. In the Chalmers--Siegel formulation the $\Psi$-field obeys the same equations as in SDYM and, in addition, $\epsilon \Psi_\pm=D_\pm \Phi_\pm$. In some sense, 'Dirac' operators $D_\pm$ factorize $\square=D_+D_-$ and this allows one to separate regular and irregular solutions. The only fully Lorentz- and gauge-invariant way to do it is via the Chalmers--Siegel action, which also links regularity to helicity.\footnote{Let us note that another way to factorize the Maxwell equations, but without separating (ir)regular solutions and helicities, is via the usual first order action where an auxiliary field $\Psi_{[\mu\nu]}$ is introduced $
    \mathcal{L}= \Psi^{\mu\nu}F_{\mu\nu}(A)-\tfrac{\epsilon}2 \Psi_{\mu\nu}\Psi^{\mu\nu}
$.}

%%%%%%%%%%%%%%%%%%%%%%%%%%%%%%%%%%%%%%%%%%%%%%%%%%%%%%%%%%%%%
\section{AdS/CFT dictionary}
\label{sec:AdSCFTdict}
%%%%%%%%%%%%%%%%%%%%%%%%%%%%%%%%%%%%%%%%%%%%%%%%%%%%%%%%%%%%%
The goal of this section is to put together the bulk and the boundary sides of the story, i.e. to discuss the AdS/CFT duality for self-dual theories (SDYM) and the relation to the parent theories they originate from (Yang--Mills theory). We first recall a number of useful facts about CFT two-point functions and the spinor-helicity language. Then, we proceed to discuss the variety of boundary conditions that are available in (SD)YM. Finally, we attempt to put SDYM into the web of AdS/CFT dualities.  

%%%%%%%%%%%%%%%%%%%%%%%%%%%%%%%%%%%%%%%%%%%%%%%%%%%%%%%%%%%%%
\subsection{Two-point functions and spinor-helicity}
\label{sec:}
%%%%%%%%%%%%%%%%%%%%%%%%%%%%%%%%%%%%%%%%%%%%%%%%%%%%%%%%%%%%%
In the paper we present all CFT structures in momentum space taking also advantage of the CFT analog of the spinor-helicity language. Momentum space has certain pros and cons. The advantage of the momentum space is that conservation and Ward identities (contact terms therein) acquire simpler forms. Among cons is that the conformal symmetry itself (conformal boosts) is no longer represented by vector fields. 

The (non-normalized) two-point function of a scalar operator $O_\Delta(k)$ with conformal dimension $\Delta$ in dimension $d$ reads in momentum space 
\begin{align}
    \langle O(k) O(-k)\rangle&\sim k^{d/2-\Delta} \,.
\end{align}
The two-point function of a conserved current $J_m(k)$ is
\begin{align}
    \langle J_m(k) J_n(-k)\rangle& \sim k^{d-2}\pi_{mn} \,, & \pi_{mn}&= \left(\delta_{mn}-\frac{k_m k_n}{k^2}\right) \,,
\end{align}
where the projector $\pi^{mn}$ ensures conservation, $k_m \pi^{mn}=0$. In three dimensions there is an additional parity odd contact contribution possible
\begin{align}
    \langle J_m(k) J_n(-k)\rangle& = a\,k\,\pi_{mn}+b \,\epsilon_{mn\lambda}k^\lambda \,,
\end{align}
where $a$, $b$ are some constants. There exists another projector $\hat{\pi}_{mn}=\epsilon_{mn\lambda}k^\lambda /k$. In the spinor language this translates into two projectors: the parity even
\begin{align}
    \Pi_{AA,BB}^e&=\epsilon^{AB}\epsilon^{AB}+\frac{ k^{AA}k^{BB}}{2k^2} \,, && k^{AA}\Pi_{AA,BB}^e=0
\end{align}
and parity odd
\begin{align}
    \Pi^\text{o}_{AA,BB}&= \frac{1}{k}\epsilon_{AB}k_{AB} \,, && k^{AA}\Pi_{AA,BB}^\text{o}=0 \,.
\end{align}
The two-point function is now
\begin{align}
    \langle J_{AA}(k) J_{BB}(-k)\rangle& = a\,k\,\Pi_{AA,BB}^\text{e}+b \,k\, \Pi^\text{o}_{AA,BB} \,.
\end{align}
The (tensor) currents are traceless, e.g. $T\fud{m}{m}=0$ for the stress-tensor, and are conserved, i.e. orthogonal to $k^m$ modulo contact terms.\footnote{One usually gets the correlation functions that are identically conserved and the contact terms' contributions can be recovered from the Ward identities.} Therefore, it is convenient to use generating functions that depend on polarization vectors $\epsilon^m(k)$ that are chosen to be null, $\epsilon(k)\cdot \epsilon(k)=0$ due to the tracelessness, and orthogonal to $k^m$, $\epsilon(k)\cdot k=0$ due to the conservation. The two-point function can be encoded as
\begin{align} \label{CFT2point}
    \epsilon_1^m(k) \epsilon_2^n(-k)  \langle J_m(k) J_n(-k)\rangle& = a \,k (\epsilon_1 \cdot \epsilon_2) + b\, \epsilon_1^m \epsilon_2^n \epsilon_{mn\lambda}k^\lambda \,.
\end{align}
Recall that the key idea of the $3d$ spinor-helicity formalism is to complete $3d$ momentum $k^m$ to a $4d$ null momentum\footnote{As the formula implies, $k\cdot k=k^2=-\tfrac12 k_{AB} k^{AB}$. Also, $\brk^C k_C=2k$ and $k_{AB} \brk^{B}=+k \brk_A$, $k^Ak_{AB}=-kk_B$. }
\begin{align}
    k_{A,B'}&=\vec{k}_{AB'}+ \epsilon_{AB'}k=k_A\brk_{B'} \,,&& k_{A,A'}k^{A,A'}=0\,,
\end{align}
which immediately implies its factorization into two spinors $k_A$ and $\brk_{A'}$. Since the $4d$ Lorentz symmetry $sl(2,\mathbb{C})$ is reduced to the $3d$ Lorentz symmetry $sl(2,\mathbb{R})$, there is no difference between primed and unprimed indices --- they transform in the fundamental of $sl(2,\mathbb{R})$.\footnote{It is convenient to introduce $\epsilon_{AA'}$ as a $4d$ vector that breaks $sl(2,\mathbb{C})$ down to $sl(2,\mathbb{R})$. } It is convenient to introduce the same polarization spinors and polarization vectors \eqref{polarizationspinors} as in the bulk
\begin{align}
    \epsilon^+_A(k)&= \frac{\brk_A}{\sqrt{2k}} \,,& \epsilon^-_A(k)&= \frac{k_A}{\sqrt{2k}} \,,& \epsilon^\pm_{AB}(k)&= \epsilon^\pm_A \epsilon^\pm_B \,,
\end{align}
where $\epsilon^{\pm}_{AB}(k)$ are both null and orthogonal to $\vec{k}$, $\epsilon_{\pm}\cdot \epsilon_{\pm}=0$, $(\epsilon_+)^A (\epsilon_-)_A=1$. The spinors $k_A$, $\brk_{A'}$ can be used to define a (complex) two-dimensional subspace of vectors orthogonal to $\vec{k}^{AB}$. It is spanned by $k^Ak^B$ and $\brk^A \brk^B$. Therefore, the parity-even and parity-odd projectors can nicely be rewritten as 
\besubeqs
\begin{align}
    \Pi^\text{e}_{AA,BB}&= \frac{1}{4k^2}(k_A k_A \brk_{B}\brk_{B}+\brk_A \brk_A k_{B} k_{B})\,,\\
    \Pi^\text{o}_{AA,BB}&= \frac{1}{4k^2}(k_A k_A \brk_{B}\brk_{B}-\brk_A \brk_A k_{B} k_{B})   \,.
\end{align}
\esubeqs
Two-point functions are special in that the momenta of the two operators are simply $\vec{k}$ and $-\vec{k}$ and the polarization vectors/spinors are related. Indeed, from 
\begin{align}
    -k_{AB}+\epsilon_{AB}k&= -(k_{BA}+k\epsilon_{BA})=-k_B\brk_A
\end{align}
we see that $k^A(-k)=-c\brk^A(k)$, $\brk^A(-k)=\bar{c}k^A(k)$ and $c\bar{c}=1$.\footnote{We will not need this fact, but $c$ , $\bar{c}$ do depend on $k$, otherwise there is a contradiction. In practice, the spinors do not appear alone, but contracted, and one can simply set $k^A(-k)=-\brk^A(k)$, $\brk^A(-k)=k^A(k)$ and, hence, $\epsilon_{\pm}^{AB}(-k)=\epsilon_{\mp}^{AB}(k)$.} In particular,
\begin{align}
    \epsilon^\pm (+k)\cdot \epsilon^\pm (-k)&=1 \,, & \epsilon_+(+k)\cdot \epsilon_-(-k)&=0 \,,
\end{align}
or, more compactly, $\epsilon^\alpha (+k)\cdot \epsilon^\beta (-k)=\delta^{\alpha,\beta}$, where the helicity is defined as $\alpha,\beta=\pm1$. In the spinor language $\epsilon_1^m \epsilon_2^n \epsilon_{mn\lambda}k^\lambda $ is $\epsilon_1\fud{A}{C} \epsilon_2^{AC} k_{AA}$. The current two-point function can now be rewritten in the spinor-helicity language as
\begin{align}
    \langle J^\alpha(k) J^\beta(-k)\rangle&= \delta^{\alpha,\beta} 4 k (a-\alpha b) \,,
\end{align}
where we introduced the projections $J^\alpha =J(k)\cdot \epsilon^\alpha(k)$ of the currents onto the components with definite helicity. In other words, the following two-point functions do not vanish
\begin{align}\label{twopointab}
       \langle J^+(k) J^+(-k)\rangle&=  k (a- b) \,, &
       \langle J^-(k) J^-(-k)\rangle&=  k (a+ b) \,.
\end{align}
We can also define $\langle 1 2\rangle= k^A(k) k_A(-k)=-k^A(k) \brk_A(k)=+2k$, $\langle \bar{1} \bar{2}\rangle= \brk^A(k) \brk_A(-k)=+2k$ and rewrite the two-point function as\footnote{Note that, as is standard in the $4d$ spinor-helicity language, $1 \rangle = k^A$ and is not normalized by $k$, whereas it is also standard to normalize polarization vectors not to have any mass dimension. }
\begin{align}
       \langle J^+(k) J^+(-k)\rangle&= \delta^{\alpha,\beta} \frac{\langle 1 2\rangle^2 }{k^2} (a- b)k \,,&
       \langle J^-(k) J^-(-k)\rangle&= \delta^{\alpha,\beta} \frac{\langle \bar 1 \bar 2\rangle^2 }{k^2} (a+ b)k \,.
\end{align}
An interesting feature of the helicity decomposition of the two-point function is that the even and odd structures contribute asymmetrically and one can tune either $\langle --\rangle $ or $\langle ++\rangle $ to vanish. 

Lastly, let us also note that, in general, contact terms can be present in two-point functions, e.g. in the parity-odd contribution to $\langle J_n J_m\rangle$ which may come from a Chern--Simons term. One typical example is the two-point function of an operator $O_\Delta$ and its shadow $\tilde O_{d-\Delta}$ with the two-point function
\begin{align}
    \langle O_\Delta(x) \tilde O_{d-\Delta}(y)\rangle&\sim \delta^d(x-y) && \Longleftrightarrow&& \langle O_\Delta(-k) \tilde O_{d-\Delta}(k)\rangle\sim 1 \,.
\end{align}
Such a contact term can appear in the case of a toy-model of the scalar field where instead of considering $\phi(0)$ and $\pl_z \phi(0)$ one can choose the shadow $|k|\phi(0)$ of the former. A similar situation is found in the case of massless spin-half fields: the conformal dimension $\Delta=3/2\pm m$ becomes exactly $d/2$, which leads to the delta function ambiguity in the two-point function. This is also important for the spin-one case since the dimension of $a_i$ is $1$ and the dimension of $J_i\sim E_i$ is $2$. Therefore, one can again expect a contact term of type $\langle a_n J_m\rangle \sim \Pi^\text{e}_{nm}$, being of order $0$, it corresponds to $\delta^3(x-y)$ in the position space. Indeed, as we will see, the bulk-to-bulk propagator $\langle \Psi \Phi\rangle$ in SDYM does have a term with $\sign(z-z')$ that leads to similar contact-term ambiguities. One subtlety of contact terms is to justify they have an invariant meaning. For example, the coefficient of the parity-odd structure can be the Chern--Simons level $k$, in which case the quantization of $k$ can make $(k\mod \mathbb{Z})$ physical.

%%%%%%%%%%%%%%%%%%%%%%%%%%%%%%%%%%%%%%%%%%%%%%%%%%%%%%%%%%%%%
\subsection{Boundary conditions}
\label{sec:}
%%%%%%%%%%%%%%%%%%%%%%%%%%%%%%%%%%%%%%%%%%%%%%%%%%%%%%%%%%%%%
Let us begin with Maxwell/YM theory. It is easier to discuss the AdS/CFT dictionary in the axial gauge $A_z=0$, where the boundary asymptotic reads
\begin{align}
    A_i&= a_i +z J_i+... \,, && J_i=E_i=F_{zi}|_{z=0}=\pl_z A_i|_{z=0} \,.
\end{align}
The Dirichlet condition corresponds to fixing $a_i$ and the Neumann one to fixing $J_i$. With these boundary conditions the bulk field is dual to a conserved current $J_i$ or to a gauge field $a_i$ in the boundary CFT. There is a gauge-invariant way to impose the boundary conditions. One notices that the magnetic field $B_i\sim \epsilon_{ijk} F^{jk}$ detects the $a_i$ asymptotic quotiented by the gauge symmetry and the electric field $E_i\sim F_{zi}$ detects the $J_i$ asymptotic. In our convention $F^{AA}\sim B^{AA}+E^{AA}$ and $\bar F^{AA}\sim B^{AA}-E^{AA}$, $B^{AA}=k\fud{A}{B}\Phi^{AB}$. Therefore, the simplest boundary conditions are identified as (also called, magnetic and electric)
\besubeqs
\begin{align}
    \text{Dirichlet}&: & 2B^{AA}=F^{AA}+\bar F^{AA}&=\text{fixed}\,,\\
    \text{Neumann}&: & 2E^{AA}=F^{AA}-\bar F^{AA}&=\text{fixed} \,.
\end{align}
\esubeqs
In general, boundary conditions should correspond to a Lagrangian submanifold of the boundary data that is compatible with regularity in the bulk. The canonical symplectic structure that comes from the Yang--Mills action is (it gives $\delta A_i \pl_z A^i \sim p_i dq^i$ as a potential)
\begin{align}
    \Omega= \delta^{ij}\,\delta A^0_i \wedge \delta A^1_j= \delta \Phi_+^0\wedge \delta \Phi_+^1+\delta \Phi_-^0\wedge \delta \Phi_-^1\,,
\end{align}
where for the future convenience we also expanded it in terms of the helicity components. It is clear that one can also impose mixed (conformally-invariant) boundary conditions \cite{Witten:2003ya,Marolf:2006nd}
\begin{align}\label{mixedbc}
    F^{AA} e^{i\gamma}+ \bar F^{AA} e^{-i\gamma}&=\text{fixed} && \Longleftrightarrow && B_i \cos \gamma +i E_i\sin \gamma=\text{fixed} \,,
\end{align}
with $\gamma=0$ being Dirichlet and $\gamma=\pi/2$ being Neumann. The $B$-term in the boundary conditions can be thought of as coming from the Chern--Simons term on the boundary (equivalently, from the theta-term in the bulk). Indeed, the variation of the Yang--Mills action gives $\delta S_{\text{YM}}=\int_{\pl M} E_i \delta A^i$, while the variation of the Chern--Simons term is $\delta S_{\text{CS}}=i\kappa \int _{\pl M} B_i \delta A^i$, $\kappa$ being the non-normalized level. Therefore, the mixed boundary conditions can also be written for the free theory (in other words, with sources turned off) as
\begin{align}\label{mixedbcEB}
    i\kappa B_i + E_i&=0 \,, && \kappa = -\cot(\gamma) \,.
\end{align}
In terms of the helicity fields, the Dirichlet and Neumann options with regularity in the bulk imposed on the free fields lead to 
\besubeqs
\begin{align}
    \text{D}&: & A_i&=e^{-k z}a_i\,, \,\,\qquad k_i a^i=0 && \Longleftrightarrow && \Phi_\pm=e^{-kz} \phi_{\pm}^0 \,.\\
    \text{N}&: & A_i&=-\tfrac1{k}e^{-k z}J_i\,, \quad k_i J^i=0 && \Longleftrightarrow && \Phi_\pm=-\tfrac1{k}e^{-kz} \phi_{\pm}^1 \,.
\end{align}
\esubeqs
The mixed boundary conditions translate into
\begin{align}\label{mixedbcphi}
    k \Phi^0_+ \cos \gamma +\Phi^1_+ i\sin \gamma&= \text{fixed}_+\,, &
    -k \Phi^0_- \cos \gamma +\Phi^1_- i\sin \gamma&= \text{fixed}_-\,.
\end{align}
Since the helicity decomposition is Lorentz-invariant on the boundary, one can play with many more options, which we won't attempt. 

The self-dual boundary conditions fix $F_{AA}=0$. Formally, one can consider the limit of $\gamma=-i\infty$ in \eqref{mixedbc} (or $\kappa=-i$ in \eqref{mixedbcEB}), which amplifies $F_{AA}$ and dumps the rest, including the source on the r.h.s. One peculiarity is that $F_{AA}=0$ is not just a boundary condition but the equation of motion that is satisfied in the bulk, see footnote \ref{ft:Feq}. In the helicity basis the self-dual boundary conditions read
\begin{align}\label{sdbcphi}
    k \Phi^0_+ +\Phi^1_+&=0 \,,&
    -k \Phi^0_-  +\Phi^1_- &= 0\,.
\end{align}
There are two types of self-duality one can consider: (1) in the original YM or (2) in its Chalmers--Siegel form. Firstly, let us discuss (1). The self-dual boundary condition in (complexified) YM implies setting the source \eqref{mixedbcphi} to zero (unless we manually amplify it by $\exp[i\gamma]$, which does not give self-duality).\footnote{For example, see also Appendix \ref{app:boundaryterms}, one can consider a mixed boundary condition $(aq+bp)|_{z=0}=Q$. The Dirichlet case corresponds to either $b=0$ or to $a=\infty$, in the latter case one has to amplify the source $Q\rightarrow a Q$ as well. This clearly does not work for $F_{AA}=0$, where the source has to be set to zero. } This kills the $\Phi_-$ degree of freedom: regularity gives $\Phi_-=e^{-kz} \phi_-^0$, which is inconsistent with \eqref{sdbcphi}. On the other hand, $\Phi_+$ can propagate since $\Phi_+=e^{-kz} \phi_+^0$ respects \eqref{sdbcphi}. However, there cannot be any cubic and higher order interaction in the bulk built of just $\Phi_+$. Indeed, passing to the chiral YM theory, the vertex always contains $F_{AA}$, which vanishes. We will also see that certain terms in the propagators blow up in the self-dual limit of type (1). As it was already discussed in Sections \ref{sec:2.5} and \ref{sec:FG}, the self-dual limit should not be taken in the original YM theory. We have identified two problems: (a) viewing self-duality as the limit of mixed boundary conditions in YM kills the source, which is essential for extracting observables on the CFT side; (b) there cannot be any Lorentz invariant theory of just a single helicity degree of freedom $\Phi_+$. It would be interesting to consider the extended YM theory of Section \ref{sec:2.5} with two coupling constants, especially if it admits a covariant formulation, as it does have SDYM as a smooth limit. 

The situation (2) is not much better at first glance since $\epsilon$, unless $\epsilon=0$, is not a genuine coupling constant. Therefore, imposing self-dual boundary conditions for $\epsilon\neq0$ does not lead anywhere. Fortunately, the degrees of freedom carried by $A_\mu \sim \Phi_{A,A'}$ are not lost at $\epsilon=0$: the positive helicity $\Phi_+$ is still there and the negative one has moved to $\Psi_{AA}$. We have $F_{AA}\rightarrow 0 $ and $F_{AA}=\epsilon \Psi_{AA}$. One can make $\epsilon$ go to $0$ as well as to detach $\Psi_{AA}$ from $F_{AA}$. At least, it is clear that one cannot consider any other but the self-dual boundary condition in $\epsilon\rightarrow 0$ limit as it would over-constrain the system. A natural proposal is to consider a double limit: $\gamma=-i\infty$, $\epsilon=0$. Note that it is impossible to have a mixed boundary condition in SDYM since it is a first order theory.

As the Fefferman-Graham analysis of Section \ref{sec:FG} shows, in SDYM the only free boundary data is related to $\Phi_+^0$ and $\Psi_-^0$, which both satisfy the $D_+$-equation, thereby respecting regularity. This is also compatible with the SDYM symplectic form
\begin{align}
    \Omega&= \delta \Psi_+ \wedge \delta \Phi_++\delta \Psi_- \wedge \delta \Phi_- \,.
\end{align}
Therefore, the only reasonable boundary conditions in SDYM are Dirichlet ones:
\begin{align}
    \Phi_+^0&=\text{fixed} \,, & \Psi_-^0=\text{fixed}\,.
\end{align}
It is interesting that one can choose the ``same boundary conditions'' in the original YM theory. Indeed, for $\Phi_+^0$ it is the usual Dirichlet and let us choose $\Phi_-^1=\text{fixed}$. The regularity gives $\Phi_-=-k^{-1} e^{-kz} \Phi_-^1$ and, hence, $\epsilon\Psi_-= D_-\Phi_-=2 e^{-kz}\Phi_-^1$, as it should ($\epsilon\Psi_-^0=2\Phi_-^1$), i.e. we are effectively fixing the same $\Phi_+$ and $\Psi_-$ as in SDYM. This means that we have Dirichlet for $\Phi_+$ and Neumann for $\Phi_-$, which explains why it has not been considered before. Let us stress that keeping the same degrees of freedom fixed on the boundary in YM and SDYM does not mean that the self-dual boundary condition is imposed ($\Psi_- \epsilon=D_-\Phi_-\neq0$, i.e. $F_{AA}\neq0$). 

Proceeding as above, from the $A_i=a_i+z E_i+...$ vantage point, we fix the positive helicity part of $a_i$ (supposed to be a source for the CFT current) and the negative helicity part of $E_i$ (supposed to be a source for the gauge field on the boundary). This might be a good interpretation of what happens in SDYM: the theory keeps $F_{AA}$ as $\Psi_{AA}$ (it satisfies the same equation as $F_{AA}$ would) and preserves half of the YM's $\Phi_{A,A'}$. Therefore, one should maintain the same interpretation for $\Phi_{A,A'}$ in (SD)YM and for $\Psi_{AA}$ one should use the same interpretation as for $F_{AA}$, the only difference is that one should apply these interpretations to the components with specific helicities.

%%%%%%%%%%%%%%%%%%%%%%%%%%%%%%%%%%%%%%%%%%%%%%%%%%%%%%%%%%%%%
\subsection{Duality}
\label{sec:}
%%%%%%%%%%%%%%%%%%%%%%%%%%%%%%%%%%%%%%%%%%%%%%%%%%%%%%%%%%%%%
Once the AdS and CFT ingredients have been introduced above, one can put them together. An action does not yet completely specify a theory on a manifold with a boundary. Boundary conditions need to be prescribed, which in its turn may require adding a boundary term to the action. Within AdS/CFT correspondence, given a theory in the bulk, a natural question to ask is what it is dual to for a specific boundary condition. Also, it is interesting to see how the dual CFT depends on the boundary conditions in the bulk. For example, how the CFT correlators computed with Dirichlet boundary conditions are related to those with Neumann ones. The dependence of AdS/CFT duality on boundary conditions is what we review below, which allows us to approach SDYM.

\paragraph{Dirichlet.} For Dirichlet boundary conditions the AdS/CFT dual CFT has a global symmetry current $J_m$ on the boundary. The Dirichlet boundary data has the interpretation of a source and the duality implies
\begin{align}
    \left\langle \exp \int d^3 x\, J^m(x) A_m(x) \right\rangle_{\text{CFT}}&=\int D\Phi\Big|_{\Phi^0_i\rightarrow A_i} \exp{S[\Phi]} \,.
\end{align}
As before, $\Phi^0_i$ denotes the leading fall-off coefficient. Realistically, the bulk action $S$ has to contain gravity and other fields to make the bulk theory quantum consistent once gravity is in, but we are insensitive to this incompleteness at the order we work. The semi-classical approximation is to compute the action on the saddle point, which is a solution of the variational problem with the boundary value $\Phi^0$ of $\Phi$ kept fixed to be the source $A$, hence, Dirichlet.

\paragraph{Neumann.} Given a CFT with a global symmetry current $J_m$, as before, one can promote the infinitesimal source $A_m$ to the full-fledged background field.\footnote{This may require additional local terms to be added to the action, e.g. $|D\phi|^2$ contains $J_m A^m$ as well as $-\phi^\dag A_m A^m \phi$ to make the coupling gauge invariant, which were dubbed Seagulls. } Let us assume that there exists a new gauge-invariant action 
\begin{align}
    S[A]&= S_{\text{CFT}}+ \int d^3 x\, J^m(x) A_m(x) +\text{Seagulls} \,.
\end{align}
Given a gauge field $A_m$ one can define a conserved current $\tilde{J}_m=\epsilon_{mnk} \pl^nA^k$.\footnote{$\tilde J=*dA$  is for $U(1)$ (one has to divide $*dA$ by $2\pi$ to be precise), otherwise one needs to dualize the Yang--Mills field strength, $\tilde J=*(dA+AA)$. We will ignore the nonabelian correction to the current. } It then makes sense to take the path integral over $A_m$ and study the correlation functions of $\tilde{J}_m$. This is called the $S$ operation \cite{Witten:2003ya}. In practice, one first gets correlators of $A_m$ itself:
\begin{align}
    \int DA\, \left\langle \exp \int d^3 x\, J^m(x) A_m(x)+E^m A_m(x) +...\right\rangle_{\text{CFT}}&=\int D\Phi\Big|_{\Phi^1_i\rightarrow E_i} \exp{S[\Phi]} \,.
\end{align}
Here, $\Phi^1_i$ denotes the second, subleading, asymptotic, i.e. corresponds to the Neumann boundary condition. For $s=1$ it is picked by $\pl_z \Phi_{AA'}$, which is the electric field in the axial gauge. Note that $\pl_i E^i=0$ (Gauss law for $U(1)$) and, hence, $E^{AA}=k\fud{A}{B}\tilde{A}^{AB}$ ($E=\slashed k \tilde{A}$ for short) for some $\tilde A^i\equiv \tilde A^{AB}$. This correctly takes into account that a gauge field $A_m$ should be contracted with a  conserved current.  Upon this substitution, $k$ can be moved to $A_m$ to yield the correlators of $\tilde{J}_m$:
\begin{align}
    E^{i_1}(k_1)\ldots \langle A_{i_1}(k_1)\ldots\rangle \quad \longrightarrow \quad \tilde{A}^{i_1}(k_1)\ldots \langle \tilde{J}_{i_1}(k_1)\ldots\rangle 
\end{align}
To the leading order one can expand the exponent to find that the effective (nonlocal) kinetic term for $A_m$ comes from the two-point function of $J_m$:
\begin{align}
    \langle \exp S[A]\rangle&= 1+\tfrac12\int d^3x\, d^3y\, A^m(x) \langle J_m(x) J_n(y)\rangle A^n(y)  +... \,.
\end{align}
The propagator $\langle A_mA_n\rangle$ for $A_m$ comes from the inverse of the two-point function, which requires some gauge-fixing, a convenient one being the Lorenz gauge $\pl_m A^m=0$. Then, $\langle J_mJ_n\rangle = a k \pi_{mn}$ in the momentum space and, hence, $\langle A_mA_n\rangle = a^{-1}k^{-1} \pi_{mn}$ in the Lorenz gauge. The two-point function of the new current is 
\begin{align}
    \langle \tilde{J}_m \tilde{J}_n\rangle&= a^{-1} k \pi_{mn} \,.
\end{align}
Perhaps, the simplest of this situation is given \cite{Witten:2003ya} by critical QED. More generally, one can start with a CFT that has a generic two-point function  
\begin{align}\label{genericJJ}
    \langle J_mJ_n\rangle =a\pi_{mn}+ b \hat{\pi}_{mn}
\end{align}
($\pi\pi=\pi$, $\pi \hat{\pi}=\hat{\pi}$, $\hat{\pi}\hat{\pi}=-\pi$). The $S$-operation gives to the leading order \cite{Witten:2003ya}
\begin{align}\label{Soperation}
    \langle \tilde{J}_m \tilde{J}_n\rangle&= \frac{a}{a^2+b^2} k \pi_{mn} + \frac{-b}{a^2+b^2} k \hat{\pi}_{mn} \,.
\end{align}

\paragraph{Dirichlet vs. Neumann.} The Neumann problem is closely related to the Dirichlet one \cite{Witten:2001ua, Klebanov:1999tb}. Suppose we have already computed the path integral with the fields satisfying the Dirichlet condition, i.e. the on-shell action in the first approximation:
\begin{align}
    e^{S_{\text{eff}}[q]}&= \int D\Phi \Big|_{\Phi^0\rightarrow q} \exp{S[\Phi]}\,.
\end{align}
Its variation gives $\delta S_{\text{eff}}[q]=\int_{\pl M} p\,\delta q$, where $p=p[q]$ is a functional of the boundary data $q$ (fixed by the equations of motion and regularity). In order to replace Neumann with Dirichlet one can add $-\int_{\pl M} q \pi$ to the action and, then, integrate over $q$.\footnote{This is sometimes being considered as $\Phi$ with free boundary conditions since we integrate over all boundary values in the end. } The saddle point approximation instructs us to solve $p[q]=\pi$, i.e. the saddle point imposes the Neumann condition and is equivalent to performing the Legendre transform. The latter has a clear diagrammatic interpretation in the bulk.

\paragraph{Diagrammatics.} There is a way to diagrammatically understand what changing boundary conditions does on the bulk and on the CFT sides \cite{Hartman:2006dy,Giombi:2011ya}. While the relations can be given for any $\Delta$ and any $d$, we concentrate on the $\Delta=1,2$ scalars in $\text{AdS}_4$, \cite{Giombi:2011ya}. For Dirichlet vs. Neumann the key relations are: (1) the boundary-to-bulk propagators for $\Delta$ and $d-\Delta$ conditions are related via two-point functions of $O_\Delta$ or $O_{d-\Delta}$ (the latter being inverse of each other); the difference between bulk-to-bulk propagators with $\Delta$ and $d-\Delta$ boundary conditions factorizes into the product of the boundary-to-bulk propagators and a two-point function. More specifically, we have ($d=3$, $K_{\Delta=2}\equiv K_D$ and $K_{\Delta=1}\equiv K_N$)
\begin{align}
    K_{D}&=K_{N} \langle O_2 O_2\rangle \,,  &
    G_{D}-G_{N}&= \frac{1}{k}e^{-k(z+z')}=K_{N}K_{N}\langle O_2 O_2\rangle=K_{D}K_{D}\langle O_1 O_1\rangle \,.
\end{align}
These relations imply that once $\Delta=1$ duality is established, the $\Delta=2$ duality follows automatically or other way around. Indeed, one can represent the AdS/CFT correlators obtained with D/N boundary conditions as a particular manipulation over the AdS/CFT correlators obtained with N/D boundary conditions, which quietly performs a Legendre transform. Therefore, one does not have to recompute AdS diagrams one by one for another choice of boundary conditions.

\paragraph{Theta-term.} There is one more operation natural on the CFT side \cite{Witten:2003ya}, which was called T-operation. Given a CFT with a global symmetry current $J_m$ one can add to the CFT action a Chern--Simons term at level $\kappa$\footnote{To be precise, $\kappa$ has to be $\kappa/(4\pi)$ and what was called $T$-operation in \cite{Witten:2003ya} is adding the Chern--Simons term at level half, which corresponds to $\theta\rightarrow\theta +2\pi$ in the bulk. Here, we are more concerned with how to reproduce such contact terms from the bulk in general, rather than with making the theory well-defined on arbitrary manifolds, hence, $\kappa$ will be assumed a free parameter. Note also, that according to \cite{Jain:2024bza,Aharony:2024nqs}, to get self-dual correlation functions one has to take the level into the complex plane (hence, not dividing it by $4\pi$ is the least of our problems). }
\begin{align}
    S[A]&= S_{\text{CFT}}+ \int d^3 x\, J^m(x) A_m(x) + {\kappa}\,\Tr \int AdA +\tfrac23 A^3\,.
\end{align}
This just shifts the coefficient of the contact term given by the parity-odd structure. From the bulk point of view, one can add the same Chern--Simons term as a boundary term, which can also be represented as the theta-term in the bulk. The action is still stationary under the Dirichlet boundary conditions and $A_m$ is the boundary value to be fixed.  

\paragraph{Mixed boundary conditions. } Adding a boundary term does not necessarily modify the boundary conditions. Indeed, given a bulk action that is already stationary under Dirichlet boundary conditions, e.g. $S_{\text{YM}}=\int_{\pl M} E_i \delta A^i$, and supplementing it with a boundary action $S_b[A]$ such that $\delta S_b[A]=\int_{\pl M} \pi_i \delta A^i$, one can continue using the Dirichlet conditions. In this case, $S_b[A]$, whether it is local or not, will be just an addon to the generating functional of correlators $W_D[A]$ obtained in the initial Dirichlet theory. If $S_b[A]$ is local, it will add certain contact terms to the correlation functions, a good example being how a Chern--Simons term adds parity-odd contact terms to $2$- and $3$-point functions. In this cases, adding $S_b[A]$ to the CFT action means the same for the bulk theory and can be understood as adding a boundary term. 

Let us come back to the Neumann/mixed boundary conditions story where the boundary value $A_m$ is a dynamical field on the CFT side. Adding a boundary term can modify the boundary conditions. As it was already mentioned above, adding the Chern--Simons term with the sources turned off can be interpreted as a mixed boundary condition (the second option is Dirichlet, $\delta A_i=0$):
\begin{align}
    \delta S=\delta A^m( i\kappa B_m + E_m)&=0 \,,  && S[A]=S_{\text{YM}}[A]+i\kappa S_{\text{CS}}[A] \,.
\end{align}
In order to turn on the source, i.e. to fix $Q_m=( i\kappa B_m + E_m)$ on the boundary, one can add a boundary term $-\int_{\pl M} ( i\kappa B_m + E_m)A^m$, essentially a Legendre transform. Alternatively, one can just add $-\int_{\pl M} A^m Q_m^0$, where $Q_m^0$ is considered a fixed boundary data. Boundary conditions for vector fields were extensively discussed in \cite{Marolf:2006nd}.

For the Dirichlet boundary condition the bulk partition function is a generating functional $W[A_i]$ of the boundary correlators $\langle J...J\rangle$. A gauge transformation of $A_i$ will lead to contact terms that are due to Ward identities that the correlators satisfy. Therefore, one expects that the result is invariant up to contact terms. For Neumann/mixed boundary conditions, one gets correlation functions of a gauge field on the boundary, which are not gauge-invariant. For the $U(1)$ case upon passing to the magnetic field, i.e. to the dual current, $\tilde J=*dA$, one should observe the same gauge independence up to contact terms as for Dirichlet. For a non-abelian gauge group, the dual current is $\tilde J=*(dA+A^2)$ and it is only covariantly conserved. Therefore, the correlation functions of $\tilde {J}$ do depend on the gauge choice, as we will see.

\paragraph{Flow of mixed boundary conditions.}
A complementary question is what happens when the mixed boundary conditions are slightly changed. The piece of the bulk-to-bulk propagator that depends on the boundary conditions is a homogeneous one, see Section \ref{sec:props}. For the jump between the Neumann and a generic, $\gamma\neq0$, mixed boundary condition we have
\besubeqs
\begin{align}
    G_{\gamma}-G_{\text{N}}&= \frac{1}{k}e^{-k(z+z')}\cos(\gamma)[\cos (\gamma) \pi +i \sin(\gamma) \hat{\pi}]\,,\\
    K_\gamma&= \sin(\gamma)[\sin (\gamma) \pi +i \cos(\gamma)\hat{\pi}]K_{N}\,.
\end{align}
\esubeqs
The same arguments as in \cite{Hartman:2006dy,Giombi:2011ya} can be applied to show that one can express the correlators $\langle A...A\rangle$ of the gauge field for any $\gamma$ in terms of those for $\gamma=\pi/2$. It looks even better for an infinitesimal change of the mixed-boundary conditions\footnote{The formulas in this section are tied to the standard form of YM in terms of $A_\mu$. As we noted, the self-dual limit requires the Chalmers--Siegel formulation. For example, the expressions here would simply diverge for $\gamma=-i\infty$. It would be interesting to adapt them in such a way as to get the results for the correlators we obtain later in the text. The easiest way is to rescale the external lines or to compute boundary limit of correlators of $\Psi_{AA}$, $\bar{F}_{A'A'}$, see Section \ref{sec:Correlators}. }
\begin{align}
    \frac{\pl}{\pl \gamma} G&= \frac{1}{k}e^{-k(z+z')} P \,, & \frac{\pl}{\pl \gamma}  K&= \frac{1}{k}e^{-kz} P \,, && P=[- \sin (2\gamma) \pi +i \cos(2\gamma) \hat{\pi}] \,.
\end{align}
\paragraph{Self-dual holography.} Based on the discussion above, let us formulate a natural proposal for self-dual holography. The free boundary data in SDYM are $\Phi_+^0$ and $\Psi_-^0$, i.e. Dirichlet. Via the Chalmers--Siegel action they can be traced back to YM, where they correspond to Dirichlet for $\Phi_+$ and Neumann for $\Phi_-$. Indeed, we see that $\Psi_{AA}$ corresponds to a conserved current on the boundary, which is exactly a Neumann data for $A_i$ and $\Phi_{AA'}$ corresponds to a gauge field on the boundary, which is a Dirichlet data for $A_i$. Bearing in mind this picture, the dual CFT has a positive helicity component of the current $J_+$ that couples to $\Phi_+$ and a negative helicity gauge field that couples to $\Psi_-$. By dualizing the gauge field's field strength, one can build $J_-$, which together with $J_+$ gives a complete current. Therefore, the AdS/CFT dual of a self-dual theory is a CFT of the same type (with a current) as for all the other boundary conditions.

A less exotic model that does not discriminate helicities can be cooked up from two Yang--Mills theories $\text{YM}_{1,2}$ coupled together via a bi-fundamental matter. One can impose Dirichlet/Neumann on $\text{YM}_{1}/\text{YM}_{2}$. The CFT dual of the $\text{YM}_{1}$-part would have a current $J_m^1$ and the dual of the $\text{YM}_{2}$-part would have a gauge field $A_m$ whose magnetic field $B=*(dA+AA)$ would play the role of the second current $J_m^2$.

%%%%%%%%%%%%%%%%%%%%%%%%%%%%%%%%%%%%%%%%%%%%%%%%%%%%%%%%%%%%%
\section{Propagators in various gauges}
\label{sec:props}
%%%%%%%%%%%%%%%%%%%%%%%%%%%%%%%%%%%%%%%%%%%%%%%%%%%%%%%%%%%%%

%%%%%%%%%%%%%%%%%%%%%%%%%%%%%%%%%%%%%%%%%%%%%%%%%%%%%%%%%%%%%
\subsection{General aspects}
\label{sec:}
%%%%%%%%%%%%%%%%%%%%%%%%%%%%%%%%%%%%%%%%%%%%%%%%%%%%%%%%%%%%%
The kinetic terms of the Yang--Mills and chiral Yang--Mills theories are the same, just written slightly differently (ignoring the boundary term). Therefore, for the purpose of deriving the propagators the theories are identical. While gauge fixing YM theory is textbook, gauge fixing SDYM is discussed below. Besides, all theories are conformally invariant and, hence, there is little need to consider the anti-de Sitter space instead of the flat space with a boundary at $z=0$.\footnote{Therefore, we define our theories already in the flat space with a boundary at $z=0$ and proceed to impose various gauge conditions. One subtlety is that the gauge fixing conditions we are going to use are not conformally-invariant, e.g. $\pl_\mu A^\mu=0$ is not mapped to $\nabla_\mu A^\mu$ in anti-de Sitter space. It is easier to impose $\pl_\mu A^\mu=0$, which leads to a modified Lorenz gauge from the AdS vantage point, see \cite{Metsaev:2008ks} for extension to all spins. There exists one gauge condition that is conformally invariant \cite{Eastwood:1985eh}, but we do not use it.}

Some general aspects that are shared by all theories include the following. The bulk-to-bulk propagator\footnote{Earlier references discussing propagators include \cite{Allen:1985wd,Allen:1986qj,DHoker:1999bve, Raju:2010by, Moga:2025gdy}.} consists of two parts --- inhomogeneous and homogeneous:
\begin{align}
    G(-k,z;k,z')&=G^\text{hom}(-k,z;k,z')+G^\text{inh}(-k,z;k,z') \,.
\end{align}
The inhomogeneous part solves the equation of motion with the delta-function $\delta(z-z')$ on the right-hand side. It can easily be obtained by the Fourier transform of the usual momentum-space propagators in flat space with respect to $k^0$ 
\begin{align}
    G(k^i,z)&= \frac{1}{2\pi}\int d\omega\, G(k^i,k^0=\omega)\, e^{i\omega z} \,.
\end{align}
The homogeneous part is needed to impose the required boundary conditions in anti-de Sitter space. The homogeneous one subdivides into two parts, in accordance with the two regions: (a) boundary conditions at the conformal boundary of AdS, i.e. $z=0$; (b) ``boundary conditions deep in the bulk'', i.e. regularity in the deep interior of AdS (the propagator should go to zero when the points are far away from each other and from the boundary).\footnote{At least it should not blow up, see e.g. an example of the (most popular) axial gauge, where it does not decay at large separation. }

The Feynman gauge leads to the simplest form of the inhomogeneous part. All other propagators can be related to the Feynman one via a gauge transformation
\begin{align}
    \begin{aligned}
        G^{\text{some gauge}}_{AA',BB'}&(-k,z;k,z')= G^{\text{F}}_{AA',BB'}(-k,z;k,z')+\\
        &+\nabla_{k,z'}^{BB'} \xi_\text{L}^{AA'}(k,z,z')+\nabla_{-k,z}^{AA'} \xi_\text{R}^{BB'}(k,z,z')+\nabla_{-k,z}^{AA'} \nabla_{k,z'}^{BB'}\xi_{\text{c}}(k,z,z') \,,
    \end{aligned}
\end{align}
which turns out to be very useful since a pure gauge term collapses the bulk integral to a boundary term. There is some redundancy among the three types of pure gauge terms above as the last one can be absorbed by either of the first two. Note that the pure gauge terms can also be split into homogeneous and inhomogeneous. The latter are Fourier transforms of the corresponding pure gauge terms in flat space.

As it was discussed in Section \ref{sec:AdSCFTdict}, it is convenient to impose boundary conditions in terms of gauge (co/in)variant quantities such as $F^{AA}$, $\bar{F}^{A'A'}$. Since $F_{AA}=0$ in SDYM, but $\Psi_{AA}$ exists, we change our notation to refer to $F_{AA}$, whenever it makes sense, as $\Psi^{AA}$, and to $\bar{F}_{A'A'}$ as $\bar{\Psi}^{A'A'}$ (just to have a symmetric notation). To this end, given a $\langle \Phi^{A,A'} \Phi^{B,B'}\rangle $-propagator for the Yang--Mills theory or $\langle \Psi^{AA} \Phi^{B,B'}\rangle $ propagator for Chalmers--Siegel theory or SDYM, we compute the gauge-invariant objects of type $\langle \Psi/\bar\Psi\, \Psi/\bar{\Psi}\rangle $. The boundary conditions are then imposed as
\besubeqs
\begin{align}
    e^{i\gamma} \langle \Psi^{AA}(-k,z\rightarrow 0) \Psi^{BB}(k,z')\rangle &=-e^{-i\gamma} \langle \bar\Psi^{AA}(-k,z\rightarrow 0) \Psi^{BB}(k,z')\rangle\,,\\
    e^{i\gamma} \langle \Psi^{AA}(-k,z\rightarrow 0) \bar\Psi^{BB}(k,z')\rangle &=-e^{-i\gamma} \langle \bar\Psi^{AA}(-k,z\rightarrow 0) \bar\Psi^{BB}(k,z')\rangle  \,.  
\end{align}
\esubeqs
It is also worth checking that the similar boundary conditions are satisfied with respect to $(k,z')$-leg since the most general ansatz for the homogeneous terms is not symmetric with respect to the left/right legs exchange. 

When imposed this way, i.e. via gauge-invariant $\Psi/\bar\Psi$, the boundary conditions are insensitive to pure gauge homogeneous $\nabla\xi$-terms that differentiate between the gauges, but the terms themselves depend on the gauge. In order to fix these terms one could try to impose the boundary condition of the form
\begin{align}
       e^{i\gamma} \langle \Psi^{AA}(-k,z\rightarrow 0) \Phi^{B,B'}(k,z')\rangle &=-e^{-i\gamma} \langle \bar\Psi^{AA}(-k,z\rightarrow 0) \Phi^{B,B}(k,z')\rangle  \,,
\end{align}
where only the leg on which the boundary condition is imposed is represented by the gauge-invariant field strengths $F\equiv\Psi$, $\bar F\equiv \bar\Psi$. However, this condition turns out to be too strong in some gauges. For example, it does not admit solutions that decay in the deep interior of AdS in the axial gauge. In addition, it may feature some discontinuity between Dirichlet and mixed boundary conditions.

Note that the homogeneous terms are not the same in all gauges: they have to satisfy the corresponding gauge conditions and to solve the (effective) equations of motion. Nevertheless, there is a part of the homogeneous terms that is the same in all gauges. It has a $\gamma$-dependence and affects only the physical polarizations. The pure gauge homogeneous terms must not spoil two-point functions on the boundary. 

Concretely, except for the axial gauge where we will also need some $z$-dependent terms without the exponent, the homogeneous terms can be parameterized as
\begin{align}\notag
    G^{\text{hom}}_{AA',BB'}&= e^{-k(z+z')} \Big(a_1 \Pi^\text{e}_{AA',BB'}+a_2 \Pi^\text{o}_{AA',BB'} +a_3 \epsilon^{AA'}\epsilon^{BB'} +a_4 \epsilon^{AB}\epsilon^{A'B'}\\
    &+ a_5 \tfrac{1}{k}(k^{AA'}\epsilon^{BB'}+k^{BB'}\epsilon^{AA'})+ a_{6} \tfrac{1}{k}(k^{AA'}\epsilon^{BB'}-k^{BB'}\epsilon^{AA'})\Big) \,.\notag
\end{align}
This is the most general ansatz in terms of $\epsilon^{AB}$ and $k^{AB}$. One can use a slightly more general basis where each two-dimensional subspace is spanned by $k_A$, $\brk_{A}$. This way, the most general ansatz reads (the actual solutions we find can be written in the form above)
\begin{align}\label{mostgeneralhomo}
    G^{\text{hom}}_{AA',BB'}&= e^{-k(z+z')} \Big(a_1 k_A k_{A'}k_B k_{B'}+\text{15 more}\Big) 
\end{align}
and we will use this ansatz for all gauges. Some of these terms are pure gauge, but the number depends on the additional gauge-fixing conditions that $\Phi_{A,A'}$ has to obey. It is easy to manipulate such pure gauge terms, since 
\begin{align}
    \nabla^{-k,z}_{AA'} e^{-kz} &=  - e^{-kz} \brk_{A} k_{A'} \,, &
    \nabla_{BB'}^{k,z'} e^{-kz'} &=   e^{-kz'} k_{B} \brk_{B'}
\end{align}
and the result modulo overall $e^{-k(z+z')}$ is
\begin{align}
    -\brk_{A} k_{A'} T^\text{L}_{BB'}+T^\text{R}_{AA'} k_{B} \brk_{B'}-\brk_{A} k_{A'}k_{B} \brk_{B'} T^\text{c} \,,
\end{align}
where $T$s parameterize the tensorial structure of $\xi$s. We will use both $\langle ...\rangle$ and $G$ notation for two-point functions. Lastly, in the main text we consider the Feynman and axial gauges, but some comments on the Lorenz and $R_\xi$ gauges can be found in Appendix \ref{app:Lorenz}, while the complete gauge is discussed in Appendix \ref{app:ham}.

%%%%%%%%%%%%%%%%%%%%%%%%%%%%%%%%%%%%%%%%%%%%%%%%%%%%%%%%%%%%%
\subsection{(Chiral) Yang--Mills theory}
\label{sec:cYMBC}
%%%%%%%%%%%%%%%%%%%%%%%%%%%%%%%%%%%%%%%%%%%%%%%%%%%%%%%%%%%%%
The gauge fixing of the Yang--Mills theory is done in the standard way, i.e. we add $\tfrac1{2\xi} G^2$ to the action, where $G$ is a gauge-fixing function. We will discuss several gauges: Feynman gauge, i.e. $G=\pl_\mu \Phi^\mu$, $\xi=1$, (for the simplicity of its propagator); axial gauge, i.e. $G=n_\mu \Phi^\mu$, $\xi=0$,  (for its AdS/CFT helicity-friendly basis); Lorenz/Landau gauge, i.e. $G=\pl_\mu \Phi^\mu$, $\xi=0$, (for fun); lastly, the $R_\xi$-gauge for completeness. Here $n_\mu$ is taken along the radial $z$-direction, which we associate with $n_{AA'}=\epsilon_{AA'}$.

%%%%%%%%%%%%%%%%%%%%%%%%%%%%%%%%%%%%%%%%%%%%%%%%%%%%%%%%%%%%%
\subsubsection{Feynman gauge}
\label{sec:Fgauge}
%%%%%%%%%%%%%%%%%%%%%%%%%%%%%%%%%%%%%%%%%%%%%%%%%%%%%%%%%%%%%
The inhomogeneous part of the second-order propagator in Feynman gauge is
\begin{align}\label{inhomoFeynman}
    G^\text{inh}_{AA',BB'}(-k,z;k,z')=\langle \Phi_{A,A'}(-k,z)\Phi_{B,B'}(k,z') \rangle^\text{inh}_{\text{F}}= -\frac{\epsilon_{AB}\epsilon_{A'B'}}{2k}e^{-k|z-z'|}\,,
\end{align}
which comes from the Fourier transform of the flat-space result (Euclidean version)
\begin{align} \label{phiphiProp}
  &  \langle A_\mu A_\nu\rangle=\frac{1}{p^2}\eta_{\mu\nu} && \Longrightarrow && \langle\Phi_{A,A'} \Phi_{B,B'}\rangle=-\epsilon_{AB}\epsilon_{A'B'} \frac{1}{p^2} \,,
\end{align}
see Appendix \ref{app:fourier} for the Fourier transforms we need. 

\paragraph{Naive Dirichlet/Neumann.} The naive Dirichlet (magnetic) and the Neumann (electric) propagators have a very simple form\footnote{Note, however, that we imposed the Feynman gauge in the half-space model of $\text{AdS}_4$, whereas the previous papers imposed it directly in $\text{AdS}_4$, e.g. in Poincare coordinates. Since the most popular gauge-fixing conditions, such the Feynman one, are not conformally invariant, our Feynman gauge differs from the usual one. This greatly simplifies the propagator (cf. the recent \cite{Moga:2025gdy})! On the other hand, the Poincare coordinate version of our Feynman gauge would give $G=\nabla_{AA'} \Phi^{A,A'} -2\Phi_{A,A'}n^{AA'}$ for the gauge fixing function. Since $A_z$ is present, this gauge-fixing breaks $\text{AdS}_4$ isometries, like an axial gauge. }
\begin{align}
    \langle \Phi_{A,A'}(-k,z)\Phi_{B,B'}(k,z') \rangle_\text{D/N}= -\frac{\epsilon_{AB}\epsilon_{A'B'}}{2k}\left( e^{-k|z-z'|}\mp e^{-k(z+z')}\right)\,,
\end{align}
where the $-$/$+$ signs correspond to Dirichlet/Neumann, respectively. They are obtained by the method of images and are just the scalar propagator times $\eta_{\mu\nu}$. Note that the image matrix is taken not to act on the vector index of $A_\mu$. In the Feynman gauge the Lagrangian, $\tfrac12A_\mu \square A^\mu$, is that of $4$ decoupled scalars and we can choose to treat them in the same way.

Let us have a look at the Dirichlet two-point function as a boundary limit of the bulk-to-bulk propagator:
\begin{align}
    \langle\Phi_{A,A'}(-k,z)\Phi_{B,B'}(k,z'\rightarrow 0) \rangle_\text{D}=-{\epsilon_{AB}\epsilon_{A'B'}}z'(1-kz+...) \,,
\end{align}
where $z'\rightarrow0$ was taken first. The two-point function is given by amputating the $z z'$-factor in accordance with the BDHM-prescription \cite{Banks:1998dd}. The leading term is $\delta^3(x-y)$, which is the source term. The ``response'' is proportional to ${\epsilon_{AB}\epsilon_{A'B'}} k$. In the Feynman gauge $\Phi_z\neq0$, which is not a problem since it is not dual to anything on the boundary (however, $\Phi_z$ is a Lagrange multiplier for Gauss' law, but it stops being so in the Feynman gauge and becomes a dynamical degree of freedom). A worrisome feature is that we do not get the conformally-invariant $\Pi^\text{e}_{AA,BB}=\delta_{ij}-\tfrac{k_i k_j}{k^2}$-structure, but just $\delta_{ij}$. It gives the correct $\langle ++\rangle $ and $\langle --\rangle $ (note that $\langle +-\rangle =0$ always by Lorentz symmetry on the boundary) two-point functions for the physical polarizations, but contains a longitudinal component. This undesirable feature cannot be fixed by an additional homogeneous term. In accordance with AdS/CFT, the boundary current is dual to the canonical momenta on the boundary, i.e. to the electric field $E_i=F_{zi}=\pl_z A_i-\pl_i A_z$. Therefore, we should extract the boundary value of $\langle F_{zi}\, F_{zi} \rangle $. Alternatively, one can compute $\Psi/\bar \Psi$ and look at
\begin{align}
    \tfrac14\langle (\Psi^{AA}-\bar\Psi^{AA}) (\Psi^{BB}-\bar{\Psi}^{BB})\rangle\Big|_{z=z'=0}&= k \epsilon^{AB}\epsilon^{AB} \sim k \delta^{ij} \,,
\end{align}
where we used $\Psi=B+E$, $\bar\Psi=B-E$.\footnote{Let us decompose $\Phi^{A,A'}=\phi^{AA'}+\epsilon^{AA'}\phi$, then (anti-)self-dual parts of the field strength read:
    $\Psi^{AA}=B^{AA}+E^{AA}$, $\bar\Psi^{AA}=B^{AA}-E^{AA}$, 
    $B^{AA}= k\fud{A}{B} \phi^{AB}$, $E^{AA}= \pl_z\phi^{AA}-k^{AA}\phi$. }
In Maxwell theory $B^i$ is always transverse and $E^i$ is so thanks to Gauss' law and, hence, $k_{AA}\Psi^{AA}=k_{AA}\bar\Psi^{AA}=0$. Via the GKPW prescription \cite{Gubser:1998bc,Witten:1998qj}, the on-shell action is $\tfrac12\int_{{\partial M}} A^\mu \pl_z A_\mu$ and also does not have the transverse projector built in. Therefore, the naive Feynman gauge is not immediately conformally friendly, but the propagator is simple and fully $4d$ Lorentz-covariant.\footnote{We recall, however, that, when mapped to AdS, our gauge is not Lorentz covariant.}

With the Neumann boundary condition we have a gauge boson propagator on the boundary $\langle A_i A_j\rangle\sim k^{-1}\delta_{ij}$, which should correspond to the Feynman gauge on the boundary. One can also extract the magnetic field, which is the same as the current dual to $A_i$, $J^i=\epsilon^{ijk}F_{jk}=B^i$. 
\begin{align}
    \tfrac14\langle (\Psi^{AA}+\bar\Psi^{AA}) (\Psi^{BB}+\bar{\Psi}^{BB})\rangle\Big|_{z=z'=0}& = -k \Pi_\text{e}^{AA,BB}\sim k \delta^{ij} \,.
\end{align}

\paragraph{Less naive Dirichlet/Neumann.} A less naive Dirichlet/Neumann propagator is
\begin{align}\label{lessnaiveF}
    \langle \Phi_{A,A'}(-k,z)\Phi_{B,B'}(k,z') \rangle_\text{D/N}= -\frac{\epsilon_{AB}\epsilon_{A'B'}}{2k} e^{-k|z-z'|}\pm e^{-k(z+z')}\frac1{2k}\left(\epsilon_{AB}\epsilon_{A'B'}-\epsilon_{AA'}\epsilon_{BB'}\right)\,,
\end{align}
where $+$ is for Dirichlet. It is obtained by the method of images, where the reflection also acts on $\mu$ of $A_\mu$. The additional tensor structure corresponds to a linear combination of the projectors: $n_\mu n_\nu/n^2$ onto the direction of $n_\mu$ and $\Pi^n_{\mu\nu}$ onto the $3$-dimensional subspace orthogonal to $n_\mu$:
\begin{align}
    \Pi^n_{\mu\nu}\equiv \eta_{\mu\nu}-n_\mu n_\nu/n^2&= \epsilon_{AB}\epsilon_{A'B'} -\tfrac12 \epsilon_{AA'}\epsilon_{BB'}\equiv \Pi^n_{AA',BB'}=\tfrac12 (\epsilon_{AB}\epsilon_{A'B'}+\epsilon_{A'B}\epsilon_{AB'}) \,.
\end{align}
The propagator contains the operator of the reflection along $n_\mu$:
\begin{align}
    \eta_{\mu\nu}-2n_\mu n_\nu/n^2&=\epsilon_{AB}\epsilon_{A'B'}-\epsilon_{AA'}\epsilon_{BB'} \,.
\end{align}
The propagator imposes the Dirichlet/Neumann condition for $A_i$ and the opposite for $A_z$. The propagator ensures that Gauss' law is satisfied. It also implies that 
\begin{align}
   D&:&&  \tfrac14\langle (\Psi^{AA}-\bar\Psi^{AA}) (\Psi^{BB}-\bar{\Psi}^{BB})\rangle\Big|_{z=z'=0}= +k \Pi_\text{e}^{AA,BB} \,, \\
    N&:&&  \tfrac14\langle (\Psi^{AA}+\bar\Psi^{AA}) (\Psi^{BB}+\bar{\Psi}^{BB})\rangle\Big|_{z=z'=0}= -k \Pi_\text{e}^{AA,BB} \,,
\end{align}
which is a gauge-invariant way to extract the correlators. For the gauge field itself one finds
\begin{align}
    \pl_z \pl_{z'} \langle \Phi_{A,A'}(-k,z)\Phi_{B,B'}(k,z') \rangle_\text{D}\Big|_{z=z'=0}&= k \Pi_n^{AA',BB'} \,,\\
     \langle \Phi_{A,A'}(-k,z)\Phi_{B,B'}(k,z') \rangle_\text{N}\Big|_{z=z'=0}&= -\frac{1}{k} \Pi_n^{AA',BB'} \,.
\end{align}
The latter is the (boundary) Feynman gauge propagator, i.e. $k^{-1} \delta_{ij}$. The former is $\langle \pl_z A_i \pl_{z'} A_j\rangle =k \delta_{ij}$ and one should not expect to find a conformally-invariant structure here since the boundary current is dual to $F_{zi}$, which receives a correction due to nonvanishing $A_z$. For the near-boundary behavior we have
\begin{align}
    \langle \Phi_{A,A'}(-k,z)\Phi_{B,B'}(k,z') \rangle_\text{D}\Big|_{z'\rightarrow0}&= -\frac{\epsilon_{AA'}\epsilon_{BB'}}{2k}- z' \Pi_n^{AA',BB'}e^{-kz}+... \,,\\
     \langle \Phi_{A,A'}(-k,z)\Phi_{B,B'}(k,z') \rangle_\text{N}\Big|_{z'\rightarrow0}&= -\frac{1}{k}\Pi_n^{AA',BB'}e^{-kz}-z'\frac{\epsilon_{AA'}\epsilon_{BB'}}{2}e^{-kz}+... \,.
\end{align}

\paragraph{$\boldsymbol{\Psi-\Psi}$ boundary conditions.} Let us use the most general ansatz with $16$ terms \eqref{mostgeneralhomo} for the homogeneous part of the propagator. They satisfy $\square \bullet=0$ due to $e^{-k(z+z')}$ and this does not constrain the tensor structure. Out of these $16$ terms we can single out $9$ of type\footnote{To be precise $\nabla$'s should have subscripts $\pm k$ and $z/z'$!}
\begin{align}
    T^{AA'}_1 \nabla^{BB'} e^{-k(z+z')}+ T_2^{BB'}\nabla^{AA'}e^{-k(z+z')}+c\nabla^{AA'}\nabla^{BB'} e^{-k(z+z')} \,,
\end{align}
where $T_{1,2}^{AA'}$ is the most general spin-tensor, i.e. it has $4$ parameters. Out of the $9$ structures, the last one appears also in the first two. Therefore, we have only $7$ pure-gauge structures, which do not affect the $\Psi-\Psi$ boundary conditions. Note, however, that the pure gauge terms cannot reproduce neither the $n_\mu n_\nu$ nor $\Pi^n_{\mu\nu}$ tensor structures. In addition, the pure gauge terms do not affect the two-point functions when $\Phi$ is contracted with the physical polarizations. 

As we are going to impose the $\Psi-\Psi$ boundary conditions, it is important to know what the inhomogeneous part of the propagator gives for this:
\besubeqs\label{gaugeinvinhF}
\begin{align}
    \langle \Psi_{AA}(-k,z) \Psi_{BB}(k,z')\rangle^{\text{inh}}&= - \Pi^n_{AA,BB} \delta(z-z')\,,\\
    \langle \Psi_{AA}(-k,z) \bar\Psi_{BB}(k,z')\rangle^{\text{inh}}&=+\Pi^n_{AA,BB} \delta(z-z')+k\left(-\Pi^\text{e}_{AA,BB}+\Pi^\text{o}_{AA,BB}\right)e^{-k|z-z'|} \,,\\
    \langle \bar\Psi_{AA}(-k,z) \Psi_{BB}(k,z')\rangle^{\text{inh}}&=+\Pi^n_{AA,BB} \delta(z-z')+k\left(-\Pi^\text{e}_{AA,BB}-\Pi^\text{o}_{AA,BB}\right)e^{-k|z-z'|} \,,\\
    \langle \bar\Psi_{AA}(-k,z) \bar\Psi_{BB}(k,z')\rangle^{\text{inh}}&=- \Pi^n_{AA,BB} \delta(z-z') \,,
\end{align}
\esubeqs
where one can use the $\Pi^n_{AA,BB}=\epsilon_{AB}\epsilon_{AB}$ form of the projector, which is just $\delta_{ij}$.

The rest of the $16-7=9$ independent coefficients are completely fixed by the mixed boundary conditions for $\gamma\neq0$. The final result, dropping the pure gauge terms, is\footnote{For the first time the mixed boundary condition propagator was discussed in \cite{Chang:2012kt} in the non-covariant complete gauge $A_{3,0}=0$. }
\begin{align}
    G^{\text{homo},\gamma}_{AA',BB'}&= \frac{e^{-k(z+z')}}{2k} \Pi^\gamma_{AA',BB'}\,,
\end{align}
where the projector $\Pi^\gamma$ has the structure of the two-point function in a parity-violating theory (to get one, one needs to multiply it by $k$). It admits two equivalent forms\footnote{Note that on the CFT side all allowed 2-point structures have the form of $kk\brk\brk$ or $\brk\brk kk$. Having either $kkkk$ or $\brk\brk\brk\brk$ would lead to $\langle\pm\mp\rangle\neq0$, i.e. to a violation of conformal symmetry. }
\begin{align} \label{Pigamma}
   \Pi^\gamma_{AA',BB'}&= \left\{\begin{aligned}
       \cos (2\gamma)\Pi^\text{e}_{AA',BB'}+i\sin (2\gamma) \Pi^\text{o}_{AA',BB'} \,, \\ \frac{1}{4k^2}\left(e^{2i\gamma}k_{A}k_{A'} \brk_B \brk_{B'} +e^{-2i\gamma} \brk_{A}\brk_{A'} k_Bk_{B'}\right) \,.
       \end{aligned} \right.
\end{align}
For the record we also give 
\besubeqs\label{gaugeinvhomoF}
\begin{align}
    \langle \Psi_{AA}(-k,z) \Psi_{BB}(k,z')\rangle^{\text{hom}}&= e^{-k(z+z')} \frac{e^{-2i\gamma}\brk^{A}\brk^A k^B k^B}{2k} \,,\label{SDYMlimit}\\
    \langle \Psi_{AA}(-k,z) \bar\Psi_{BB}(k,z')\rangle^{\text{hom}}&=0 \,,\\
    \langle \bar\Psi_{AA}(-k,z) \Psi_{BB}(k,z')\rangle^{\text{hom}}&=0 \,,\\
    \langle \bar\Psi_{AA}(-k,z) \bar\Psi_{BB}(k,z')\rangle^{\text{hom}}&=e^{-k(z+z')} \frac{e^{+2i\gamma}k^{A}k^A \brk^B \brk^B}{2k} \,,
\end{align}
\esubeqs
where it can be useful to note that
\begin{align}
    \frac{\brk_{A}\brk_A k_B k_B}{2k^2}&= (\Pi^\text{e}_{AA,BB}-\Pi^\text{o}_{AA,BB}) \,,&  \frac{k_{A}k_A \brk_B \brk_B}{2k^2}&= (\Pi^\text{e}_{AA,BB}+\Pi^\text{o}_{AA,BB}) \,.
\end{align}
That $\langle \Psi \Psi \rangle \neq0$ tells us that the $\langle \Psi \Phi\rangle$ two-point function does not satisfy the SDYM equation on the $\Phi$-leg unless $\gamma=-i \infty$.

It is worth noting that Gauss' law is satisfied for all values of $\gamma$. For $\gamma=0$, i.e. Dirichlet/Magnetic, there is some discontinuity since the linear systems for $\gamma=0$ and $\gamma\neq0$ have different rank (some of the equations are proportional to $\sin(\gamma)$). The linear structure that is released for $\gamma=0$ is $k^A\brk^{A'}\brk^{B}k^{B'}$. For this structure the magnetic field vanishes because it corresponds to $A_i\sim k_i$ (however, $A_\mu$ is not pure gauge). The structure violates Gauss' law and we eliminate it. With this in mind the solution at $\gamma=0$ coincides with the limit $\gamma\rightarrow0$ of the general solution. The homogeneous part of the $\gamma=0$ propagator reads (the inhomogeneous is always the same)
\begin{align}\label{homogamma}
    G^{\text{homo},\gamma=0}_{AA',BB'}&= \frac{e^{-k(z+z')}}{2k} \Pi^\text{e}_{AA',BB'} \,.
\end{align}
This differs from the less naive Dirichlet propagator by $k_ik_j/k^2$ and by $n_\mu n_\nu$ terms
\begin{align}\label{diffdgamma}
G^{\text{homo},D}_{AA',BB'}-G^{\text{homo},\gamma=0}_{AA',BB'}&=-\frac{1}{4k}\left(\frac{k_{AA'}k_{BB'}}{k^2}+\epsilon_{AA'}\epsilon_{BB'}\right)e^{-k(z+z')} \,,
\end{align}
which are pure gauge, of course, since the gauge-invariant part has already been fixed:
\begin{align}
    \eqref{diffdgamma}&=\nabla^{AA'}\xi_\text{R}^{BB'}+\nabla^{BB'}\xi_L^{AA'} \,, && \begin{aligned}
        \xi_\text{R}^{BB'}&=+\frac{1}{8k^3}e^{-k(z+z')}\brk^Bk^{B'} \,, \\ \xi_\text{L}^{AA'}&=-\frac{1}{8k^3}e^{-k(z+z')}k^A\brk^{A'} \,.
    \end{aligned}
\end{align}
An alternative form
\begin{align}
    \eqref{diffdgamma}&=\nabla^{AA'}\xi_\text{R}^{BB'}+\nabla^{BB'}\xi_\text{L}^{AA'} +\nabla^{AA'}\nabla^{BB'}\xi_\text{c} \,,&& \begin{aligned}
        \xi_\text{R}^{BB'}&=-\frac{1}{4k^2}e^{-k(z+z')}\epsilon^{BB'} \,, \\ \xi_\text{L}^{AA'}&=-\frac{1}{4k^2}e^{-k(z+z')}\epsilon^{AA'}\,, \\
        \xi_\text{c}&=+\frac{1}{4k^3}e^{-k(z+z')}\,,
    \end{aligned}
\end{align}
looks more complicated, but will prove to be useful when discussing gauge independence of physical observables. 
Since the additional terms are pure gauge the gauge-invariant two-point functions stay the same. They are given by adding \eqref{gaugeinvinhF} and \eqref{gaugeinvhomoF}. The homogeneous term for the Neumann condition is the negative of that for the Dirichlet one:
\begin{align}\label{diffdgammaA}
G^{\text{homo},\text{N}}_{AA',BB'}-G^{\text{homo},\gamma=\pi/2}_{AA',BB'}&=-\eqref{diffdgamma} \,.
\end{align}
Therefore, the pure gauge term is the same up to a sign. The boundary limit for $\gamma=0$ is
\begin{align}
    \langle \Phi_{A,A'}(-k,z)\Phi_{B,B'}(k,z') \rangle_\text{D}\Big|_{z=z'=0}&= +\tfrac{1}{2k}(\Pi^\text{e}_{AA',BB'}- \epsilon_{AB}\epsilon_{A'B'}) \,.
\end{align}
This corresponds to $A_i$ of the form $\tfrac{1}{2k}\frac{k_ik_j}{k^2}$, i.e. with vanishing magnetic field. Therefore, the leading asymptotics is effectively absent, as it should. The boundary-to-bulk propagator is
\begin{align}
    \pl_{z'} \langle \Phi_{A,A'}(-k,z)\Phi_{B,B'}(k,z') \rangle_\text{D}\Big|_{z'=0}&= -\tfrac{e^{-kz}}{2}(\Pi^\text{e}_{AA',BB'}+ \epsilon_{AB}\epsilon_{A'B'}) \,.
\end{align}
The two-point function, which comes from the subleading term, is
\begin{align}
    \pl_z \pl_{z'} \langle \Phi_{A,A'}(-k,z)\Phi_{B,B'}(k,z') \rangle_\text{D}\Big|_{z=z'=0}&= \tfrac{k}{2}(\Pi^\text{e}_{AA',BB'}+ \epsilon_{AB}\epsilon_{A'B'}) \,.
\end{align}
For generic $\gamma$ we find for the two-point function of the gauge field and for the boundary-to-bulk propagator
\begin{align}\notag
        \langle \Phi_{A,A'}(-k,z)\Phi_{B,B'}(k,z') \rangle\Big|_{z=z'=0}&= +\tfrac{1}{2k}(\Pi^\gamma_{AA',BB'}- \epsilon_{AB}\epsilon_{A'B'}) \,, \\ \notag
    \langle \Phi_{A,A'}(-k,z)\Phi_{B,B'}(k,z') \rangle\Big|_{z'=0}&=e^{-kz} \tfrac{1}{2k}(\Pi^\gamma_{AA',BB'}- \epsilon_{AB}\epsilon_{A'B'}) \,.
\end{align}
In writing the expressions above we have dropped all pure gauge terms that can be added in the bulk. 

As discussed in Appendix \ref{app:boundaryterms}, the mixed/Robin boundary condition on the initial variables is equivalent to the Dirichlet boundary condition on certain new canonical variables. It is interesting to compute the two-point function for arbitrary $\gamma$ in a gauge-invariant way in terms of the new variables. The canonical variable we fix on the boundary is $Q$
\begin{align}
    Q^{AA}&= \tfrac{1}{\sqrt{2}}\left[ +\Psi^{AA} e^{i\gamma}+ \bar\Psi^{AA}e^{-i\gamma} \right] \,, &
    P^{AA}&= \tfrac{1}{\sqrt{2}}\left[ -\Psi^{AA} e^{i\gamma}+ \bar\Psi^{AA}e^{-i\gamma} \right] \,,
\end{align}
where $P$ is the canonically conjugate momentum. The matrix above belongs to $Sp(2)$.  For free theory the bulk Lagrangian is $\tfrac12(\pl_z Q)^2+\text{spatial derivatives}$, whose on-shell value upon integration by parts is $\tfrac12 P  Q|_{z=0}$. Therefore, the two-point function is 
\begin{align}
    \langle  P_{AA}(-k,0) P_{BB}(k,0)\rangle&= k \Pi^\text{e}_{AA,BB} \,,
\end{align}
which is obvious since a canonical transformation does not change the symplectic potential $p\, dq$. 

\paragraph{$\boldsymbol{\Psi-\Phi}$ boundary conditions.}
We can try to impose $\Psi-\Phi$ boundary conditions with respect to both legs, which is important since the ansatz is not symmetric with respect to $z-z'$. We find exactly the same answer as above for the gauge-invariant part $G^{\text{homo},\gamma}_{AA',BB'}$, of course. In addition, all pure gauge terms are also fixed (not by hand as above, but by the boundary conditions) except for the doubly pure-gauge term 
\begin{align}
    a \tfrac{1}{2k^2} \nabla^{AA'}_{-k,z}\nabla^{BB'}_{k,z'} e^{-k(z+z')} \,.
\end{align}
Nicely, the pure gauge terms, which we found to account for the difference \eqref{diffdgammaA} between the Neumann propagator obtained via images and the $\gamma=\pi/2$ propagator with the pure gauge terms dropped, are now fixed. These terms do not depend on $\gamma$ and are the same as for the Neumann gauge. All other pure gauge terms are now fixed to zero.  

We again note that the system degenerates slightly for $\gamma=0$ and, in addition to the doubly pure-gauge term, the same two pure gauge terms 
\begin{align}
    &\nabla^{AA'}\brk^Bk^{B'}e^{-k(z+z')} \,, &&\nabla^{BB'}k^A\brk^{A'}e^{-k(z+z')}\,,
\end{align}
as appeared in the discussion of the Dirichlet/Neumann conditions above, cannot be fixed. For $\gamma\neq0$ they do not depend on $\gamma$, actually, and give the right pure gauge term for the Neumann boundary condition. For $\gamma=0$ they are released and can be set to minus that of the Neumann's, which is the right structure for the Dirichlet.

To summarize, for the Dirichlet/Neumann boundary conditions our best candidate is \eqref{lessnaiveF}, while for the mixed boundary conditions one should use the homogeneous term \eqref{homogamma} completed by the pure gauge terms \eqref{diffdgamma} and \eqref{diffdgammaA}:
\begin{align}
    \text{Dirichlet}&: && G_{AA',BB'}^{\text{D}}= \eqref{lessnaiveF}_{+}=G^\text{inh}_{AA',BB'}+G^{\text{homo},\gamma=0}_{AA',BB'}+\eqref{diffdgamma} \,,\\
    \text{Neumann}&: && G_{AA',BB'}^{\text{N}}= \eqref{lessnaiveF}_{-}=G^\text{inh}_{AA',BB'}+G^{\text{homo},\gamma=\pi/2}_{AA',BB'}-\eqref{diffdgamma} \,,\\
    \text{generic}, \gamma\neq0&: && G_{AA',BB'}^{\gamma}=G^\text{inh}_{AA',BB'}+G^{\text{homo},\gamma}_{AA',BB'}-\eqref{diffdgamma} \,, \label{genericgamma}
\end{align}
where we note that \eqref{diffdgamma} can be represented in a pure gauge form in more than one way. 

There is a strong motivation to consider \eqref{genericgamma} for mixed boundary conditions as it is uniquely fixed by the $\Psi-\Phi$ boundary conditions, which also fixes pure gauge terms. As is easy to see, \eqref{diffdgamma} affects boundary conditions for the unphysical components, $\Phi_z$ and $\Phi_k=k_i\Phi^i$. Indeed, for the Dirichlet conditions, $\Phi_k$ is fixed and $\pl_z\Phi_z$ is also fixed (as a consequence of the gauge fixing). The latter implies that the Dirichlet propagator obeys the Dirichlet condition for $\Phi_i$ and Neumann one for $\Phi_z$. For the Neumann propagator it is the opposite. Via the BRST-transformations these conditions are directly related to ghosts and, hence, to whether or not the gauge symmetries survive in the boundary theory. Both options are available.  Therefore, one can keep them in mind by defining 
\begin{align}
    \text{generic}, \gamma\neq0&: && G_{AA',BB'}^{\gamma}=G^\text{inh}_{AA',BB'}+G^{\text{homo},\gamma}_{AA',BB'}+\sigma \eqref{diffdgamma} \,,
\end{align}
where $\sigma=\pm1$ ($\sigma=1$ can be dubbed Dirichlet-like and $\sigma=-1$ can be dubbed Neumann-like). We will find that for $\sigma=1$ the results in different gauges agree, i.e. they are (bulk)gauge independent.

%%%%%%%%%%%%%%%%%%%%%%%%%%%%%%%%%%%%%%%%%%%%%%%%%%%%%%%%%%%%%
\subsubsection{Axial gauge}
\label{sec:}
%%%%%%%%%%%%%%%%%%%%%%%%%%%%%%%%%%%%%%%%%%%%%%%%%%%%%%%%%%%%%
The inhomogeneous part of the AdS bulk-to-bulk propagator in the axial gauge is obtained by Fourier transforming the well-known flat-space result \cite{Leibbrandt:1987qv}
\begin{align} \label{flatphiphiaxial}
        G_{\mu\nu}&=\frac{1}{p^2}\left(\eta_{\mu\nu}- \frac{p_\mu n_\nu+n_\mu p_\nu}{(n\cdot p)}+ p_\mu p_\nu \frac{1}{(n \cdot p)^2} (\xi p^2+n^2)\right) \,,
\end{align}
where $n$ is taken to be the radial direction $n_{AA'}=\epsilon_{AA'}$ and we set $\xi=0$ to have the strict axial gauge. The Fourier transform gives
\begin{align}
    \begin{aligned}
    &G^\text{inh, A}_{AA';BB'}(-k,z;k,z')=\langle \Phi_{AA'}(-k,z)\Phi_{BB'}(k,z') \rangle^\text{inh}_\text{A} =\\
    &\qquad=\frac{-1}{2k}\big(\epsilon_{AB}\epsilon_{A'B'}-\tfrac{1}{2}\epsilon_{AA'}\epsilon_{BB'}+\frac{k_{AA'}k_{BB'}}{2k^2}\big)e^{-k|z-z'|}-\frac{1}{4k^2}|z-z'|k_{AA'}k_{BB'} \,.
    \end{aligned}
\end{align}
The first bracket is the same structure as the parity even two-point function
\begin{align}
    \epsilon_{AB}\epsilon_{A'B'}-\tfrac{1}{2}\epsilon_{AA'}\epsilon_{BB'}+\frac{k_{AA'}k_{BB'}}{2k^2}&= \Pi_{AA',BB'}^\text{e}\,,
\end{align}
which is just not written in the manifestly symmetric form. One might be surprised by the last linear term. It comes from the regularized $\mathrm{PV}\tfrac{1}{\omega^2}$, $\omega= (p\cdot n)$. Such additional poles are a general feature of noncovariant gauges and require some care \cite{Leibbrandt:1987qv}. The propagator satisfies 
\begin{align}
    \square_z G^\text{inh, A}_{AA';BB'}   - \tfrac12 k_{AA'} k^{CC'}G^\text{inh, A}_{CC';BB'}&=\delta(z-z')\Pi^n_{AA',BB'}\,.
\end{align}
Here, $\square_z=\pl^2_z-k^2$. The propagator is in fact symmetric in $AA'$ and $BB'$ separately since the transversality with $n_{AA'}=\epsilon_{AA'}$ just implies the symmetry. Therefore, we can as well write $\Phi^{AA}$ to stress that $A_z=0$.

One can also represent the difference between the propagators in the axial and Feynman gauge as a pure gauge term:
\begin{align} \label{axialnablaxi}
\begin{aligned}
       \langle \Phi_{AA'}(-k,z)\Phi_{BB'}(k,z') \rangle^\text{inh}_\text{A} &=\langle \Phi_{A,A'}(-k,z)\Phi_{B,B'}(k,z')\rangle^\text{inh}_\text{F}+\\&+
    \nabla_{k,z'}^{BB'} \xi_\text{L}^{AA'}+\nabla_{-k,z}^{AA'} \xi_\text{R}^{BB'}+\nabla_{-k,z}^{AA'} \nabla_{k,z'}^{BB'}\xi_{\text{c}} \,, 
\end{aligned}
\end{align}
where
\besubeqs\label{puregaugeaxial}
\begin{align}
    \xi_\text{L}^{AA'}&=-\epsilon^{AA'} \xi_{\text{L,R}} \,, && \xi_\text{R}^{BB'}=+\epsilon^{BB'} \xi_{\text{L,R}} \,,\\
    \xi_{\text{L,R}}&= \frac{1}{4k^2}\sign(z-z') (-1+e^{-k|z-z'|}) \,,\\
     \xi_\text{c}&= \frac{1}{4k^3}(e^{-k|z-z'|}+k|z-z'|) \,.
\end{align}
\esubeqs
Note that none of the terms on the r.h.s. is orthogonal to $n_\mu$ by itself, but all together they are. This is stressed by writing $\Phi_{A,A'}$ instead of $\Phi_{AA}$.

\paragraph{Homogeneous terms.} 
For the homogeneous terms we take the ansatz that contains the most general structure for $\exp{[-k(z+z')]}$. It also has to contain the linear term $(z+z')$ in order to ensure the decay of the propagator in the deep bulk of AdS, which is now spoiled by the $|z-z'|$-term in the inhomogeneous part. The ansatz reads
\begin{align}\notag
    G^{\text{hom}}_{AA',BB'}&= e^{-k(z+z')} \Big(a_1 k_A k_{A'}k_B k_{B'}+\text{15 more}\Big) + (z+z') \times \Big(b_1 k_A k_{A'}k_B k_{B'}+\text{15 more}\Big) +\\&\qquad\qquad +\Big(c_1 k_A k_{A'}k_B k_{B'}+\text{15 more}\Big) \,.\notag
\end{align}
Firstly, we constrain the ansatz by imposing the axial gauge and the free equations of motion
\begin{align}
    A_z&=0 \,, & \square A_i - \pl_i \pl^k A_k&=0 \,.
\end{align}
Very few survive as a result:
\begin{align}\label{axhom}
    (c+b (z+z') )k^{AA}k^{BB} +\sum_{\alpha,\beta=\pm} e^{-k(z+z')} \epsilon_{ \alpha}^{AA}\epsilon_{ \beta}^{BB} \,,
\end{align}
where $\epsilon^{AA}_\alpha= \{k^A k^A, \brk^A \brk^A\}$ are the polarization vectors of the free physical states. Indeed, the product of the boundary-to-bulk propagators is always an admissible homogeneous solution, which is not pure gauge. The other terms are pure gauge. Secondly, we impose the boundary conditions on the complete propagator. 

\paragraph{$\boldsymbol{\Psi-\Psi}$ boundary conditions plus regularity.} The linear piece of \eqref{axhom} disappears from the boundary conditions. Having the linear term free is crucial for being able to impose the regularity in the deep bulk. The final answer is 
\begin{align}
    G^{\text{hom, A}}_{AA',BB'}&= G^{\text{hom},\gamma}_{AA',BB'} +\frac{1}{4k^2}(z+z')k_{AA'}k_{BB'} \,.
\end{align}
All coefficients of the exponential terms turn out to be fixed by the boundary conditions. The linear term is designed to cancel the leading behavior of $|z-z'|$-term in the inhomogeneous part. Note, however, that at $z\rightarrow\infty$ the leftover, $|z-z'|-(z+z')=-2z'$ is finite. In other words, the propagator does not decay to zero at large separation.

The first term, $G^{\text{hom},\gamma}_{AA',BB'}$, is the same as in the Feynman gauge and will be the same in all gauges. In the axial gauge it can be recognized as the products of the boundary-to-bulk propagators: 
\begin{align}\notag
    e^{-2i\gamma} \brk^{A}\brk^{A'} k^Bk^{B'}+e^{+2i\gamma} k^{A}k^{A'} \brk^B \brk^{B'} \sim  e^{-2i\gamma} \Phi^{AA'}_-(-k,z) \Phi_-^{BB'}(k,z')+e^{+2i\gamma} \Phi^{AA'}_+(-k,z) \Phi_+^{BB'}(k,z') \,.
\end{align}
We also note that there is no discontinuity for $\gamma=0$. Finally, the complete propagator reads\footnote{For $\gamma=0$ this agrees with \cite{Raju:2011mp}, where the propagator was represented as
\begin{align}
    \langle A_i(z)A_j(z') \rangle =\int_0^\infty dp p \sqrt{zz'}\frac{J_{\frac{1}{2}}(p z)J_{\frac{1}{2}}(p z')}{k^2+p^2}\mathcal{T}_{ij}\,,
\end{align}
where $\mathcal{T}_{ij}=\eta_{ij}+\frac{k_ik_j}{p^2}$, i.e. it becomes the projector onto the direction transverse to $k_i$ when $p^2=-k^2$.}
\begin{align}
    \begin{aligned}
    \langle \Phi_{AA'}(-k,z)\Phi_{BB'}(k,z') \rangle_{\text{A}}&=-\frac{1}{2k}\Pi^\text{e}_{AA',BB'} e^{-k|z-z'|}+G^{\text{hom},\gamma}_{AA',BB'}+\\& -\frac{1}{4k^2}|z-z'|k_{AA'}k_{BB'}  +\frac{1}{4k^2}(z+z')k_{AA'}k_{BB'}\,.
    \end{aligned}
\end{align}
Since the difference between any two gauges is pure gauge, it is convenient to represent the last piece in a pure gauge form as
\begin{align}
    \frac{k_{AA'}k_{BB'}}{4k^2}(z+z')&= -\frac{1}{4k^2}\Big[\nabla^{AA'}\nabla^{BB'} [(z+z')]+\nabla^{AA'}\epsilon^{BB'}+\nabla^{BB'}\epsilon^{AA'}\Big] \,,
\end{align}
which is somewhat reminiscent of the pure gauge terms for the inhomogeneous part. For practical applications it is useful to set the inhomogeneous part of the Feynman propagator as the reference point
\begin{align}\label{axialYMpropagator}
\begin{aligned}
    &\langle \Phi_{AA'}(-k,z)\Phi_{BB'}(k,z') \rangle_{\text{A}}=\langle \Phi_{A,A'}(-k,z)\Phi_{B,B'}(k,z') \rangle^\text{inh}_{\text{F}}+G^{\text{hom},\gamma}_{AA',BB'}+\\
    &\quad+\nabla_{BB'}^{k,z'} \xi^\text{L}_{AA'}+\nabla_{AA'}^{-k,z} \xi^\text{R}_{BB'}+\nabla_{AA'}^{-k,z}\nabla_{BB'}^{k,z'}\xi_{\text{c}}\,,
\end{aligned}
\end{align}
where
\besubeqs\label{puregaugeaxialFull}
\begin{align}
    \xi_\text{L}^{AA'}&=\epsilon^{AA'} \frac{\sign\left(z-z'\right) \left(1-e^{-k \left| z-z'\right| }\right)-1}{4 k^2} \,,\\
    \xi_\text{R}^{AA'}&=\epsilon^{AA'} \frac{\sign\left(z-z'\right) \left(e^{-k \left| z-z'\right| }-1\right)-1}{4 k^2}\,,\\
     \xi_\text{c}&= \frac{k \left| z-z'\right| +e^{-k \left| z-z'\right| }-k \left(z+z'\right)}{4 k^3}\,.
\end{align}
\esubeqs
In the diagrams the pure gauge terms are immediately reduced to the boundary terms. Therefore, we will need only 
\begin{align}\label{forthecomment}
    \xi_\text{L}^{AA'}\big|_{z'=0}&=-\epsilon^{AA'} \frac{e^{-k z}}{4 k^2} \,, & \xi_\text{R}^{BB'}\big|_{z=0}&=-\epsilon^{BB'}\frac{e^{-k z'}}{4 k^2}\,, &
     \xi_\text{c}\big|_{z=z'=0}&= \frac{1}{4 k^3} \,.
\end{align}
Note that, for instance, $\xi_\text{R}$ enters as $-\pl_z \epsilon^{AA'}$ into the integral. Since $z=0$ is the lower limit, we have $\epsilon^{AA'}\xi_\text{R}^{BB'}|_{z=0}$, etc. In fact, all of the boundary terms above contribute with $\epsilon^{AA'}\epsilon^{BB'}$ tensor structure, which is $n_\mu n_\nu$. This means that they can be dropped, but this is not always useful. Note that none of the terms on the r.h.s. of \eqref{axialYMpropagator} is orthogonal to $n_\mu$ by itself. Nevertheless, such representation of the propagator is useful in practice because it isolates energy poles of different type. We will see on the example of an exchange diagram that the Feynman piece has the leading pole and is easy to match with the flat-space amplitude. If instead we choose to drop the pure gauge terms at the price of symmetrizing the Feynman propagator in $AA'$ and $BB'$ (this is equivalent to making it orthogonal to $n_{AA'}$), it will be hard to isolate the leading pole and make contact with the flat limit (the symmerization contains 4 terms, one of which produces the poles and others conspire not to have it when summed, which is difficult to see).

\paragraph{$\boldsymbol{\Psi-\Phi}$ boundary conditions.} Here the exponential homogeneous terms have the same solution, of course. However, some of the linear terms turn out to be also fixed. In particular, we have 
\begin{align}
G^{\text{hom}}_{AA',BB'}&=G^{\text{homo},\gamma}_{AA',BB'}-\frac{1}{4k^2}(z+z')k_{AA'}k_{BB'}\,,
\end{align}
with $-$ instead of $+$. This violates the regularity in the deep bulk. On top of that there is a discontinuity between mixed boundary conditions and the Dirichlet one. Therefore, the $\Psi-\Phi$ boundary conditions seem to be too strong.

\paragraph{Boundary-to-bulk propagator, two-point function.} Let us first take $z'=0$ to get the boundary-to-bulk propagator. For generic $\gamma\neq0$ the leading coefficient does not vanish
\begin{align}\label{btobAxial}
    \langle \Phi_{AA'}(-k,z)\Phi_{BB'}(k,0) \rangle&=\frac{1}{2k}\Big[ \Pi^\gamma_{AA',BB'} -\Pi^\text{e}_{AA',BB'}\Big]e^{-kz} \,,
\end{align}
which is the boundary-to-bulk propagator for the mixed boundary conditions. For the Dirichlet condition, $\gamma=0$, we have to look at the subleading term
\begin{align}
    \pl_{z'}\langle \Phi_{AA'}(-k,z)\Phi_{BB'}(k,0) \rangle&=- \Pi^\text{e}_{AA',BB'}e^{-kz}+\frac{1}{2} \frac{k_{AA'}k_{BB'}}{k^2} \,.
\end{align}
The last term vanishes when its boundary indices are contracted with the transverse polarization vector. From the expressions above we can get the two-point functions. For generic $\gamma\neq0$ we have
\begin{align}\label{twopointAxial}
    \langle \Phi_{AA'}(-k,0)\Phi_{BB'}(k,0) \rangle&=\frac{1}{2k}\Big[ \Pi^\gamma_{AA',BB'} -\Pi^\text{e}_{AA',BB'}\Big] \,,
\end{align}
which is a two-point function of a gauge field on the boundary. For the Dirichlet condition we expand
\begin{align}
    \pl_{z'}\langle \Phi_{AA'}(-k,z)\Phi_{BB'}(k,0) \rangle&=-\Pi^n_{AA',BB'} +\mathcal{O}(z)
\end{align}
to find the $\delta^3(x)$-term on the boundary.\footnote{The scalar analog is $\exp[-kz]=1-kz+...$, where the first term is the usual $\delta^3(x)$ on the boundary and the next term gives the two-point function.} The subleading term is the two-point function:
\begin{align}
    \pl_z\pl_{z'}\langle \Phi_{AA'}(-k,0)\Phi_{BB'}(k,0) \rangle&=k \,\Pi^\text{e}_{AA',BB'} \,,
\end{align}
which is the right conformally-invariant expression. We can also have a look at the gauge-invariant observables on the boundary
\begin{align}
    \tfrac14\langle (\Psi^{AA}-\bar\Psi^{AA}) (\Psi^{BB}-\bar{\Psi}^{BB})\rangle\Big|_{z=z'=0}&=\frac{k}2\Big[ \Pi^\gamma_{AA',BB'} +\Pi^\text{e}_{AA',BB'}\Big] \,,\\
    \tfrac14\langle (\Psi^{AA}+\bar\Psi^{AA}) (\Psi^{BB}+\bar{\Psi}^{BB})\rangle\Big|_{z=z'=0}&=\frac{k}2\Big[ \Pi^\gamma_{AA',BB'} -\Pi^\text{e}_{AA',BB'}\Big]\,.
\end{align}

%%%%%%%%%%%%%%%%%%%%%%%%%%%%%%%%%%%%%%%%%%%%%%%%%%%%%%%%%%%%%
\subsection{Self-dual Yang--Mills theory (SDYM)}
\label{sec:SDYMprops}
%%%%%%%%%%%%%%%%%%%%%%%%%%%%%%%%%%%%%%%%%%%%%%%%%%%%%%%%%%%%%
Since SDYM is not usually found in textbooks, it is instructive to discuss how to gauge fix this theory, see e.g. \cite{Krasnov:2016emc} for more details and for examples with loop diagrams. 
\paragraph{Gauge fixing.} Let us begin with the component action (we keep the full Chalmers--Siegel action for the time being)
\begin{align}
   \mathcal{L}&= \Tr \int \Psi^{AA} (\pl_{AC'} \Phi\fdu{A,}{C'}+\Phi_{A,C'}\Phi\fdu{A,}{C'}- \tfrac{\epsilon}2 \Psi_{AA}) \,.
\end{align}
The BRST transformations are defined as usual for the Yang--Mills theory (below we use the gauge-covariant derivative $D_{AA'}=\pl_{AA'}\bullet +[\Phi_{A,A'}, \bullet]$)
\begin{align}
    \delta \Phi^{A,A'}&= D^{AA'} c^{}\,, &
    \delta \Psi^{AA}&=[c,\Psi^{AA}]\,, &
    \delta c^{}&=\tfrac12[c,c]\,, &
    \delta \bar{c}^{}&=B^{}\,, &
    \delta B^{}&=0\,,
\end{align}
and $\Psi$ is treated as a matter field in the adjoint. 
Here, $c$, $\bar{c}$ is the (anti)ghost and $B$ is the Nakanishi-Lautrup field, all borrowed from the standard Yang--Mills theory. We can try the usual choice for the gauge fixing fermion
\begin{align}
    O&= \bar{c}\left(\tfrac{\xi}{2} B- G(\Phi)\right)\,, & sO&=B\left(\tfrac{\xi}{2} B- G(\Phi)\right)+\text{ghosts} \,,
\end{align}
where $G=\pl_{AA'}\Phi^{A,A'}$ for Lorenz gauge and $G=n_{AA'}\Phi^{A,A'}$ for the axial one. We uniformize them as $G=q_{AA'}\Phi^{AA'}$, where $q_{AA'}$ is either $\pl_{AA'}$ or $n_{AA'}$. Now, several options are available. One can integrate out $B$, as usual, getting $\tfrac{1}{2\xi} G^2$, where it is tempting to take the $\xi\rightarrow0$ limit. In the limit we find $q_{AA'}\Phi^{A,A'}=0$ and, hence, the kinetic term is invertible since we have $3(\Psi^{AB})+3(\Phi^{A,A'})$ off-shell degrees of freedom (one degree of freedom removed from $\Phi$ now). Alternatively, one can set $\xi=0$ from the start and notice that $B$ can be combined with $\Psi^{AA}$ into a ``traceful'' $\Psi^{A,B}$
\begin{align}
    \Psi^{AB}\pl_{AA'}\Phi\fdu{B}{A'} - \epsilon^{AB}B q_{AA'}\Phi\fdu{B}{A'} \,.
\end{align}
In the case $q_{AA'}=\pl_{AA'}$, the free SDYM action acquires the most suggestive form
\begin{align}
    & \mathcal{L}=\Psi^{A, B}\pl_{AA'}\Phi\fdu{B,}{A'} \,,&& \Psi^{A,B}=\Psi^{AB} - \epsilon^{AB}B \,.
\end{align}
This can be called the Feynman gauge in the sense that no actual gauge condition is imposed on $\Phi_{A,A'}$. This is also justified by the fact that the gauge-fixed Chalmers--Siegel action, upon integrating out $\Psi^{A,B}$, gives the expected $\tfrac12\Phi_{A,A'} \square \Phi^{A,A'}$ kinetic term. More generally, one finds
\begin{align}
    & \mathcal{L}=\Psi^{A, B}\Pi_{A,B|CC'} \Phi^{C,C'} \,, & \Pi_{A,B|CC'}&= \tfrac12(\pl_{AC'}\epsilon_{CB}+\pl_{BC'}\epsilon_{CA}+\epsilon_{AB} q_{CC'}) \,.
\end{align}
With $B$ joining $\Psi^{AA}$ the set of fields is balanced again, $4(\Psi^{A,B})+4(\Phi^{A,A'})$, and the kinetic term is invertible. Since the vertices are still made of $\Psi^{AA}$, which is symmetric, the anti-symmetric part of $\Psi^{A,B}$ will not contribute.\footnote{As a side remark, we can start with the cYM action in the first order form, $\mathcal{L} \sim F_{AB} \Psi^{AB}-\tfrac{\epsilon}{2} \Psi_{AB}\Psi^{AB}$, relate $\epsilon$ to $\xi$ to get $\mathcal{L} \sim  \Psi^{A,B}D_{AC'}\Phi\fdu{B,}{C'}-\tfrac{\epsilon}{2} \Psi_{A,B}\Psi^{A,B}+\text{ghosts}$, which also requires $G=D_{AA'}\Phi^{A,A'}$ gauge fixing term. }

%%%%%%%%%%%%%%%%%%%%%%%%%%%%%%%%%%%%%%%%%%%%%%%%%%%%%%%%%%%%%
\subsubsection{Lorenz/Feynman gauge}
\label{sec:}
%%%%%%%%%%%%%%%%%%%%%%%%%%%%%%%%%%%%%%%%%%%%%%%%%%%%%%%%%%%%%
In the Feynman gauge we find the following propagator in flat space 
\begin{align}
    -p_{AB'}\langle \phi\fdu{B}{B'}(-p)\phi_{C,C'}(p)\rangle_{\text{2nd order}} = \langle \psi_{A,B}(-p) \phi_{C,C'}(p)\rangle_{\text{1st order}}=-\frac{\epsilon_{BC}p_{AC'}}{p^2} \,,
\end{align}
which is related in the expected way to the second order propagator of the Yang--Mills theory. Note that with indices symmetrized, i.e. with $B$ integrated out and the Lorenz gauge imposed, we simply find
\begin{align}
    -p_{AB'}\langle \phi\fdu{A}{,B'}(-p)\phi_{C,C'}(p)\rangle_{\text{2nd order}} = \langle \psi_{AA} (-p)\phi_{C,C'}(p)\rangle_{\text{1st order}}=-\frac{\epsilon_{AC}p_{AC'}}{p^2} \,.
\end{align}
Note, however, that with $K_{AA;BB'}=p_{AB'}\epsilon_{BA}$, the symmetrized propagator satisfies 
\begin{align}
    G^{AA;BB'}K_{AA;CC'}&=\epsilon\fdu{C}{B}\epsilon\fdu{C'}{B'}-p\fdu{C}{B'}p\fud{B}{C'}/p^2 \,,&G^{AA;CC'}K_{BB;CC'}&=-\epsilon\fdu{B}{A}\epsilon\fdu{B}{A}\,,
\end{align}
i.e. with the $p$-transverse projector on the $\Phi$-leg. The effective equations of motion in the Feynman gauge are
\begin{align}
    \nabla\fud{A}{C'}\Phi^{B,C'}-\epsilon \Psi^{A,B}&=0 \,, && \nabla\fdu{A}{A'}\Psi^{A,B}=0 \,.
\end{align}
In the Lorenz gauge we have instead (valid for all values of $\epsilon$) 
\begin{align}
   \nabla_{AA'}\Phi^{A,A'}&=0 \,, & \nabla\fud{A}{C'}\Phi^{A,C'}-\epsilon \Psi^{AA}&=0 \,, && \nabla\fdu{C}{A'}\Psi^{AC}=\nabla^{AA'} \alpha \,,
\end{align}
for some $\alpha$ (this corresponds to the fact that only the $p$-transverse part of $p\fdu{C}{A'}\Psi^{AC}$ needs to vanish, i.e. the expression has to be proportional to $p^{AA'}$). In anti-de Sitter space, either directly or via the Fourier transform, we have (we use $\Psi^{A,B}=\nabla\fud{A}{C'}\Phi^{B,C'}$)
\begin{align} 
    \begin{aligned}
        \langle \Psi_{A,B}(-k,z)\Phi_{C,C'}(k,z')\rangle^{\text{inh}}_{\text{F}} &= (\nabla_{-k,z})_{AA'}\langle \Phi\fdu{B,}{A'}(-k,z)\Phi_{C,C'}(k,z') \rangle^{\text{inh}}_{\text{F}} =\\
        &=-\frac{\epsilon_{BC}}{2k}(k_{AC'}-\text{sign}(z-z')k\epsilon_{AC'})e^{-k|z-z'|} \,.
    \end{aligned}
\end{align}
This is just the spin-half field propagator \eqref{spinhalfpropag} times $\epsilon_{AB}$. The propagator in the Lorenz gauge is the symmetrized part of the above:
\begin{align} 
    \begin{aligned}
        &\langle \Psi_{AA}(-k,z)\Phi_{B,B'}(k,z')\rangle^{\text{inh}}_{\text{L}} = 
        -\frac{\epsilon_{AB}}{2k}(k_{AB'}-\text{sign}(z-z')k\epsilon_{AB'})e^{-k|z-z'|} \,.
    \end{aligned}
\end{align}
Here, the $\Phi-\Phi$ propagator can be taken either in the Feynman, $R_\xi$ or Lorenz gauge with the same result at the end (the $p_\mu p_\nu$-term will disappear). The $\Psi$-indices are always effectively symmetrized since the vertices contain only the symmetric part of $\Psi^{A,B}$ anyway. As a consequence of the equations of motion for $\Phi$, the propagator satisfies
\begin{align}
    (\nabla_{k,z'})\fud{B}{B'}\langle \Psi^{AA}(-k,z) \Phi^{B,B'}(k,z')\rangle&=-\delta(z-z') \epsilon^{AB}\epsilon^{AB}\,.
\end{align}

\paragraph{Homogeneous terms.} Naively, taking the derivative of the Yang--Mills propagator in the Feynman gauge, as was done for the inhomogeneous terms, we find the following homogeneous term (for the $\Psi-\Psi$ boundary conditions and we keep the pure gauge terms)
\begin{align}
    G^{AA;BB'}_{\text{hom}}&= \frac{1}{4k^2 }e^{-k(z+z')} \Big[-e^{-2i\gamma} \brk^A\brk^{A} k^B k^{B'} +c \brk^A \brk^{A} k^B \brk^{B'} + c' k^{AA} k^B \brk^{B'}\Big] \,.
\end{align}
Note that the free terms are proportional to $k^B \brk^{B'}$, which is just $\nabla^{BB'}e^{-k(z+z')}$, i.e. they are pure gauge. The $c'$-term breaks Gauss' law on the $\Psi=B+E$ leg (but the other leg is still $\Phi$, i.e. this is not a gauge-invariant statement). The SDYM equation on the right leg is satisfied only if $\gamma=-i\infty$, i.e. the self-dual (SD) boundary condition $\Psi^{AA}\equiv\nabla\fud{A}{A'}\Phi^{A,A'}=0$ at $z=0$, cf. also \eqref{SDYMlimit},
\begin{align}
   -\nabla_{k,z'}^{BB'}G\fud{AA';B}{B'}_{\text{hom}} &= \frac{1}{2k}e^{-k(z+z')} e^{-2i\gamma} \brk^A\brk^{A'} k^B k^B \,.
\end{align}
The SDYM equation on the left leg implies
\begin{align}
    -\nabla^{AA'} G_{\text{hom}}\fudu{A}{A',}{BB'}&= \frac{c'}{4k }e^{-k(z+z')} \brk^{A}k^{A'} k^B \brk^{B'} \,,
\end{align}
which is of the form $\nabla^{AA'}\alpha^{BB'}$ (in fact, it is doubly pure-gauge). Therefore, the equations of motion for $\Psi$ are satisfied in the Lorenz gauge. To summarize, the leftover pure gauge terms tolerate self-duality, but $\gamma$ needs to be set to $-i\infty$, i.e. the SDYM propagator wants the SD boundary conditions, which is not surprising. 

In the Feynman gauge our best candidate for the propagator is the one fixed by the $\Psi-\Phi$ boundary conditions. In this case we find
\begin{align}
    G^{AA;BB'}_{\text{hom}}&= -\frac{1}{4k^2 }e^{-k(z+z')} \Big[e^{-2i\gamma} \brk^A\brk^{A} k^B k^{B'} + k^{AA} k^B \brk^{B'}\Big]=\\
    &=-\frac{1}{4k^2 }e^{-k(z+z')} e^{-2i\gamma} \brk^A\brk^{A} k^B k^{B'}-\frac{k^{AA}}{4k^2} \nabla^{BB'}e^{-k(z+z')} \,,
\end{align}
where the coefficient of the second, pure gauge-term, is fixed (by the Neumann conditions, in fact). The SDYM equations of motion are satisfied for $\gamma=-i\infty$, as above. Since the $\Psi-\Phi$ boundary conditions extrapolate nicely from the Neumann point, the final result for the SDYM propagator is its inhomogeneous part plus the Neumann correction
\begin{align} 
    \begin{aligned}
        &\langle \Psi_{AA}(-k,z)\Phi_{B,B'}(k,z')\rangle_{\text{L}}=\langle \Psi_{AA}(-k,z)\Phi_{B,B'}(k,z')\rangle^{\text{inh}}_{\text{L}} -\frac{k_{AA}}{4k^2} \nabla_{BB'}^{k,z'}e^{-k(z+z')}= \\
        &=-\frac{\epsilon_{AB}}{2k}(k_{AB'}-\sign(z-z')k\epsilon_{AB'})e^{-k|z-z'|} +\sigma\frac{k_{AA}}{4k^2} \nabla_{BB'}^{k,z'}e^{-k(z+z')}\,.
    \end{aligned}
\end{align}
Here, we reinstated $\sigma=\pm1$ and the Neumann option is $\sigma=-1$. 
Since the last term is pure gauge, only its boundary value 
\begin{align}
    \sigma\frac{k^{AA}}{4k^2} \epsilon^{BB'}e^{-k(z+z')}
\end{align}
will be needed. The boundary-to-bulk propagators follow:
\begin{align}
    \langle \Psi_{AA}(-k,0)\Phi_{B,B'}(k,z')\rangle_{\text{L}}&=-\frac{k^A k^A \brk^{B} \brk^{B'}}{4k^2} e^{-kz'} \,,\\
    \langle \Psi_{AA}(-k,z)\Phi_{B,B'}(k,0)\rangle_{\text{L}}&=+\frac{\brk^A \brk^A k^{B} k^{B'}}{4k^2} e^{-kz}-\frac{k^{AA}k^{BB'}}{2k^2} e^{-kz} \,.
\end{align}
The very last term will cancel against the physical polarization vector we choose. When the legs on the boundary are contracted with the physical polarizations we get the same results as for the axial gauge, cf. Section \ref{sec:axial}. 

%%%%%%%%%%%%%%%%%%%%%%%%%%%%%%%%%%%%%%%%%%%%%%%%%%%%%%%%%%%%%
\subsubsection{Axial gauge}
\label{sec:axial}
%%%%%%%%%%%%%%%%%%%%%%%%%%%%%%%%%%%%%%%%%%%%%%%%%%%%%%%%%%%%%
In the strict axial gauge, $n_{AA'}\Phi^{A,A'}=0$, the kinetic operator $D$ needs to be projected onto the space orthogonal to $n$:
\begin{align}
    D^\perp_{AA'|BB}&=D_{AA'|BB}-n^{-2} n_{AA'} n^{CC'}D_{CC'|BB}= \epsilon_{BA}p_{BA'}+n^{-2}p_{BC'} n\fdu{B}{C'} n_{AA'} \,,
\end{align}
where $n^2=n\cdot n$ and $(n\cdot p)=n_{CC'}p^{CC'}$ and $D_{AA'|BB}= \epsilon_{BA}p_{BA'}$. 
The propagator is the inverse of $D^\perp$ above and reads
\begin{align} \label{axialProp}
    G^{BB|CC'}&=\frac{2}{p\cdot p} \left(\epsilon^{BC} p^{BC'}+\frac{1}{(p\cdot n)} p\fud{B}{M'}n^{BM'} p^{CC'}\right) \,,
\end{align}
which satisfies
\begin{align}
    G^{CC|AA'}D_{AA'|BB}^\perp&= \epsilon\fdu{B}{C}\epsilon\fdu{B}{C}\,, &
    D_{AA'|BB}^\perp G^{BB|CC'}&= \epsilon\fdu{A}{C}\epsilon\fdu{A'}{C'}-n^{-2} n_{AA'}n^{CC'}\,.
\end{align}
We also have the familiar relation
\begin{align}
    -p_{AB'}\langle \phi\fdu{A}{B'}\phi_{CC'}\rangle_{\text{2nd order}}^{\text{A}} = \langle \psi_{AA} \phi_{CC'}\rangle_{\text{1st order}}^{\text{A}}\,,
\end{align}
now for the axial gauge. The Fourier transform gives the following inhomogeneous bulk-to-bulk propagator, where it is convenient to use the Lorenz gauge as a reference point,
\begin{align}
    &\langle \Psi_{AA}(-k,z)\Phi_{BB'}(k,z')\rangle_{\text{A}}^\text{inh}=\langle \Psi_{AA}(-k,z)\Phi_{B,B'}(k,z')\rangle_{\text{F}}^\text{inh}+\nabla^{k,z'}_{BB'}\xi_{AA}^\text{L} \,.
\end{align}
Here, the pure gauge term comes from \eqref{puregaugeaxial}
\begin{align}
    \xi_\text{L}^{AA}&=k^{AA} \xi_{\text{L,R}} \,, &
    \xi_{\text{L,R}}&= \frac{1}{4k^2}\sign(z-z') (-1+e^{-k|z-z'|})\,.
\end{align}
There is another form, where the indices are explicitly symmetrized on the Feynman propagator
\begin{align}\notag
\begin{aligned}
         \langle \Psi_{AA}(-k,z)\Phi_{BB}(k,z')\rangle^{\text{inh}}_{\text{A}} &= -\frac{\epsilon_{AB}}{2\tilde{a}k}\big(k_{AB}-\text{sign}(z-z')k\epsilon_{AB}\big)e^{-k|z-z'|}+\xi_{\text{L,R}}k_{AA}k_{BB}=\\
     &=\langle \Psi_{AA}(-k,z)\Phi_{BB}(k,z')\rangle^{\text{inh}}_{\text{F}}+\xi_{\text{L,R}} k_{AA}k_{BB}\,.
\end{aligned}
\end{align}
The propagator satisfies
\begin{align}
    -\nabla^{BB'}_{k,z'}\langle \Psi^{AA}(-k,z) \Phi\fud{B}{B'}(k,z')\rangle&=-\delta(z-z') \epsilon^{AB}\epsilon^{AB}\,,\\
    -\nabla^{MA}_{-k,z}\langle \Psi\fud{A}{M}(-k,z) \Phi^{BB}(k,z')\rangle&=-\delta(z-z') \epsilon^{AB}\epsilon^{AB}\,.
\end{align}
In the second line the equations of motion for $\Psi$ are projected onto the $n$-transverse direction via the symmetrization, $-\nabla^{MA}\Psi\fud{A}{M}$. Note that the effective equations of motion read
\begin{align}
   \epsilon_{AA'}\Phi^{AA'}&=0 \,,& \nabla\fud{A}{C'}\Phi^{AC'}-\epsilon \Psi^{AA}&=0\,, && \nabla\fdu{C}{A'}\Psi^{AC}=\epsilon^{AA'} \alpha \,,
\end{align}
for some $\alpha$. The latter means that $\nabla\fdu{C}{A}\Psi^{AC}=0$.

\paragraph{Homogeneous terms.}  As before, we try to obtain the homogeneous terms from those of Yang--Mills theory, to find
\begin{align}
    G_{AA;BB}^{\text{hom}}&= -\frac{1}{4k^2 }e^{-k(z+z')} e^{-2i\gamma} \brk_A\brk_{A} k_B k_{B} +\frac{1}{4k^2} k_{AA}k_{BB}\,.
\end{align}
The linear term $(z+z')$ turned into the last constant term. The inhomogeneous part of the propagator features a similar term 
\begin{align}
    G_{AA;BB}^{\text{inh}}&\ni -\frac{1}{4k^2} k_{AA}k_{BB} \sign(z-z')\,.
\end{align}
Similarly to the Yang--Mills case it all came from, we see that there is a cancellation for $z\rightarrow\infty$ between the two terms and they double for $z'\rightarrow\infty$, leaving some constant background value: 
\begin{align}
    G_{AA;BB}&\rightarrow \theta(z'-z)\frac{1}{2k^2} k_{AA}k_{BB}\,, && z \gg z' \text{ or } z'\gg z\,.
\end{align}
The equation of motion for the $\Phi$-leg,
\begin{align}
   \nabla_{k,z'}\fud{B}{B'}G^{AA;BB'}_{\text{hom}} &= \frac{1}{2k}e^{-k(z+z')} e^{-2i\gamma} \brk^A\brk^{A} k^B k^B \,,
\end{align}
is satisfied in the self-dual limit, $\gamma\rightarrow -i \infty$. The (symmetrized) equation of motion for the $\Psi$-leg is also satisfied 
\begin{align}
   -\nabla_{-k,z}^{MA}G_{\text{hom}}\fudu{A}{M;}{BB}&= 0 \,.
\end{align}
Therefore, in the axial gauge the SDYM propagator is a smooth limit $\gamma\rightarrow -i \infty$ of the YM propagator, as expected. In this limit, the complete SDYM propagator reads
\begin{align}\label{SDYMAxialLDifference}
        &\langle \Psi_{AA}(-k,z)\Phi_{BB'}(k,z')\rangle_{\text{A}}^\text{inh}=\langle \Psi_{AA}(-k,z)\Phi_{B,B'}(k,z')\rangle_{\text{F}}^\text{inh}+\nabla^{k,z'}_{BB'}\hat{\xi}_{AA} \,,
\end{align}
where 
\begin{align}
    \hat\xi_{AA}&=-\frac{k_{AA}}{4 k^2}\left(\sign\left(z-z'\right) \left(1-e^{-k \left| z-z'\right| }\right)-1\right) \,.
\end{align}
The last $(-1)$ in the bracket is the contribution of the constant homogeneous term. The boundary limit of this gives
\begin{align}
    \epsilon_{BB'}k_{AA} e^{-kz}\frac{1}{4k^2}\,.
\end{align}
Note that this is not symmetric since $\nabla_{BB'}$ is not. The Feynman part gives the opposite contribution to the propagator and the complete expression is symmetric. One can make a similar comment to the one below \eqref{forthecomment}, i.e. this split may be helpful to reveal the pole structure and flat limit.

%%%%%%%%%%%%%%%%%%%%%%%%%%%%%%%%%%%%%%%%%%%%%%%%%%%%%%%%%%%%%
\subsection{Chalmers--Siegel's amendments}
\label{sec:ChSiAmend}
%%%%%%%%%%%%%%%%%%%%%%%%%%%%%%%%%%%%%%%%%%%%%%%%%%%%%%%%%%%%%
For the purpose of gauge fixing and deriving the propagators the Chalmers--Siegel action can be taken in the form \eqref{SDYMexpandedPert} (dropping the topological and other boundary terms, the latter, of course, must support the boundary conditions) 
\begin{align}\label{SDYMexpandedPertB}
    S_{\text{Ch.Si.}}&=\Tr\int \Psi^{AA}(\nabla_{AC'}\Phi\fdu{A,}{C'} +g\,\Phi_{A,C'}\Phi\fdu{A,}{C'})-\frac{\epsilon}{2}\Psi^{AB}\Psi_{AB}\,.
\end{align}
Since the $\Phi^{A,A'}$-field is the same as in the Yang--Mills and SDYM theories and $\Psi$ is auxiliary in the former, most of the results can be borrowed. One can use the same gauge-fixing options as in YM and SDYM. The $\langle \Phi \Phi\rangle$-propagator will be the same, except that it is multiplied by $\epsilon$ if we stick to \eqref{SDYMexpandedPertB}. The $\langle \Psi \Phi\rangle$-propagator can be derived via the $\Psi$ equation of motion. A surprise is awaiting at the $\langle \Psi \Psi\rangle$-case. Here, the two-point function and the propagator turn out to be not the same thing. Let us consider the axial gauge. On one hand, the two-point function can be derived by using the $\Psi$ equation once again
\begin{align}\label{psispsi2pt}
\begin{aligned}
    \langle \Psi^{AA}(-k,z) \Psi^{BB}(k,z')\rangle&=(\nabla_{-k,z})\fud{A}{A'} (\nabla_{k,z'})\fud{B}{B'}\langle \Phi^{AA'}(-k,z) \Phi^{BB'}(k,z')\rangle=\\
    &=-\delta(z-z') \epsilon^{AB}\epsilon^{AB}+e^{-2i\gamma} e^{-k(z+z')} \frac{\brk^A\brk^A k^Bk^B}{2k}\,.
\end{aligned}
\end{align}
This two-point function should be denoted $ \langle F^{AA}(-k,z) F^{BB}(k,z')\rangle$ judging on how it was constructed. 
On the other hand, in inverting the block matrix of the kinetic term we find for the $\Psi-\Psi$ piece of the propagator 
\begin{align}
    \nabla\fdu{A}{A'}\langle \Psi^{AA}(-k,z) \Psi^{BB}\,,(k,z')\rangle&=0
\end{align}
which is clearly not satisfied by the two-point function \eqref{psispsi2pt}. On the third hand, the $\Psi-\Psi$ propagator must not vanish since it is the $\Psi-\Psi$ two-point function that we used to impose the boundary conditions. The resolution of the apparent paradox is that \eqref{psispsi2pt} is correct as the two-point function of a descendant. The $\Psi-\Psi$ propagator consists of the homogeneous part of \eqref{psispsi2pt}:
\begin{align}
    \langle \Psi^{AA}(-k,z) \Psi^{BB}(k,z')\rangle&=+e^{-2i\gamma} e^{-k(z+z')} \frac{\brk^A\brk^A k^Bk^B}{2k}\,.
\end{align}
Therefore, we can write
\begin{align}\label{FFvsPsiPsi}
    \langle F^{AA}(-k,z) F^{BB}(k,z')\rangle&= -\delta(z-z')\epsilon^{AB}\epsilon^{AB} +\langle \Psi^{AA}(-k,z) \Psi^{BB}(k,z')\rangle\,.
\end{align}
Note that the inhomogeneous part does not affect the boundary conditions since $z\neq z'$ (one point being in the bulk and another on the boundary). The fact that the $\Psi-\Psi$ piece does not vanish will have important consequences: it will assure that the YM and the Chalmers--Siegel actions give the same four-point function. That $\langle \Psi\Psi\rangle \neq0$ is similar to the spin-half case discussed in Appendix \ref{app:boundaryterms}, where it also satisfies a homogeneous equation and is there to support the boundary condition. Lastly, in the self-dual limit the homogeneous part vanishes, which implies that in SDYM $\langle \Psi\Psi\rangle = 0$ (as the propagator).

%%%%%%%%%%%%%%%%%%%%%%%%%%%%%%%%%%%%%%%%%%%%%%%%%%%%%%%%%%%%%
\subsection{Boundary limits, helicity structure}
\label{sec:bdyLim}
%%%%%%%%%%%%%%%%%%%%%%%%%%%%%%%%%%%%%%%%%%%%%%%%%%%%%%%%%%%%%
In principle, the information below is contained in the form of the propagators just discussed. However, given that different helicities play very different roles in the self-dual limit, it is instructive to see what is the helicity structure of the propagators and two-point functions. The polarization vectors in axial gauge are chosen as follows (they are also an admissible choice in any other gauge we discussed):
\begin{align}
    \begin{aligned}
        \epsilon^{+AA}(k)&=+\frac{\bar{k}^A\bar{k}^A}{2k} \,, & \epsilon^{+AA}(-k)&=+\frac{k^Ak^A}{2k} \,,\\
        \epsilon^{-AA}(k)&=+\frac{k^Ak^A}{2k} \,, & \epsilon^{-AA}(-k)&=+\frac{\bar{k}^A\bar{k}^A}{2k} \,.
    \end{aligned}
\end{align}
Note that $\epsilon^\pm(-k) \cdot \epsilon^\pm(k)=1$ and $\epsilon^\pm(-k) \cdot \epsilon^\mp(k)=0$. Below, we will project the two-point functions on $\pm$ helicity structures. It should be noted that $\langle \pm \mp\rangle=0$ by Lorentz invariance and, hence, there is no need to write them down. A typical expression we compute is 
\begin{align}
    \epsilon^{+AA}(k)\langle \Phi_{AA}(-k,0)\Phi_{BB}(k,0)\rangle\epsilon^{+BB}(-k) \,,
\end{align}
or with just one side contracted onto a polarization vector. Here, it is important that the polarization vector on the left has momentum $+k$, while the one on the right has $-k$. It makes sense to contract with $\epsilon$'s the $\Psi$-legs and those legs $\Phi$ that do not have $A_z$-component. Due to the latter, we restrict ourselves to the most reliable axial gauge.

Due to the great number of expressions below, a rule of thumb to keep in mind is that, after decomposing everything into components of definite helicity, the theories we consider reduce to ``scalar fields''. For scalar fields the Dirichlet propagator is $e^{-kz}$ and the Neumann one is $-k^{-1}e^{-kz}$. In practice this means that Dirichlet propagators have prefactors of weight-zero in $k^A$ (identity map on the corresponding space). Also, the fields that satisfy first order equations can only have Dirichlet boundary condition imposed, hence, the structure has to be of the same $e^{-kz}$ form times a weight-zero expression. 

\paragraph{$\boldsymbol{\Phi-\Phi}$.} In the axial gauge for $\gamma\neq0$ we have for the two-point functions
\begin{align} \label{PhiPhi2}
    \begin{aligned}
        \epsilon^{+AA}(k)\langle \Phi_{AA}(-k,0)\Phi_{BB}(k,0)\rangle\epsilon^{+BB}(-k) &= -\frac{1}{2k}(1-e^{+2i\gamma})\,,\\
        \epsilon^{-AA}(k)\langle \Phi_{AA}(-k,0)\Phi_{BB}(k,0)\rangle\epsilon^{-BB}(-k) &= -\frac{1}{2k}(1-e^{-2i\gamma})\,.
    \end{aligned}
\end{align}
It also makes sense to look at the full bulk-to-bulk propagator projected onto the definite helicity states
\begin{align} 
    \begin{aligned}
        \epsilon^{+AA}(k)\langle \Phi_{AA}(-k,z)\Phi_{BB}(k,z')\rangle\epsilon^{+BB}(-k) &= -\frac{1}{2k}\left(e^{-k|z-z'|}-e^{-k(z+z')}e^{+2i\gamma}\right)\,,\\
        \epsilon^{-AA}(k)\langle \Phi_{AA}(-k,z)\Phi_{BB}(k,z')\rangle\epsilon^{-BB}(-k) &= -\frac{1}{2k}\left(e^{-k|z-z'|}-e^{-k(z+z')}e^{-2i\gamma}\right) \,,
    \end{aligned}
\end{align}
which are exactly the scalar propagators for the mixed boundary conditions that involve the shadow operator, see Appendix \ref{app:boundaryterms}, or the propagators for the components of definite helicity discussed in Section \ref{sec:boundaryterms}. For the boundary-to-bulk propagator we find
\begin{align} \label{PhiPhi1}
    \begin{aligned}
        \epsilon^{+AA}(k)\langle \Phi_{AA}(-k,0)\Phi_{BB}(k,z')\rangle &= -\frac{1}{4k^2}(1-e^{+2i\gamma})\brk_B\brk_Be^{-kz'} \,,\\
        \epsilon^{-AA}(k)\langle \Phi_{AA}(-k,0)\Phi_{BB}(k,z')\rangle &= -\frac{1}{4k^2}(1-e^{-2i\gamma}){k}_B{k}_Be^{-kz'}\,.
    \end{aligned}
\end{align}
The uncontracted boundary-to-bulk propagator is
\begin{align}
    \langle \Phi_{AA}(-k,0)\Phi_{BB}(k,z')\rangle_{\gamma\neq0}&= \frac{1}{k}e^{-kz'}\sin(\gamma)[-\sin(\gamma)\Pi^\text{e}_{AA,BB}+i\cos(\gamma)\Pi^\text{o}_{AA,BB}]\,,
\end{align}
from which the boundary-to-boundary propagator follows by dropping $e^{-kz'}$. This is exactly the two-point function we find in Section \ref{sec:boundaryterms}. For the Neumann case we simply have 
\begin{align}
   \langle \Phi_{AA}(-k,0)\Phi_{BB}(k,0)\rangle&= -\frac{1}{k}\Pi^\text{e}_{AA,BB}\,.
\end{align}
At the Dirichlet point $\gamma=0$, we have for the two-point functions 
\begin{align} \label{PhiPhi3}
    \begin{aligned}
        \epsilon^{+AA}(k)\pl_z\pl_{z'}\langle \Phi_{AA}(-k,0)\Phi_{BB}(k,0)\rangle\epsilon^{+BB}(k) &= k\,,\\
        \epsilon^{-AA}(k)\pl_z\pl_{z'}\langle \Phi_{AA}(-k,0)\Phi_{BB}(k,0)\rangle\epsilon^{-BB}(k) &= k \,.
    \end{aligned}
\end{align}
In other words, the complete Dirichlet two-point function is
\begin{align}
    \pl_z\pl_{z'}\langle \Phi_{AA}(-k,0)\Phi_{BB}(k,0)\rangle &= k \Pi^\text{e}_{AA,BB}\,.
\end{align}
For the boundary-to-bulk $\gamma=0$ propagator we have
\begin{align} \label{PhiPhi4}
    \begin{aligned}
        \epsilon^{+AA}(k)\pl_z\langle \Phi_{AA}(-k,0)\Phi_{BB}(k,z')\rangle &= -\frac{1}{2k}\brk_B\brk_Be^{-kz'} \,,\\
        \epsilon^{-AA}(k)\pl_z\langle \Phi_{AA}(-k,0)\Phi_{BB}(k,z')\rangle &= -\frac{1}{2k}{k}_B {k}_Be^{-kz'}\,.
    \end{aligned}
\end{align}

\paragraph{$\boldsymbol{\Psi-\Phi}$.} First note that the $\Psi-\Phi$ propagator, which is, basically, the same as for spin-half, contains $\sign(z-z')$. Therefore, the double boundary limit seems to depend on the order. This term, $\epsilon_{AB}\epsilon_{AB}, \sign(z-z')$ is analytic in $k$ and, hence, the difference between the two ways to take the limit is $\delta^3(x)$ on the boundary. This is similar to the scalar field Dirichlet propagator $\exp[-kz]$ having $\delta^3(x)$ as the leading term, but no $\sign(z-z')$. The sign-function shows up in the two-point function $\langle \pl_z \phi \phi\rangle$, which is of type $\langle p q\rangle$.\footnote{This $\sign$-function is present in any first order theory's propagator and is also responsible for $[p,q]\neq 0$.} 

The bulk-to-bulk propagator sandwiched in between two definite helicity states gives
\begin{align} \notag
    \begin{aligned}
        \epsilon^{+AA}(k)\langle \Psi_{AA}(-k,z)\Phi_{BB}(k,z')\rangle\epsilon^{+BB}(-k) &= \frac12 e^{-k|z-z'|}[\sign(z-z')-1]\,,\\
        \epsilon^{-AA}(k)\langle \Psi_{AA}(-k,z)\Phi_{BB}(k,z')\rangle\epsilon^{-BB}(-k) &=   \frac12 e^{-k|z-z'|}[\sign(z-z')+1] -e^{-2i\gamma} e^{-k(z+z')}\,.
    \end{aligned}
\end{align}
In the self-dual limit the $\gamma$-term in the first line dies off and the propagators are exactly of the type discussed in Appendix \ref{app:boundaryterms} for the first order toy-model with scalar fields, $\mathcal{L}\sim \psi D_{\pm}\phi$. The same expressions result from projecting the spin-half propagators onto the definite helicity components. Note that $\epsilon^\pm \cdot \epsilon^\pm=0$. Therefore, in terms of the helicity decomposition the first line selects $\langle \Psi_+ \Phi_+\rangle$ and the second line gives $\langle \Psi_- \Phi_-\rangle$. 

For the boundary-to-bulk propagator we find (the last term in the first line gives zero when contracted with polarization vectors on the boundary)
\begin{align}
    \begin{aligned}
        \langle \Psi_{AA}(-k,0)\Phi_{BB}(k,z')\rangle &= -\frac{1}{4k^2}\Big(k_Ak_A\bar{k}_B\bar{k}_B+e^{-2i\gamma}\bar{k}_A\bar{k}_Ak_Bk_B\Big)e^{-kz'}+\frac{1}{2}\frac{k_{AA}k_{BB}}{2k^2} \,,\\
    \langle\Psi_{AA}(-k,z)\Phi_{BB}(k,0)\rangle &=  \frac{1}{4k^2}\big(1-e^{-2i\gamma}\big)\bar{k}_A\bar{k}_Ak_Bk_Be^{-kz}\,.
    \end{aligned}
\end{align}
The projection onto the definite helicity components reads
\begin{align} 
    \begin{aligned}
        \epsilon^{+AA}(k)\langle \Psi_{AA}(-k,0)\Phi_{BB}(k,z')\rangle &= -\frac{1}{2k}\bar{k}_B\bar{k}_Be^{-kz'} \,,\\
        \epsilon^{-AA}(k)\langle \Psi_{AA}(-k,0)\Phi_{BB}(k,z')\rangle &= -\frac{1}{2k}k_Bk_Be^{-2i\gamma}e^{-kz'}\,,
    \end{aligned}\\
    \begin{aligned}
        \epsilon^{+AA}(k)\langle \Phi_{AA}(-k,0)\Psi_{BB}(k,z')\rangle &= 0 \,,\\
        \epsilon^{-AA}(k)\langle \Phi_{AA}(-k,0)\Psi_{BB}(k,z')\rangle &= +\frac{1}{2k}(1-e^{-2i\gamma})k_B k_B e^{-kz'}\,.
    \end{aligned}
\end{align}
One can also think of replacing $\Phi$ on the boundary with $\bar\Psi\equiv \bar{F}$, which is gauge covariant, to find
\begin{align} 
    \begin{aligned}
        \epsilon^{+AA}(k)\langle \bar\Psi_{AA}(-k,0)\Phi_{BB}(k,z')\rangle &= +\frac{1}{2k}\bar{k}_B\bar{k}_B e^{+2i\gamma} e^{-kz'} \,,\\
        \epsilon^{-AA}(k)\langle \bar\Psi_{AA}(-k,0)\Phi_{BB}(k,z')\rangle &= +\frac{1}{2k}k_Bk_Be^{-kz'}\,,
    \end{aligned}
\end{align}
For completeness, the two-point functions for the two different orders of the limits
\begin{align} 
    \begin{aligned}
        \epsilon^{+AA}(k)\langle \Psi_{AA}(-k,0)\Phi_{BB}(k,0^2)\rangle\epsilon^{+BB}(-k) &= 0\,,\\
        \epsilon^{-AA}(k)\langle \Psi_{AA}(-k,0)\Phi_{BB}(k,0^2)\rangle\epsilon^{-BB}(-k) &= 1-e^{-2i\gamma} \,,
    \end{aligned}
\end{align}
where $0^2\ll 0$ for the order of the limits. 
\begin{align} 
    \begin{aligned}
        \epsilon^{+AA}(k)\langle \Psi_{AA}(-k,0^2)\Phi_{BB}(k,0)\rangle\epsilon^{+BB}(-k) &= -1\,,\\
        \epsilon^{-AA}(k)\langle \Psi_{AA}(-k,0^2)\Phi_{BB}(k,0)\rangle\epsilon^{-BB}(-k) &= -e^{-2i\gamma} \,.
    \end{aligned}
\end{align}
Here, we note again that the difference between the two two-point functions is $\langle\pm\pm\rangle=1$, which is what $\delta_{ij} \delta^3(x-y)$ would give.

At the Dirichlet point $\gamma=0$ for the boundary-to-bulk propagator we have
\begin{align} \label{PhiPhi4}
    \begin{aligned}
        \epsilon^{+AA}(k)\pl_z\langle \Phi_{AA}(-k,0)\Psi_{BB}(k,z')\rangle &= 0 \,,\\
        \epsilon^{-AA}(k)\pl_z\langle \Phi_{AA}(-k,0)\Psi_{BB}(k,z')\rangle &= k_Bk_Be^{-kz'}\,.
    \end{aligned}
\end{align}

\paragraph{$\boldsymbol{\Psi-\Psi}$.} This type of two-point functions contains $\delta(z-z')$, homogeneous terms and $\sign(z-z')$ sometimes. One easily finds
\begin{align} 
    \begin{aligned}
        \epsilon^{+AA}(k)\langle \Psi_{AA}(-k,0)\Psi_{BB}(k,0)\rangle\epsilon^{+BB}(-k) &= 0\,,\\
        \epsilon^{-AA}(k)\langle \Psi_{AA}(-k,0)\Psi_{BB}(k,0)\rangle\epsilon^{-BB}(-k) &= 2 e^{-2i\gamma} k\,,
    \end{aligned}\\
    \begin{aligned}
        \epsilon^{+AA}(k)\langle \bar\Psi_{AA}(-k,0)\bar\Psi_{BB}(k,0)\rangle\epsilon^{+BB}(-k) &= 2 e^{2i\gamma} k\,,\\
        \epsilon^{-AA}(k)\langle \bar\Psi_{AA}(-k,0)\bar\Psi_{BB}(k,0)\rangle\epsilon^{-BB}(-k) &= 0 \,,
    \end{aligned}
\end{align}
which do not depend on the order of the limits. The following ones do, however, depend on the order
\begin{align} 
    &\begin{aligned}
        \epsilon^{+AA}(k)\langle \Psi_{AA}(-k,0^2)\bar\Psi_{BB}(k,0)\rangle\epsilon^{+BB}(-k) &=-2k \,,\\
        \epsilon^{-AA}(k)\langle \Psi_{AA}(-k,0^2)\bar\Psi_{BB}(k,0)\rangle\epsilon^{-BB}(-k) &= 0 \,,
    \end{aligned}\\
    &\begin{aligned}
        \epsilon^{+AA}(k)\langle \Psi_{AA}(-k,0)\bar\Psi_{BB}(k,0^2)\rangle\epsilon^{+BB}(-k) &=0 \,,\\
        \epsilon^{-AA}(k)\langle \Psi_{AA}(-k,0)\bar\Psi_{BB}(k,0^2)\rangle\epsilon^{-BB}(-k) &= -2k \,.
    \end{aligned}
\end{align}
Two-point functions with one leg in the bulk are also useful
\begin{align} 
    \begin{aligned}
        \epsilon^{+AA}(k)\langle \Psi_{AA}(-k,z)\Psi_{BB}(k,0)\rangle &= 0\,,\\
        \epsilon^{-AA}(k)\langle \Psi_{AA}(-k,z)\Psi_{BB}(k,0)\rangle &= e^{-kz} e^{-2i\gamma} k_B k_{B}
    \end{aligned}
\end{align}
and 
\begin{align} 
    \begin{aligned}
        \epsilon^{+AA}(k)\langle \Psi_{AA}(-k,z)\bar\Psi_{BB}(k,0)\rangle &= -e^{-kz}  \brk_B \brk_{B}\,,\\
        \epsilon^{-AA}(k)\langle \Psi_{AA}(-k,z)\bar\Psi_{BB}(k,0)\rangle &= 0\,,\\
        \epsilon^{+AA}(k)\langle \bar\Psi_{AA}(-k,z)\Psi_{BB}(k,0)\rangle &= 0\,,\\
        \epsilon^{-AA}(k)\langle \bar\Psi_{AA}(-k,z)\Psi_{BB}(k,0)\rangle &= -e^{-kz}  k_B k_{B} \,.
    \end{aligned}
\end{align}

%%%%%%%%%%%%%%%%%%%%%%%%%%%%%%%%%%%%%%%%%%%%%%%%%%%%%%%%%%%%%
\subsection{Boundary terms, two-point functions from the action}
\label{sec:boundaryterms}
%%%%%%%%%%%%%%%%%%%%%%%%%%%%%%%%%%%%%%%%%%%%%%%%%%%%%%%%%%%%%

While three- and higher-point correlators are usually computed directly via Witten diagrams, it may be instructive to see how two-point functions can emerge from the on-shell action. This would require appropriate boundary terms to be added to the action to make the variational principle well-defined. 

Let us recall that the starting point is the Yang--Mills action with the theta-term \eqref{YMthetaAction} and its Chalmers--Siegel counterpart \eqref{CSthetaAction}. The free on-shell action is\footnote{Free on-shell action $S(\phi)$ is just a boundary term. Indeed, since it is bilinear in the fields, $\phi\pl_\phi S=2S$ up to boundary terms. On the other hand $\pl_\phi S$ are the bulk equations of motion, i.e. $S=\tfrac12 \phi\pl_\phi S$ picks only the boundary terms and does it with the factor $\tfrac12$. The minus originates from $\delta F_{AA}=\nabla_{AA'}\delta\Phi\fdu{A}{A'}\ni -\pl_z \epsilon_{AA'}\delta \Phi\fdu{A}{A'}=\pl_z \delta\Phi_{AA}$, $\delta F_{A'A'}\ni -\pl_z \epsilon_{BA'}\delta\Phi\fud{B}{A'}=-\pl_z\delta\Phi_{A'A'}$ and the fact that $z=0$ is the lower limit. } 
\begin{align}\label{YMactionOnshell}
    -S_{\text{YM},\theta}\Big|_{\text{on-shell}}&= \tfrac14\int_{\pl M}(\alpha+\beta) F^{AA}\Phi_{AA}-(\alpha-\beta) \bar{F}^{AA}\Phi_{AA} \,. 
\end{align}
The Chalmers--Siegel action has the same on-shell value, of course, but $F^{AA}$ comes originally as $\epsilon \Psi^{AA}$. Recalling that $F=B+E$ and $\bar{F}=B-E$, we have 
\begin{align}
     -S_{\text{YM},\theta}\Big|_{\text{on-shell}}&= \tfrac12\int_{\pl M} (\beta B + \alpha E )^i\Phi_i \,.
\end{align}

\paragraph{Dirichlet two-point function.} The Yang--Mills action is well-defined with Dirichlet boundary conditions for any value of the theta-term. Let us see what the net effect is. The on-shell action of the pure Yang--Mills theory is just the integral of $\tfrac12A^i \pl_zA_i=\tfrac12A^i E_i$ over the boundary, which gives the parity-even contribution. Activating $\beta$ brings the magnetic field $B$, which comes from the Chern--Simons term $A dA$. This leads to the parity-odd contribution:
\begin{align}
    \langle J_{AA} J_{BB}\rangle&= \alpha k \Pi^\text{e}_{AA,BB}+\beta k\Pi^\text{o}_{AA,BB} \,.
\end{align}
Therefore, with just Dirichlet boundary condition it is easy to get any ratio $\alpha/\beta$ between the parity-even and parity-odd contributions.

\paragraph{Mixed two-point function.} To consider the generic case $\gamma\neq0$ we start with the variation, which is almost the same as the on-shell action, \eqref{YMactionOnshell}, but without $\tfrac12$:
\begin{align}\label{YMactionVar}
    -\delta S_{\text{YM},\theta}&=\tfrac12 \Tr\int_{\pl M}[(\alpha+\beta) F^{AA}-(\alpha-\beta) \bar{F}^{AA}]\delta \Phi_{AA} =\Tr\int_{\pl M}[ \beta B^{AA}+\alpha E^{AA}]\delta \Phi_{AA} \,.
\end{align}
We follow closely Appendix \ref{app:boundaryterms}. Firstly, we absorb the contact term with $\beta$ into the boundary term. Effectively, we just drop it. We introduce the source (or the new canonical coordinate)
\begin{align}
    J^{AA}=Q^{AA}=\tfrac12 [e^{i\gamma}F^{AA}+e^{-i\gamma} \bar{F}^{AA}]=\cos (\gamma) B^{AA}+ i \sin(\gamma) E^{AA} \,,
\end{align}
the combination we would like to keep fixed on the boundary, i.e. $\delta Q^{AA}=0$. Since $B^{AA}=(\slashed k \Phi)^{AA}=k\fud{A}{B}\Phi^{AB}$, in terms of the Appendix \ref{app:boundaryterms}, we have $a=\cos(\gamma)\slashed k$ and $b=i \sin(\gamma)$. The action is stationary, provided one adds a boundary term\footnote{One could just derive propagators for a given boundary condition and proceed to computing Witten diagrams with the bulk action. If the boundary conditions are nonlinear ($F$ contains $A^2$), one can take this fact into account perturbatively. Effectively, this corresponds to adding the same boundary vertices to the action that would make the action stationary for the chosen boundary conditions. Therefore, it is hard to bypass the variational principle. }
\begin{align}
    -S_{\text{bndry}}&= -\alpha \Tr\int _{\pl M}\left[ E\cdot \Phi +\frac{\cos(\gamma)}{2i \sin(\gamma)} \Phi \cdot B\right]\,.%=-\alpha \Tr\int _{\pl M}\left[ \tfrac12 E\cdot \Phi +\frac{1}{2i \sin(\gamma)} \Phi \cdot Q\right] \,.
\end{align}
The first term $E\cdot \Phi$ is to implement a ``Legendre transform'' that switches the Dirichlet problem to the Neumann one (note that generic mixed boundary conditions are ``closer'' to Neumann ones than to Dirichlet). Note that $\Phi \cdot B = \Phi \cdot(\slashed  k\Phi)$, which is the free Chern--Simons term $AdA$. Therefore, the mixed boundary conditions require the coefficient of the Chern--Simons term to be $-i \alpha \cot(\gamma)$. Next, we need the boundary-to-bulk propagator \eqref{btobAxial} to get $\Phi\equiv q(z)$ in the bulk that is sourced by source $J^{AA}$ at the boundary:\footnote{Note that we chose to have $G(-k,z;k,z')$, hence, $G(-k,0;k,z')^{AA,BB}=\frac{1}{2k}[\Pi^{\gamma} -\Pi^\text{e}]^{AA,BB} e^{-kz'}$ and swapping the two sides costs us $\Pi^\text{o}\rightarrow-\Pi^\text{o}$. In addition, one can see that $-b^{-1}[\cos(\gamma) k\fud{B}{C}G(-k,0;k,z')^{AA,BC}+i \sin(\gamma)\pl_{z'}G(-k,0;k,z')^{AA,BB}]=-e^{-kz'}\Pi_\text{e}^{AA,BB}$. In other words, the solution induced via the boundary-to-bulk propagator does satisfy the mixed boundary condition except at the same point as the source. }
\begin{align}
    \Phi^{AA}(z)&=-\frac{1}{b}K^{AA,BB}(z)J_{BB}\,, &&K^{AA,BB}(k,z)=\frac{1}{2k}[\Pi^{-\gamma} -\Pi^\text{e}]^{AA,BB} e^{-kz} \,.
\end{align}
Since $Q^{AA}(z)$ satisfies the Dirichlet condition we have (as well as the direct calculation shows)
\begin{align}
    Q^{AA}(z)&= -\Pi^{AA,BB}_\text{e} J_{BB} e^{-kz} \,,
\end{align}
where $\Pi^{AA,BB}_e$ is just the identity map on the space of transverse $3d$ vectors, cf. the scalar case where $Q=\exp[-kz]J$. Finally, the on-shell action is
\begin{align}
    S_{\text{on-shell}}&= \Tr\int_{\pl M} -\frac{1}{2b} q Q \longrightarrow -\frac{\alpha}{2b^2} J_{AA}K^{AA,BB}(z=0) J_{BB} \,,
\end{align}
where, in principle, we have a freedom of adding a contact term, but we drop it right away to find the generating functional on the right. We see that the two-point function coincides with the one obtained directly as $\langle \Phi^{AA}(-k,0)\Phi^{BB}(k,0)\rangle$, see \eqref{twopointAxial}, modulo $1/b$ per leg, which is a factor that accompanies every external leg within the standard perturbation theory, but has no physical meaning, see Appendix \ref{app:boundaryterms}. Note that the fact the two ways give the same answer is not a coincidence: $Q$, being canonical with the Dirichlet boundary condition, gives $Q=J$ near the boundary, whereas $q$ is related to $J$ via the boundary-to-bulk propagator $K$, whose $z=0$ limit is the same as $\langle \Phi^{AA}(-k,0)\Phi^{BB}(k,0)\rangle$.

Lastly, we can promote the correlator $\langle A_i A_j\rangle$ of the gauge field on the boundary just computed to the correlation function of the associated currents $\tilde{J}=B$ ($B$ is the magnetic field for $A_i$). One needs to use $\slashed k \Pi^{\text{e,o}}=-k \Pi^{\text{o,e}}$ twice (on every leg) to get 
\begin{align}
    \langle J^{AA} J^{BB}\rangle&= k\fud{A}{C} k\fud{B}{M} \langle A^{AC} A^{BM}\rangle=  k \sin(\gamma)[ - \sin(\gamma)\Pi_\text{e}^{AA,BB}+i\cos (\gamma) \Pi_\text{o}^{AA,BB}] \,.
\end{align}
Therefore, mixed boundary conditions allow us to get any value for the ratio, $-\tan(\gamma)$, of the parity-even to the parity-odd structures. We have to conclude that two-point functions cannot yet make a difference between the mixed and Dirichlet boundary conditions and one can get any ratio of $\langle ++\rangle$ to $\langle -- \rangle$ in either way.

There seems to be one more free parameter. Upon coupling the dual current $\tilde{J}$ to a source $\tilde{A}_i$, one can consider adding a Chern--Simons term for $\tilde{A}$, which would give an additional contribution to the parity-odd parts of two- and three-point functions of $\tilde{J}$. 

\paragraph{Spin-half interlude.} As it was already noted, SDYM is very close in spirit to a Weyl fermion, see also Appendix \ref{app:boundaryterms}. Let us recall, that the expansion near the boundary for a Dirac spinor reads $\Psi^\alpha= z^{\Delta_- } \psi_-^\alpha +... +z^{\Delta_+ } \psi_+^\alpha+...$. One should interpret $\psi_-\sim q$ and $\psi_+\sim p$, meaning that the Dirichlet boundary condition should fix $q$. The variation of the Dirac action gives 
\begin{align}
    \delta S_\text{D}&= \text{EOM}+ \int_{\pl M} \bar{q}\delta p+ \delta \bar{p} q \,.
\end{align}
The action is not stationary with the Dirichlet boundary conditions. However, the leftover is a variation of a boundary term
\begin{align}
    S_b&= \int_{\pl M} \bar{q} p+ \bar{p} q
\end{align}
if we take into account that $\delta q=\delta\bar{q}=0$. Therefore, the correct action for the AdS/CFT applications is $S_\text{D}-S_b$. Since the bulk action vanishes on solutions, the two-point function is generated by the added boundary term.

Now, we consider a Weyl fermion: we drop the Dirac conjugate $\bar{\chi}^\alpha$ used to write down an action and proceed with $\chi^\alpha=(\Phi^{A'},\Psi^A)$, where $\Psi^A=(\Phi^{A'})^*$ is the reality condition. The best action is
\begin{align}
    S_W&=\tfrac12\int |e|\, \Psi^{A} \overleftrightarrow{\nabla}_{AA'}\Phi^{A'} \,.
\end{align}
The use of $\overleftrightarrow{\nabla}$ makes it Hermitian in Minkowski signature without boundary terms (with just $\nabla$ the action is Hermitian up to a boundary term). It also makes it Hermitian in Euclidean signature where $x^{AA'}$ does not obey any simple Hermiticity rules. The kinetic term is
\begin{align}
    S_W& \ni-\tfrac12\int  \Psi_{A} \pl_z \Phi^{A'}- \pl_z\Psi_{A}\Phi^{A'} \,.
\end{align}
It will generate a boundary term
\begin{align}
    \delta S_W&=\tfrac12\int_{\pl M}  \Psi_{A} \delta \Phi^{A}- \delta\Psi_{A}\Phi^{A} \,,
\end{align}
which breaks the bulk Lorentz symmetry down to the boundary one, hence, the contracted indices can all be $A$s. The action is not stationary unless we require $\delta \Phi=\delta \Psi=0$, which is too strong. In the second order Maxwell/Yang--Mills theory one can impose Dirichlet/Neumann or mixed boundary conditions. In the first order theory the only option is Dirichlet. Options for boundary conditions are not unrelated to the requirement of regularity in the bulk, the latter being related to helicity now. In terms of the helicity components the action can be rewritten as ($D_\pm= \pl_z \pm |k|$ and $(f,g)\equiv f(-\vec k,z)g(\vec k,z)$):
\begin{align}
    S&=\int_M (\psi_+,D_+ \phi_+)+(\psi_-,D_- \phi_-)\,.
\end{align}
The variation of the action is simply
\begin{align}
    \delta S&=-\int_{\pl M} (\psi_+, \delta \phi_+)+(\psi_-,\delta \phi_-)\,.
\end{align}
Since the only solutions for $\phi_-$ are irregular, we cannot let it fluctuate, but $\psi_-$ satisfies $D_+\psi_-=0$ and can fluctuate. Therefore, the boundary data that can be freely fixed without having to violate the regularity in the bulk are $\phi_+$ and $\psi_-$ (they are related by the reality condition in the Minkowski signature). At the same time, $\phi_-=\psi_+=0$ at $\pl M$. The variational principle is well-defined now, but we can still add boundary terms, of which the only nonvanishing, when evaluated close to the boundary, is $(\psi_-,\phi_+)$. Note that the conformal weight $\Delta=3/2\pm m$ becomes critical $\Delta=3/2$ at $m=0$ in the sense that the position space two-point function scales as $1/x^d$ and the momentum space two-point function is of weight $0$, i.e. $\slashed k/|k|$.

\paragraph{SDYM: Boundary terms and boundary conditions.} The Fefferman-Graham analysis of Section \ref{sec:FG} revealed that the boundary data needs to be split into helicity eigenstates. We will use the same polarization vectors $\epsilon_{\pm}^{AA}$ as before. In the physical gauge (i.e. transverse $k_i A^i=0$ and $A_z=0$) one can write
\begin{align}
    \Phi^{AA}(\vec k,z)&= \epsilon_+^{AA}(\vec k) \Phi_+(\vec k,z)+\epsilon_-^{AA}(\vec k) \Phi_-(\vec k,z)\,, \quad \text{same for } \quad \Psi^{AA}(\vec k,z)=...
\end{align}
Introducing $D_\pm=\pl_z \pm k$, the free action of SDYM/Chalmers--Siegel can be written as
\begin{align}
    S&= \int_M (\Psi_+, D_+ \Phi_+)+(\Psi_-, D_- \Phi_-)-\tfrac{\epsilon}2[(\Psi_+,\Psi_+)+(\Psi_-,\Psi_-)]
\end{align}
where $(f,g)\equiv f(-\vec k,z) g(+\vec k,z) $ is induced by contracting the polarization vectors. Note that there is little to no changes between the SDYM example here and the case of spin-half. The variation produces a boundary term, from which the presymplectic potential can be read off:
\begin{align}
   \delta S&= -(\Psi_+, \delta \Phi_+)-(\Psi_-, \delta \Phi_-)\,.
\end{align}
In the helicity basis regular fields satisfy $D_+ \bullet=0$, which gives the scalar Dirichlet propagator $e^{-kz}$, and irregular ones satisfy $D_-\bullet=0$, which gives $e^{+kz}$. With this in mind, we see that $\Phi_+$ and $\Psi_-$ are allowed to fluctuate, while $\Phi_-$ and $\Psi_+$ must be frozen, i.e. set to zero. With these conditions the action is stationary, but one still supplement the action with $S_{\text{bndry}}= +(\Psi_-, \Phi_-)$. The total variation looks now more canonical 
\begin{align}
   \delta (S+S_{\text{bndry}})&= -(\Psi_+, \delta \Phi_+)+(\delta\Psi_-,\Phi_-)\,,
\end{align}
as it has $\delta \bullet$ for the fields that can fluctuate. 
One can also add $(\Psi_-,\Phi_+)$, which is the only one not to vanish close to the boundary. Indeed, it is $\Psi_-$ and $\Phi_+$ that are allowed to propagate into the bulk as $e^{-kz}$. Therefore, $(\Psi_-,\Phi_+)\neq0$ when one steps $\epsilon$ into the bulk. In any case, the two-point function turns out to be a contact term and, hence, additional arguments are needed to fix its value. This may not be too surprising given that $A_i\sim \Phi_{AA}$ and $J_i \sim \Psi_{AA}$ have dual dimensions, which allows for a non-vanishing contact two-point function. Since the presymplectic potential is entirely generated by the free kinetic term, no further boundary terms are needed when interactions are turned on. 

In the Chalmers--Siegel case, upon solving for $\Psi_\pm$ one gets just an action for two scalar fields
\begin{align}
    S&= \epsilon^{-1}\int (D_+\Phi_+, D_+ \Phi_+)+(D_-\Phi_-, D_- \Phi_-)\,.
\end{align}
Note that $(D_{\pm})^T=-D_{\pm}$ upon integration by parts. The Klein-Gordon operator is $D_+D_-=\pl^2_z-k^2$. Now, with the help of Appendix \ref{app:boundaryterms}, one can impose generic mixed boundary conditions, which requires certain boundary terms.

Lastly, let us consider the Chern--Simons term on the boundary. At the free level or for $U(1)$, it is helicity-friendly
\begin{align}
    \int A dA \sim k(A_+,A_+)-k(A_-,A_-)\,.
\end{align}
However, the $A^3$-term mixes up all helicities together and its variation cannot be made to vanish unless both $\delta A_+=\delta A_-=0$ or the field strength $F(A)$ is constrained on the boundary.\footnote{A potentially useful axial gauge choice that makes the $A^3$-term vanish is impossible in the helicity basis.} Therefore, it seems that we have to drop the Chern--Simons term in SDYM. Nevertheless, let us note that the $AdA$-structure leads to $\langle ++\rangle$ two-point function if we evaluate it as the limit $z\rightarrow0$ of $AdA$ close to the boundary where $\Phi(z)=A(z)$ comes from $\epsilon^+\langle\Psi(0) \Phi(z)\rangle$.

\paragraph{From SDYM to Chalmers--Siegel.} Since SDYM is a first order theory, we do not have the Neumann option. However, in the YM/Chalmers--Siegel theories we can impose any mixed-boundary condition. The latter affects the bulk-to-bulk propagator. Can we still interpret the Chalmers--Siegel action as a perturbation of SDYM and where are the mixed boundary conditions coming from? Let us interpret $\tfrac{\epsilon}2 \Psi^2 $ as an interaction vertex. The $\Phi-\Phi$ propagator results from integrating two $\Psi-\Phi$ propagators sewn via the quadratic vertex:
\begin{align}\notag
    \epsilon\int d\tilde z\,  \langle \Psi^{CC}(-k,\tilde z) \Phi^{BB}(k,z')\rangle \, \langle \Phi^{AA}(-k,z) \Psi_{CC}(k,\tilde z)\rangle &= -\epsilon\langle \Phi^{AA}(-k,z)  \Phi^{BB}(k,z')\rangle \Big|_{\gamma=0}\,.
\end{align}
For $\gamma=0$ this correctly reproduces the Dirichlet propagator. The Chalmers--Siegel action is stationary for the Dirichlet problem, but requires a boundary term for mixed boundary conditions. This term can be rewritten as
\begin{align}
    S_{\text{bndry}}&= -\epsilon \tfrac14(1-e^{2i\gamma})\int \frac{1}{k} \Psi^{AA} \Pi^\text{o}_{AA,BB} \Psi^{BB}\,.
\end{align}
When interpreted as an additional boundary vertex, it does reproduce the correction that takes the Dirichlet propagator into the mixed one. Note that the boundary term looks non-local. In the Chalmers--Siegel theory $\epsilon^{-1}\Psi^{AA}=\nabla\fud{A}{A'}\Phi^{AA'}$ and the nonlocal term is $\pl_z \Phi \pl_z \Phi$, i.e. a total derivative on the boundary and the boundary term turns out to be local in disguise. At the end of Appendix \ref{app:boundaryterms} we consider a toy model that illustrates some of the features discussed above with the help of scalar fields.

%%%%%%%%%%%%%%%%%%%%%%%%%%%%%%%%%%%%%%%%%%%%%%%%%%%%%%%%%%%%%
\section{AdS/CFT correlators}
\label{sec:Correlators}
%%%%%%%%%%%%%%%%%%%%%%%%%%%%%%%%%%%%%%%%%%%%%%%%%%%%%%%%%%%%%
With all preparatory steps collected in the previous sections and Appendix \ref{app:boundaryterms}, we are ready to compute AdS/CFT correlators in the Yang--Mills theory (plus the theta-term) with the most general mixed boundary conditions and in SDYM. Following Section \ref{sec:2.5}, the Yang--Mills theory can be rewritten in the first order Chalmers--Siegel form plus the theta-term (with a shifted value of the coefficient). The latter is beneficial for several reasons: (i) the Chalmers--Siegel action does not have the quartic term, which saves time; (ii) the anti-MHV vertex is absorbed to give twice the MHV plus the topological term, which simplifies cubic interactions as well; (iii) the topological term leads to a boundary vertex that does not generate genuine bulk singularities, which makes the structure of the final answer more transparent; (iv) YM, in its original form, does not have SDYM as a limit, it is the Chalmers--Siegel action that does.   

Our main goal, apart from exploring the holography for the most general boundary conditions, is to identify the subsector of YM that is captured by SDYM. It is also interesting to see to which extent SDYM appears to be a limit of YM. 

We begin with the three-point functions (the two-point functions, which can be obtained in several different ways, were discussed in Section \ref{sec:boundaryterms} and \ref{sec:bdyLim}). Here, there is already a detectable difference between YM and SDYM. At the four-point level there are quite a few diagrams present in YM on top of what one finds in SDYM. The SDYM/YM four-point function of type $(+++-)$ should not have the energy pole since the corresponding flat space amplitude vanishes, see also \cite{Gomez:2026yno} for the recent progress on the structure of the pole expansion. 

A few general comments before the start. (1) The two-point function $\langle \Phi (-k,0)\Phi(k,0)\rangle$ scales as $1/k$ for non-Dirichlet boundary conditions (as that of a Chern--Simons gauge field on the boundary). For the helicity configuration we could choose the boundary-to-bulk propagator for the positive helicity to be $\Phi^{AA}_+\sim k^{-1} \epsilon_+^{AA} e^{-kz}\sim \brk^A\brk^A/k^2 e^{-kz}$. This is dual to a gauge field $a_i$ on the boundary. A better observable is the corresponding magnetic field $B=*(da+g a\wedge a)$ (we added the nonabelian piece). $*da$ corresponds to $\slashed k a$, for which $\epsilon^{AA}_\pm$ are eigen-vectors, $\slashed k \epsilon^{AA}_\pm=\pm k \epsilon^{AA}_\pm$. To save time, we choose the wave function $\Phi^{AA}\sim \epsilon_+^{AA} e^{-kz}$, i.e. we drop $k$ like if $*da$ has already been applied. In other words, we directly compute the correlators of the magnetic field (and do not mention this explicitly below). This applies to the non-Dirichlet cases. For the Dirichlet case the propagator already has this form $\Phi^{AA}_\pm\sim \epsilon_\pm^{AA} e^{-kz}$, i.e. without $k^{-1}$. (2) We also fix $\Psi^{AA}(k,z)= k^Ak^A e^{-kz}$, as in the Dirichlet case. The difference between this and non-Dirichlet is $k^{-1}$ and a numerical factor. Here, we do not have an excuse for $\gamma\neq0$, since $\Psi$ is already gauge-covariant. However, this choice facilitates comparison with the flat space results. (3) For mixed boundary conditions the boundary-to-bulk propagators have additional purely numerical $\gamma$-dependent factors, which we drop since they just rescale the basis of the dual operators (unless these factors vanish or blow up, the cases we discuss separately). These shortcuts are made since the external lines do not depend much on the boundary conditions. 

If all factors are kept, the external lines are saturated by the boundary-to-bulk propagators, see Section \ref{sec:bdyLim}, whose functional form is the same, but a pre-factor depends on the boundary conditions chosen:
\begin{align}
    \Psi_{AA}(k,z) &=e^{-kz}k_Ak_A
    \begin{cases}
        +\frac{1}{2k}(1-e^{-2i\gamma}) \,, & \gamma\neq0\,,\\
        1\,, & \gamma =0 \,,
    \end{cases}
\end{align}
all of which are obtained by contracting $\epsilon^{+BB}(k)$ with the $\Phi$-leg on the boundary. For the $\Phi$-leg in the bulk we have
\begin{align}
    \Phi_{AA}(k,z) &=-e^{-kz} \epsilon^{\pm}_{AA}
    \begin{cases}
        \frac{1}{2k}(1-e^{ \pm 2i\gamma})\,, & \gamma\neq0\,,\\
        1\,, & \gamma =0 \,,
    \end{cases}
\end{align}
where we recall that $\epsilon_+^{AA}(k)=\brk^A \brk^A/(2k)$, $\epsilon_-^{AA}(k)=k^A k^A/(2k)$ in the axial gauge. We will also use Feynman gauge a few times. 

The boundary-to-bulk propagators above originate from $\epsilon^{AA}(k)\langle \Phi_{AA}(-k,0)\,\, \bullet(k,z)\rangle$, i.e. from the bulk-to-bulk propagators with the $\Phi$-leg on the boundary. The bulk helicity configuration is the same as the $\epsilon^{AA}(k)$ that the boundary $\Phi(-k,0)$ has been contracted with. In other words, $\epsilon_\pm(k)\langle \Phi(-k,0) \bullet \rangle \sim \epsilon_{\pm}(k)$. This choice of the external lines means that for $\gamma\neq0$ we compute $\langle \Phi_{AA}(k_1,0)\ldots \rangle$ correlators, i.e. of the gauge field $a_i$ on the boundary. One can later compute its magnetic field as a better observable. Alternatively one can compute the boundary limit of bulk correlators where the boundary $\Phi$-leg is replaced by $F_{AA}\sim\Psi_{AA}$ (in SDYM we have $\Psi_{AA}$ while $F_{AA}=0$, which is why $\Psi_{AA}$ is a better notation for all cases) or by $\bar{\Psi}_{AA}(k_1,0)\equiv\bar F_{AA}(k_1,0)$. These options just change the overall coefficient by $k$ times a numerical factor, which sometimes can vanish or diverge in the self-dual limit:
\besubeqs
\begin{align}
    \text{from }\Psi(0): & & \Phi_{AA}(k,z) &=-e^{-kz} 
    \begin{cases}
       \epsilon^{-}_{AA} e^{ -2i\gamma}\\
        \epsilon^{+}_{AA}\,, 
    \end{cases}
\\
    \text{from }\bar\Psi(0)\equiv\bar F(0): & & \Phi_{AA}(k,z) &=+e^{-kz} 
    \begin{cases}
       \epsilon^{-}_{AA} \\
        \epsilon^{+}_{AA}e^{ +2i\gamma}\,, 
    \end{cases}
\\ \label{fromF}
    \text{from }\Psi(0): & & \Psi_{AA}(k,z) &=+e^{-kz} 
    \begin{cases}
       \epsilon^{-}_{AA} 2k e^{ -2i\gamma}\\
        0\times \epsilon^+_{AA}\,, 
    \end{cases}
\\
    \text{from }\bar\Psi(0)\equiv\bar F(0): & & \Psi_{AA}(k,z) &=-e^{-kz} 
    \begin{cases}
        \epsilon^{-}_{AA}2k \phantom{e^{ -2i\gamma}}\\
               0\times \epsilon^+_{AA} \,.\\
    \end{cases}
\end{align}
\esubeqs
$\Psi(z)$ always satisfies the first order equation and, hence, carries just one helicity state, which can propagate from $\Phi(0)$, or $\bar{F}(0)$ or even from $F(0)$ unless we are in the self-dual limit. One option in AdS/CFT is to just compute the on-shell action according to GKPW \cite{Gubser:1998bc,Witten:1998qj}. On the other hand, following BDHM \cite{Banks:1998dd}, one can compute boundary limits of the bulk correlators (with $z^\Delta$-factors amputated). The two recipes give the same answer once in BDHM we take the fundamental fields themselves. BDHM provides us with more options as one can directly compute, say, the magnetic field of the boundary gauge field by taking $F+\bar{F}$ instead of $\Phi$. Since the choice of the boundary observable affects only the external lines and in a simple way described above, we will not address this issue in what follows. Note that when $\Phi(z)$ propagates from $\Phi(0)$ or $\bar{F}(0)$ on the boundary, the $\epsilon_+$ state diverges. To get a smooth SDYM limit for external lines one can induce $\Phi(z) \sim \epsilon_+$ from $F_{AA}(0)$ on the boundary, while $\Phi(z)$ in the bulk can be induced by $\Phi(0)$ or $\bar{F}(0)$ and not by $\Psi(0)$ as the latter vanishes in the self-dual limit.  

We will carefully check in all cases the fact that the leading $1/E$-pole gives the flat space amplitude. For Dirichlet boundary conditions this is literally true. For Neumann/mixed boundary conditions, $\gamma\neq0$, one has to remember that the external legs carry the boundary-to-bulk propagators that have additional $1/k$ times a numerical factor, which has to be taken into account. For example, for the Neumann case the statement is
\begin{align}
    k_1...k_n \times \text{leading $E$ pole of AdS correlator}&= \text{flat space amplitude}\,.
\end{align}
Since we always use fixed (Dirichlet) boundary-to-bulk propagators, no additional $k$-factors will be needed. 

We will also check that the Dirichlet results do not depend on the gauge choice. This should give more confidence in the non-Dirichlet correlators. The latter do depend on the gauge for $\sigma=-1$ and do not for $\sigma=+1$. In principle, the correlators correspond either to a non-abelian gauge field $a_i$ on the boundary or to its magnetic field $B=*(da+g\,a\wedge a)$, which is gauge covariant rather than invariant. The composite operator contribution $a\wedge a$ is considered in Appendix \ref{app:sdymleftovers}.

%%%%%%%%%%%%%%%%%%%%%%%%%%%%%%%%%%%%%%%%%%%%%%%%%%%%%%%%%%%%%
\subsection{Three point}
\label{sec:}
%%%%%%%%%%%%%%%%%%%%%%%%%%%%%%%%%%%%%%%%%%%%%%%%%%%%%%%%%%%%%
We begin with the SDYM case as it is the one having fewer diagrams and then proceed to the Yang--Mills theory in the chiral form. The latter is easier to compare to the Chalmers--Siegel action.

%%%%%%%%%%%%%%%%%%%%%%%%%%%%%%%%%%%%%%%%%%%%%%%%%%%%%%%%%%%%%
\subsubsection{SDYM}
\label{sec:SDYM}
%%%%%%%%%%%%%%%%%%%%%%%%%%%%%%%%%%%%%%%%%%%%%%%%%%%%%%%%%%%%%
Recall that the SDYM action in $\text{AdS}_4$ is
\begin{align}
    S=\Tr\int d^3xdz\,\Psi^{AA}H_{AA}F=\Tr\int d^3x\frac{dz}{z^4}\,\Psi^{AA}F_{AA}\,.
\end{align}
We perform a conformal rescaling of the fields 
\begin{align}
    \Psi^{AA} & \rightarrow z^2\,\Psi^{AA} \,, & \Phi_{A,A'} &\rightarrow z\, \Phi_{A,A'}\,.
\end{align}
This brings the action to the same form as the flat-space action, \eqref{SDYMaction}. Consequently, the interaction vertex in $\text{AdS}_4$ is the same as in flat space.
\begin{figure}[h]
\centering
\begin{tabular}{c c c}
    % Boundary-to-bulk propagator (+,-)
    \begin{tikzpicture}
        % Define external points for propagator
        \coordinate (a) at (-2, 2);
        \coordinate (b) at (0, 2);
        \coordinate (boundary) at (-3.5,2);
        
        % Draw boundary-to-bulk propagator
        \draw[decorate,decoration={snake},thick] (a) -- (b);
        
        % Draw arc on the boundary
        \draw[thick] (-1.9,2.7) arc[start angle=160, end angle=200, radius=2cm];
        
        % Labels for fields
        \node[left] at (a) {\( \epsilon^-_i \)};
        \node[right] at (b) {\( (+,AA') \)};
    \end{tikzpicture}
    & \quad \raisebox{0mm}{\quad : \quad} &
    \quad \raisebox{0mm}{\( \Phi^{A,A'}(k_i,z)=\frac{q_i^{A}\bar{i}^{A'}}{\langle q_ii \rangle}e^{-kz} \)}
    \\[1.5cm]
    % Boundary-to-bulk propagator (-,+)
    \begin{tikzpicture}
        % Define external points for propagator
        \coordinate (a) at (-2, -2);
        \coordinate (b) at (0, -2);
        \coordinate (boundary) at (-3.5,-2);
        
        % Draw boundary-to-bulk propagator
        \draw[decorate,decoration={snake},thick] (a) -- (b);
        
        % Draw arc on the boundary
        \draw[thick] (-1.9,-1.2) arc[start angle=160, end angle=200, radius=2cm];
        
        % Labels for fields
        \node[left] at (a) {\( \epsilon^+_i \)};
        \node[right] at (b) {\( (-,AA) \)};
    \end{tikzpicture}
    & \quad \raisebox{0mm}{\quad : \quad} &
    \quad \raisebox{0mm}{\( \Psi^{AA}(k_i,z)=i^Ai^A e^{-kz} \)}
\end{tabular}
\caption{SDYM Feynman rules in $\text{AdS}_4$ (Feynman gauge).}
\label{fig:feynmanRulesAdS}
\end{figure}

The $\text{AdS}_4$ correlator differs from the flat-space amplitude in three key ways. Firstly, the $\text{AdS}_4$ correlator contains a bulk integral over the radial coordinate $z$. Secondly, instead of a polarization vector, we attach boundary-to-bulk propagators, see Figure \ref{fig:feynmanRulesAdS}.\footnote{Naming the boundary-to-bulk propagators is ambiguous, as the $\Phi$- and $\Psi$-field propagate into each other. We choose to name the boundary-to-bulk propagators after the nature of the corresponding field in the bulk.} Lastly, in $\text{AdS}_4$ only three-momentum is conserved,
\begin{align}
    \sum_{i=1}^n \vec{k}^i_{AA'}=0 \,.
\end{align}
One can introduce a virtual null-momentum associated to the boundary three-momenta of the fields in a correlation function and measure the failure of its four-momentum conservation. We will refer to this failure as the total energy $E$.\footnote{We use the same notation for the energy in the case of the three-point function and four-point function in later sections. We assume the distinction is clear from context.} We recall that this can be expressed by constructing a null-vector $k_i^{AA'}+k_i\epsilon^{AA'}=i^A\bar{i}^{A'}$ from a given three-momentum $k_i^{AA'}$. For the three-point correlation function we get
\begin{align} \label{AdSMomCons}
    1^{A}\bar{1}^{A'}+2^A\bar{2}^{A'}+3^A\bar{3}^{A'}= (k_1+k_2+k_3)\epsilon^{AA'}\equiv E\epsilon^{AA'}\,.
\end{align}
Another comment about correlation functions is that they can be computed for generic reference spinors in the Feynman gauge, but in order to ensure a definite boundary helicity structure, one should choose $q_i^A=i^A$.\footnote{For example, some amplitudes in SDYM and SGDR were computed in the light-cone gauge in \cite{Skvortsov:2018uru} (as a part of a more general chiral higher-spin gravity result) and in \cite{Chowdhury:2024dcy}. In \cite{Chowdhury:2024dcy}, these amplitudes were covariantized in such a way that there is a single reference spinor shared by all legs. However, in order to project onto the definite helicity states reference spinors need to be chosen as $q_i^A=\bar{i}^A$ (as in the axial gauge).}

\begin{figure}[h!]
\centering
\begin{tabular}{c c}
    % S-channel Witten diagram
    \begin{tikzpicture}
        % Define the boundary circle
        \draw[thick] (0,0) circle (2.1cm);
        
        % Define the vertices inside the bulk
        \coordinate (v) at (0,0);
        
        % Define the external legs on the boundary
        \coordinate (a) at (-1.48,1.48);
        \coordinate (b) at (1.48,1.48);
        \coordinate (c) at (0,-2.1);
        
        % Draw the gauge boson lines
        \draw[decorate,decoration={snake},thick] (a) -- (v);
        \draw[decorate,decoration={snake},thick] (b) -- (v);
        \draw[decorate,decoration={snake},thick] (c) -- (v);
        
        % Draw parallel momentum arrows
        \draw[-{Latex},thick] (-1.3,1.0) -- (-0.9,0.6) node[midway,below left] {\( k_1 \)};
        \draw[-{Latex},thick] (0.2,-1.6) -- (0.2,-1) node[midway, right] {\( k_2 \)};
        \draw[{Latex}-,thick] (0.9,0.6) -- (1.3,1) node[midway,below right] {\( k_3 \)};
        
        % Add labels for fields next to endpoints
        \node[above left] at (a) {\( \epsilon^+_1 \)};
        \node[above right] at (b) {\( \epsilon^-_3 \)};
        \node[below] at (c) {\( \epsilon^+_2 \)};
        
        % Add labels for vertices
        \node[below right] at (v) {\( z \)};        
    \end{tikzpicture}
\end{tabular}
\caption{SDYM Witten diagram for the three-point correlator.}
\label{fig:SDYMthreepoint}
\end{figure}
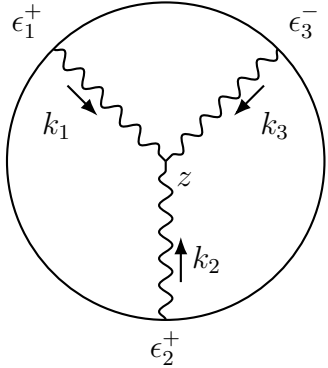

The Witten diagram for the SDYM three-point $(++-)$ correlation function is shown in Figure \ref{fig:SDYMthreepoint}. Due to conformal invariance of the theory, the three-point vertex is the same as in flat space -- see Appendix \ref{sec:flatSDYM3pt} for more details --,
\begin{align}
    V_{AA',BB',CC}=2g\epsilon_{AC}\epsilon_{BC}\epsilon_{A'B'} \,.
\end{align}
The three-point correlation function then gives
\begin{align}
    \begin{aligned}\label{SDYM3pt}
        \mathcal{W}_3^{\text{SDYM}} &= \int_0^{\infty}dz\, V_{AA',BB',CC}\Phi^{AA'}(k_1)\Phi^{BB'}(k_2)\Psi^{CC}(k_3)=\frac{1}{E}\mathcal{A}_3=\\
        &=-\frac{2g}{E}\frac{\langle q_13 \rangle \langle q_23 \rangle}{\langle q_11 \rangle \langle q_22 \rangle} \langle \bar{1}\bar{2} \rangle \,.
    \end{aligned}
\end{align}
Comparing this with the flat space amplitude $\mathcal{A}_3$ that is given in \eqref{SDYM3ptRefSpin}, it is clear that the residue of the energy pole is the flat-space amplitude, i.e.
\begin{align}
    \lim_{E\rightarrow 0} E \, \mathcal{W}_3^{\text{SDYM}} = \mathcal{A}_3\,.
\end{align}
In flat space one can use certain spinor-helicity identities, see \eqref{refSpinors}, to eliminate the reference spinors. In AdS these relations are modified by $\mathcal{O}(E)$ corrections in $\text{AdS}_4$ due to the failure of four-momentum conservation \eqref{AdSMomCons} and read
\begin{align} \label{momConsCorrections}
    \langle q_13 \rangle &= \frac{\langle q_11 \rangle \langle \bar{1}\bar{2} \rangle-E\langle q_1\bar{2} \rangle}{\langle \bar{2}\bar{3} \rangle} \,, & \langle q_23 \rangle &= -\frac{\langle q_22\rangle \langle \bar{1}\bar{2} \rangle+E\langle q_2\bar{1} \rangle}{\langle \bar{1}\bar{3} \rangle}\,.
\end{align}
The correlation function can then be written as
\begin{align}
    \mathcal{W}_3^{\text{SDYM}} &= \frac{2g}{E}\frac{\langle \bar{1}\bar{2} \rangle}{\langle \bar{1}\bar{3}\rangle\langle \bar{2}\bar{3} \rangle}\Big(\langle \bar{1}\bar{2}\rangle^2 + E \langle \bar{1}\bar{2} \rangle \frac{\langle q_11\rangle \langle q_2\bar{1}\rangle - \langle q_1 \bar{2} \rangle \langle q_2 2 \rangle}{\langle q_11 \rangle \langle q_2 2 \rangle}-E^2\frac{\langle q_1\bar{2} \rangle \langle q_2\bar{1} \rangle}{\langle q_11 \rangle \langle q_22 \rangle}\Big) \,,
\end{align}
where we recognize the familiar $q$-independent flat space amplitude \eqref{3ptSDYM} residing in the residue of the energy pole, supplemented by higher order corrections in energy.

The dependence on $q$ represents the fact that the current has other components than with the definite helicity, which are redundant as they can be reconstructed from the conservation. When choosing a definite helicity structure on the boundary, we obtain from \eqref{SDYM3pt}
\begin{align}\label{SDYM3ptAxial}
    \mathcal{W}_3^{\text{SDYM}} = -\frac{g}{2E}\frac{\langle \bar{1}3 \rangle \langle \bar{2}3 \rangle}{k_1 k_2} \langle \bar{1}\bar{2} \rangle\,.
\end{align}

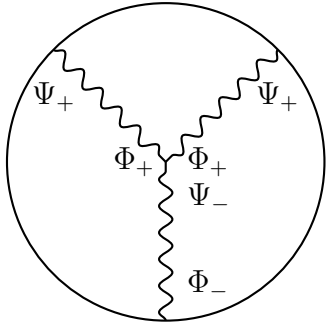
\begin{figure}[h!]
\centering
\begin{tabular}{c c}
    % S-channel Witten diagram
    \begin{tikzpicture}
        % Define the boundary circle
        \draw[thick] (0,0) circle (2.1cm);
        
        % Define the vertices inside the bulk
        \coordinate (v) at (0,0);
        
        % Define the external legs on the boundary
        \coordinate (a) at (-1.48,1.48);
        \coordinate (b) at (1.48,1.48);
        \coordinate (c) at (0,-2.1);
        
        % Draw the gauge boson lines
        \draw[decorate,decoration={snake},thick] (a) -- (v);
        \draw[decorate,decoration={snake},thick] (b) -- (v);
        \draw[decorate,decoration={snake},thick] (c) -- (v);

        \node[below=8pt] at (a) {\( \Psi_+ \)};
        \node[below=8pt] at (b) {\( \Psi_+ \)};
        \node[above right=4pt] at (c) {\( \Phi_- \)};

        \node[left] at (v) {\( \Phi_+ \)};
        \node[right=4pt] at (v) {\( \Phi_+ \)};
        \node[below right=4pt] at (v) {\( \Psi_- \)};
        
        % Add labels for fields next to endpoints
        % \node[above left] at (a) {\( \epsilon^+_1 \)};
        % \node[above right] at (b) {\( \epsilon^-_3 \)};
        % \node[below] at (c) {\( \epsilon^+_2 \)};     
    \end{tikzpicture}
\end{tabular}
\caption{The three-point function in terms of the helicity fields. The vertex allows to change the pair from $\Psi_+-\Phi_+$ to $\Psi_--\Phi_-$.}
\label{fig:helicity}
\end{figure}

\paragraph{Helicity decomposition.} One can compute the same three-point function in terms of the scalars $\Psi_\pm$, $\Phi_\pm$ that define the helicity decomposition of $\Psi^{AA}$ and $\Phi^{A,A'}$ in the complete physical gauge. The diagram in Figure \ref{fig:helicity} takes two $\Phi_+$ off the boundary (they have $D_+$ as the kinetic operator and can propagate).\footnote{Note that due to the off-diagonal kinetic term one is using $\phi(z')=\int dz\, J_\phi(z)\, \langle \psi(z)\phi(z')\rangle $ for iterating the bulk equations and with $J_\phi$ on the boundary $J_\phi=\delta(z) \phi_0$ we find the usual Fefferman-Graham's $\phi(z')=e^{-kz} \phi_0$. This can make an impression that it is $\psi$ that propagates from the boundary, but the boundary value is given to $\phi$. } Then it merges them via $\Psi_-(\Phi_+\Phi_+)$-component of the cubic vertex and propagates to the boundary via $\langle \Psi_-(z)\Phi_-(0)\rangle$ which does not vanish as different from $\langle \Psi_+(z)\Phi_+(0)\rangle=0$. Note that $\Psi_-(\Phi_+\Phi_+)$ does not vanish: explicitly it reads $\epsilon^-_{AA}(k_1)\epsilon_+^{AB}(k_2)\epsilon_+^{AC}(k_3)\epsilon_{BC}\,\Psi_-\Phi_+\Phi_+$. 

%%%%%%%%%%%%%%%%%%%%%%%%%%%%%%%%%%%%%%%%%%%%%%%%%%%%%%%%%%%%%
\subsubsection{(Chiral) Yang--Mills}
\label{sec:cYMAdS}
%%%%%%%%%%%%%%%%%%%%%%%%%%%%%%%%%%%%%%%%%%%%%%%%%%%%%%%%%%%%%
In Section \ref{sec:2.5} we obtain cYM by adding a topological Chern--Simons term to the YM and in Appendix \ref{sec:YMFlat} we derive the associated Feynman rules and scattering amplitudes. This is perturbatively equivalent to YM in flat space. In AdS, however, the topological term remains due to the presence of a boundary and it therefore alters both the propagators and the interaction vertex. 

In Section \ref{sec:YMFlat} and \ref{sec:cYMFlat}, we observe significant simplifications of both the three-point amplitude and the $s$-channel diagram in flat space. The anti-self-dual (ASD) vertex was converted into the SD one by adding a boundary term, which vanishes in flat space. In AdS, this conversion introduces a boundary vertex. We remind ourselves that the topological term can be related to the Chern--Simons term $\text{C}S(A)=AdA+\frac{2}{3}A^3$, which reads
\begin{align}
    \frac{6}{v}\Tr\int d\text{CS}(A) =-\Tr\int \partial_z\,\Big(\Phi^{AA'}k\fdu{A}{A'}\Phi_{AA'}+\frac{2}{3}\Phi^{AA}\Phi\fdu{A}{A'}\Phi_{AA'}\Big) \,.
\end{align}
In the latter expression $\Phi^{AA'}$ is the symmetric, i.e. orthogonal to $n_\mu$, component of $\Phi^{A,A'}$, i.e. $a_i=A_i|_{z=0}$ on the boundary.

Analogously to SDYM, one constructs the boundary-to-bulk propagators for cYM in AdS from the flat-space cYM polarization vectors, see Figure \ref{fig:boundary-to-bulkcYM}.\footnote{Note that we present the correlation functions for the most natural normalization, given in Figure \ref{fig:boundary-to-bulkcYM}. However, one must keep in mind that the normalization should be adjusted for the chosen boundary conditions, as explained at the beginning of this Section.} Due to the conformal invariance of the theory, the SD and ASD vertices are the same as in flat space, see \eqref{vertex1}, and \eqref{vertex2} and Figure \ref{fig:YM&top1}, \ref{fig:YM&top2}, \ref{fig:cYMSD}, and \ref{fig:cYMASD}.
\begin{figure}[h]
\centering
\begin{tabular}{c c c}
    % Boundary-to-bulk propagator (+,-)
    \begin{tikzpicture}
        % Define external points for propagator
        \coordinate (a) at (-2, 2);
        \coordinate (b) at (0, 2);
        \coordinate (boundary) at (-3.5,2);
        
        % Draw boundary-to-bulk propagator
        \draw[decorate,decoration={snake},thick] (a) -- (b);
        
        % Draw arc on the boundary
        \draw[thick] (-1.9,2.7) arc[start angle=160, end angle=200, radius=2cm];
        
        % Labels for fields
        \node[right] at (b) {\( \Phi_+^{A,A'} \)};
    \end{tikzpicture}
    & \quad \raisebox{6mm}{\quad : \quad} &
    \quad \raisebox{6mm}{\( \Phi_+^{A,A'}(k_i,z)=\frac{q_{i}^A\bar{i}^{A'}}{\langle q_i i\rangle}e^{-k_iz} \)}
    \\[1.5cm]
    % Boundary-to-bulk propagator (-,+)
    \begin{tikzpicture}
        % Define external points for propagator
        \coordinate (a) at (-2, -2);
        \coordinate (b) at (0, -2);
        \coordinate (boundary) at (-3.5,-2);
        
        % Draw boundary-to-bulk propagator
        \draw[decorate,decoration={snake},thick] (a) -- (b);
        
        % Draw arc on the boundary
        \draw[thick] (-1.9,-1.2) arc[start angle=160, end angle=200, radius=2cm];
        
        % Labels for fields
        \node[right] at (b) {\( \Phi_-^{A,A'} \)};
    \end{tikzpicture}
    & \quad \raisebox{6mm}{\quad : \quad} &
    \quad \raisebox{6mm}{\( \Phi_-^{A,A'}(k_i,z)=\frac{i^A\bar{q}_i^{A'}}{\langle \bar{q}\bar{i}\rangle} e^{-k_iz} \)}
\end{tabular}
\caption{YM boundary-to-bulk propagators in $\text{AdS}_4$.}
\label{fig:boundary-to-bulkcYM}
\end{figure}

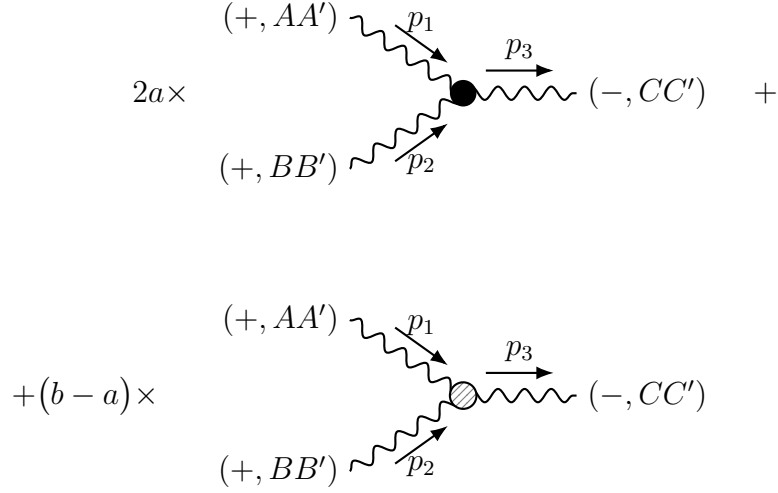
\begin{figure}[h!]
\centering
\begin{tikzpicture}
          
    % Equal sign between top diagrams
    
        \coordinate (v) at (0,0);
        
        % Define external legs
        \coordinate (a) at (-1.5,1);
        \coordinate (b) at (-1.5,-1);
        \coordinate (c) at (1.5,0);
        
        % Draw gauge boson lines 
        \draw[decorate,decoration={snake},thick] (a) -- (v);
        \draw[decorate,decoration={snake},thick] (b) -- (v);
        \draw[decorate,decoration={snake},thick] (v) -- (c);
        
        % Draw parallel momentum arrows 
        \draw[-{Latex},thick] (-0.9,0.9) -- (-0.2,0.4) node[midway,above] {\( p_1 \)};
        \draw[-{Latex},thick] (-0.9,-0.9) -- (-0.2,-0.4) node[midway,below] {\( p_2 \)};
        \draw[-{Latex},thick] (0.3,0.3) --  (1.2,0.3) node[midway,above] {\( p_3 \)};
        
        % Add labels for fields next to endpoints
        \node[left] at (a) {\( (+,AA') \)};
        \node[left] at (b) {\( (+,BB') \)};
        \node[right] at (c) {\( (-,CC') \)};
        
        \fill (v) circle (5pt);
        \node at (-4,0) {$2a\times$};
        \node at (4,0) {$+$};

    \begin{scope}[yshift=-4cm]
        \coordinate (v) at (0,0);

        \node at (-5,0) {$+\big(b-a\big)\times$};
        
        % Define external legs
        \coordinate (a) at (-1.5,1);
        \coordinate (b) at (-1.5,-1);
        \coordinate (c) at (1.5,0);
        
        % Draw gauge boson lines 
        \draw[decorate,decoration={snake},thick] (a) -- (v);
        \draw[decorate,decoration={snake},thick] (b) -- (v);
        \draw[decorate,decoration={snake},thick] (v) -- (c);
        
        % Draw parallel momentum arrows 
        \draw[-{Latex},thick] (-0.9,0.9) -- (-0.2,0.4) node[midway,above] {\( p_1 \)};
        \draw[-{Latex},thick] (-0.9,-0.9) -- (-0.2,-0.4) node[midway,below] {\( p_2 \)};
        \draw[-{Latex},thick] (0.3,0.3) --  (1.2,0.3) node[midway,above] {\( p_3 \)};
        
        % Add labels for fields next to endpoints
        \node[left] at (a) {\( (+,AA') \)};
        \node[left] at (b) {\( (+,BB') \)};
        \node[right] at (c) {\( (-,CC') \)};
        
        \fill[white] (0,0) circle (5pt);
        \draw[black, thick]
        (v) node[draw, circle, minimum size=10pt, inner sep=0pt,
              pattern=north east lines, pattern color=gray] {} ;
    \end{scope}
\end{tikzpicture}
\caption{The full vertex is a linear combination of the SD vertex and the topological one.}
\label{fig:2SDVertexAdS}
\end{figure}

\noindent We begin with the following action
\begin{align} \label{AdSAction}
    S = \frac{a}{4}\,\Tr\int \left(F_{AB}F^{AB} +\bar{F}_{A'B'} \bar{F}^{A'B'}\right)+\frac{b}{4} \,\Tr\int \left(F_{AB}F^{AB} -\bar{F}_{A'B'} \bar{F}^{A'B'}\right) \,,
\end{align}
with arbitrary coefficients $a$ and $b$. Note that we make no distinction between YM and cYM; one can go between the theories by changing the coefficient in front of the topological term. Following Section \ref{sec:2.5}, the action can be decomposed into its self-dual and anti-self-dual components,
\begin{align}\label{actiondecomp1}
    S=\frac{(a+b)}{4}\Tr\int F_{AB}F^{AB}+\frac{(a-b)}{4}\Tr\int \bar{F}_{A'B'} \bar{F}^{A'B'}\,,
\end{align}
or it can be expressed in a more chiral form,
\begin{align}\label{actiondecomp2}
    \begin{aligned}
        S &= \frac{a}{2} \Tr\int F_{AB}F^{AB} + \frac{(b-a)}{4} \,\Tr\int \left(F_{AB}F^{AB} -\bar{F}_{A'B'} \bar{F}^{A'B'}\right)\,.
    \end{aligned}
\end{align}
The former leads to a decomposition of the vertex of the form
\begin{align}
    V = (a+b)V_{\text{SD}}+(a-b)V_{\text{ASD}}
\end{align}
and the latter of the form
\begin{align} \label{abVertex}
    V&= 2aV_{\text{SD}}  + (b-a) V_{\text{top}} \,,
\end{align}
which is also illustrated in Figure \ref{fig:2SDVertexAdS}. The self-dual vertex $V_{\text{SD}}$ originates from the $F^2$ term in the action and the anti-self-dual vertex comes from the $\bar{F}^2$ term. Due to conformal invariance of the theory, the vertices are the same as in flat space, see Appendix \ref{sec:YMFlat} for more details. The above vertices read
\begin{align}
    \begin{aligned}
        V^{\text{SD}}_{AA',BB',CC'}&= \frac{g}{2}\Big[\epsilon_{B'C'}(\epsilon_{AC}p^1_{BA'}+\epsilon_{AB}p^1_{CA'})+\epsilon_{C'A'}(\epsilon_{BA}p^2_{CB'}+\epsilon_{BC}p^2_{AB'})+\\
        &+\epsilon_{A'B'}(\epsilon_{CB}p^3_{AC'}+\epsilon_{CA}p^3_{BC'})\Big] \,,\\
        V^{\text{ASD}}_{AA',BB',CC'}&= \frac{g}{2} \Big[\epsilon_{BC}(\epsilon_{A'C'}p^1_{AB'}+\epsilon_{A'B'}p^1_{AC'})+\epsilon_{CA}(\epsilon_{B'A'}p^2_{BC'}+\epsilon_{B'C'}p^2_{BA'})+\\
        &+\epsilon_{AB}(\epsilon_{C'B'}p^3_{CA'}+\epsilon_{C'A'}p^3_{CB'})\Big] \,.
    \end{aligned}
\end{align}
\noindent The topological vertex is given by
\begin{align}\label{topVertex}
    V^{\text{top}}_{AA',BB',CC'}=V^{\text{SD}}_{AA',BB',CC'}-V^{\text{ASD}}_{AA',BB',CC'}=g(\epsilon_{AC}\epsilon_{A'B'}\epsilon_{BC'}-\epsilon_{AB}\epsilon_{A'C'}\epsilon_{CB'})\partial_z\,.
\end{align}
Together with the bulk integral and the remainder of the diagram represented by $f(z)$, integration by parts is used to observe that the vertex is placed on the boundary as
\begin{align}
    \int_0^\infty dz\,  V_{AA',BB',CC'}^{\text{top}} f(z)= -gf(0)(\epsilon_{AC}\epsilon_{A'B'}\epsilon_{BC'}-\epsilon_{AB}\epsilon_{A'C'}\epsilon_{CB'}) \,,
\end{align}
if $f(z)$ vanishes for $z\rightarrow\infty$.

Despite the additional layer of complexity introduced by the boundary term, following similar steps as in flat space still simplifies the three-point vertex -- albeit in a slightly modified form. Originally, the Feynman rule for the three-point vertex obtained from \eqref{AdSAction} can be interpreted in two ways: i) the total vertex is a linear combination of the YM vertex, which itself contains the SD and ASD vertex, and the topological vertex; ii) upon decomposing the topological vertex, the total vertex is just a linear combination of the SD and ASD vertex.

The second interpretation is clearly the simplest of the two. However, after the rewriting of the action as in \eqref{actiondecomp2}, the total vertex becomes a linear combination of the SD and topological vertex. At first glance, this might not seem advantageous compared to ii), since both decompositions contain two components. Nevertheless, there are a few reasons to favor the topological vertex over the ASD one.

Firstly, the topological vertex is a total derivative and eliminates the need for a bulk integral; secondly, as a result, it does not contribute to the energy pole of the correlator -- which is significant, as it is well-known \cite{Maldacena:2011nz} that the flat-space amplitude is contained in the leading order energy pole of the AdS correlator -- and this decomposition explicitly shows that the AdS correction to the vertex cannot spoil this; third, while the ASD vertex decomposes into three distinct terms depending on the placement of the derivative, the topological vertex is derivative-free and therefore structurally simpler.

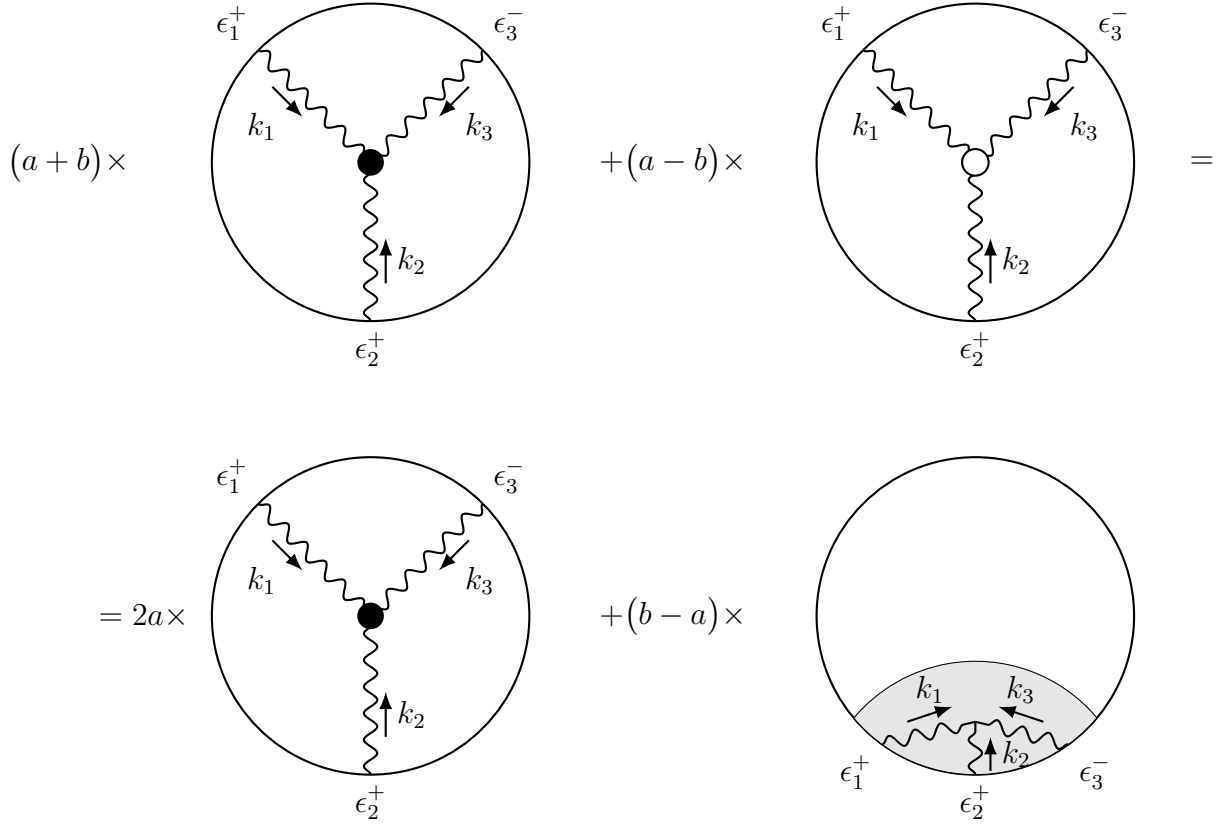
\begin{figure}[h!]
\centering
\begin{tabular}{c c}
    % S-channel Witten diagram
    \begin{tikzpicture}
        
            \begin{scope}[xshift=-8cm,yshift=-6cm]
            \node at (-4,0) {$\big(a+b\big)\times$};
                % Define the boundary circle
            \draw[thick] (0,0) circle (2.1cm);
            
            % Define the vertices inside the bulk
            \coordinate (v) at (0,0);
            
            % Define the external legs on the boundary
            \coordinate (a) at (-1.48,1.48);
            \coordinate (b) at (1.48,1.48);
            \coordinate (c) at (0,-2.1);
            
            % Draw the gauge boson lines
            \draw[decorate,decoration={snake},thick] (a) -- (v);
            \draw[decorate,decoration={snake},thick] (b) -- (v);
            \draw[decorate,decoration={snake},thick] (c) -- (v);
            
            % Draw parallel momentum arrows
            \draw[-{Latex},thick] (-1.3,1.0) -- (-0.9,0.6) node[midway,below left] {\( k_1 \)};
            \draw[-{Latex},thick] (0.2,-1.6) -- (0.2,-1) node[midway, right] {\( k_2 \)};
            \draw[{Latex}-,thick] (0.9,0.6) -- (1.3,1) node[midway,below right] {\( k_3 \)};
            
            % Add labels for fields next to endpoints
            \node[above left] at (a) {\( \epsilon^+_1 \)};
            \node[above right] at (b) {\( \epsilon^-_3 \)};
            \node[below] at (c) {\( \epsilon^+_2 \)};

            \node at (4,0) {$+\big(a-b\big)\times$};
            \fill (0,0) circle (5pt);
            \end{scope}
            \begin{scope}[yshift=-6cm]
            % Define the boundary circle
            \draw[thick] (0,0) circle (2.1cm);
            
            % Define the vertices inside the bulk
            \coordinate (v) at (0,0);
            
            % Define the external legs on the boundary
            \coordinate (a) at (-1.48,1.48);
            \coordinate (b) at (1.48,1.48);
            \coordinate (c) at (0,-2.1);
            
            % Draw the gauge boson lines
            \draw[decorate,decoration={snake},thick] (a) -- (v);
            \draw[decorate,decoration={snake},thick] (b) -- (v);
            \draw[decorate,decoration={snake},thick] (c) -- (v);
            
            % Draw parallel momentum arrows
            \draw[-{Latex},thick] (-1.3,1.0) -- (-0.9,0.6) node[midway,below left] {\( k_1 \)};
            \draw[-{Latex},thick] (0.2,-1.6) -- (0.2,-1) node[midway, right] {\( k_2 \)};
            \draw[{Latex}-,thick] (0.9,0.6) -- (1.3,1) node[midway,below right] {\( k_3 \)};
            
            % Add labels for fields next to endpoints
            \node[above left] at (a) {\( \epsilon^+_1 \)};
            \node[above right] at (b) {\( \epsilon^-_3 \)};
            \node[below] at (c) {\( \epsilon^+_2 \)}; 
            \fill[white] (0,0) circle (5pt);
            \draw[thick] (0,0) circle (5pt);
            \node at (3,0) {$=$};
            \end{scope}

            \begin{scope}[yshift=-6cm]
                \begin{scope}[xshift=-8cm,yshift=-6cm]
            \node at (-3,0) {$=2a\times$};
                % Define the boundary circle
            \draw[thick] (0,0) circle (2.1cm);
            
            % Define the vertices inside the bulk
            \coordinate (v) at (0,0);
            
            % Define the external legs on the boundary
            \coordinate (a) at (-1.48,1.48);
            \coordinate (b) at (1.48,1.48);
            \coordinate (c) at (0,-2.1);
            
            % Draw the gauge boson lines
            \draw[decorate,decoration={snake},thick] (a) -- (v);
            \draw[decorate,decoration={snake},thick] (b) -- (v);
            \draw[decorate,decoration={snake},thick] (c) -- (v);
            
            % Draw parallel momentum arrows
            \draw[-{Latex},thick] (-1.3,1.0) -- (-0.9,0.6) node[midway,below left] {\( k_1 \)};
            \draw[-{Latex},thick] (0.2,-1.6) -- (0.2,-1) node[midway, right] {\( k_2 \)};
            \draw[{Latex}-,thick] (0.9,0.6) -- (1.3,1) node[midway,below right] {\( k_3 \)};
            
            % Add labels for fields next to endpoints
            \node[above left] at (a) {\( \epsilon^+_1 \)};
            \node[above right] at (b) {\( \epsilon^-_3 \)};
            \node[below] at (c) {\( \epsilon^+_2 \)};

            \node at (4,0) {$+\big(b-a\big)\times$};
            \fill (0,0) circle (5pt);
            \end{scope}
            \begin{scope}[yshift=-6cm]
            % Define the boundary circle
            \draw[thick] (0,0) circle (2.1cm);

            \filldraw[fill=gray!20, draw=black] 
            (1.61,-1.35) arc[start angle=40, end angle=140, radius=2.1cm];
            \filldraw[fill=gray!20, draw=black] 
            (-1.61,-1.35) arc[start angle=220, end angle=320, radius=2.1cm];
            
            % Define the vertices inside the bulk
            \coordinate (v) at (0,-1.4);
            
            % Define the external legs on the boundary
            \coordinate (a) at (-1.23,-1.7);
            \coordinate (b) at (0,-2.1);
            \coordinate (c) at (1.23,-1.7);
            
            % Draw the gauge boson lines
            \draw[decorate,decoration={snake},thick] (a) -- (v);
            \draw[decorate,decoration={snake},thick] (b) -- (v);
            \draw[decorate,decoration={snake},thick] (c) -- (v);
            
            % Add labels for fields next to endpoints
            \node[below left] at (a) {\( \epsilon^+_1 \)};
            \node[below] at (b) {\( \epsilon^+_2 \)};
            \node[below right] at (c) {\( \epsilon^-_3 \)};

             % Draw parallel momentum arrows
            \draw[-{Latex},thick] (-0.9,-1.4) -- (-0.3,-1.2) node[midway,above] {\( k_1 \)};
            \draw[-{Latex},thick] (0.2,-2.05) -- (0.2,-1.6) node[midway, right] {\( k_2 \)};
            \draw[-{Latex},thick] (0.9,-1.4) -- (0.3,-1.2) node[midway,above] {\( k_3 \)};

            \end{scope}
            \end{scope}
    \end{tikzpicture}
\end{tabular}
\caption{The YM three-point correlator is a linear combination of the SD and ASD vertex, see top row. Alternatively, it can be expressed as a linear combination of the SD and topological vertex, as shown in the bottom row. Here, the topological vertex is placed on the boundary, which is visually extended to the gray region.}
\label{fig:YMthreepoint}
\end{figure}

Despite the advantage of computing the full YM correlator from only the SD and topological vertex, we will derive it using both strategies and verify that they agree. The three-point correlation function is represented schematically in Figure \ref{fig:YMthreepoint}. Since the topological vertex takes place on the boundary, it is placed on the gray region that visually extends the AdS boundary. The previously used visual distinction at the junction of the three legs is omitted, since it should be clear that only the topological vertex lies on the boundary. The boundary-to-boundary propagators that are connected to the topological vertex, are just the associated polarization vectors, which are simply the boundary-to-bulk propagators with $z=0$.

\noindent The correlation function computed with the SD vertex yields
\begin{align}
    \begin{aligned}
        \mathcal{W}_3^{\text{SD}} &= \int_0^\infty dz V^{\text{SD}}_{AA',BB',CC'}\Phi_+^{AA'}(k_1)\Phi_+^{BB'}(k_2)\Phi_-^{CC'}(k_3)=-\frac{g}{E}\frac{\langle q_13\rangle \langle q_23 \rangle}{\langle q_11 \rangle \langle q_22 \rangle} \langle \bar{1}\bar{2} \rangle\,,
    \end{aligned}
\end{align}
which is just $2W^{\text{SDYM}}$, \eqref{SDYM3pt}. The ASD one gives
\begin{align}
    \begin{aligned}
                \mathcal{W}_3^{\text{ASD}} &= \int_0^\infty dz V^{\text{ASD}}_{AA',BB',CC'}\Phi_+^{AA'}(k_1)\Phi_+^{BB'}(k_2)\Phi_-^{CC'}(k_3)\\
                &=g\frac{\langle \bar{1}\bar{2} \rangle}{E}\frac{\langle q_11\rangle \langle q_23 \rangle \langle \bar{q}_3\bar{1} \rangle + \langle q_13 \rangle \langle q_22 \rangle \langle \bar{q}_3\bar{2} \rangle}{\langle q_11\rangle \langle q_22\rangle\langle\bar{q}_3\bar{3}\rangle} \,.
    \end{aligned}
\end{align}
Similar to the SDYM correlation function, we observe that the reference spinors cannot be completely removed, as the four-momentum is not conserved in AdS. We again use the identities \eqref{momConsCorrections} to massage the self-dual and anti-self-dual correlation functions into
\begin{align}
    \begin{aligned}
        \mathcal{W}_3^{\text{SD}} &= \frac{g}{E}\frac{\langle \bar{1}\bar{2} \rangle}{\langle \bar{1}\bar{3}\rangle\langle \bar{2}\bar{3} \rangle}\Big(\langle \bar{1}\bar{2}\rangle^2 + E \langle \bar{1}\bar{2} \rangle \frac{\langle q_11\rangle \langle q_2\bar{1}\rangle - \langle q_1 \bar{2} \rangle \langle q_2 2 \rangle}{\langle q_11 \rangle \langle q_2 2 \rangle}-E^2\frac{\langle q_1\bar{2} \rangle \langle q_2\bar{1} \rangle}{\langle q_11 \rangle \langle q_22 \rangle}\Big)\,,\\
        \mathcal{W}_3^{\text{ASD}} &= \frac{g}{E}\frac{\langle \bar{1}\bar{2} \rangle}{\langle \bar{1}\bar{3} \rangle \langle \bar{2}\bar{3} \rangle}\Big(\langle \bar{1}\bar{2} \rangle^2 - E\frac{\langle q_11 \rangle \langle q_2\bar{1} \rangle \langle \bar{q}_3\bar{1}\rangle \langle \bar{2}\bar{3} \rangle + \langle q_1\bar{2} \rangle \langle q_22 \rangle \langle \bar{q}_3\bar{2} \rangle \langle \bar{1}\bar{3} \rangle}{\langle q_11 \rangle \langle q_22 \rangle \langle \bar{q}_3 \bar{3} \rangle }\Big) \,.
    \end{aligned}
\end{align}
This form reveals the $E$-expansion and the flat limit as the leading pole. Meanwhile, the diagram containing the topological vertex yields
\begin{align}\label{Wtop}
    \begin{aligned}
        \mathcal{W}_3^{\text{top}} &=\int_0^\infty dz V^{\text{top}}_{AA',BB',CC'}\Phi_+^{AA'}(k_1)\Phi_+^{BB'}(k_2)\Phi_-^{CC'}(k_3) =\\
        &=g\frac{\langle q_13\rangle \langle \bar{1}\bar{2} \rangle \langle q_2\bar{q}_3 \rangle - \langle q_1q_2 \rangle \langle \bar{q}_3\bar{1}\rangle \langle \bar{2}3 \rangle}{\langle q_11\rangle \langle q_22 \rangle \langle \bar{q}_3\bar{3} \rangle}\rightarrow-\frac{g}{4}\frac{\langle \bar{1}\bar{2} \rangle \langle \bar{1}3 \rangle \langle \bar{2}3 \rangle}{k_1k_2k_3}\,,
    \end{aligned}
\end{align}
where the last form is obtained by using the reference spinors of the axial gauge. 
As a consistency check, it was confirmed that $\mathcal{W}^{\text{SD}}_3-\mathcal{W}^{\text{ASD}}_3=\mathcal{W}^{\text{top}}_3$, which requires applying the Fierz identity and momentum conservation several times. With this, the correlation function for YM theory can be obtained in two equivalent ways, as follows from \eqref{actiondecomp1} and \eqref{actiondecomp2}. They read
\begin{align}
    \begin{aligned}
        \mathcal{W}_3 &= (a+b)\mathcal{W}_3^{\text{SD}} + (a-b)\mathcal{W}_3^{\text{ASD}} = 2a\mathcal{W}_3^{\text{SD}}+(b-a)\mathcal{W}_3^{\text{top}}\,, 
    \end{aligned}
\end{align}
respectively. In addition, $\mathcal{W}_3^{\text{SD}}=2\mathcal{W}^{\text{SDYM}}$ in our normalization. Now, it is easy to see that the residue of the energy pole gives the flat-space amplitude, while the topological term contains no energy pole, i.e.
\begin{align}
    \lim_{E \rightarrow 0} E\mathcal{W}_3 &= a\mathcal{A}_3 \,, & \lim_{E \rightarrow 0} E\mathcal{W}_{\text{top}} &= 0 \,.
\end{align}
This is expected as the topological term is a boundary contribution and has no analogue in flat space. Choosing a definite helicity structure, the SD and ASD correlators become
\begin{align}
        \mathcal{W}_3^{\text{SD}} &= -\frac{g}{4E}\frac{k_3}{k_1k_2} \langle \bar{1}\bar{2} \rangle\langle \bar{1}3\rangle \langle \bar{2}3 \rangle\,, &
        \mathcal{W}_3^{\text{ASD}} &=\frac{g}{4E}\frac{k_1+k_2}{k_1k_2k_3}\langle \bar{1}\bar{2} \rangle \langle \bar{1}3\rangle \langle \bar{2}3 \rangle \,.
\end{align}
Using the definition of the energy, \eqref{AdSMomCons}, it is straightforward to see that the difference $\mathcal{W}_3^{\text{SD}}-\mathcal{W}_3^{\text{ASD}}$ does not contribute to the energy pole:
\begin{align}
    \mathcal{W}_3^{\text{SD}}-\mathcal{W}_3^{\text{ASD}} = -\frac{g}{4}\frac{\langle \bar{1}\bar{2}\rangle \langle \bar{1}3\rangle \langle \bar{2}3 \rangle}{k_1k_2k_3} \,.
\end{align}
As a consistency check, we note that this is in agreement with the result \eqref{Wtop}, which is computed directly using the topological vertex.

%%%%%%%%%%%%%%%%%%%%%%%%%%%%%%%%%%%%%%%%%%%%%%%%%%%%%%%%%%%%%
\subsection{Four point}
\label{sec:fourAdS}
%%%%%%%%%%%%%%%%%%%%%%%%%%%%%%%%%%%%%%%%%%%%%%%%%%%%%%%%%%%%%
In this section, we compute the four-point correlation functions for SDYM, cYM and YM in $\text{AdS}_4$. We begin with the SDYM correlation function in Lorenz/Feynman gauge,\footnote{The Lorenz and Feynman gauge are really different at the YM level, but the $p_\mu p_\nu/p^2$-term disappears when passing to the $\Psi-\Phi$-propagator.} which serves as a convenient reference point to express the results in the other theories. To keep the notation streamlined, we will omit labels indicating which channel an exchange diagram is evaluated in, unless specified otherwise. By default, we study the $s$-channel diagram from which the $t$- and $u$-channel diagrams can be obtained by permutation of the external states. More details on color factors can be found in Appendix \ref{app:color}. In each of the mentioned theories we consider diagrams with and without boundary vertices, which we refer to as topological and non-topological diagrams, respectively.

%%%%%%%%%%%%%%%%%%%%%%%%%%%%%%%%%%%%%%%%%%%%%%%%%%%%%%%%%%%%%
\subsubsection{SDYM Lorenz}
\label{sec:4ptSDYM}
%%%%%%%%%%%%%%%%%%%%%%%%%%%%%%%%%%%%%%%%%%%%%%%%%%%%%%%%%%%%%
The four-point function for SDYM consists of just the exchange diagram, since SDYM contains no four-point vertex. The $s$-channel exchange diagram is depicted in Figure \ref{fig:schannelAdS}. On top of the Feynman rules described in Figure \ref{fig:feynmanRulesAdS}, this diagram requires the bulk-to-bulk propagator, which is displayed in Figure \ref{fig:Lorenzbulktobulk}. The bulk-to-bulk propagator in Lorenz gauge reads
\begin{align} \notag
  %  \begin{aligned}
        \langle \Psi_{AA}(-k,z)\Phi_{B,B'}(k,z')\rangle_{\text{L}}
        &=-\frac{\epsilon_{AB}}{2k}(k_{AB'}-\sign(z-z')k\epsilon_{AB'})e^{-k|z-z'|}+\sigma\frac{k_{AA}}{4k^2} \nabla_{BB'}^{k,z'}e^{-k(z+z')}
  %  \end{aligned}
\end{align}
and it admits the self-dual boundary condition by construction. We recall that $\sigma=\pm1$ and encodes boundary conditions on the unphysical components.

\begin{figure}[h!]
\centering
\begin{tabular}{c c}
    % S-channel Witten diagram
    \begin{tikzpicture}
        % Define the boundary circle
        \draw[thick] (0,0) circle (2.1cm);
        
        % Define the vertices inside the bulk
        \coordinate (v1) at (-0.7,0);
        \coordinate (v2) at (0.7,0);
        
        % Define the external legs on the boundary
        \coordinate (a) at (-1.7,1.2);
        \coordinate (b) at (-1.7,-1.2);
        \coordinate (c) at (1.7,1.2);
        \coordinate (d) at (1.7,-1.2);
        
        % Draw the gauge boson lines
        \draw[decorate,decoration={snake},thick] (a) -- (v1);
            \draw[decorate,decoration={snake},thick] (b) -- (v1);
        \draw[decorate,decoration={snake},thick] (c) -- (v2);
        \draw[decorate,decoration={snake},thick] (d) -- (v2);
        
        % Draw the internal propagator 
        \draw[decorate,decoration={snake},thick] (v1) -- (v2);
        
        % Draw parallel momentum arrows
        \draw[-{Latex},thick] (-1.3,1.0) -- (-0.9,0.6) node[midway,above right] {\( k_1 \)};
        \draw[-{Latex},thick] (-1.3,-1) -- (-0.9,-0.6) node[midway,below right] {\( k_2 \)};
        \draw[{Latex}-,thick] (0.9,0.6) -- (1.3,1) node[midway,above left] {\( k_4 \)};
        \draw[{Latex}-,thick] (0.9,-0.6) -- (1.3,-1) node[midway,below left] {\( k_3 \)};
        \draw[-{Latex},thick] (-0.2,0.2) -- (0.2,0.2) node[midway,above] {\( k_1+k_2 \)};
        
        % Add labels for fields next to endpoints
        \node[left] at (a) {\( \epsilon^+_1 \)};
        \node[left] at (b) {\( \epsilon^+_2 \)};
        \node[right] at (c) {\( \epsilon^-_4 \)};
        \node[right] at (d) {\( \epsilon^+_3 \)};
        
        % Add labels for vertices
        \node[below] at (v1) {\( z \)};
        \node[shift={(-1mm,-2mm)}] at (v2) {\( z' \)};
        
    \end{tikzpicture}
    & \raisebox{20mm}{\(\quad : \quad\)}
\end{tabular}

\vspace{0.5cm}

\[
\begin{aligned}
\mathcal{W}_s = \int_{0}^{\infty} dz\int_0^\infty dz' & \Phi^{A,A'}(k_1,z)\Phi^{B,B'}(k_2,z)V\fdu{AA',BB',}{EE} \, \langle \Psi_{EE}(-k,z)\Phi_{F,F'}(k,z')\rangle_\text{L} \times\, \\
&\times V\fud{FF'}{,CC',DD}\Phi^{C,C'}(k_3,z')\Psi^{DD}(k_4,z') 
\end{aligned}
\]

\caption{SDYM Lorenz gauge $\text{AdS}_4$ $s$-channel Witten diagram.}
\label{fig:schannelAdS}
\end{figure}

\begin{figure}[h]
\centering
\begin{tabular}{c c c}
    % Bulk-to-bulk propagator
    \begin{tikzpicture}
        \coordinate (a) at (-1.5,0);
        \coordinate (b) at (1.5,0);
        
        % Draw wavy propagator line
        \draw[decorate,decoration={snake},thick] (a) -- (b);

        % Draw momentum arrow
        \draw[-{Latex},thick] (-0.6,0.3) -- (0.6,0.3) node[midway,above] {\( k \)};
        
        % Labels for fields
        \node[left] at (a) {\( (-,AB) \)};
        \node[right] at (b) {\( (+,CC') \)};
    \end{tikzpicture}
    & \raisebox{3mm}{\quad : \quad} &
    \raisebox{3mm}{\( 
    \begin{aligned}
    &\langle \Psi_{AA}(-k,z) \Phi_{B,B'}(k,z')\rangle_\text{L}
    \end{aligned}\)}
\end{tabular}
\caption{SDYM Lorenz gauge bulk-to-bulk propagator in $\text{AdS}_4$.}
\label{fig:Lorenzbulktobulk}
\end{figure}

One crucial difference between flat space and $\text{AdS}_4$ is that propagators in the latter depend on the radial coordinate and one must perform an integral over it. In fact, the $s$-channel diagram contains the radial coordinates of two vertices, $z$ and $z'$. In order to relate the correlation function to the flat-space amplitude, it is convenient to first perform the relevant integrals in the diagram. Thus, we integrate the bulk-to-bulk propagator multiplied by the exponentials from the boundary-to-bulk propagators. The relevant integrals are listed in Appendix \ref{app:fourier}.
Here the energy $E$ is defined through\footnote{Although the same notation for the energy $E$ is used for the three-point and four-point function, their difference should be clear from the context.}
\begin{align}
    1^A\bar{1}^{A'}+2^A\bar{2}^{A'}+3^A\bar{3}^{A'}+4^A\bar{4}^{A'}=E\epsilon^{AA'}
\end{align}
and
\begin{align}
    E_\text{L} &= k_1+k_2+k \,, & E_\text{R} &= k_3 +k_4 +k  && k=|\vec k_1+\vec k_2|=|\vec k_3+\vec k_4|\,.
\end{align}
We find
\begin{align}\label{int2}
    \begin{aligned}
        &\int_0^\infty dz \int_0^\infty dz' e^{-(k_1+k_2)z}  \langle \Psi_{AA}(-k,z)\Phi_{B,B'}(k,z')\rangle_\text{L} e^{-(k_3+k_4)z'} =\\
        &= \frac{\epsilon_{AB}}{EE_\text{L}E_\text{R}}(3_A\bar{3}_{B'}+4_A\bar{4}_{B'})-\frac{k_Ak_A\bar{k}_B\bar{k}_{B'}}{4k^2E_\text{L}E_\text{R}}+(1+\sigma)\frac{k_A\bar{k}_Ak_B\bar{k}_{B'}}{4k^2E_\text{L}E_\text{R}}\,.
    \end{aligned}
\end{align}
Here, we recognize the physical momentum of the internal line in flat space. Note that due to the energy non-conservation, $(3_A\bar{3}_{B'}+4_A\bar{4}_{B'})$ is not equal to $-(1_A\bar{1}_{B'}+2_A\bar{2}_{B'})$. It is worth mentioning that in order to get such ``reminiscent of flat space'' expressions both the inhomogeneous and homogeneous parts of the propagator conspired nicely together, see Appendix \ref{app:fourier}. With the relation
\begin{align} \label{ELER}
    E_\text{L}E_\text{R}=-\langle 12 \rangle \langle \bar{1}\bar{2} \rangle +EE_\text{L} \,,
\end{align}
one easily observes that the leading energy pole of \eqref{int2} contains the flat-space propagator in the flat limit,
\begin{align}
    \lim_{E \rightarrow 0} E\int e^{-(k_1+k_2)z} \langle \Psi_{AA}(-k,z)\Phi_{B,B'}(k,z')\rangle_{\text{L}} e^{-(k_3+k_4)z'} = \langle \Psi_{AA}(-p)\Phi_{B,B'}(p)\rangle_{\text{flat}}\,.
\end{align}
Note that $-\langle 12 \rangle \langle \bar{1}\bar{2} \rangle$ is the Mandelstam $s$ in flat space. Also note that due to $E_\text{L}E_\text{R}$ appearing in the denominator, there are infinitely many $E^{\bullet>0}$-terms in the expansion. Since the vertices and polarizations are the same as in flat space, it then immediately follows that
\begin{align}
    \lim_{E \rightarrow 0}E \, \mathcal{W}_s^{\text{SDYM}}=\mathcal{A}_s^{\text{SDYM}} \,,
\end{align}
which holds channel by channel and we remember that the $(+++-)$-amplitude vanishes in flat space. At arbitrary energy, the $s$-channel of the $\text{AdS}_4$ correlation function reads 
\begin{align} \label{WSDYM}
    \begin{aligned}
        \mathcal{W}_{s,\text{L}}^{\text{SDYM}} &= -\frac{4g^2}{EE_\text{L}E_\text{R}}\frac{\langle \bar{1}\bar{2}\rangle \langle q_34 \rangle}{\langle q_11\rangle \langle q_2 2\rangle \langle q_3 3\rangle}\Big[\langle q_14 \rangle \langle q_24 \rangle \langle \bar{3}\bar{4} \rangle +\\
        &+\frac{E}{4k^2}\Big(\langle q_1q_2|kk\bar{k}\bar{k}|\bar{3}4\rangle-\frac{1+\sigma}{2}\big(\langle q_1q_2|k\bar{k}\bar{k}k|\bar{3}4\rangle+\langle q_1q_2|\bar{k}k\bar{k}k|\bar{3}4\rangle\big)\Big)\Big] \,,
    \end{aligned}
\end{align}
where we defined
\begin{align}\label{abmnrscd}
    \langle ab|mnrs|cd\rangle=a^Ab^Bc^Cd^Dm_An_Br_Cs_D\,.
\end{align}
Specializing to a definite helicity structure, i.e. choosing $q_i^A=\bar{i}^A$, one obtains
\begin{align}\label{SDYM4ptLorenzAx}
    \begin{aligned}
        \mathcal{W}_{s,\text{L}}^{\text{SDYM}} &= -\frac{g^2}{2}\frac{1}{EE_\text{L}E_\text{R}}\frac{\langle \bar{1}\bar{2}\rangle \langle \bar{3}4 \rangle}{k_1k_2k_3}\Big[\langle \bar{1}4 \rangle \langle \bar{2}4 \rangle \langle \bar{3}\bar{4} \rangle +\frac{E}{4k^2}\langle \bar{1}\bar{2}|kk\bar{k}\bar{k}|\bar{3}4\rangle\Big] +\\
        &-(1+\sigma)\frac{g^2}{8E_\text{L}}\frac{\langle \bar{1}\bar{2} \rangle^2\langle \bar{3}4\rangle^2}{k_1k_2k_3}\frac{k_1-k_2}{k^2}\,.
    \end{aligned}
\end{align}
The last term on the first line can be expressed in terms of only boundary momentum spinors through\footnote{This follows from some algebra after writing $\langle \bar{1}\bar{2}|kk\bar{k}\bar{k}|\bar{3}4\rangle=\bar{1}^A\bar{2}^A\bar{3}^B4^B(k^1_{AB}+k^2_{AB}+k\epsilon_{AB})(k^1_{AB}+k^2_{AB}+k\epsilon_{AB})$, where $k_{A}\bar{k}_{A'}=k_{AA'}+k\epsilon_{AA'}$ was used.}
\begin{align} \label{bdymomspin}
    \begin{aligned}
        \langle \bar{1}\bar{2} | kk\bar{k}\bar{k} |\bar{3}4 \rangle &= k \Big(\langle \bar{1}\bar{3} \rangle \big(\langle \bar{2}\bar{3} \rangle \langle 34 \rangle - 2\langle 1 \bar{2} \rangle \langle \bar{1}4 \rangle\big) - \langle \bar{2}4 \rangle \big(\langle \bar{1}4 \rangle \langle \bar{3} \bar{4} \rangle - 2\langle \bar{1}2 \rangle \langle \bar{2}\bar{3} \rangle\big) \Big)+ \\
        &+k^2 \big(\langle \bar{1}4 \rangle \langle \bar{2}\bar{3} \rangle + \langle \bar{1}\bar{3} \rangle \langle \bar{2} 4 \rangle\big) - (k_1-k_2)(k_3+k_4)\langle \bar{1}\bar{2} \rangle \langle \bar{3}4 \rangle \,.
    \end{aligned}
\end{align}
The $t$-channel diagram is obtained by swapping $1 \leftrightarrow 3$, $\bar{1} \leftrightarrow \bar{3}$ and $q_1 \leftrightarrow q_3$. Reinstating labels that denote which channel various quantities belong to, the total correlation function is then found to be
\begin{align}
    \begin{aligned}
       & \mathcal{W}_{s,\text{L}}^{\text{SDYM}}+\mathcal{W}_{t,\text{L}}^{\text{SDYM}} =\\
        &= \frac{4g^2}{E_\text{L}^sE_\text{R}^sE_\text{L}^tE_\text{R}^t}\frac{1}{\langle q_11\rangle \langle q_2 2\rangle \langle q_3 3\rangle}\Big[\langle q_14 \rangle \langle q_24 \rangle \langle q_34 \rangle \Big(\langle \bar{1}\bar{2} \rangle \langle \bar{2}\bar{3} \rangle \frac{\langle 23 \rangle \langle \bar{3}\bar{4} \rangle - \langle 12 \rangle \langle \bar{1}\bar{4} \rangle}{E}+\\
        &\qquad\qquad\qquad\qquad\qquad\qquad\qquad\qquad\qquad\qquad\quad+E_\text{L}^s\langle \bar{1}\bar{4} \rangle \langle \bar{2}\bar{3} \rangle -E_\text{L}^t\langle \bar{1}\bar{2} \rangle \langle \bar{3}\bar{4} \rangle  \Big)+\\
        &+\frac{E_\text{L}^sE_\text{R}^s}{4k_t^2}\langle q_14 \rangle \langle \bar{2}\bar{3} \rangle \Big(\langle q_2q_3|k_tk_t\bar{k}_t\bar{k}_t|\bar{1}4\rangle-\frac{1+\sigma}{2}\big(\langle q_2q_3|\bar{k}_tk_t\bar{k}_tk_t|\bar{1}4\rangle+\langle q_2q_3|k_t\bar{k}_t\bar{k}_tk_t|\bar{1}4\rangle\big)\Big)+\\
        &-\frac{E_\text{L}^tE_\text{R}^t}{4k_s^2}\langle \bar{1}\bar{2} \rangle \langle q_34 \rangle\Big( \langle q_1q_2|k_sk_s\bar{k}_s\bar{k}_s|\bar{3}4\rangle-\frac{1+\sigma}{2} \big(\langle q_1q_2|k_s\bar{k}_s\bar{k}_sk_s|\bar{3}4 \rangle +\langle q_1q_2|\bar{k}_sk_s\bar{k}_sk_s|\bar{3}4 \rangle \big)\Big)\Big]\,.
    \end{aligned}
\end{align}
Only the first term contains an energy pole and we see that its residue is the flat space amplitude \eqref{As+At} in the flat limit $E\rightarrow 0$. All other terms are AdS corrections and are subleading in energy. In flat space the sum of the $s$-channel and $t$-channel amplitude vanishes, which can be seen by using momentum conservation, c.f. \eqref{pcons}. In AdS this does not vanish and it adds a power of $E$, which mixes the flat space amplitude with the AdS corrections. Applying energy momentum then yields

\begin{align}\label{SDYM4ptComplete}
    \begin{aligned}
       & \mathcal{W}_{s,\text{L}}^{\text{SDYM}}+\mathcal{W}_{t,\text{L}}^{\text{SDYM}} =\\
        &= \frac{4g^2}{E_\text{L}^sE_\text{R}^sE_\text{L}^tE_\text{R}^t}\frac{1}{\langle q_11\rangle \langle q_2 2\rangle \langle q_3 3\rangle}\Big[\langle q_14 \rangle \langle q_24 \rangle \langle q_34 \rangle \Big(\langle \bar{1}\bar{2} \rangle \langle \bar{2}\bar{3} \rangle \langle 2\bar{4} \rangle+E_\text{L}^s\langle \bar{1}\bar{4} \rangle \langle \bar{2}\bar{3} \rangle -E_\text{L}^t\langle \bar{1}\bar{2} \rangle \langle \bar{3}\bar{4} \rangle  \Big)+\\
        &+\frac{E_\text{L}^sE_\text{R}^s}{4k_t^2}\langle q_14 \rangle \langle \bar{2}\bar{3} \rangle \Big(\langle q_2q_3|k_tk_t\bar{k}_t\bar{k}_t|\bar{1}4\rangle-\frac{1+\sigma}{2}\big(\langle q_2q_3|\bar{k}_tk_t\bar{k}_tk_t|\bar{1}4\rangle+\langle q_2q_3|k_t\bar{k}_t\bar{k}_tk_t|\bar{1}4\rangle\big)\Big)+\\
        &-\frac{E_\text{L}^tE_\text{R}^t}{4k_s^2}\langle \bar{1}\bar{2} \rangle \langle q_34 \rangle\Big( \langle q_1q_2|k_sk_s\bar{k}_s\bar{k}_s|\bar{3}4\rangle-\frac{1+\sigma}{2} \big(\langle q_1q_2|k_s\bar{k}_s\bar{k}_sk_s|\bar{3}4 \rangle +\langle q_1q_2|\bar{k}_sk_s\bar{k}_sk_s|\bar{3}4 \rangle \big)\Big)\Big] \,.
    \end{aligned}
\end{align}
Note that in order to add the $s$-channel and $t$-channel correlators their denominators must be matched and subsequently \eqref{ELER} is used to obtain the above expression. The term in \eqref{ELER} that survives in the flat space limit is responsible for the first term in the total correlator. The second and third term in the first line arise from the energy correction in \eqref{ELER}. We directly observe that the correlation function contains no energy pole, which agrees with the vanishing of the flat-space amplitude. Choosing a definite helicity structure, one finds
\begin{align}
    \begin{aligned}
       & \mathcal{W}_{s,\text{L}}^{\text{SDYM}}+\mathcal{W}_{t,\text{L}}^{\text{SDYM}} =\\
        &= \frac{g^2}{2E_\text{L}^sE_\text{R}^sE_\text{L}^tE_\text{R}^t}\frac{1}{k_1k_2k_3}\Big[\langle \bar{1}4 \rangle \langle \bar{2}4 \rangle \langle \bar{3}4 \rangle \Big(\langle \bar{1}\bar{2} \rangle \langle \bar{2}\bar{3} \rangle \langle 2\bar{4}\rangle+E_\text{L}^s\langle \bar{1}\bar{4} \rangle \langle \bar{2}\bar{3} \rangle -E_\text{L}^t\langle \bar{1}\bar{2} \rangle \langle \bar{3}\bar{4} \rangle  \Big)+\\
        &+\frac{E_\text{L}^sE_\text{R}^s}{4k_t^2}\langle \bar{1}4 \rangle \langle \bar{2}\bar{3} \rangle \Big(\langle \bar{2}\bar{3}|k_tk_t\bar{k}_t\bar{k}_t|\bar{1}4\rangle+(1+\sigma)E_\text{R}^t\langle\bar{2}\bar{3}\rangle\langle \bar{1}4\rangle(k_2-k_3)\Big)+\\
        &-\frac{E_\text{L}^tE_\text{R}^t}{4k_s^2}\langle \bar{1}\bar{2} \rangle \langle \bar{3}4 \rangle\Big( \langle \bar{1}\bar{2}|k_sk_s\bar{k}_s\bar{k}_s|\bar{3}4\rangle+(1+\sigma) E_\text{R}^s\langle \bar{1}\bar{2} \rangle \langle \bar{3}4 \rangle(k_1-k_2) \Big)\Big]\,.
    \end{aligned}
\end{align}

%%%%%%%%%%%%%%%%%%%%%%%%%%%%%%%%%%%%%%%%%%%%%%%%%%%%%%%%%%%%%
\subsubsection{SDYM Axial}
%%%%%%%%%%%%%%%%%%%%%%%%%%%%%%%%%%%%%%%%%%%%%%%%%%%%%%%%%%%%%
The Feynman rules in axial gauge are shown in Figure \ref{fig:axial}. In particular, the bulk-to-bulk propagator reads
\begin{align} \label{SDYMPropDecomp}
    \begin{aligned}
        \langle \Psi_{AA}(-k,z) \Phi_{BB'}(k,z') \rangle^{\text{SDYM}}_{\text{A}} &= \langle \Psi_{AA}(-k,z)\Phi_{B,B'}(k,z')\rangle_{\text{L}}^{\text{SDYM}}+\nabla_{BB'}^{k,z'}\Delta\eta^\text{L}_{AA}\,,
    \end{aligned}
\end{align}
see \eqref{deltadeltanablaxi1} and \eqref{deltadeltanablaxi2} for more details. It is convenient to use the Feynman/Lorenz gauge propagator as the reference point, the difference with the axial one being pure gauge. 
\begin{figure}[h]
\centering
\begin{tabular}{c c c}
    % Bulk-to-bulk propagator
    \begin{tikzpicture}
        \coordinate (a) at (-1.5,0);
        \coordinate (b) at (1.5,0);
        
        % Draw wavy propagator line
        \draw[decorate,decoration={snake},thick] (a) -- (b);

        % Draw momentum arrow
        \draw[-{Latex},thick] (-0.6,0.3) -- (0.6,0.3) node[midway,above] {\( k \)};
        
        % Labels for fields
        \node[left] at (a) {\( (-,AA) \)};
        \node[right] at (b) {\( (+,BB') \)};
    \end{tikzpicture}
    & \raisebox{3mm}{\quad : \quad} &
    \raisebox{3mm}{\( 
    \begin{aligned}
    &\langle \Psi_{AA}(-k,z)\Phi_{BB'}(k,z')\rangle_{A}=\\
    &=\langle \Psi_{AA}(-k,z)\Phi_{B,B'}(k,z')\rangle_{\text{L}}+\nabla_{BB'}\Delta\eta^\text{L}_{AA}
    \end{aligned}\)}
    \\[1.5cm]
    % Vertex
    \begin{tikzpicture}
        \coordinate (v) at (0,0);
        
        % Define external legs
        \coordinate (a) at (-1.5,1);
        \coordinate (b) at (-1.5,-1);
        \coordinate (c) at (1.5,0);
        
        % Draw gauge boson lines 
        \draw[decorate,decoration={snake},thick] (a) -- (v);
        \draw[decorate,decoration={snake},thick] (b) -- (v);
        \draw[decorate,decoration={snake},thick] (v) -- (c);
        
        % Draw parallel momentum arrows 
        \draw[-{Latex},thick] (-0.9,0.9) -- (-0.2,0.4) node[midway,above] {\( p_1 \)};
        \draw[-{Latex},thick] (-0.9,-0.9) -- (-0.2,-0.4) node[midway,below] {\( p_2 \)};
        \draw[-{Latex},thick] (0.3,0.3) --  (1.2,0.3) node[midway,above] {\( p_3 \)};
        
        % Add labels for fields next to endpoints
        \node[left] at (a) {\( (+,AA') \)};
        \node[left] at (b) {\( (+,BB') \)};
        \node[right] at (c) {\( (-,CD) \)};
    \end{tikzpicture}
    & \raisebox{12mm}{\quad : \quad} &
    \raisebox{12mm}{\( 
    V_{AA,BB,CC}= 2g\,\epsilon_{A'B'}\epsilon_{AC}\epsilon_{BC}
    \)}
    \\[2.0cm]
    % Boundary-to-bulk propagator (+,-)
    \begin{tikzpicture}
        % Define external points for propagator
        \coordinate (a) at (-2, 2);
        \coordinate (b) at (0, 2);
        \coordinate (boundary) at (-3.5,2);
        
        % Draw boundary-to-bulk propagator
        \draw[decorate,decoration={snake},thick] (a) -- (b);
        
        % Draw arc on the boundary
        \draw[thick] (-1.9,2.7) arc[start angle=160, end angle=200, radius=2cm];
        
        % Labels for fields
        \node[left] at (a) {\( \epsilon^-_i \)};
        \node[right] at (b) {\( (+,AA) \)};
    \end{tikzpicture}
    & \quad \raisebox{6mm}{\quad : \quad} &
    \quad \raisebox{6mm}{\( \Phi^{AA}_+(k_i,z)=\frac{\bar{i}^A\bar{i}^{A}}{2k_i}e^{-kz} \)}
    \\[1.5cm]
    % Boundary-to-bulk propagator (-,+)
    \begin{tikzpicture}
        % Define external points for propagator
        \coordinate (a) at (-2, -2);
        \coordinate (b) at (0, -2);
        \coordinate (boundary) at (-3.5,-2);
        
        % Draw boundary-to-bulk propagator
        \draw[decorate,decoration={snake},thick] (a) -- (b);
        
        % Draw arc on the boundary
        \draw[thick] (-1.9,-1.2) arc[start angle=160, end angle=200, radius=2cm];
        
        % Labels for fields
        \node[left] at (a) {\( \epsilon^+_i \)};
        \node[right] at (b) {\( (-,AA) \)};
    \end{tikzpicture}
    & \quad \raisebox{6mm}{\quad : \quad} &
    \quad \raisebox{6mm}{\( \Psi^{AA}_-(k_i,z)=i^Ai^{A} e^{-kz} \)}
\end{tabular}
\caption{SDYM axial gauge Feynman rules in $\text{AdS}_4$.}
\label{fig:axial}
\end{figure}
Likewise, we decompose the correlator into its contributions coming from the various pieces of the bulk-to-bulk propagator in \eqref{SDYMPropDecomp},
\begin{align}\label{SDYMAxialcorr}
    \mathcal{W}^{\text{SDYM}}_{s,\text{A}} = \mathcal{W}_{s,\text{L}}^{\text{SDYM}} + \mathcal{W}_{s,\text{gauge}}^{\text{SDYM}}\,,
\end{align}
where $\mathcal{W}_{s,\text{L}}^{\text{SDYM}}$ is \eqref{WSDYM} with $q_i^A=\bar{i}^{A}$ for $i=1,2,3$, i.e. \eqref{SDYM4ptLorenzAx}. Using
\begin{align} \label{intbyparts}
    \begin{aligned}
        \int &dz'\, V_{AA',BB',CC}[\nabla_{k,z'}^{AA'}\Delta\eta^{DD}]\phi^{B,B'}(z',k_3)\psi^{CC}(z',k_4)= \\
        &=-V_{AA',BB',CC}[-\epsilon^{AA'}\Delta\eta^{DD}]\phi^{B,B'}(0,k_3)\psi^{CC}(0,k_4)\Big|_{z'=0}+\text{eom}\,,
    \end{aligned}
\end{align}
we obtain (recall that $\mathcal{W}_{s,\text{L}}^{\text{SDYM}}=\eqref{SDYM4ptLorenzAx}$)
\begin{align} \label{SDYMaxialcorr}
    \begin{aligned}
        \mathcal{W}_{s,\text{L}}^{\text{SDYM}} &= -\frac{g^2}{2}\frac{1}{EE_\text{L}E_\text{R}}\frac{\langle \bar{1}\bar{2}\rangle \langle \bar{3}4 \rangle}{k_1k_2k_3}\Big[\langle \bar{1}4 \rangle \langle \bar{2}4 \rangle \langle \bar{3}\bar{4} \rangle +\frac{E}{4k^2}\langle \bar{1}\bar{2}|kk\bar{k}\bar{k}|\bar{3}4\rangle\Big] +\\
        &-(1+\sigma)\frac{g^2}{8E_\text{L}}\frac{\langle \bar{1}\bar{2} \rangle^2\langle \bar{3}4\rangle^2}{k_1k_2k_3}\frac{k_1-k_2}{k^2}\,,\\
        \mathcal{W}_{s,\text{gauge}}^{\text{SDYM}} &= -(1-\sigma)\frac{g^2}{8E_\text{L}}\frac{\langle \bar{1}\bar{2} \rangle^2 \langle \bar{3}4 \rangle^2}{k_1k_2k_3}\frac{k_1-k_2}{k^2}\,.
    \end{aligned}
\end{align}
The non-vanishing $\mathcal{W}_{s,\text{gauge}}^{\text{SDYM}}$ implies that the correlation function is gauge dependent for the Neumann-like propagator ($\sigma=-1$), while for the Dirichlet-like propagator ($\sigma=+1$), the gauge variation vanishes. Because $\sigma$ is purely a parameter of the Feynman/Lorenz gauge propagator and not the axial gauge one, we also observe that after adding $\mathcal{W}_{s,\text{L}}^{\text{SDYM}} + \mathcal{W}_{s,\text{gauge}}^{\text{SDYM}}$ the $\sigma$-dependence disappears in the axial gauge correlator.

%%%%%%%%%%%%%%%%%%%%%%%%%%%%%%%%%%%%%%%%%%%%%%%%%%%%%%%%%%%%%
\subsubsection{cYM Feynman}
\label{subsec:cYM}
%%%%%%%%%%%%%%%%%%%%%%%%%%%%%%%%%%%%%%%%%%%%%%%%%%%%%%%%%%%%%
In Appendix \ref{sec:cYMFlat} we discuss how the flat-space YM four-point amplitude arising from the exchange and contact diagrams is equivalent to the exchange diagrams in SDYM. As was discussed in Section \ref{sec:cYMAdS}, see for instance \eqref{abVertex}, the full three-point vertex can be expressed as a linear combination of the SD and the topological vertex. As a result, the exchange diagrams are naturally decomposed into i) diagrams with only the SD vertex; ii) diagrams with one or two topological vertices. Moreover, the boundary conditions in AdS call for homogeneous solutions to be added to the bulk-to-bulk propagator, a feature absent in flat space. This homogeneous contribution will enter both types of diagrams. As for the three-point function, we aim to identify the leading energy pole of the correlator with the flat-space amplitude. This implies that only the diagrams of type i) evaluated with the inhomogeneous part of the bulk-to-bulk propagator and the contact diagram may contribute to the leading energy pole, since both the homogeneous and topological contributions are absent in flat space.

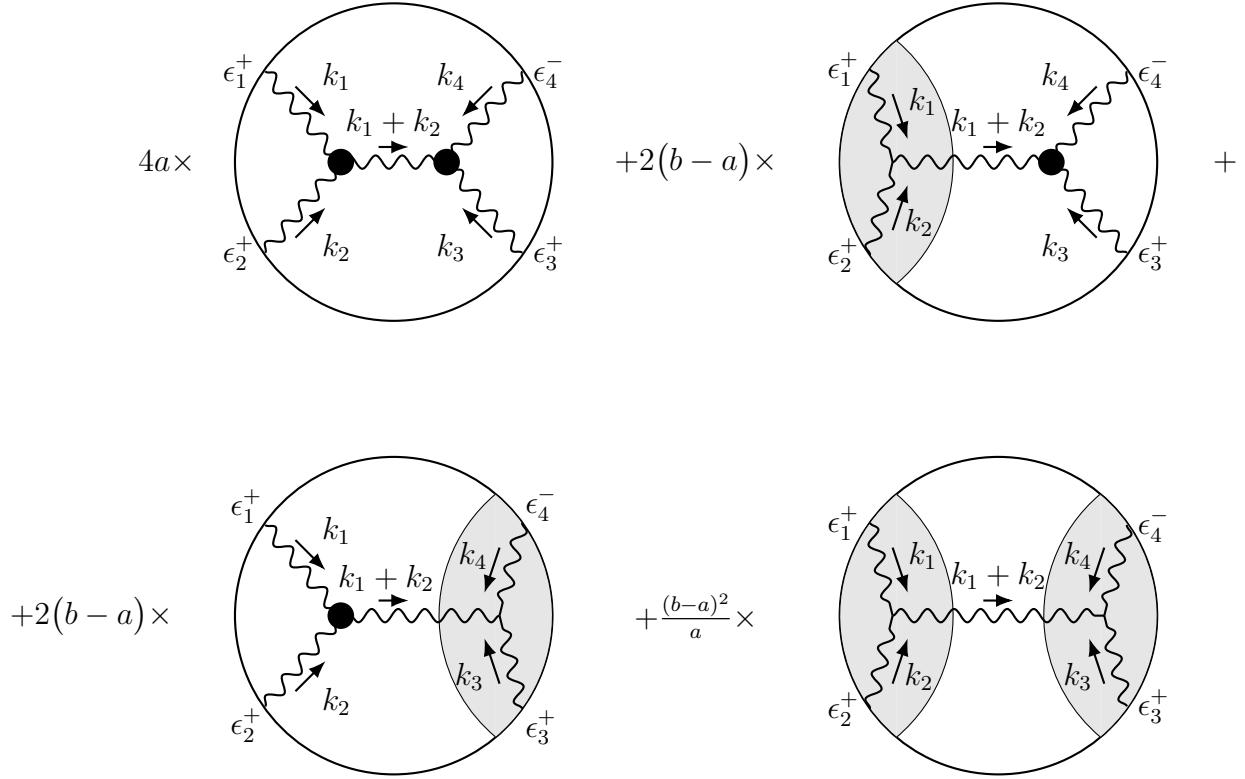
\begin{figure}[h!]
\centering
    % S-channel Witten diagram
    \begin{tikzpicture}
        \begin{scope}[xshift=-8cm]
        \node at (-3,0) {$4a\times$};
        % Define the boundary circle
        \draw[thick] (0,0) circle (2.1cm);
        
        % Define the vertices inside the bulk
        \coordinate (v1) at (-0.7,0);
        \coordinate (v2) at (0.7,0);
        
        % Define the external legs on the boundary
        \coordinate (a) at (-1.7,1.2);
        \coordinate (b) at (-1.7,-1.2);
        \coordinate (c) at (1.7,1.2);
        \coordinate (d) at (1.7,-1.2);
        
        % Draw the gauge boson lines
        \draw[decorate,decoration={snake},thick] (a) -- (v1);
        \draw[decorate,decoration={snake},thick] (b) -- (v1);
        \draw[decorate,decoration={snake},thick] (c) -- (v2);
        \draw[decorate,decoration={snake},thick] (d) -- (v2);
        
        % Draw the internal propagator 
        \draw[decorate,decoration={snake},thick] (v1) -- (v2);
        
        % Draw parallel momentum arrows
        \draw[-{Latex},thick] (-1.3,1.0) -- (-0.9,0.6) node[midway,above right] {\( k_1 \)};
        \draw[-{Latex},thick] (-1.3,-1) -- (-0.9,-0.6) node[midway,below right] {\( k_2 \)};
        \draw[{Latex}-,thick] (0.9,0.6) -- (1.3,1) node[midway,above left] {\( k_4 \)};
        \draw[{Latex}-,thick] (0.9,-0.6) -- (1.3,-1) node[midway,below left] {\( k_3 \)};
        \draw[-{Latex},thick] (-0.2,0.2) -- (0.2,0.2) node[midway,above] {\( k_1+k_2 \)};
        
        % Add labels for fields next to endpoints
        \node[left] at (a) {\( \epsilon^+_1 \)};
        \node[left] at (b) {\( \epsilon^+_2 \)};
        \node[right] at (c) {\( \epsilon^-_4 \)};
        \node[right] at (d) {\( \epsilon^+_3 \)};

        \fill (v1) circle (5pt);
        \fill (v2) circle (5pt);
        \end{scope}

        \begin{scope}
        \node at (-4,0) {$+2\big(b-a\big)\times$};

            % Define the boundary circle
        \draw[thick] (0,0) circle (2.1cm);
        
        % Define the vertices inside the bulk
        \coordinate (v1) at (-1.4,0);
        \coordinate (v2) at (0.7,0);
        
        % Define the external legs on the boundary
        \coordinate (a) at (-1.7,1.23);
        \coordinate (b) at (-1.7,-1.23);
        \coordinate (c) at (1.7,1.2);
        \coordinate (d) at (1.7,-1.2);

        \filldraw[fill=gray!20, draw=black,rotate=-90] 
            (1.61,-1.35) arc[start angle=40, end angle=140, radius=2.1cm];
        \filldraw[fill=gray!20, draw=black,rotate=-90] 
        (-1.61,-1.35) arc[start angle=220, end angle=320, radius=2.1cm];

                % Draw parallel momentum arrows
            \draw[-{Latex},thick,rotate=-90] (-0.9,-1.4) -- (-0.3,-1.2) node[midway,xshift=0.3 cm,yshift=0.2 cm] {\( k_1 \)};
            \draw[-{Latex},thick,rotate=-90] (0.9,-1.4) -- (0.3,-1.2) node[midway,xshift=0.25 cm,yshift=-0.2 cm] {\( k_2 \)};
        
        % Draw the gauge boson lines
        \draw[decorate,decoration={snake},thick] (a) -- (v1);
        \draw[decorate,decoration={snake},thick] (b) -- (v1);
        \draw[decorate,decoration={snake},thick] (c) -- (v2);
        \draw[decorate,decoration={snake},thick] (d) -- (v2);
        
        % Draw the internal propagator 
        \draw[decorate,decoration={snake},thick] (v1) -- (v2);
        
        % Draw parallel momentum arrows
        \draw[{Latex}-,thick] (0.9,0.6) -- (1.3,1) node[midway,above left] {\( k_4 \)};
        \draw[{Latex}-,thick] (0.9,-0.6) -- (1.3,-1) node[midway,below left] {\( k_3 \)};
        \draw[-{Latex},thick] (-0.2,0.2) -- (0.2,0.2) node[midway,above] {\( k_1+k_2 \)};
        
        % Add labels for fields next to endpoints
        \node[left] at (a) {\( \epsilon^+_1 \)};
        \node[left] at (b) {\( \epsilon^+_2 \)};
        \node[right] at (c) {\( \epsilon^-_4 \)};
        \node[right] at (d) {\( \epsilon^+_3 \)};

        \fill (v2) circle (5pt);
            \node at (3,0) {$+$};
        \end{scope}
        
        \begin{scope}[yshift=-6cm]
            \begin{scope}[xshift=-8cm,rotate=180]
                    \node at (-4,0) {$+\frac{(b-a)^2}{a}\times$};
        
                    % Define the boundary circle
                \draw[thick] (0,0) circle (2.1cm);
                
                % Define the vertices inside the bulk
                \coordinate (v1) at (-1.4,0);
                \coordinate (v2) at (0.7,0);
                
                % Define the external legs on the boundary
                \coordinate (a) at (-1.7,1.23);
                \coordinate (b) at (-1.7,-1.23);
                \coordinate (c) at (1.7,1.2);
                \coordinate (d) at (1.7,-1.2);

                \filldraw[fill=gray!20, draw=black,rotate=-90] 
                    (1.61,-1.35) arc[start angle=40, end angle=140, radius=2.1cm];
                \filldraw[fill=gray!20, draw=black,rotate=-90] 
                (-1.61,-1.35) arc[start angle=220, end angle=320, radius=2.1cm];
        
                        % Draw parallel momentum arrows
                    \draw[-{Latex},thick,rotate=-90] (-0.9,-1.4) -- (-0.3,-1.2) node[midway,xshift=-0.3 cm,yshift=-0.2 cm] {\( k_3 \)};
                    \draw[-{Latex},thick,rotate=-90] (0.9,-1.4) -- (0.3,-1.2) node[midway,xshift=-0.25 cm,yshift=0.2 cm] {\( k_4 \)};
                
                % Draw the gauge boson lines
                \draw[decorate,decoration={snake},thick] (a) -- (v1);
                \draw[decorate,decoration={snake},thick] (b) -- (v1);
                \draw[decorate,decoration={snake},thick] (c) -- (v2);
                \draw[decorate,decoration={snake},thick] (d) -- (v2);
                
                % Draw the internal propagator 
                \draw[decorate,decoration={snake},thick] (v1) -- (v2);
                
                % Draw parallel momentum arrows
                \draw[{Latex}-,thick] (0.9,0.6) -- (1.3,1) node[midway,below right] {\( k_2 \)};
                \draw[{Latex}-,thick] (0.9,-0.6) -- (1.3,-1) node[midway,above right] {\( k_1 \)};
                \draw[-{Latex},thick]  (0.2,-0.2) -- (-0.2,-0.2)node[midway,xshift=-0.1 cm,yshift=0.3 cm] {\( k_1+k_2 \)};
                
                % Add labels for fields next to endpoints
                \node[xshift=0.25cm,yshift=-0.25cm] at (a) {\( \epsilon^+_3 \)};
                \node[xshift=0.25cm,yshift=0.25cm] at (b) {\( \epsilon^-_4 \)};
                \node[xshift=-0.25cm,yshift=-0.25cm] at (c) {\( \epsilon^+_2 \)};
                \node[xshift=-0.25cm,yshift=0.25cm] at (d) {\( \epsilon^+_1 \)};

                \fill (v2) circle (5pt);
                    \node at (4,0) {$+ 2\big(b-a\big)\times$};
                \end{scope}
                \begin{scope}
                                % Define the boundary circle
                    \draw[thick] (0,0) circle (2.1cm);

                    \filldraw[fill=gray!20, draw=black,rotate=90] 
                    (1.61,-1.35) arc[start angle=40, end angle=140, radius=2.1cm];
                    \filldraw[fill=gray!20, draw=black,rotate=90] 
                    (-1.61,-1.35) arc[start angle=220, end angle=320, radius=2.1cm];
                    
                    % Define the vertices inside the bulk
                    \coordinate (v1) at (-1.4,0);
                    \coordinate (v2) at (1.4,0);
                    
                    % Define the external legs on the boundary
                    \coordinate (a) at (-1.7,1.23);
                    \coordinate (b) at (-1.7,-1.23);
                    \coordinate (c) at (1.7,1.2);
                    \coordinate (d) at (1.7,-1.2);

                    \filldraw[fill=gray!20, draw=black,rotate=-90] 
                        (1.61,-1.35) arc[start angle=40, end angle=140, radius=2.1cm];
                    \filldraw[fill=gray!20, draw=black,rotate=-90] 
                    (-1.61,-1.35) arc[start angle=220, end angle=320, radius=2.1cm];
            
                            % Draw parallel momentum arrows
                        \draw[-{Latex},thick,rotate=-90] (-0.9,-1.4) -- (-0.3,-1.2) node[midway,xshift=0.3 cm,yshift=0.2 cm] {\( k_1 \)};
                        \draw[-{Latex},thick,rotate=-90] (0.9,-1.4) -- (0.3,-1.2) node[midway,xshift=0.25 cm,yshift=-0.2 cm] {\( k_2 \)};
                    
                    % Draw the gauge boson lines
                    \draw[decorate,decoration={snake},thick] (a) -- (v1);
                    \draw[decorate,decoration={snake},thick] (b) -- (v1);
                    \draw[decorate,decoration={snake},thick] (c) -- (v2);
                    \draw[decorate,decoration={snake},thick] (d) -- (v2);
                    
                    % Draw the internal propagator 
                    \draw[decorate,decoration={snake},thick] (v1) -- (v2);
                    
                    % Draw parallel momentum arrows
                    \draw[-{Latex},thick,rotate=90] (-0.9,-1.4) -- (-0.3,-1.2) node[midway,xshift=-0.25 cm,yshift=-0.2 cm] {\( k_3 \)};
                    \draw[-{Latex},thick,rotate=90] (0.9,-1.4) -- (0.3,-1.2) node[midway,xshift=-0.25 cm,yshift=0.2 cm] {\( k_4 \)};
                    \draw[-{Latex},thick] (-0.2,0.2) -- (0.2,0.2) node[midway,above] {\( k_1+k_2 \)};
                    
                    % Add labels for fields next to endpoints
                    \node[left] at (a) {\( \epsilon^+_1 \)};
                    \node[left] at (b) {\( \epsilon^+_2 \)};
                    \node[right] at (c) {\( \epsilon^-_4 \)};
                    \node[right] at (d) {\( \epsilon^+_3 \)};
                \end{scope}
        \end{scope}
        
    \end{tikzpicture}
\caption{The full exchange diagram is a linear combination of four diagrams: one with only the SD vertex (top left), which resembles the flat-space diagram and is denoted by $\mathcal{W}^{\text{SD}}$; one with the left vertex replaced by the topological vertex (top right), denoted by $\mathcal{T}^{\text{L}}$; one with the right vertex replaced by the topological vertex (bottom left), denoted by $\mathcal{T}^{\text{R}}$; and one with both vertices replaced by the topological vertex (bottom right), denoted by $\mathcal{W}^{\text{L,R}}$.}
\label{fig:exchangeDecomp}
\end{figure}

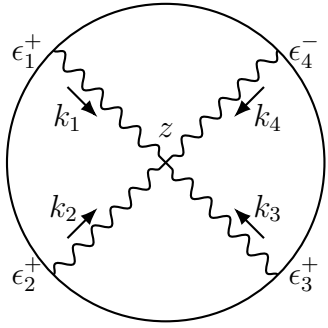
\begin{figure}[h!]
\centering
\begin{tabular}{c c}
    % S-channel Witten diagram
    \begin{tikzpicture}
        % Define the boundary circle
        \draw[thick] (0,0) circle (2.1cm);
        
        % Define the vertices inside the bulk
        \coordinate (v) at (0,0);
        
        % Define the external legs on the boundary
        \coordinate (a) at (-1.48,1.48);
        \coordinate (b) at (-1.48,-1.48);
        \coordinate (c) at (1.48,1.48);
        \coordinate (d) at (1.48,-1.48);
        
        % Draw the gauge boson lines
        \draw[decorate,decoration={snake},thick] (a) -- (v);
        \draw[decorate,decoration={snake},thick] (b) -- (v);
        \draw[decorate,decoration={snake},thick] (c) -- (v);
        \draw[decorate,decoration={snake},thick] (d) -- (v);

        % Draw parallel momentum arrows
        \draw[-{Latex},thick] (-1.3,1.0) -- (-0.9,0.6) node[midway,xshift=-0.2cm,yshift=-0.2cm] {\( k_1 \)};
        \draw[-{Latex},thick] (-1.3,-1) -- (-0.9,-0.6) node[midway,xshift=-0.25cm,yshift=0.2cm] {\( k_2 \)};
        \draw[{Latex}-,thick] (0.9,0.6) -- (1.3,1) node[midway,xshift=0.25cm,yshift=-0.2cm] {\( k_4 \)};
        \draw[{Latex}-,thick] (0.9,-0.6) -- (1.3,-1) node[midway,xshift=0.25cm,yshift=0.2cm] {\( k_3 \)};
        
        % Add labels for fields next to endpoints
        \node[left] at (a) {\( \epsilon^+_1 \)};
        \node[left] at (b) {\( \epsilon^+_2 \)};
        \node[right] at (c) {\( \epsilon^-_4 \)};
        \node[right] at (d) {\( \epsilon^+_3 \)};
        
        % Add labels for vertices
        \node[above] at (0,0.2) {\( z \)};
        
    \end{tikzpicture}
\end{tabular}

\caption{YM four-point contact diagram.}
\label{fig:contactAdS}
\end{figure}

Using the decomposition of the cubic vertex as provided in Figure \ref{fig:2SDVertexAdS}, the $s$-channel decomposes as illustrated in Figure \ref{fig:exchangeDecomp}. As is shown, the exchange diagram can be decomposed into four types of diagrams. Together with the contact diagram, the total four-point correlator can be written as\footnote{The YM action came with an overall factor $a$, which would cause the bulk-to-bulk propagator to be proportional to $\frac{1}{a}$. However, we chose to extract this factor and incorporate it directly into the normalization of the diagrams. One should also have $1/a$ for every external leg, but we absorb it into the normalization of $\epsilon_\pm$.}
\begin{align}\label{W4Decomp}
    \mathcal{W}_s = 4a\mathcal{W}_s^{\text{SD}}+2\big(b-a\big)\Big(\mathcal{T}_s^\text{L}+\mathcal{T}_s^\text{R}\Big)+\frac{\big(b-a\big)^2}{a}\mathcal{T}_s^{\text{L,R}}+a\mathcal{W}_{s,\text{contact}}\,.
\end{align}
$\mathcal{W}_s^{\text{SD}}$ is the diagram containing two SD vertices, $\mathcal{T}_s^{\text{L}}$ and $\mathcal{T}_s^{\text{R}}$ are the diagrams with the left and right vertex replaced by the topological vertex, respectively, and $\mathcal{T}_s^{\text{L,R}}$ is the one with both vertices being the topological vertex. We will refer to $\mathcal{W}_s^\text{SD}$ as non-topological diagrams, while we call the other topological diagrams. Figure \ref{fig:contactAdS} illustrates the contact diagram.

All diagrams will be evaluated separately and the calculation will be decomposed into contributions from various types of bulk-to-bulk propagators and components thereof. For instance, the correlation function will be computed for the Dirichlet and mixed boundary conditions both in Feynman gauge and in axial gauge. We stress that Lorenz gauge and Feynman gauge can be used interchangeably in SDYM, because the SDYM vertex symmetrizes the indices of the $\Psi$-leg in the $\langle\Psi\Phi\rangle$ propagator, but this is not always true in YM. With the help of the derivative present in the YM vertex, it produces a $\langle \Psi \Phi \rangle$ propagator and symmetrizes $\Psi$ when acting on $\langle \Phi \Phi \rangle$. However, when considering certain topological diagrams, a genuine $\langle \Phi\Phi \rangle$ propagator must be used. The Lorenz gauge $\langle \Phi\Phi \rangle$ propagator is given in Appendix \ref{app:Lorenz} and it can be seen that it differs significantly from the Feynman gauge propagator. Since the latter is the simplest of the two, we will use the Feynman gauge for the topological diagram, and also for the non-topological ones for consistency. In Appendix \ref{app:fourier} we collected all bulk integrals relevant in this section.

The correlation function will be derived using the symmetrized polarization vectors from axial gauge, i.e. we choose $q_i^A=\bar{i}^A$ and $\bar{q}_i^{A'}=i^{A'}$. While the correlation functions can be derived for arbitrary reference spinors, it is only for this specific choice that Feynman gauge and axial gauge can be compared.

\begin{figure}[h!]
\centering
    % S-channel Witten diagram
    \begin{tikzpicture}
        \begin{scope}[xshift=-3cm]
        % Define the boundary circle
        \draw[thick] (0,0) circle (2.1cm);
        
        % Define the vertices inside the bulk
        \coordinate (v1) at (-0.7,0);
        \coordinate (v2) at (0.7,0);
        
        % Define the external legs on the boundary
        \coordinate (a) at (-1.7,1.2);
        \coordinate (b) at (-1.7,-1.2);
        \coordinate (c) at (1.7,1.2);
        \coordinate (d) at (1.7,-1.2);
        
        % Draw the gauge boson lines
        \draw[decorate,decoration={snake},thick] (a) -- (v1);
        \draw[decorate,decoration={snake},thick] (b) -- (v1);
        \draw[decorate,decoration={snake},thick] (c) -- (v2);
        \draw[decorate,decoration={snake},thick] (d) -- (v2);
        
        % Draw the internal propagator 
        \draw[decorate,decoration={snake},thick] (v1) -- (v2);
        
        % Draw parallel momentum arrows
        \draw[-{Latex},thick] (-1.3,1.0) -- (-0.9,0.6) node[midway,above right] {\( k_1 \)};
        \draw[-{Latex},thick] (-1.3,-1) -- (-0.9,-0.6) node[midway,below right] {\( k_2 \)};
        \draw[{Latex}-,thick] (0.9,0.6) -- (1.3,1) node[midway,above left] {\( k_4 \)};
        \draw[{Latex}-,thick] (0.9,-0.6) -- (1.3,-1) node[midway,below left] {\( k_3 \)};
        \draw[-{Latex},thick] (-0.2,0.2) -- (0.2,0.2) node[midway,above] {\( k_1+k_2 \)};
        
        % Add labels for fields next to endpoints
        \node[left] at (a) {\( \epsilon^+_1 \)};
        \node[left] at (b) {\( \epsilon^+_2 \)};
        \node[right] at (c) {\( \epsilon^-_4 \)};
        \node[right] at (d) {\( \epsilon^+_3 \)};
        \node at (3,0) {$=$};

        \fill (v1) circle (5pt);
        \fill (v2) circle (5pt);
        \end{scope}
        
        \begin{scope}[xshift=-6cm,yshift=-6cm]
            % Define the boundary circle
        \draw[thick] (0,0) circle (2.1cm);
        
        % Define the vertices inside the bulk
        \coordinate (v1) at (-0.7,0);
        \coordinate (v2) at (0.7,0);
        
        % Define the external legs on the boundary
        \coordinate (a) at (-1.7,1.2);
        \coordinate (b) at (-1.7,-1.2);
        \coordinate (c) at (1.7,1.2);
        \coordinate (d) at (1.7,-1.2);
        
        % Draw the gauge boson lines
        \draw[decorate,decoration={snake},thick] (a) -- (v1);
        \draw[decorate,decoration={snake},thick] (b) -- (v1);
        \draw[decorate,decoration={snake},thick] (c) -- (v2);
        \draw[decorate,decoration={snake},thick] (d) -- (v2);
        
        % Draw the internal propagator 
        \draw[decorate,decoration={snake},thick] (v1) -- (v2);
        
        % Draw parallel momentum arrows
        \draw[-{Latex},thick] (-1.3,1.0) -- (-0.9,0.6) node[midway,above right] {\( k_1 \)};
        \draw[-{Latex},thick] (-1.3,-1) -- (-0.9,-0.6) node[midway,below right] {\( k_2 \)};
        \draw[{Latex}-,thick] (0.9,0.6) -- (1.3,1) node[midway,above left] {\( k_4 \)};
        \draw[{Latex}-,thick] (0.9,-0.6) -- (1.3,-1) node[midway,below left] {\( k_3 \)};
        \draw[-{Latex},thick] (-0.2,0.2) -- (0.2,0.2) node[midway,above] {\( k_1+k_2 \)};
        
        % Add labels for fields next to endpoints
        \node[left] at (a) {\( \epsilon^+_1 \)};
        \node[left] at (b) {\( \epsilon^+_2 \)};
        \node[right] at (c) {\( \epsilon^-_4 \)};
        \node[right] at (d) {\( \epsilon^+_3 \)};

        \fill[blue] (-0.4,0) circle (5pt);
        \fill[blue] (0.9,0.2) circle (5pt);
        
        \node at (3,0) {$+$}; 
        \node at (-3,0) {$=$}; 
        
        \end{scope}
        \begin{scope}[yshift=-6cm]
            % Define the boundary circle
        \draw[thick] (0,0) circle (2.1cm);
        
        % Define the vertices inside the bulk
        \coordinate (v1) at (-0.7,0);
        \coordinate (v2) at (0.7,0);
        
        % Define the external legs on the boundary
        \coordinate (a) at (-1.7,1.2);
        \coordinate (b) at (-1.7,-1.2);
        \coordinate (c) at (1.7,1.2);
        \coordinate (d) at (1.7,-1.2);
        
        % Draw the gauge boson lines
        \draw[decorate,decoration={snake},thick] (a) -- (v1);
        \draw[decorate,decoration={snake},thick] (b) -- (v1);
        \draw[decorate,decoration={snake},thick] (c) -- (v2);
        \draw[decorate,decoration={snake},thick] (d) -- (v2);
        
        % Draw the internal propagator 
        \draw[decorate,decoration={snake},thick] (v1) -- (v2);
        
        % Draw parallel momentum arrows
        \draw[-{Latex},thick] (-1.3,1.0) -- (-0.9,0.6) node[midway,above right] {\( k_1 \)};
        \draw[-{Latex},thick] (-1.3,-1) -- (-0.9,-0.6) node[midway,below right] {\( k_2 \)};
        \draw[{Latex}-,thick] (0.9,0.6) -- (1.3,1) node[midway,above left] {\( k_4 \)};
        \draw[{Latex}-,thick] (0.9,-0.6) -- (1.3,-1) node[midway,below left] {\( k_3 \)};
        \draw[-{Latex},thick] (-0.2,0.2) -- (0.2,0.2) node[midway,above] {\( k_1+k_2 \)};
        
        % Add labels for fields next to endpoints
        \node[left] at (a) {\( \epsilon^+_1 \)};
        \node[left] at (b) {\( \epsilon^+_2 \)};
        \node[right] at (c) {\( \epsilon^-_4 \)};
        \node[right] at (d) {\( \epsilon^+_3 \)};

        \fill[blue] (-0.4,0) circle (5pt);
        \fill[blue] (0.5,0) circle (5pt);
        \end{scope}
        
    \end{tikzpicture}

\caption{The YM $s$-channel Witten diagram is composed of two diagrams. The left diagram is of the same form as the diagram in SDYM and is denoted by $\mathcal{W}_s^1$, while the right diagram -- absent in SDYM -- is labeled by $\mathcal{W}_s^2$.}
\label{fig:cYM1}
\end{figure}
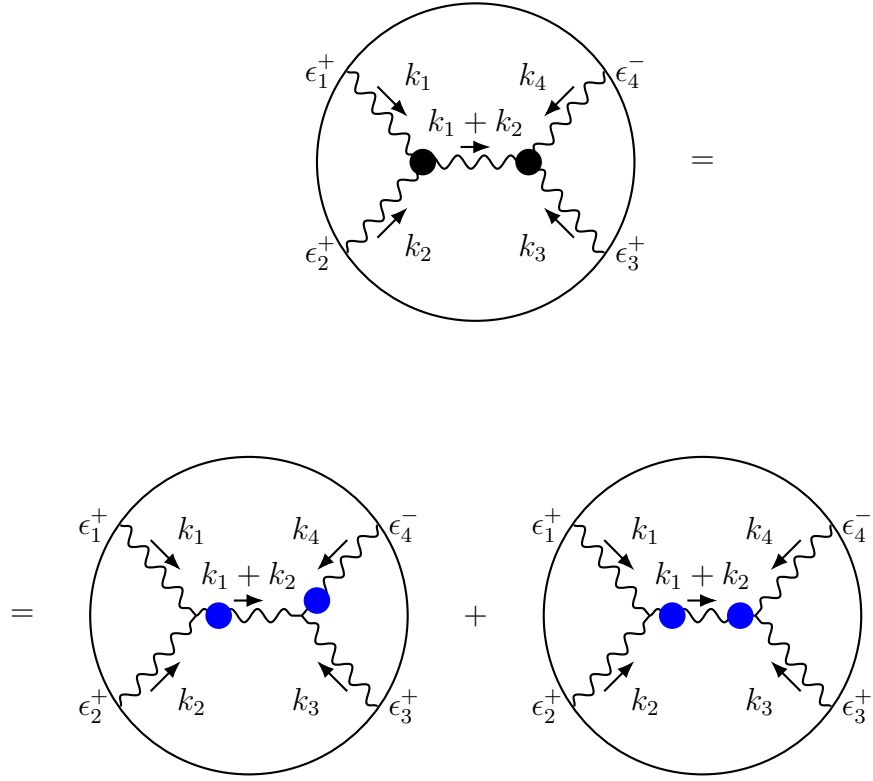

\paragraph{Non-topological diagrams.}
The YM diagrams are richer than the SDYM diagrams, both in multitude and complexity. While the SDYM bulk-to-bulk propagator had to admit the self-dual boundary condition, mixed boundary conditions, including Dirichlet and Neumann, can be chosen for the YM propagator. Moreover, the $s$-channel diagram picks up a contribution with a derivative from one vertex acting on the boundary-to-bulk propagator and the other one on the bulk-to-bulk propagator. We will denote this contribution to the correlator $\mathcal{W}^{1}_s$. There will also be a contribution with the derivative from both vertices acting on the bulk-to-bulk propagator and we will call its contribution to the correlator $\mathcal{W}^{2}_s$. The total (non-topological) $s$-channel contribution to the correlator is then given by
\begin{align} \label{Wdecomp}       \mathcal{W}_s^{\text{SD}}=\mathcal{W}_s^{1}+\mathcal{W}_s^{2} \,,
\end{align}
which is also depicted in Figure \ref{fig:cYM1}, with $\mathcal{W}_s^{1}$ and $\mathcal{W}_s^{2}$ corresponding to the left and right diagram, respectively. Note that only the former diagram is present in SDYM.

The $s$-channel contribution to the correlator will be computed from
\begin{align}
    \begin{aligned}
        \mathcal{W}_s^{\text{SD}}&=\int_0^\infty dz\int_0^\infty dz' \Phi_+^{A,A}(k_1,z)\Phi_+^{B,B'}(k_2,z)V\fdu{AA',BB'}{EE'}\langle \Phi_{EE'}(-k,z)\Phi_{FF'}(k,z') \rangle \times\\
        &\times V\fud{FF'}{CC',DD'}\Phi_+^{C,C'}(k_3,z')\Phi_-^{D,D'}(k_4,z') \,.
    \end{aligned}
\end{align}
The subdiagrams $\mathcal{W}^{1}$ and $\mathcal{W}^{2}$ read
\begin{align} \label{w1w2}
    \begin{aligned}
       \mathcal{W}_s^{1} &= -g^2\int\Phi_{+}^{A,A'}(k_1,z)\Phi_+\fud{A,}{A'}(k_2,z)\langle \Psi_{AA}(-k,z)\Phi_{B,B'}(k,z') \rangle \Phi\fdu{+B,}{B'}(k_3,z')\Psi_-^{BB}(k_4,z')\,,\\
        \mathcal{W}_s^{2} &= g^2\int\Phi_{+}^{AA'}(k_1,z)\Phi_+\fud{A,}{A'}(k_2,z)\langle F_{AA}(-k,z)F_{BB}(k,z') \rangle \Phi_+^{BB'}(k_3,z')\Phi\fdud{-}{B,}{B'}(k_4,z') \,.
    \end{aligned}
\end{align}
Here, we would like to draw the reader's attention to two points: (a) in $\mathcal{W}_s^{1}$ the derivative hits the $\langle \Phi\Phi\rangle$-propagator and gives the two-point function $\langle F \Phi \rangle $, which coincides with the $\langle \Psi\Phi\rangle$-propagator in Chalmers--Siegel theory, hence, the notation; (b) in $\mathcal{W}_s^{2}$ the two derivatives produce $\langle F F \rangle $ out of $\langle \Phi\Phi\rangle$ and it is very important to distinguish it from the $\langle \Psi\Psi\rangle$-propagator in Chalmers--Siegel theory, see Section \ref{sec:ChSiAmend}. The contact diagram, illustrated in Figure \ref{fig:contactAdS}, requires the four-point vertex,\footnote{Note that the quartic vertex contains terms with the color factors of the $s$-channel, $t$-channel and $u$-channel, see Appendix \ref{app:color} for more details. Here we only use the term with the color factors of the $s$-channel.}
\begin{align}
    V^{\text{quartic}}_{AA',BB',CC',DD'}&= 2g^2\epsilon_{A'B'}\epsilon_{C'D'}(\epsilon_{AD}\epsilon_{BC}+\epsilon    _{AC}\epsilon_{BD})\,.
\end{align}
The diagram is computed from
\begin{align}
    \begin{aligned}
        \mathcal{W}_{s,\text{contact}}&=\int \Phi_+^{A,A'}(k_1,z)\Phi_+^{B,B'}(k_2,z)V^{\text{quartic}}_{AA',BB',CC',DD'}\Phi_+^{C,C'}(k_3,z)\Phi_-^{D,D'}(k_4,z) \,.
    \end{aligned}
\end{align}

\paragraph{Dirichlet boundary conditions.} For Dirichlet boundary conditions, the four-point correlation function is captured by only the non-topological diagrams, as any propagator with Dirichlet boundary conditions vanishes when the $\Phi$-leg is placed on the boundary. The diagrams yield
\begin{align}
    \begin{aligned}
        \mathcal{W}^1_{s,\text{D}} &= -\frac{g^2}{8EE_\text{L}E_\text{R}}\frac{\langle \bar{1}\bar{2}\rangle \langle \bar{3}4\rangle}{k_1k_2k_3}\Big(\langle \bar{1}4\rangle\langle\bar{2}4\rangle\langle \bar{3}\bar{4}\rangle-\frac{E}{4k}\big(\langle \bar{1}4 \rangle \langle \bar{2}|k\bar{k}|\bar{3}\rangle+\langle \bar{2}4 \rangle \langle \bar{1}|k\bar{k}|\bar{3}\rangle+\\
        &\qquad\qquad\qquad\qquad\qquad\qquad\qquad\qquad\qquad\qquad-\langle \bar{1}\bar{3} \rangle \langle 4|k\bar{k}|\bar{2}\rangle-\langle \bar{2}\bar{3} \rangle \langle 4|k\bar{k}|\bar{1}\rangle\big)\Big) \,,\\
        \mathcal{W}^2_{s,\text{D}} &= \frac{g^2}{32}\frac{\langle \bar{1}\bar{2}\rangle \langle \bar{3}4\rangle}{k_1k_2k_3k_4}\Big(\underbrace{\frac{\langle \bar{1}\bar{3} \rangle \langle \bar{2}4\rangle +\langle \bar{1}4 \rangle\langle \bar{2}\bar{3}\rangle}{E}}_{\text{contact}}  \underbrace{-\frac{\langle \bar{1}\bar{2}|\bar{k}\bar{k}kk|\bar{3}4\rangle}{kE_\text{L}E_\text{R}}}_{\langle\Psi\Psi\rangle}\Big) \,.
    \end{aligned}
\end{align}
Here we used \eqref{abmnrscd} and we defined 
\begin{align}
    \begin{aligned}
    \langle a|mn|b\rangle &= a^Am_Am_Bb^B\,.
    \end{aligned}
\end{align}
We have also split the contribution from the $\langle F F \rangle $ two-point function into $\langle \Psi\Psi\rangle$ and a contact piece in accordance with \eqref{FFvsPsiPsi}. The contact diagram, which, due to the absence of a bulk-to-bulk propagator, is the same for all boundary conditions, reads
\begin{align}\label{cont}
    \mathcal{W}_{s}^{\text{contact}} &=-\frac{g^2}{8}\frac{\langle \bar{1}\bar{2}\rangle \langle \bar{3}4\rangle}{k_1k_2k_3k_4}\frac{\langle \bar{1}\bar{3} \rangle \langle \bar{2}4\rangle +\langle \bar{1}4 \rangle\langle \bar{2}\bar{3}\rangle}{E} \,.
\end{align}
We observe that $\mathcal{W}^2_{s,\text{D}}$-diagram cancels the energy pole in the contact diagram when the coefficients in \eqref{W4Decomp} are taken into account. In Appendix \ref{sec:reltoflat} we prove that the Dirichlet correlator is gauge-invariant, so the above results hold for both Feynman  and axial gauges.

The correlator can be written exclusively in boundary momentum spinors using
\begin{align}
    \begin{aligned}
        \langle \bar{1}\bar{2}|\bar{k}\bar{k}kk|\bar{3}4\rangle &= k\big(\langle \bar{1}\bar{3}\rangle \langle \bar{2}\bar{3} \rangle \langle 34 \rangle - \langle \bar{1}4\rangle \langle \bar{2}4\rangle \langle \bar{3}\bar{4} \rangle\big) +k^2\big(\langle \bar{1}4\rangle \langle \bar{2}\bar{3} \rangle + \langle \bar{1}\bar{3} \rangle \langle \bar{2}4 \rangle\big)+\\
        &- (k_1-k_2)(k_3+k_4) \langle \bar{1}\bar{2} \rangle \langle \bar{3}4 \rangle \,,\\
        \langle \bar{1}4\rangle \langle \bar{2}|k\bar{k}|\bar{3} \rangle &+ \langle \bar{2}4 \rangle \langle \bar{1}|k\bar{k} |\bar{3}\rangle - \langle \bar{1}\bar{3} \rangle \langle 4|k\bar{k}|\bar{2} \rangle - \langle \bar{2}\bar{3} \rangle \langle 4|k\bar{k}|\bar{1} \rangle =\\
        &=-2(k_1-k_2)\langle \bar{1}\bar{2} \rangle \langle \bar{3}4\rangle -2k\big(\langle\bar{1}\bar{3}\rangle \langle \bar{2}4\rangle + \langle \bar{1}4\rangle \langle \bar{2}\bar{3} \rangle\big) \,.
    \end{aligned}
\end{align}
It then reads
\begin{align}
    \begin{aligned}
        \mathcal{W}_{s,\text{D}} = -\frac{g^2}{32E_\text{L}E_\text{R}}\frac{\langle \bar{1}\bar{2} \rangle \langle \bar{3} 4 \rangle}{k_1k_2k_3k_4}\Big(&\frac{4k_4}{E}\langle \bar{1}4 \rangle \langle \bar{2}4\rangle \langle \bar{3}\bar{4} \rangle +\big(\langle \bar{1}\bar{3} \rangle \langle \bar{2} \bar{3} \rangle \langle 34 \rangle - \langle \bar{1}4 \rangle \langle \bar{2} 4 \rangle \langle \bar{3} \bar{4} \rangle\big) +\\
        &+ (k+2k_4)\big(\langle \bar{1}4 \rangle \langle \bar{2}\bar{3} \rangle + \langle \bar{1} \bar{3} \rangle \langle \bar{2}4 \rangle\big) - \frac{(k_1-k_2)(k_3-k_4)}{k}\langle \bar{1}\bar{2} \rangle \langle \bar{3} 4 \rangle\Big) \,.
    \end{aligned}
\end{align}
The Dirichlet four-point correlator for the $(+++-)$ helicity configuration was first obtained in \cite{Armstrong:2020woi}. Replacing factors of $i$ by $-1$ and changing overall sign of their correlator agrees with our result.

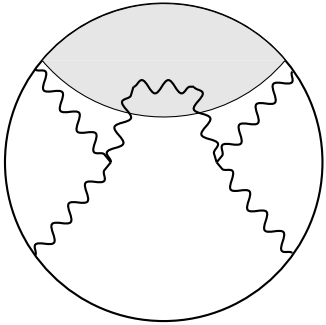
\begin{figure}[h!]
\centering
    \begin{tikzpicture}        
                    % Define the boundary circle
                \draw[thick] (0,0) circle (2.1cm);
                
                % Define the vertices inside the bulk
                \coordinate (v1) at (-0.7,0);
                \coordinate (v2) at (0.7,0);

                \coordinate (v3) at (-0.4,1);
                \coordinate (v4) at (0.4,1);
                
                % Define the external legs on the boundary
                \coordinate (a) at (-1.7,1.23);
                \coordinate (b) at (-1.7,-1.23);
                \coordinate (c) at (1.7,1.2);
                \coordinate (d) at (1.7,-1.2);

                \filldraw[fill=gray!20, draw=black,rotate=180] 
                    (1.61,-1.35) arc[start angle=40, end angle=140, radius=2.1cm];
                \filldraw[fill=gray!20, draw=black,rotate=-180] 
                (-1.61,-1.35) arc[start angle=220, end angle=320, radius=2.1cm];

                % Draw the gauge boson lines
                \draw[decorate,decoration={snake},thick] (a) -- (v1);
                \draw[decorate,decoration={snake},thick] (b) -- (v1);
                \draw[decorate,decoration={snake},thick] (c) -- (v2);
                \draw[decorate,decoration={snake},thick] (d) -- (v2);
                
                % Draw the internal propagator 
                \draw[decorate,decoration={snake},thick] (v1) -- (v3);

                \draw[decorate,decoration={snake},thick] (v4) -- (v2);

                \draw[decorate,decoration={snake},thick] (v3) -- (v4);
    \end{tikzpicture}
\caption{The figure illustrates that the difference between Dirichlet and Neumann propagators yields a product of two boundary-to-bulk propagators, glued by a ``two-point'' function along the boundary.}
\label{fig:NDdiff}
\end{figure}

\paragraph{Neumann boundary conditions.} For the Neumann boundary condition in Feynman gauge we find
\begin{align}
    \begin{aligned}
        \mathcal{W}^1_{s,\text{N}} &= -\frac{g^2}{8EE_\text{L}E_\text{R}}\frac{\langle \bar{1}\bar{2}\rangle \langle \bar{3}4\rangle}{k_1k_2k_3}\Big(\langle \bar{1}4\rangle\langle\bar{2}4\rangle\langle \bar{3}\bar{4}\rangle-\frac{E}{4k}\big(\langle \bar{1}4 \rangle \langle \bar{2}|k\bar{k}|\bar{3}\rangle+\langle \bar{2}4 \rangle \langle \bar{1}|k\bar{k}|\bar{3}\rangle+\\
        &\qquad\qquad\qquad\qquad\qquad\qquad\qquad\qquad\qquad\qquad+\underline{\langle \bar{1}\bar{3} \rangle \langle 4|k\bar{k}|\bar{2}\rangle+\langle \bar{2}\bar{3} \rangle \langle 4|k\bar{k}|\bar{1}\rangle}\big)\Big) \,,\\
        \mathcal{W}^2_{s,\text{N}} &= \frac{g^2}{32}\frac{\langle \bar{1}\bar{2}\rangle \langle \bar{3}4\rangle}{k_1k_2k_3k_4}\Big(\frac{\langle \bar{1}\bar{3} \rangle \langle \bar{2}4\rangle +\langle \bar{1}4 \rangle\langle \bar{2}\bar{3}\rangle}{E} +\underline{\frac{\langle \bar{1}\bar{2}|\bar{k}\bar{k}kk|\bar{3}4\rangle}{kE_\text{L}E_\text{R}}}\Big) \,,
    \end{aligned}
\end{align}
where
\begin{align}
    \langle \bar{1}4 \rangle \langle \bar{2}|k\bar{k}|\bar{3}\rangle+\langle \bar{2}4 \rangle \langle \bar{1}|k\bar{k}|\bar{3}\rangle+\langle \bar{1}\bar{3} \rangle \langle 4|k\bar{k}|\bar{2}\rangle+\langle \bar{2}\bar{3} \rangle \langle 4|k\bar{k}|\bar{1}\rangle = 2\big(\langle \bar{1}4 \rangle \langle\bar{1}\bar{3}\rangle\langle 1\bar{2}\rangle - \langle \bar{1}2 \rangle \langle \bar{2}\bar{3} \rangle \langle \bar{2}4 \rangle\big) \,.
\end{align}
Again, the energy poles of the contact diagram and $\mathcal{W}^2_{s,\text{N}}$ cancel. We have underlined the terms that change sign when going from Dirichlet to Neumann, the rest remains the same. The difference $\mathcal{W}_{s,\text{N}-\text{D}}\equiv \mathcal{W}_{s,\text{N}}-\mathcal{W}_{s,\text{D}}$ between the Neumann and Dirichlet correlator for the diagram in \eqref{w1w2} reads
\begin{align}
    \begin{aligned}
        \mathcal{W}^1_{s,\text{N}-\text{D}} &= \frac{g^2}{16kE_\text{L}E_\text{R}}\frac{\langle \bar{1}\bar{2}\rangle\langle \bar{3}4\rangle}{k_1k_2k_3}\Big(\langle \bar{1}\bar{3} \rangle \langle 4|k\bar{k}|\bar{2}\rangle+\langle\bar{2}\bar{3}\rangle\langle4|k\bar{k}|\bar{1}\rangle\Big) \,,\\
        \mathcal{W}^2_{s,\text{N}-\text{D}} &= \frac{g^2}{16kE_\text{L}E_\text{R}}\frac{\langle \bar{1}\bar{2}\rangle\langle \bar{3}4\rangle}{k_1k_2k_3k_4}\langle \bar{1}\bar{2}|\bar{k}\bar{k}kk|\bar{3}4\rangle \,,
    \end{aligned}
\end{align}
which can again be expressed in only boundary momentum spinors using \eqref{bdymomspin} and
\begin{align}
    \langle \bar{1}\bar{3}\rangle \langle 4|k\bar{k}|\bar{2} \rangle + \langle \bar{2}\bar{3} \rangle \langle 4 |k\bar{k} | \bar{1} \rangle = \langle \bar{1} \bar{2} \rangle \big(\langle \bar{1}\bar{3} \rangle \langle 14\rangle - \langle \bar{2}\bar{3} \rangle \langle 24 \rangle\big)+(k-k_1-k_2)\big(\langle \bar{1}\bar{3} \rangle \langle \bar{2}4 \rangle + \langle 14 \rangle \langle \bar{2}\bar{3} \rangle\big) \,.
\end{align}
As it was discussed in Section \ref{sec:AdSCFTdict}, this difference has a very special structure, see Figure \ref{fig:NDdiff}. Basically, it is the product of two three-point functions glued along one pair of legs. This structure is clearly visible from the formula above: it has $E_\text{L}$, $E_\text{R}$ poles characteristic of three-point functions.

\paragraph{Mixed boundary conditions.} The bulk-to-bulk propagator with mixed boundary conditions can be decomposed as in \eqref{Gdecomposed}. We will compute the correlators for each component separately and write
\begin{align}
    \mathcal{W}=\mathcal{W}_\text{inh}+\mathcal{W}_\text{pure gauge}+\mathcal{W}_\gamma\,.
\end{align}
Keeping in mind that the contact diagram is given by \eqref{cont}, as it is insensitive to boundary conditions, the other non-topological diagrams are given by
\begin{align}
    \begin{aligned}
        \mathcal{W}^1_{s,\text{inh}} &= -\frac{g^2}{8EE_\text{L}E_\text{R}}\frac{\langle \bar{1}\bar{2} \rangle \langle \bar{3}4\rangle}{k_1k_2k_3}\Big(\langle \bar{1}4\rangle \langle \bar{2} 4\rangle \langle \bar{3}4\rangle-\frac{E}{4k}\big(\langle \bar{1}4 \rangle\langle \bar{2}|k\bar{k}|\bar{3}\rangle+\langle \bar{2}4 \rangle\langle \bar{1}|k\bar{k}|\bar{3}\rangle\big)\Big) \,,\\
        \mathcal{W}_{s,\text{pure gauge}}^1 &=-\sigma\frac{g^2}{32k^2E_\text{L}}\frac{\langle \bar{1}\bar{2}\rangle^2\langle \bar{3}4\rangle^2}{k_1k_2k_3}(k_1-k_2) \,,\\
        \mathcal{W}^1_{s,\gamma} &= -e^{-2i\gamma}\frac{g^2}{32k^2E_\text{L}E_\text{R}}\frac{\langle \bar{1}\bar{2}\rangle \langle \bar{3}4\rangle}{k_1k_2k_3}\langle \bar{1}\bar{2}|\bar{k}\bar{k}kk|\bar{3}4\rangle\,,\\
        \mathcal{W}^2_{s,\text{inh}} &= \frac{g^2}{32E}\frac{\langle \bar{1}\bar{2}\rangle \langle \bar{3}4\rangle}{k_1k_2k_3k_4}\big(\langle \bar{1}\bar{3}\rangle \langle \bar{2}4\rangle +\langle \bar{1}4\rangle \langle \bar{2}\bar{3}\rangle\big)\,,\\
        \mathcal{W}^2_{s,\text{pure gauge}}&= 0\,,\\
        \mathcal{W}^2_{s,\gamma} &= -e^{-2i\gamma}\frac{g^2}{64kE_\text{L}E_\text{R}}\frac{\langle \bar{1}\bar{2} \rangle \langle \bar{3}4\rangle}{k_1k_2k_3k_4}\langle \bar{1}\bar{2}|\bar{k}\bar{k}kk|\bar{3}4\rangle\,.
    \end{aligned}
\end{align}
To compare to the SDYM result, we note that the total $\mathcal{W}^2_{s}=\mathcal{W}^2_{s,\text{inh}}+\mathcal{W}^2_{s,\text{pure gauge}}+\mathcal{W}^2_{s,\gamma}+\mathcal{W}_s^\text{contact}$ vanishes in the self-dual limit. The remainder becomes
\begin{align}
    \begin{aligned}
        &\big(\mathcal{W}^1_{s,\text{inh}} + \mathcal{W}_{s,\text{pure gauge}} +\mathcal{W}^1_{s,\gamma}\big)\Big|_{\gamma\rightarrow-i\infty} =\\
        &=-\frac{g^2}{8EE_\text{L}E_\text{R}}\frac{\langle \bar{1}\bar{2} \rangle \langle \bar{3}4\rangle}{k_1k_2k_3}\Big(\langle \bar{1}4\rangle \langle \bar{2} 4\rangle \langle \bar{3}4\rangle-\frac{E}{4k}\big(\langle \bar{1}4 \rangle\langle \bar{2}|k\bar{k}|\bar{3}\rangle+\langle \bar{2}4 \rangle\langle \bar{1}|k\bar{k}|\bar{3}\rangle\big)\Big)+\\
        &-\sigma\frac{g^2}{32k^2E_\text{L}}\frac{\langle \bar{1}\bar{2}\rangle^2\langle \bar{3}4\rangle^2}{k_1k_2k_3}(k_1-k_2) \,,
    \end{aligned}
\end{align}
where we observe that $\mathcal{W}^{1/2}_{s,\gamma}$ vanishes in the limit. It is not immediately obvious that this agrees with the SDYM correlator \eqref{SDYM4ptLorenzAx}. However, this is easily confirmed by writing
\begin{align}
    \begin{aligned}
        \frac{1}{2}\big(\langle \bar{1}4\rangle \langle \bar{2}|k\bar{k}|\bar{3}\rangle + \langle \bar{2}4\rangle \langle \bar{1}|k\bar{k}|\bar{3}\rangle\big) &= -\bar{1}^A\bar{2}^A\bar{3}^{B'}4^B\epsilon_{AB}k_A\bar{k}_{B'} =\\
        &= -\frac{1}{2k}\langle \bar{1}\bar{2}|kk\bar{k}\bar{k}|\bar{3}4\rangle -\frac{E_\text{R}}{2k}\langle \bar{1}\bar{2}\rangle \langle \bar{3}4\rangle (k_1-k_2) \,,
    \end{aligned}
\end{align}
for which we used $\epsilon_{AB}=\frac{1}{2k}(k_A\bar{k}_B-\bar{k}_Ak_B)$.

\paragraph{Gauge dependence.} The gauge variation of the mixed boundary conditions correlation functions resides only inside the diagram $\mathcal{W}_s^1$: in $\mathcal{W}_s^2$ we use a totally gauge-invariant two-point function $\langle F F\rangle$ and $\mathcal{W}^{\text{contact}}_s$ contains no bulk-to-bulk propagator. Moreover, the $\gamma$-dependent component of the propagators are the same in all gauges, so it suffices to only consider the Neumann component. To simplify calculations, we recall that the difference between the Dirichlet and Neumann propagators are expressed as $\nabla\xi_\text{D}$ and $\nabla\xi_\text{N}$, respectively, see \eqref{deltaphiphi} and \eqref{deltapsiphietaN} for explicit expressions. Since $\nabla\xi_\text{D}$ does not lead to a change in the correlator (for it vanishes on the boundary), $\nabla\xi_\text{N}$ can effectively be rewritten as the much simpler $\nabla\Delta\xi\equiv \nabla\xi_\text{N}-\nabla\xi_\text{D}$, given in \eqref{deltanablaxi2}, when evaluated in the exchange diagram. This is illustrated in Figure \ref{fig:gaugevariation}. See Appendix \ref{sec:reltoflat} for more details. The gauge variation reads
\begin{align}
    \Delta\mathcal{W}_{s,\gamma}=\Delta\mathcal{W}^1_{s,\text{N}} = -\frac{g^2}{16E_\text{L}}\frac{\langle \bar{1}\bar{2} \rangle^2\langle \bar{3}4\rangle^2}{k_1k_2k_3}\frac{k_1-k_2}{k^2} \,,
\end{align}
which, again taking into account the coefficients in \eqref{W4Decomp}, agrees with the gauge variation in SDYM for the Neumann-like propagator, see \eqref{SDYMaxialcorr}.

The results are gauge independent for the Dirichlet propagator and, hence, for the same reason, the correlators are gauge-independent for the Dirichlet-like propagator with $\sigma=+1$.  

\begin{figure}
\[ \begin{tikzcd}[column sep = 20ex]
\mathcal{W}_\text{D,F} \arrow[leftrightarrow]{r}{\nabla\xi_\text{D} \rightarrow 0}& \mathcal{W}_{\text{D,A}} \\%
\mathcal{W}_{\text{N,F}} \arrow[leftrightarrow]{r}{\nabla\xi_\text{N} \rightarrow \nabla\xi_\text{N}-\nabla\xi_\text{D}\equiv \nabla\Delta \xi}& \mathcal{W}_{\text{N,A}}
\end{tikzcd}
\]
\label{fig:gaugevariation}
\caption{In the Dirichlet case, the correlation function is the same in Feynman gauge and axial gauge. It is therefore easier to evaluate the gauge variation for Neumann by subtracting the Dirichlet gauge variation to obtain the simpler gauge variation $\nabla\Delta\xi$.}
\end{figure}

\paragraph{Topological diagrams.}

\begin{figure}[h!]
\centering
    \begin{tikzpicture}
        \begin{scope}[xshift=-3cm]
            % Define the boundary circle
        \draw[thick] (0,0) circle (2.1cm);
        
        % Define the vertices inside the bulk
        \coordinate (v1) at (-1.4,0);
        \coordinate (v2) at (0.7,0);
        
        % Define the external legs on the boundary
        \coordinate (a) at (-1.7,1.23);
        \coordinate (b) at (-1.7,-1.23);
        \coordinate (c) at (1.7,1.2);
        \coordinate (d) at (1.7,-1.2);

        \filldraw[fill=gray!20, draw=black,rotate=-90] 
            (1.61,-1.35) arc[start angle=40, end angle=140, radius=2.1cm];
        \filldraw[fill=gray!20, draw=black,rotate=-90] 
        (-1.61,-1.35) arc[start angle=220, end angle=320, radius=2.1cm];

                % Draw parallel momentum arrows
            \draw[-{Latex},thick,rotate=-90] (-0.9,-1.4) -- (-0.3,-1.2) node[midway,xshift=0.3 cm,yshift=0.2 cm] {\( k_1 \)};
            \draw[-{Latex},thick,rotate=-90] (0.9,-1.4) -- (0.3,-1.2) node[midway,xshift=0.25 cm,yshift=-0.2 cm] {\( k_2 \)};
        
        % Draw the gauge boson lines
        \draw[decorate,decoration={snake},thick] (a) -- (v1);
        \draw[decorate,decoration={snake},thick] (b) -- (v1);
        \draw[decorate,decoration={snake},thick] (c) -- (v2);
        \draw[decorate,decoration={snake},thick] (d) -- (v2);
        
        % Draw the internal propagator 
        \draw[decorate,decoration={snake},thick] (v1) -- (v2);
        
        % Draw parallel momentum arrows
        \draw[{Latex}-,thick] (0.9,0.6) -- (1.3,1) node[midway,above left] {\( k_4 \)};
        \draw[{Latex}-,thick] (0.9,-0.6) -- (1.3,-1) node[midway,below left] {\( k_3 \)};
        \draw[-{Latex},thick] (-0.2,0.2) -- (0.2,0.2) node[midway,above] {\( k_1+k_2 \)};
        
        % Add labels for fields next to endpoints
        \node[left] at (a) {\( \epsilon^+_1 \)};
        \node[left] at (b) {\( \epsilon^+_2 \)};
        \node[right] at (c) {\( \epsilon^-_4 \)};
        \node[right] at (d) {\( \epsilon^+_3 \)};

        \fill (v2) circle (5pt);
            \node at (3,0) {$=$};
        \end{scope}

            \begin{scope}[yshift=-6cm]
            \begin{scope}[xshift=-6cm]
                    % Define the boundary circle
            \draw[thick] (0,0) circle (2.1cm);
            
            % Define the vertices inside the bulk
            \coordinate (v1) at (-1.4,0);
            \coordinate (v2) at (0.7,0);
            
            % Define the external legs on the boundary
            \coordinate (a) at (-1.7,1.23);
            \coordinate (b) at (-1.7,-1.23);
            \coordinate (c) at (1.7,1.2);
            \coordinate (d) at (1.7,-1.2);

            \filldraw[fill=gray!20, draw=black,rotate=-90] 
                (1.61,-1.35) arc[start angle=40, end angle=140, radius=2.1cm];
            \filldraw[fill=gray!20, draw=black,rotate=-90] 
            (-1.61,-1.35) arc[start angle=220, end angle=320, radius=2.1cm];
    
                    % Draw parallel momentum arrows
                \draw[-{Latex},thick,rotate=-90] (-0.9,-1.4) -- (-0.3,-1.2) node[midway,xshift=0.3 cm,yshift=0.2 cm] {\( k_1 \)};
                \draw[-{Latex},thick,rotate=-90] (0.9,-1.4) -- (0.3,-1.2) node[midway,xshift=0.25 cm,yshift=-0.2 cm] {\( k_2 \)};
            
            % Draw the gauge boson lines
            \draw[decorate,decoration={snake},thick] (a) -- (v1);
            \draw[decorate,decoration={snake},thick] (b) -- (v1);
            \draw[decorate,decoration={snake},thick] (c) -- (v2);
            \draw[decorate,decoration={snake},thick] (d) -- (v2);
            
            % Draw the internal propagator 
            \draw[decorate,decoration={snake},thick] (v1) -- (v2);
            
            % Draw parallel momentum arrows
            \draw[{Latex}-,thick] (0.9,0.6) -- (1.3,1) node[midway,above left] {\( k_4 \)};
            \draw[{Latex}-,thick] (0.9,-0.6) -- (1.3,-1) node[midway,below left] {\( k_3 \)};
            \draw[-{Latex},thick] (-0.2,0.2) -- (0.2,0.2) node[midway,above] {\( k_1+k_2 \)};
            
            % Add labels for fields next to endpoints
            \node[left] at (a) {\( \epsilon^+_1 \)};
            \node[left] at (b) {\( \epsilon^+_2 \)};
            \node[right] at (c) {\( \epsilon^-_4 \)};
            \node[right] at (d) {\( \epsilon^+_3 \)};

            \fill[blue] (0.9,0.2) circle (5pt);
                \node at (3,0) {$+$};
            \end{scope}
            \begin{scope}
                % Define the boundary circle
        \draw[thick] (0,0) circle (2.1cm);
        
        % Define the vertices inside the bulk
        \coordinate (v1) at (-1.4,0);
        \coordinate (v2) at (0.7,0);
        
        % Define the external legs on the boundary
        \coordinate (a) at (-1.7,1.23);
        \coordinate (b) at (-1.7,-1.23);
        \coordinate (c) at (1.7,1.2);
        \coordinate (d) at (1.7,-1.2);

        \filldraw[fill=gray!20, draw=black,rotate=-90] 
            (1.61,-1.35) arc[start angle=40, end angle=140, radius=2.1cm];
        \filldraw[fill=gray!20, draw=black,rotate=-90] 
        (-1.61,-1.35) arc[start angle=220, end angle=320, radius=2.1cm];

                % Draw parallel momentum arrows
            \draw[-{Latex},thick,rotate=-90] (-0.9,-1.4) -- (-0.3,-1.2) node[midway,xshift=0.3 cm,yshift=0.2 cm] {\( k_1 \)};
            \draw[-{Latex},thick,rotate=-90] (0.9,-1.4) -- (0.3,-1.2) node[midway,xshift=0.25 cm,yshift=-0.2 cm] {\( k_2 \)};
        
        % Draw the gauge boson lines
        \draw[decorate,decoration={snake},thick] (a) -- (v1);
        \draw[decorate,decoration={snake},thick] (b) -- (v1);
        \draw[decorate,decoration={snake},thick] (c) -- (v2);
        \draw[decorate,decoration={snake},thick] (d) -- (v2);
        
        % Draw the internal propagator 
        \draw[decorate,decoration={snake},thick] (v1) -- (v2);
        
        % Draw parallel momentum arrows
        \draw[{Latex}-,thick] (0.9,0.6) -- (1.3,1) node[midway,above left] {\( k_4 \)};
        \draw[{Latex}-,thick] (0.9,-0.6) -- (1.3,-1) node[midway,below left] {\( k_3 \)};
        \draw[-{Latex},thick] (-0.2,0.2) -- (0.2,0.2) node[midway,above] {\( k_1+k_2 \)};
        
        % Add labels for fields next to endpoints
        \node[left] at (a) {\( \epsilon^+_1 \)};
        \node[left] at (b) {\( \epsilon^+_2 \)};
        \node[right] at (c) {\( \epsilon^-_4 \)};
        \node[right] at (d) {\( \epsilon^+_3 \)};

        \fill[blue] (0.5,0) circle (5pt);
            \end{scope}
            \end{scope}
        
    \end{tikzpicture}
\caption{The diagram $\mathcal{T}_s^{\text{L}}$ is expressed as two diagrams with different placements of the derivative in the SD vertex. The left diagram is denoted by $\mathcal{T}_s^{\text{L},1}$ and the right one by $\mathcal{T}_s^{\text{L},2}$.}
\label{fig:TLDecomp}
\end{figure}
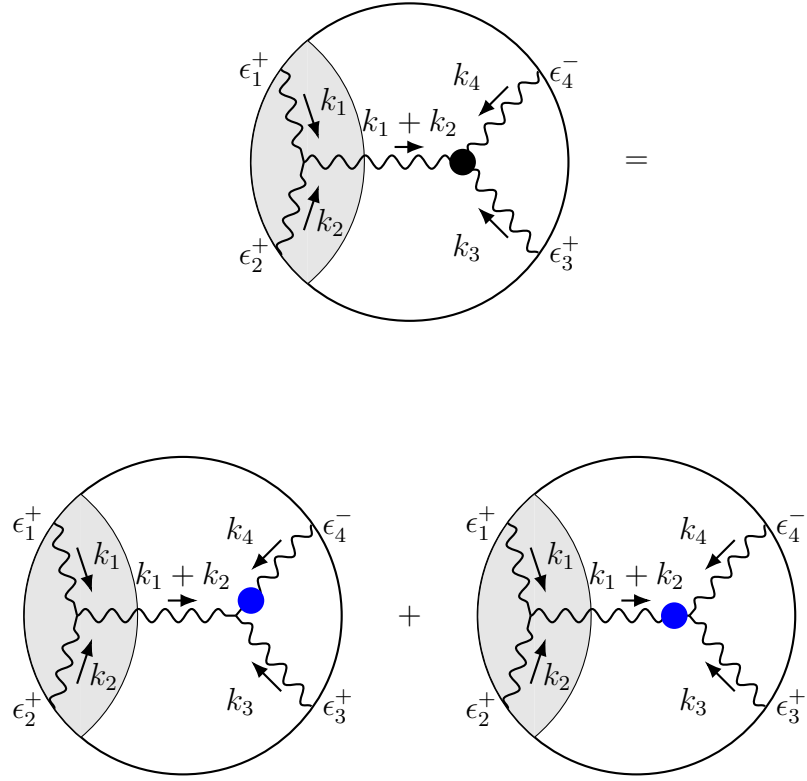

The distribution of derivatives in the diagram for $\mathcal{T}_s^{\text{L}}$ is displayed in Figure \ref{fig:TLDecomp}. It leads to the further decomposition of $\mathcal{T}_s^{\text{L}}$ into
\begin{align} \label{TDecomp}
    \mathcal{T}_s^{\text{L}} = \mathcal{T}_s^{\text{L},1}+\mathcal{T}_s^{\text{L},2} \,,
\end{align}
with $\mathcal{T}_s^{\text{L},1}$ and $\mathcal{T}_s^{\text{L},2}$ corresponding to the left and right diagram, respectively. For the diagram with the left vertex replaced by the topological vertex, $\mathcal{T}_s^\text{L}$, we have two contributions to the correlator, $\mathcal{T}_s^{\text{L},1}$ and $\mathcal{T}_s^{\text{L},2}$, as shown in Figure \ref{fig:TLDecomp}. They are given by
\begin{align}
    \begin{aligned}
        \mathcal{T}_s^{\text{L},1} = 2g^2\int \Phi_+^{AA'}(k_1,0)\Phi\fdud{+}{A}{A'}(k_2,0) \langle \Phi_{AA}(-k,0) \Phi_{BB'}(k,z') \rangle \Phi\fudu{+}{B,}{B'}(k_3,z') \Psi_-^{BB}(k_4,z') \,,\\
        \mathcal{T}_s^{\text{L},2} = -2g^2\int \Phi^{+AA'}(k_1,0)\Phi\fdud{+}{A}{A'}(k_2,0) \langle \Phi_{AA}(-k,0) \Psi_{BB'}(k,z') \rangle \Phi_+^{B,B'}(k_3,z') \Phi\fdud{-}{B}{B'}(k_4,z')\,.
    \end{aligned}
\end{align}

\begin{figure}[h!]
\centering
    \begin{tikzpicture}
        \begin{scope}[xshift=-6cm,rotate=180]
            \node at (-3,0) {$=$};
        
                    % Define the boundary circle
                \draw[thick] (0,0) circle (2.1cm);
                
                % Define the vertices inside the bulk
                \coordinate (v1) at (-1.4,0);
                \coordinate (v2) at (0.7,0);
                
                % Define the external legs on the boundary
                \coordinate (a) at (-1.7,1.23);
                \coordinate (b) at (-1.7,-1.23);
                \coordinate (c) at (1.7,1.2);
                \coordinate (d) at (1.7,-1.2);

                \filldraw[fill=gray!20, draw=black,rotate=-90] 
                    (1.61,-1.35) arc[start angle=40, end angle=140, radius=2.1cm];
                \filldraw[fill=gray!20, draw=black,rotate=-90] 
                (-1.61,-1.35) arc[start angle=220, end angle=320, radius=2.1cm];
        
                        % Draw parallel momentum arrows
                    \draw[-{Latex},thick,rotate=-90] (-0.9,-1.4) -- (-0.3,-1.2) node[midway,xshift=-0.3 cm,yshift=-0.2 cm] {\( k_3 \)};
                    \draw[-{Latex},thick,rotate=-90] (0.9,-1.4) -- (0.3,-1.2) node[midway,xshift=-0.25 cm,yshift=0.2 cm] {\( k_4 \)};
                
                % Draw the gauge boson lines
                \draw[decorate,decoration={snake},thick] (a) -- (v1);
                \draw[decorate,decoration={snake},thick] (b) -- (v1);
                \draw[decorate,decoration={snake},thick] (c) -- (v2);
                \draw[decorate,decoration={snake},thick] (d) -- (v2);
                
                % Draw the internal propagator 
                \draw[decorate,decoration={snake},thick] (v1) -- (v2);
                
                % Draw parallel momentum arrows
                \draw[{Latex}-,thick] (0.9,0.6) -- (1.3,1) node[midway,below right] {\( k_2 \)};
                \draw[{Latex}-,thick] (0.9,-0.6) -- (1.3,-1) node[midway,above right] {\( k_1 \)};
                \draw[-{Latex},thick]  (0.2,-0.2) -- (-0.2,-0.2)node[midway,xshift=-0.1 cm,yshift=0.3 cm] {\( k_1+k_2 \)};
                
                % Add labels for fields next to endpoints
                \node[xshift=0.25cm,yshift=-0.25cm] at (a) {\( \epsilon^+_3 \)};
                \node[xshift=0.25cm,yshift=0.25cm] at (b) {\( \epsilon^-_4 \)};
                \node[xshift=-0.25cm,yshift=-0.25cm] at (c) {\( \epsilon^+_2 \)};
                \node[xshift=-0.25cm,yshift=0.25cm] at (d) {\( \epsilon^+_1 \)};

                \fill (v2) circle (5pt);
        \end{scope}

                \begin{scope}[rotate=180]
                    % Define the boundary circle
                \draw[thick] (0,0) circle (2.1cm);
                
                % Define the vertices inside the bulk
                \coordinate (v1) at (-1.4,0);
                \coordinate (v2) at (0.7,0);
                
                % Define the external legs on the boundary
                \coordinate (a) at (-1.7,1.23);
                \coordinate (b) at (-1.7,-1.23);
                \coordinate (c) at (1.7,1.2);
                \coordinate (d) at (1.7,-1.2);

                \filldraw[fill=gray!20, draw=black,rotate=-90] 
                    (1.61,-1.35) arc[start angle=40, end angle=140, radius=2.1cm];
                \filldraw[fill=gray!20, draw=black,rotate=-90] 
                (-1.61,-1.35) arc[start angle=220, end angle=320, radius=2.1cm];
        
                        % Draw parallel momentum arrows
                    \draw[-{Latex},thick,rotate=-90] (-0.9,-1.4) -- (-0.3,-1.2) node[midway,xshift=-0.3 cm,yshift=-0.2 cm] {\( k_3 \)};
                    \draw[-{Latex},thick,rotate=-90] (0.9,-1.4) -- (0.3,-1.2) node[midway,xshift=-0.25 cm,yshift=0.2 cm] {\( k_4 \)};
                
                % Draw the gauge boson lines
                \draw[decorate,decoration={snake},thick] (a) -- (v1);
                \draw[decorate,decoration={snake},thick] (b) -- (v1);
                \draw[decorate,decoration={snake},thick] (c) -- (v2);
                \draw[decorate,decoration={snake},thick] (d) -- (v2);
                
                % Draw the internal propagator 
                \draw[decorate,decoration={snake},thick] (v1) -- (v2);
                
                % Draw parallel momentum arrows
                \draw[{Latex}-,thick] (0.9,0.6) -- (1.3,1) node[midway,below right] {\( k_2 \)};
                \draw[{Latex}-,thick] (0.9,-0.6) -- (1.3,-1) node[midway,above right] {\( k_1 \)};
                \draw[-{Latex},thick]  (0.2,-0.2) -- (-0.2,-0.2)node[midway,xshift=-0.1 cm,yshift=0.3 cm] {\( k_1+k_2 \)};
                
                % Add labels for fields next to endpoints
                \node[xshift=0.25cm,yshift=-0.25cm] at (a) {\( \epsilon^+_3 \)};
                \node[xshift=0.25cm,yshift=0.25cm] at (b) {\( \epsilon^-_4 \)};
                \node[xshift=-0.25cm,yshift=-0.25cm] at (c) {\( \epsilon^+_2 \)};
                \node[xshift=-0.25cm,yshift=0.25cm] at (d) {\( \epsilon^+_1 \)};
        
                \fill[blue] (0.5,0) circle (5pt);
                \end{scope}
    \end{tikzpicture}
\caption{The derivative from the SD vertex in the diagram $\mathcal{T}_s^\text{R}$ only gives a nonzero contribution if it acts on the internal line.}
\label{fig:TRDecomp}
\end{figure}
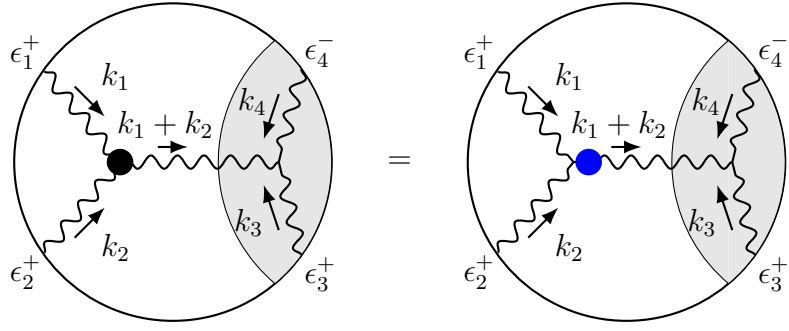
We denote the diagram with the right vertex replaced by the topological vertex $\mathcal{T}_s^\text{R}$. This contributes with only one subdiagram, see Figure \ref{fig:TRDecomp}. The partial correlator for $\mathcal{T}_s^\text{R}$ is computed from
\begin{align}
    \mathcal{T}_s^\text{R} = -2g^2\int_0^\infty dz \Phi_+^{AA'}(k_1,z)\Phi\fdud{+}{A}{A'}(k_2,z) \langle \Psi_{AA}(-k,z) \Phi_{BB}(k,0) \rangle \Phi_+^{BB'}(k_3,0) \Phi\fdud{-}{B}{B'}(k_4,0) \,.
\end{align}
The last partial correlator, $\mathcal{T}_s^{\text{L,R}}$, which has both vertices replaced by the topological vertex, is given by
\begin{align}
    \mathcal{T}_s^{\text{L,R}} = 4g^2\Phi_+^{AA'}(k_1,0)\Phi\fdud{+}{A}{A'}(k_2,0)\langle \Phi_{AA}(-k,0)\Phi_{BB}(k,0) \rangle\Phi_+^{BB'}(k_3,0)\Phi\fdud{-}{B}{B'}(k_4,0)\,. 
\end{align}
As previously mentioned, the topological diagrams are only relevant for mixed boundary conditions, including Neumann. We will present both cases separately. For the Neumann bulk-to-bulk propagator, one finds
\begin{align}
    \begin{aligned}
        \mathcal{T}^{\text{L},1}_{s,\text{N}} &= \frac{g^2}{8kE_\text{R}}\frac{\langle \bar{1}\bar{2}\rangle \langle \bar{3}4 \rangle}{k_1k_2k_3} \Big(\langle \bar{1}\bar{3} \rangle \langle \bar{2}4 \rangle + \langle \bar{1}4 \rangle \langle \bar{2}\bar{3}\rangle\Big) \,,\\
        \mathcal{T}^{\text{L},2}_{s,\text{N}} &= \frac{g^2}{32kE_\text{R}} \frac{\langle \bar{1}\bar{2} \rangle \langle \bar{3}4 \rangle}{k_1k_2k_3k_4}\Big(\langle \bar{1}\bar{3}\rangle \langle \bar{2}|\bar{k}k|4\rangle + \langle \bar{1}4\rangle \langle \bar{2}|\bar{k}k|\bar{3}\rangle + \langle \bar{2}\bar{3}\rangle \langle \bar{1}|\bar{k}k|4\rangle + \langle \bar{2}4\rangle \langle \bar{1}|\bar{k}k|\bar{3}\rangle\Big) \,, \\
        \mathcal{T}^{\text{R}}_{s,\text{N}} &= \frac{g^2}{32kE_\text{L}} \frac{\langle \bar{1}\bar{2} \rangle \langle \bar{3}4 \rangle}{k_1k_2k_3k_4}\Big(\langle \bar{1}\bar{3}\rangle \langle \bar{2}|\bar{k}k|4\rangle + \langle \bar{1}4\rangle \langle \bar{2}|\bar{k}k|\bar{3}\rangle + \langle \bar{2}\bar{3}\rangle \langle \bar{1}|\bar{k}k|4\rangle + \langle \bar{2}4\rangle \langle \bar{1}|\bar{k}k|\bar{3}\rangle\Big) \,,\\
        \mathcal{T}^{\text{L,R}}_{s,\text{N}} &= \frac{g^2}{16k} \frac{\langle \bar{1}\bar{2} \rangle \langle \bar{3}4 \rangle}{k_1k_2k_3k_4} \Big(\langle \bar{1}\bar{3} \rangle \langle \bar{2}4\rangle + \langle \bar{1}4 \rangle\langle \bar{2}\bar{3}\rangle\Big) \,.
    \end{aligned}
\end{align}

For mixed boundary conditions we obtain
{\allowdisplaybreaks\begin{align}\label{topDiag}
        \mathcal{T}^{\text{L},1}_{\text{inh}} &= \frac{g^2}{16kE_\text{R}}\frac{\langle \bar{1}\bar{2} \rangle \langle \bar{3}4\rangle}{k_1k_2k_3}\big(\langle \bar{1}\bar{3}\rangle \langle \bar{2}4\rangle+\langle \bar{1}4\rangle \langle \bar{2}\bar{3} \rangle\big)\,,\notag\\
        \mathcal{T}^{\text{L},1}_{\text{pure gauge}} &= -\sigma\frac{g^2}{16k^3E_\text{R}}\frac{\langle \bar{1}\bar{2}\rangle^2\langle \bar{3}4\rangle^2}{k_1k_2k_3}(k_1-k_2)(k_3+k_4)\,,\notag\\
        \mathcal{T}^{\text{L},1}_{\gamma} &= -\frac{g^2}{32k^3E_\text{R}} \frac{\langle \bar{1}\bar{2} \rangle \langle \bar{3}4\rangle }{k_1k_2k_3} \Big(e^{2i\gamma}\langle \bar{1}\bar{2}|kk\bar{k}\bar{k}|\bar{3}4\rangle + e^{-2i\gamma}\langle \bar{1}\bar{2}|\bar{k}\bar{k}kk|\bar{3}4\rangle\Big) \,,\notag\\
        \mathcal{T}^{\text{L},2}_{\text{inh}} &= \frac{g^2}{64kE_\text{R}}\frac{\langle \bar{1}\bar{2}\rangle\langle \bar{3}4\rangle}{k_1k_2k_3k_4}\Big(\langle \bar{1}\bar{3} \rangle \langle \bar{2}|\bar{k}k|4\rangle + \langle \bar{1}4 \rangle \langle \bar{2}|\bar{k}k|\bar{3}\rangle + \langle \bar{2}\bar{3} \rangle \langle \bar{1}|\bar{k}k|4\rangle + \langle \bar{2}4 \rangle \langle \bar{1}|\bar{k}k|\bar{3}\rangle\Big)\,,\notag\\
        \mathcal{T}^{\text{L},2}_{\text{pure gauge}} &= -\sigma \frac{g^2}{32k^2E_\text{R}}\frac{\langle \bar{1}\bar{2}\rangle^2\langle \bar{3}4\rangle^2}{k_1k_2k_3k_4}(k_1-k_2)(k_3+k_4)\,,\\
        \mathcal{T}^{\text{L},2}_{\gamma} &= -e^{-2i\gamma}\frac{g^2}{32k^2E_\text{R}} \frac{\langle \bar{1}\bar{2} \rangle \langle \bar{3}4 \rangle}{k_1k_2k_3k_4} \langle \bar{1}\bar{2}|\bar{k}\bar{k}kk|\bar{3}4\rangle\,,\notag\\
        \mathcal{T}^\text{R}_\text{inh} &= \frac{g^2}{64kE_\text{L}}\frac{\langle \bar{1}\bar{2}\rangle\langle \bar{3}4\rangle}{k_1k_2k_3k_4}\Big(\langle \bar{1}\bar{3} \rangle \langle \bar{2}|\bar{k}k|4\rangle + \langle \bar{1}4 \rangle \langle \bar{2}|\bar{k}k|\bar{3}\rangle + \langle \bar{2}\bar{3} \rangle \langle \bar{1}|\bar{k}k|4\rangle + \langle \bar{2}4 \rangle \langle \bar{1}|\bar{k}k|\bar{3}\rangle\Big)\,,\notag\\
        \mathcal{T}^\text{R}_\text{pure gauge} &= -\sigma \frac{g^2}{32k^2E_\text{L}}\frac{\langle \bar{1}\bar{2}\rangle^2\langle \bar{3}4\rangle^2}{k_1k_2k_3k_4}(k_1-k_2)(k_3+k_4) \,,\notag\\
        \mathcal{T}^{\text{R}}_{\gamma} &= -e^{-2i\gamma}\frac{g^2}{32k^2E_\text{L}} \frac{\langle \bar{1}\bar{2} \rangle \langle \bar{3}4 \rangle}{k_1k_2k_3k_4} \langle \bar{1}\bar{2}|\bar{k}\bar{k}kk|\bar{3}4\rangle\,,\notag\\
        \mathcal{T}^{\text{L,R}}_\text{inh} &= \frac{g^2}{16k}\frac{\langle \bar{1}\bar{2}\rangle \langle \bar{3}4\rangle}{k_1k_2k_3k_4}\big(\langle \bar{1}\bar{3} \rangle \langle \bar{2}4\rangle + \langle \bar{1}4\rangle \langle \bar{2}\bar{3} \rangle)\,,\notag\\
        \mathcal{T}^\text{L,R}_{\text{pure gauge}} &= -\sigma \frac{g^2}{32k^3}\frac{\langle \bar{1}\bar{2}\rangle^2\langle \bar{3}4\rangle^2}{k_1k_2k_3k_4}(k_1-k_2)(k_3+k_4)\,,\notag\\
        \mathcal{T}^{\text{L,R}}_{s,\gamma} &= -\frac{g^2}{32k^3} \frac{\langle \bar{1}\bar{2} \rangle \langle \bar{3}4 \rangle}{k_1k_2k_3k_4}\Big(e^{2i\gamma}\langle \bar{1}\bar{2}|kk\bar{k}\bar{k}|\bar{3}4\rangle + e^{-2i\gamma}\langle \bar{1}\bar{2}|\bar{k}\bar{k}kk|\bar{3}4\rangle\Big) \notag\,.
\end{align}}%
The gauge variation of the bulk-to-bulk propagator between the Feynman and axial gauge can be expressed as pure gauge terms, e.g. \eqref{deltaphiphi}. For Dirichlet boundary conditions, they vanish in the appropriate boundary limits and so they do not contribute to boundary diagrams. For mixed boundary conditions, the gauge parameters do not vanish in the respective boundary limits. However, we find that the total gauge variation becomes proportional to $\epsilon_{AA'}\epsilon_{BB'}$ after integrating by parts. We observe that
\begin{align}
    \epsilon^{CC'}V^{\text{top}}_{AA',BB',CC'}=0
\end{align}
and we conclude that topological diagrams are insensitive to the gauge variation. Let us point out that $\mathcal{T}^{\text{L},1}_{\gamma}$, when the factor of $2(b-a)$ is taken into account, survives in the self-dual limit. We analyze this issue shortly in Section \ref{subsec:cYM}.

%%%%%%%%%%%%%%%%%%%%%%%%%%%%%%%%%%%%%%%%%%%%%%%%%%%%%%%%%%%%%
\subsection{YM, cYM, Chalmers--Siegel and SDYM}
\label{subsec:cYM}
%%%%%%%%%%%%%%%%%%%%%%%%%%%%%%%%%%%%%%%%%%%%%%%%%%%%%%%%%%%%%
Let us discuss the results and the relation between them.  Before that let us recall the relation between the theories:
\begin{align}\notag 
   a F^2 +b F\wedge F \rightarrow 2a (F_+)^2+(b-a) d\text{CS}(A) \rightarrow 2\epsilon a(\Psi F_+-\tfrac{\epsilon}2\Psi^2) +(b-a) d\text{CS}(A) \rightarrow \Psi F \,.
\end{align}
Here, we start from YM plus a theta-term and move all bulk interactions into $(F_+)^2$, which is cYM. The latter can be massaged into the Chalmers--Siegel action. On dropping $\Psi^2$ and the theta-term we end up with SDYM. 

\paragraph{Chalmers--Siegel vs. YM.} First of all, Chalmers--Siegel theory and YM give the same results. This is not surprising, but there are subtleties. In general, the derivatives in the cubic vertex of cYM produce exactly the Chalmers--Siegel vertex. One genuine AdS/boundary effect is that the exchange diagram with two derivatives hitting the bulk-to-bulk propagator is not fully canceled by the quartic vertex (as they do in flat space). The two derivatives produce the $\langle F F\rangle$ two-point function. Its contact piece (inherited from the flat space) cancels the quartic contact diagram. The rest makes exactly the $\langle \Psi \Psi\rangle$-propagator of the Chalmers--Siegel theory, which vanishes in flat space. We have witnessed this effect for the four-point example, but the statement is general. 

The Chalmers--Siegel action has the theta-term even if it was zero in YM. The corresponding topological diagrams, which only make sense for non-Dirichlet boundary conditions, have milder singularities as compared to the terms inherited from flat space, which always have $1/E$-poles individually. Therefore, cYM, and its Chalmers--Siegel form, is an efficient way to separate terms by their analytic properties. Since the $(+++-)$-amplitude vanishes in flat space, the leading $1/E$-pole is absent in the amplitude.

\paragraph{Self-dual limit in YM.} As the ``flow'' of theories above illustrates, the YM theory with the theta-term fine-tuned to give cYM without the theta-term (equivalently, Chalmers--Siegel without the theta-term) is the right starting point to transmute into SDYM. This is exactly the value of the theta-term we obtained in the self-dual limit of mixed boundary conditions, see Section \ref{sec:boundaryterms}: $b=-ia\cot(\gamma)\rightarrow a$. As we also argued in Section \ref{sec:boundaryterms}, SDYM does not tolerate the theta-term. 

The two-point function in YM \eqref{PhiPhi2} with mixed boundary conditions gives the propagator for a gauge field $a_i=A_i|_{z=0}$ on the boundary. The corresponding two-point functions
\begin{align}
     \langle ++\rangle &= -\frac{1}{2k}(1-e^{-2i\gamma})\,, &
        \langle --\rangle &= -\frac{1}{2k}(1-e^{+2i\gamma})\,.
\end{align}
blow up in the self-dual limit (as many other quantities in YM theory that are absent in SDYM). This is not a big problem as external states can always be rotated to eliminate the divergent factors, see e.g. \cite{Jain:2024bza, Aharony:2024nqs}. Better behaved quantities belong to the Chalmers--Siegel theory, e.g. the $\langle \Psi \Phi\rangle$-propagator has a smooth self-dual limit, which leads to a two-point function of a contact type on the boundary. Note that in YM or Chalmers--Siegel theory $\Phi_+$ has a purely numerical divergent prefactor $(1-e^{+2i\gamma})$ or $e^{+2i\gamma}$, depending on whether it comes from $\Phi$ or $\bar{F}$ on the boundary. In SDYM $\Phi$ is fixed and finite. $\Psi$ behaves smoothly and approaches the SDYM $\Psi$ when $\gamma=-i\infty$. Due to the fact that the two-point function is contact, more information is needed to argue its coefficient has physical meaning.

The pure YM and SDYM are clearly different at the three-point level, where our result can be summarized as
\begin{align}
        \mathcal{W}_3 & = 2a\mathcal{W}_3^{\text{SD}}+(b-a)\mathcal{W}_3^{\text{top}}\,, && \mathcal{W}_3^{\text{SD}}=\tfrac12 W^{\text{SDYM}}\,.
\end{align}
The difference is exactly due to the theta/Chern--Simons term unless $b=a$, i.e. we have pure cYM. Therefore, YM/cYM/Chalmers--Siegel do reproduce the SDYM three-point function in the self-dual limit of $b=a$ and are different otherwise. Let us note that $b=a$ is possible in YM with Dirichlet boundary conditions as well.

The four-point correlator in SDYM does not have the $1/E$-pole, but it does depend on the gauge
\begin{align}\label{SDYMAxialcorrA}
    \mathcal{W}^{\text{SDYM}}_{s,\text{A}} = \mathcal{W}_{s,\text{L}}^{\text{SDYM}} + \mathcal{W}_{s,\text{gauge}}^{\text{SDYM}}\,.
\end{align}
One observes that due to the gauge dependence residing in total derivatives, the gauge dependence leads to a milder singularity $1/E_\text{L}$ as compared to the main part of the diagram that has $1/(E_\text{L} E_\text{R})$. 

For generic value of the theta-term cYM has additional topological diagrams, which are clearly absent in SDYM. It is easier to compare the Chalmers--Siegel form of cYM to SDYM as there is no contact diagram. As it was already noted, the part of the exchange with $\Psi-\Psi$-propagator vanishes in the self-dual limit. Therefore, there is a single $s$-channel exchange diagram that involves the $\Psi-\Phi$ propagator which has a smooth SDYM limit. 
\begin{align}
\mathcal{W}^1_{s,\gamma=-i\infty}=\mathcal{W}^1_{s,\text{inh}}+\mathcal{W}^{1}_{s,\text{pure gauge}}+\mathcal{W}^1_{s,\gamma=-i\infty}=\mathcal{W}^{\text{SDYM}}_{s} \,.
\end{align}
Therefore, we conclude here that SDYM does emerge in the limit of YM/cYM/Chalmers--Siegel theories provided the boundary terms are also taken into account. A word of warning, see the beginning of this section, is that there is an ambiguity in what external states to consider that may or may not require rescaling of the external lines by $\gamma$-dependent factors.

\paragraph{Composite operators.} As it was already mentioned, the gauge covariant dual current on the boundary is $J=*(da+ga\wedge a)$. In the main part of the text we have computed correlators of the gauge field $a$, which were adjusted to give those of $*da$ on the fly. If the correlators were amplitudes we could stop here. However, the $a\wedge a$-part is important for gauge covariance. From the QFT point of view, this is a typical composite operator. Its contribution is suppressed by $g$ and, hence, it is given by the diagrams made of two- and three-point functions. Due to the multitude of diagrams that need to be computed, we analyze the $a\wedge a$-part in the Appendix \ref{app:sdymleftovers}. Here, we summarize the main result in the self-dual limit. The topological diagram $\mathcal{T}^{\text{L},1}_{\gamma}$ that survives in the self-dual limit gets canceled by the composite operator contribution in the limit. Therefore, SDYM and YM agree in the self-dual limit. 

\paragraph{Comments on the light-cone gauge.} The light-cone gauge has always been instrumental in the study of self-dual theories. There has already been some work done for self-dual theories in the light-cone gauge within the (A)dS/CFT correspondence \cite{Skvortsov:2018uru, Neiman:2023bkq, Neiman:2024vit, Chowdhury:2024dcy,Kozaki:2025jrj}. However, we would like to draw one's attention to several features that make it more difficult to use in holographic applications. 

Firstly, when going into the light-cone gauge one has to drop a number of boundary terms that can potentially modify the result. For example, the theta-term does produce detectable corrections to the holographic correlators. This was discussed recently in \cite{Chowdhury:2024dcy}. Secondly, the free action in the light-cone gauge is the same for YM, Chalmers--Siegel and SDYM. It is just $\phi \square \bar \phi$. In the covariant approach, the two-point function depends a lot on the boundary terms and one can tune, for example, $\langle -- \rangle =0 $. Perhaps, in the light-cone gauge similar options are available provided one treats the shadow of $\phi(0)$ and $\pl_z \phi(0)$ as $q$, $p$ to impose conformally-invariant mixed boundary conditions and to adjust contact terms, as discussed in Appendix \ref{app:boundaryterms}. Thirdly, the light-cone basis is different from the helicity basis we use in momentum space, which can prevent one from comparing the results. In flat space there is a well-defined recipe to covariantize light-cone expressions, see e.g. \cite{Boels:2013bi}. It was applied in \cite{Chowdhury:2024dcy} to the covariantization of the amplitudes computed in SDYM. However, there is a single reference spinor shared by all legs, whereas in the helicity basis the reference spinor has to be the conjugate of the corresponding momentum spinor. We see that while our results for the three-point functions agree with \cite{Chowdhury:2024dcy} (modulo the topological term), the four-point amplitude has additional terms, which can, perhaps, be attributed to the gauge dependence and to the fact that certain boundary terms are kept that go to zero in the SDYM limit.\footnote{We are grateful to Arthur Lipstein for clarifying remarks about \cite{Chowdhury:2024dcy}. }

%%%%%%%%%%%%%%%%%%%%%%%%%%%%%%%%%%%%%%%%%%%%%%%%%%%%%%%%%%%%%
\section{Discussion and Conclusions}
\label{sec:conclusions}
%%%%%%%%%%%%%%%%%%%%%%%%%%%%%%%%%%%%%%%%%%%%%%%%%%%%%%%%%%%%%
Let us briefly summarize the main results, see also \cite{Skvortsov:2026gtq}. We have proposed the AdS/CFT dictionary for SDYM and computed the three- and four-point correlators in YM, Chalmers--Siegel theory and SDYM. It is shown that in the limit of self-dual boundary conditions the correlators of YM approach those of SDYM (the worst case scenario would have been that SDYM reproduces only a part of the corresponding YM correlator when restricted to a given helicity configuration and loop order). Several points are important to stress. Firstly, the results agree in several gauges. In particular, everything agrees with the complete physical gauge, which is considered in Appendix \ref{app:ham}. Secondly, there are additional topological diagrams in YM that are absent in SDYM and those contribution cancel against that of the composite operators.  

In Appendix \ref{app:collinear} we also consider the collinear limit of the three- and four-point correlators, following the recent \cite{Guevara:2026qwa} and \cite{Skvortsov:2026gtq}. To be precise, we massage the correlators as to reveal the corrections to the collinearity due to the energy nonconservation. 

Some obvious extensions of the results include the generalization to the higher-spin case, which is far from being trivial and has already begun in \cite{Skvortsov:2026ofl} with the example of HS-SDYM \cite{Ponomarev:2017nrr,Krasnov:2021nsq}. The maximal self-dual theory --- Chiral higher-spin gravity --- should be dual to a closed subsector of Chern--Simons vector models \cite{Sharapov:2022awp,Jain:2024bza, Aharony:2024nqs}.\footnote{Strictly speaking, in \cite{Jain:2024bza, Aharony:2024nqs} the case of Chern--Simons vector models without nonabelian global symmetries was considered, i.e. at most there can be $U(1)$ spin-one currents. The leading $1/E$-contribution to the $(++-)$-correlators agrees with our results.}  The definition of this closed subsector is yet to be completed starting from the four-point correlators and it would be interesting to compare the results of the present paper and of \cite{Skvortsov:2026ofl} with the CFT side in the future.

%%%%%%%%%%%%%%%%%%%%%%%%%%%%%%%%%%%%%%%%%%%%%%%%%%%%%%%%%%%%%
\section*{Acknowledgment}
%%%%%%%%%%%%%%%%%%%%%%%%%%%%%%%%%%%%%%%%%%%%%%%%%%%%%%%%%%%%%
We are indebted to Simone Giombi for a lot of very useful discussions regarding this project. 
E.S. is grateful to Dhruva K.S., Sachin Jain, Euihun Joung, Arthur Lipstein and Alexey Sharapov for useful discussions on the topics related to this paper. The work of E. S. and R. van D. was partially supported by the European Research Council (ERC) under the European Union’s Horizon 2020 research and innovation programme (grant agreement No 101002551).

%%%%%%%%%%%%%%%%%%%%%%%%%%%%%%%%%%%%%%%%%%%%%%%%%%%%%%%%%%%%%
\appendix
%%%%%%%%%%%%%%%%%%%%%%%%%%%%%%%%%%%%%%%%%%%%%%%%%%%%%%%%%%%%%
\section{Conventions}
\label{app:notation}
%%%%%%%%%%%%%%%%%%%%%%%%%%%%%%%%%%%%%%%%%%%%%%%%%%%%%%%%%%%%%
It can be useful to collect all notation at one place, which is where we are now. We use the Penrose-Rindler rules to raise and lower two-component indices, e.g. $\xi^A=\epsilon^{AB}\xi_B$ and $\xi^B\epsilon_{BA}=\xi_A$. These rules are applied to all objects, e.g. to derivatives, $\pl^A=\epsilon^{AB}\pl_B$, which makes $\pl^A\neq \pl/\pl y^A$. It is also true that $\epsilon^{12}=\epsilon_{12}=1$, $\epsilon_{AC}\epsilon^{AB}\equiv \epsilon\fdu{C}{B}\equiv \delta\fdu{C}{B}=-\epsilon\fud{B}{C}$. Two indices in which a tensor is symmetric or to be symmetrized can be denoted by the same letter, e.g. $\Psi^{AA}\equiv \Psi^{A_1A_2}$ and means that $\Psi$ is symmetric. The symmetrization is defined to be a projector, i.e. one has to divide by the number of permutations, e.g. $\xi^A\eta^A\equiv \tfrac12 (\xi^{A_1}\eta^{A_2}+\xi^{A_2}\eta^{A_1})$.

$4d$ vectors, in particular coordinates, are bi-spinors, $\mathrm{x}^{AA'}$. We define $\mathrm{x}^2\equiv -\tfrac12 \mathrm{x}_{AA'}\mathrm{x}^{AA'}= -\det \mathrm{x}^{AA'}$, $\mathrm{x}_{AA'}\mathrm{x}^{AA'}=2\det \mathrm{x}^{AA'}$. In Poincare coordinates or the flat half-space model of $\text{AdS}_4$ we split $\mathrm{x}^{AA'}=x^{AA'}+i \epsilon^{AA'} z$, where $x^{AA'}=x^{A'A}$ are along the boundary $z=0$. Note that $\pl_{AA'}\mathrm{x}^{BB'}=\epsilon\fdu{A}{B}\epsilon\fdu{A'}{B'}$, giving $\pl_{AA'}\mathrm{x}^{AA'}=4$. To get the Euclidean signature we can define
\begin{align}
    x^{AA'}&=\begin{pmatrix}
        w & x_3 \\ x_3& -\bar{w} 
    \end{pmatrix} \,,&
    \mathrm{x}^{AA'}&=\begin{pmatrix}
        w & x_3+iz \\ x_3-iz& -\bar{w} 
    \end{pmatrix}\,, & \mathrm{x}^2&=w\bar{w}+x_3^2+z^2\,,
\end{align}
where $w=x_1+ix_2$. The Hermiticity conditions for the Euclidean signature are $(x^{AA'})^\dag|_{x_3\rightarrow -x_3}=x_{AA'}$. In flat space the vierbein can be taken $e^{AA'}=d\mathrm{x}^{AA'}=dx^{AA'}+i \epsilon^{AA'} dz$. Regardless, the choice of coordinates we define the basis of (anti)self-dual two-forms
\begin{align}
    H^{AB}&= e\fud{A}{C'}\wedge e^{BC'}\,, & \bar{H}^{A'B'}&=e\fdu{C}{A'}\wedge e^{CB'}\,,
\end{align}
that satisfy 
\begin{align}
    e^{AA'}\wedge e^{BB'}&= \tfrac12 \epsilon^{AB} \bar{H}^{A'B'}+\tfrac12 \epsilon^{A'B'} {H}^{AB}\,.
\end{align}
We also need $H^{AB}\wedge \bar{H}^{A'B'}\equiv0$ and $H^{AA}\wedge H^{BB}=\tfrac13 \epsilon^{AB}\epsilon^{AB} \mathrm{vol}$, where $\mathrm{vol}\equiv H^{AB}\wedge H_{AB}= - \bar{H}_{A'B'}\wedge \bar{H}^{A'B'}$ is a unique (up to a multiple) four-form. In practice, $H_{AB}\wedge H^{AB}=v\, d^4x$. In the Cartesian coordinates we have
\begin{align}
    H^{AA}&= d^2x^{AA}+2i dx^{AA}\wedge dz\,, & H_{AB}\wedge H^{AB}&=4 i\, d^2x_{AA}\wedge dx^{AA} \wedge dz
\end{align}
and $d^2x_{AA}\wedge dx^{AA}=6 dw\wedge d\bar{w} \wedge db=-12 i d^3x$. Therefore, $v=48$. To properly normalize the kinetic term in the chiral form, let us massage its free part as
\begin{align}\notag
    F_{AB}F^{AB}&=\pl_{AC'}A\fdu{A}{C'}\pl\fud{A}{M'}A^{AM'}=A\fdu{A}{C'}p_{AC'}p\fud{A}{M'}A^{AM'} = -p^2 A_{MM'}A^{MM'} = 2p^2 A_\mu A^\mu\,,
\end{align}
where we temporarily imposed the Lorenz gauge condition $p\cdot A=0$. This implies $a=1/8$ for the action of the form $S_{\text{cYM}}= \frac{2a}{g^2}\,\Tr\int F_{AB}F^{AB} $.

%%%%%%%%%%%%%%%%%%%%%%%%%%%%%%%%%%%%%%%%%%%%%%%%%%%%%%%%%%%%%
\section{Lorenz gauge}
\label{app:Lorenz}
%%%%%%%%%%%%%%%%%%%%%%%%%%%%%%%%%%%%%%%%%%%%%%%%%%%%%%%%%%%%%
For completeness, let us also consider Lorenz gauge. The flat-space propagator is 
\begin{align}
        G_{\mu\nu}&=\frac{1}{p^2}\left(\eta_{\mu\nu}- \frac{p_\mu p_\nu}{p^2}\right) \,. 
\end{align}
Its direct Fourier transform leads to 
\begin{align}
    \begin{aligned}
    &G^\text{inh}_{AA';BB'}(-k,z;k,z')=\langle \Phi_{A,A'}(-k,z)\Phi_{B,B'}(k,z') \rangle^\text{inh} =\\
    &=\frac{e^{-k |z-z'|}}{k}\Big[ \epsilon_{AB}\epsilon_{A'B'} -\tfrac14 \epsilon_{AA'}\epsilon_{BB'}(1-k|z-z'|) -\tfrac{z}4 (k_{AA'} \epsilon_{BB'} +k_{BB'} \epsilon_{AA'})+\\
    &\qquad + \frac{1}{4k^2} (1+k|z-z'|) k_{AA'}k_{BB'}\Big] \,.
    \end{aligned}
\end{align}
As before, this can be linked to the Feynman gauge via a pure gauge term
\begin{align}
    \langle \Phi_{A,A'}(-k,z)\Phi_{B,B'}(k,z') \rangle^\text{inh}_\text{L} &=\langle \Phi_{A,A'}(-k,z)\Phi_{B,B'}(k,z') \rangle^\text{inh}_\text{F}+\nabla_{-k,z}^{AA'} \nabla_{k,z'}^{BB'}\xi_{\text{c}} \,,
\end{align}
where
\begin{align}\label{doublyL}
    \xi_\text{c}&= -\frac{1}{4k^3}e^{-k|z-z'|}(1+k|z-z'|) \,.
\end{align}
Since the propagator differs by a double-gauge term, it produces the same contributions to $\Psi-\Psi$ and $\Psi-\Phi$ boundary behavior as the Feynman gauge. The inhomogeneous part satisfies 
\begin{align}
    \begin{aligned}
        \square \langle \Phi_{A,A'}(-k,z) \Phi_{B,B'}(k,0) \rangle&=\delta(z)(\epsilon^{AA'}\epsilon^{BB'}-2\epsilon^{AB}\epsilon^{A'B'})+\\
        &-\frac{e^{-k|z|}}{2k}(k^{AA'}-\epsilon^{AA'}k \sign(z))(k^{BB'}-\epsilon^{BB'}k \sign(z)) \,,
    \end{aligned}
\end{align}
where the expression on the r.h.s. is divergence-free on both legs, i.e. it is a Fourier transform of $(\eta_{\mu\nu}-p_\mu p_\nu/p^2)$. The $\xi=0$ equations of motion (EOM) give 
\begin{align}
    \text{EOM}_{\xi=0} \langle \Phi_{A,A'}(-k,z) \Phi_{B,B'}(k,0) \rangle&=\delta(z)(\epsilon^{AA'}\epsilon^{BB'}-2\epsilon^{AB}\epsilon^{A'B'}) \,,
\end{align}
which is equivalent of $\text{EOM}_{\xi=0} G_{ij}=\delta(z-z') \delta_{ij}$ in the usual vector language. 

\paragraph{Homogeneous terms.} We can start with the most general ansatz of type $e^{-k(z+z')} T_{AA';BB'}$, featuring the most general tensor structure that can be at most linear in $z$ and $z'$, bearing in mind that the inhomogeneous part also has at most linear terms in front of the exponent. However, such linear terms are inconsistent with the free equations of motion, $\square \Phi^{A,A'}=0$, and transversality $\nabla_{AA'} \Phi^{A,A'}=0$. Therefore, $T_{AA';BB'}$ does not depend on $z$, $z'$. There are $9$ linearly independent structures that survive in $T_{AA';BB'}$. However, $5$ of them can be represented in a pure gauge form with the help of
\begin{align}
    T^{AA'}_1 \nabla^{BB'} e^{-k(z+z')}+ T_2^{BB'}\nabla^{AA'}e^{-k(z+z')}+c\nabla^{AA'}\nabla^{BB'} e^{-k(z+z')} \,.
\end{align}
The ansatz above is also assumed to be constrained by the transversality, which leaves exactly the $5$ free parameters. Since the inhomogeneous part decays exponentially as the points run away from each other, the homogeneous component only serves to impose the boundary conditions and there is no issue with regularity (as different from the axial gauge). 

\paragraph{$\boldsymbol{\Psi-\Psi}$ boundary conditions. } As the counting here-above shows, we are left with a $4$-parameter family of homogeneous terms that survive up to the $\Psi-\Psi$ boundary conditions. There is no degeneracy for $\gamma=0$ (Dirichlet) and the mixed boundary conditions fix the homogeneous term to be
\begin{align}
G^{\text{hom}}_{AA',BB'}&=G^{\text{homo},\gamma}_{AA',BB'}+\text{pure gauge} \,.
\end{align}

\paragraph{$\boldsymbol{\Psi-\Phi}$ boundary conditions.} In the Lorenz gauge this simply fails. The linear pre-exponent terms present in the inhomogeneous solution produce contributions that cannot be canceled by the homogeneous ones except for Dirichlet boundary conditions. This clearly shows that the $\Psi-\Phi$ boundary conditions can be too strong. Lastly, the doubly pure gauge term \eqref{doublyL} should be compensated by a homogeneous one as for $R_\xi$-gauge.

\paragraph{$R_\xi$-gauge.} $R_\xi$ can also be useful to check that the physical results do not depend on $\xi$. In the flat space the difference between the Feynman and the $R_\xi$ propagators is a doubly pure-gauge term of the form
\begin{align}
    (\xi -1) \frac{p_\mu p_\nu}{p^2 p^2} \,.
\end{align}
Therefore, in order to get the $R_\xi$-propagator we just need to add 
\begin{align}
    \nabla^{AA'}_{-k,z}\nabla^{BB'}_{k,z'}\xi_\text{c} \,,& && \xi_\text{c}=(\xi -1)\frac{e^{-k|z-z'|}}{4k^3}(1+k |z-z'|) \,.
\end{align}
Since it is a doubly pure-gauge term, it leads to the same results for the homogeneous terms that are fixed by the boundary conditions. Let us also note that in the double boundary limit  $\xi_\text{c}\neq0$ and one can expect a contribution to the four-point function coming from the integration by parts and it is a nontrivial consistency check that the final result is $\xi$-independent. The integration by parts gives $\epsilon^{AA'}\epsilon^{BB'}(4k^3)^{-1}$. Therefore, it is clear that one needs to add the following homogeneous pure gauge term
\begin{align}
    \nabla^{AA'}_{-k,z}\nabla^{BB'}_{k,z'}\tilde{\xi}_\text{c} \,,& && \tilde{\xi}_\text{c}=-(\xi -1)\frac{e^{-k(z+z')}}{4k^3}(1+k (z+z')) \,.
\end{align}

%%%%%%%%%%%%%%%%%%%%%%%%%%%%%%%%%%%%%%%%%%%%%%%%%%%%%%%%%%%%%
\section{Fourier transforms, propagators and integrals}
\label{app:fourier}
%%%%%%%%%%%%%%%%%%%%%%%%%%%%%%%%%%%%%%%%%%%%%%%%%%%%%%%%%%%%%
\paragraph{Fourier transforms.}
All Fourier transforms can be done with the help of the following 
\begin{align}
    \mathrm{Pv}\frac{1}{\omega}&= \frac{\omega}{(\omega^2+\mu^2)}\,, & \mathrm{Pv}\frac{1}{\omega^2}&= \frac{\omega^2-\mu^2}{(\omega^2+\mu^2)^2} \,.
\end{align}
Therefore, we get the following integrals
\begin{align}
    \begin{aligned}
        \frac{1}{\pi}\int e^{i\omega z} \frac{1}{(\omega^2+k^2)}&=e^{-k |z|}\frac{1}{k} \,,\\
        \frac{1}{\pi}\int e^{i\omega z} \frac{1}{(\omega^2+k^2)^2}&=e^{-k |z|}\frac{(1+k|z|)}{2k^3} \,,\\
        \frac{1}{\pi}\int e^{i\omega z} \frac{i \omega}{(\omega^2+k^2)(\omega^2+\mu^2)}&=(e^{-k|z|}-1)\frac{\sign(z)}{k^2} \,,\\
        \frac{1}{\pi}\int e^{i\omega z} \frac{1}{(\omega^2+k^2)}\frac{\omega^2-\mu^2}{(\omega^2+\mu^2)^2}&= -\frac{1}{k^3}(e^{-k|z|}+k|z|) \,,\\
        \frac{1}{\pi}\int e^{i\omega z} \frac{\omega}{(\omega^2+\mu^2)}&=-\sign z \,,\\
        \frac{1}{\pi}\int e^{i\omega z} \frac{\omega^2-\mu^2}{(\omega^2+\mu^2)^2}&=-|z| \,.
     \end{aligned}
\end{align}
\paragraph{Feynman gauge propagators.} The Dirichlet and Neumann propagators in Feynman gauge are
\begin{align}
    \begin{aligned}
        \langle \Phi_{A,A'}(-k,z)\Phi_{B,B'}(k,z') \rangle_\text{D} &= -\frac{\epsilon_{AB}\epsilon_{A'B'}}{2k}\Big(e^{-k|z-z'|}-e^{-k(z+z')}\Big)-\frac{\epsilon_{AA'}\epsilon_{BB'}}{2k}e^{-k(z+z')} \,,\\
        \langle \Phi_{A,A'}(-k,z)\Phi_{B,B'}(k,z') \rangle_\text{N} &= -\frac{\epsilon_{AB}\epsilon_{A'B'}}{2k}\Big(e^{-k|z-z'|}+e^{-k(z+z')}\Big)+\frac{\epsilon_{AA'}\epsilon_{BB'}}{2k}e^{-k(z+z')} \,,
    \end{aligned}
\end{align}
respectively. For mixed boundary conditions we use the decomposition
\begin{align} \label{Gdecomposed}
    G=G_\text{inh}+G_\text{pure gauge}+G_\gamma
\end{align}
and the various components read
\begin{align}
    \begin{aligned}
        \langle \Phi_{A,A'}(-k,z)\Phi_{B,B'}(k,z') \rangle_\text{inh} &= -\frac{\epsilon_{AB}\epsilon_{A'B'}}{2k}e^{-k|z-z'|}\,,\\
        \langle \Phi_{A,A'}(-k,z)\Phi_{B,B'}(k,z') \rangle_\text{pure gauge} &= -\frac{\sigma}{4k}\Big(\frac{k_{AA'}k_{BB'}}{k^2}+\epsilon_{AA'}\epsilon_{BB'}\Big)e^{-k(z+z')}\,,\\
        \langle \Phi_{A,A'}(-k,z)\Phi_{B,B'}(k,z') \rangle_\gamma &= \frac{e^{-k(z+z')}}{2k} \Pi^\gamma_{AA',BB'}\,,
    \end{aligned}
\end{align}
see \eqref{Pigamma} for the definition of $\Pi^\gamma$. 
The derivatives of the Dirichlet and Neumann propagators read
\begin{align}
    \begin{aligned}
        \langle \Psi_{AA}(-k,z)\Phi_{B,B'}(k,z')\rangle_\text{D} &= -\frac{\epsilon_{AB}}{2k}\big(k_{AB'}-\text{sign}(z-z')k\epsilon_{AB'}\big)e^{-k|z-z'|}+\frac{\epsilon_{AB'}}{2k}\bar{k}_Ak_Be^{-k(z+z')} \,,\\
        \langle \Psi_{AA}(-k,z)\Phi_{B,B'}(k,z')\rangle_\text{N} &= -\frac{\epsilon_{AB}}{2k}\big(k_{AB'}-\text{sign}(z-z')k\epsilon_{AB'}\big)e^{-k|z-z'|}-\frac{\epsilon_{AB'}}{2k}\bar{k}_Ak_Be^{-k(z+z')} 
    \end{aligned}
\end{align}
and the propagators with derivatives on both legs are given by
\begin{align}
    \begin{aligned}
        \langle F_{AA}(-k,z)F_{BB}(k,z')\rangle_\text{D} &= -\epsilon_{AB}\epsilon_{AB}\delta(z-z')+\frac{\bar{k}_A\bar{k}_Ak_Bk_B}{2k}e^{-k(z+z')} \,,\\
        \langle F_{AA}(-k,z)F_{BB}(k,z')\rangle_\text{N} &= -\epsilon_{AB}\epsilon_{AB}\delta(z-z')-\frac{\bar{k}_A\bar{k}_Ak_Bk_B}{2k}e^{-k(z+z')} \,.
    \end{aligned}
\end{align}
The single derivatives of the components of the propagator with mixed boundary conditions are
\begin{align}
    \begin{aligned}
        \langle \Psi_{AA}(-k,z)\Phi_{B,B'}(k,z')\rangle^{\text{inh}} &= -\frac{\epsilon_{AB}}{2k}\big(k_{AB'}-\text{sign}(z-z')k\epsilon_{AB'}\big)e^{-k|z-z'|} \,,\\
        \langle \Psi_{AA}(-k,z)\Phi_{B,B'}(k,z')\rangle^{\text{pure gauge}} &= \sigma\frac{k_{AA}}{4k^2}\nabla_{BB'}^{k,z'}e^{-k(z+z')}\,,\\
        \langle \Psi_{AA}(-k,z)\Phi_{B,B'}(k,z')\rangle^{\text{homo,}\gamma} &= -\frac{e^{-2i\gamma}}{4k^2}\bar{k}_A\bar{k}_Ak_Bk_{B'}e^{-k(z+z')}\,.
    \end{aligned}
\end{align}
and their double derivatives are given by
\begin{align}
    \begin{aligned}
        \langle F_{AA}(-k,z)F_{BB}(k,z')\rangle^\text{inh} &= -\epsilon_{AB}\epsilon_{AB}\delta(z-z')\,,\\
        \langle \Psi_{AA}(-k,z)\Psi_{BB}(k,z')\rangle^\text{pure gauge} &= 0\,,\\
        \langle \Psi_{AA}(-k,z)\Psi_{BB}(k,z')\rangle^{\text{homo},\gamma} &= \frac{e^{-2i\gamma}}{2k}\bar{k}_A\bar{k}_Ak_Bk_Be^{-k(z+z')} \,.
    \end{aligned}
\end{align}

\paragraph{Axial gauge propagators.} In Appendix \ref{sec:DirYM}, we define the gauge variation between axial gauge and Feynman gauge, of the difference between the Neumann and Dirichlet propagator, $\Delta\Delta\langle \Phi_{AA'}(-k,z)\Phi_{B,B'}(k,z')\rangle$, see \eqref{deltanablaxi1} and \eqref{deltanablaxi2}. Alternatively, it can be expressed as
\begin{align}
    \Delta\Delta\langle \Phi_{AA'}(-k,z)\Phi_{B,B'}(k,z')\rangle &= -\frac{1}{4k^3}(k_A\bar{k}_{A'}k_B\bar{k}_{B'}+\bar{k}_Ak_{A'}\bar{k}_Bk_{B'})e^{-k(z+z')} \,.
\end{align}
Its single derivatives read
\begin{align}
    \begin{aligned}
        \Delta\Delta\langle \Psi_{AA}(-k,z)\Phi_{B,B'}(k,z')\rangle &= \frac{1}{2k^2}k_{AA}k_B\bar{k}_{B'}e^{-k(z+z')} \,,\\
        \Delta\Delta\langle \Phi_{A,A'}(-k,z)\Psi_{BB}(k,z')\rangle &= \frac{1}{2k^2}\bar{k}_Ak_{A'}k_{BB}e^{-k(z+z')}
    \end{aligned}
\end{align}
and its double derivative is given by
\begin{align}
    \begin{aligned}
        \Delta\Delta\langle \Psi_{AA}(-k,z)\Psi_{BB}(k,z')\rangle &= 0\,.
    \end{aligned}
\end{align}

\paragraph{Integrals.} It is useful to collect the following integrals
\begin{align}
    \int dz \, e^{-(k_1+k_2+k_3) z } &= \frac{1}{k_1+k_2+k_3} \,,\\
    \int dz\,dz'\, e^{-(k_1+k_2) z }  e^{-k(z+z')} e^{-(k_3+k_4) z' }&= \frac{1}{E_\text{L} E_\text{R}} \,,\\
    \int dz\,dz'\, e^{-(k_1+k_2) z }  e^{-k|z-z'|} e^{-(k_3+k_4) z' }&= \frac{1}{E} \left(\frac{1}{E_\text{L}}+\frac{1}{E_\text{R}}\right) \,,\\
    \int dz\,dz'\, e^{-(k_1+k_2) z }  \sign(z-z') e^{-k|z-z'|} e^{-(k_3+k_4) z' }&= \frac{1}{E} \left(\frac{1}{E_\text{L}}-\frac{1}{E_\text{R}}\right) \,, \\
    \int dz\,dz'\, e^{-(k_1+k_2) z }  \sign(z-z')  e^{-(k_3+k_4) z' }&=   \frac{1}{E} \left(\frac{1}{k_1+k_2}-\frac{1}{k_3+k_4}\right) \,,\\
    \int dz dz' e^{-(k_1+k_2)z}e^{-(k_3+k_4)z'} &= \frac{1}{(k_1+k_2)(k_3+k_4)} \,, \\
    \int dz\,dz'\, e^{-(k_1+k_2) z }  |z-z'|  e^{-(k_3+k_4) z' }&=   \frac{1}{E} \left(\frac{1}{(k_1+k_2)^2}+\frac{1}{(k_3+k_4)^2}\right) \,,\\
    \int dz\,dz'\, e^{-(k_1+k_2) z }  (z+z')  e^{-(k_3+k_4) z' }&=   \frac{E}{(k_1+k_2)^2(k_3+k_4)^2} \,,
\end{align}
where $E=k_1+k_2+k_3+k_4$, $E_\text{L}=k_1+k_2+k$, $E_\text{R}=k_3+k_4+k$, $k=|k_1+k_2|=|k_3+k_4|$. 

\paragraph{Propagator integrals.} With the above integrals, we will evaluate the bulk integrals that appear in the four-point diagrams in Lorenz/Feynman gauge, i.e. we only consider terms in the propagator that contain an exponential. These integrals consist of bulk-to-bulk propagators and the exponentials stemming from the boundary-to-bulk propagators. Since we keep the notation minimal, it is worth mentioning once that we integrate $z$ and $z'$ from $0$ to $\infty$.

The integral for the $\langle\Phi\Phi\rangle$ propagators for Dirichlet and Neumann boundary conditions read
\begin{align}
    \begin{aligned}
        \int e^{-(k_1+k_2)z}&\langle\Phi_{A,A'}(-k,z)\Phi_{B,B'}(k,z')\rangle_\text{D}\, e^{-(k_3+k_4)z'} = -\frac{1}{EE_\text{L}E_\text{R}}\big(\epsilon_{AB}\epsilon_{A'B'}+\frac{E}{2k}\epsilon_{AA'}\epsilon_{BB'}\big) \,,\\
        \int e^{-(k_1+k_2)z}&\langle\Phi_{A,A'}(-k,z)\Phi_{B,B'}(k,z')\rangle_\text{N}\, e^{-(k_3+k_4)z'} =\\
        &= -\frac{1}{EE_\text{L}E_\text{R}}\big(\epsilon_{AB}\epsilon_{A'B'}-\frac{E}{2k}\big(\epsilon_{AA'}\epsilon_{BB'}-2\epsilon_{AB}\epsilon_{A'B'}\big)\big)\,.
    \end{aligned}
\end{align}
For mixed boundary conditions we provide
\begin{align}
    \begin{aligned}
        \int e^{-(k_1+k_2)z}\langle \Phi_{AA'}(-k,z)\Phi_{BB'}(k,z')\rangle^\text{inh} e^{-(k_3+k_4)z'} &= -\frac{\epsilon_{AB}\epsilon_{A'B'}}{EE_\text{L}E_\text{R}}\big(1+\frac{E}{2k}\big)\,,\\
        \int e^{-(k_1+k_2)z}e^{-k(z+z')} e^{-(k_3+k_4)z'} &=\frac{1}{E_\text{L}E_\text{R}} \,.
    \end{aligned}
\end{align}
The latter integral is relevant for all homogeneous terms, i.e. $\langle \Phi\Phi\rangle^\text{pure gauge}$, $\langle \Phi\Phi \rangle^{\gamma}$, and their descendants. 
We observe that the residue of the energy pole is independent of the chosen boundary conditions. Moreover, it turns out to be proportional to the flat space $\langle \Phi\Phi \rangle$ propagator.

The flat space propagator for $\langle\Psi\Phi\rangle$ can also be obtained in the flat limit after performing the bulk integral of the single derivative propagator. Only the inhomogeneous component of the propagator yields the flat space propagator, while the homogeneous terms provide higher energy corrections. Let us therefore demonstrate the integral of the former:
\begin{align}
    \begin{aligned}
        &\int e^{-(k_1+k_2)z}\langle \Psi_{AA}(-k,z)\Phi_{B,B'}(k,z')\rangle^\text{inh}e^{-(k_3+k_4)z'} =\\
        = &-\frac{\epsilon_{AB}}{2kEE_\text{L}E_\text{R}}\Big(\big(E_\text{L}+E_\text{R}\big)k_{AB'}-k\big(E_\text{R}-E_\text{L}\big)\epsilon_{AB'}\Big) = \\
        = &-\frac{\epsilon_{AB}}{2kEE_\text{L}E_\text{R}}\Big((E+2k)k_{AB'}-\big(2(k_3+k_4)-E\big)k\epsilon_{AB'}\Big) =\\
        =& \frac{\epsilon_{AB}}{EE_\text{L}E_\text{R}}\big(3_A\bar{3}_{B'}+4_A\bar{4}_{B'}-\frac{E}{2k}k_A\bar{k}_{B'}\big) \,.
    \end{aligned}
\end{align}
The integrals of the single derivatives of the Dirichlet and Neumann propagators are\footnote{One can replace $3\bar{3}+4\bar{4}$ with $-(1\bar{1}+2\bar{2})$, but one needs to remember that the indices are not symmetrized (there is no full momentum conservation), hence, there will be a term of order $E$ as well.}
\begin{align} \label{DNints}
    \begin{aligned}
        \int e^{-(k_1+k_2)z}&\langle\Psi_{AA}(-k,z)\Phi_{B,B'}(k,z')\rangle_\text{D}\,e^{-(k_3+k_4)z'} =\\
        &=\frac{\epsilon_{AB}}{EE_\text{L}E_\text{R}}\big(3_A\bar{3}_{B'}+4_A\bar{4}_{B'}\big)-\frac{1}{2kE_\text{L}E_\text{R}}\Big(\epsilon_{AB}k_A\bar{k}_{B'}-\epsilon_{AB'}\bar{k}_Ak_B\Big) \,,\\
        \int e^{-(k_1+k_2)z}&\langle\Psi_{AA}(-k,z)\Phi_{B,B'}(k,z')\rangle_\text{N}\,e^{-(k_3+k_4)z'} =\\
        &=\frac{\epsilon_{AB}}{EE_\text{L}E_\text{R}}\big(3_A\bar{3}_{B'}+4_A\bar{4}_{B'}\big)-\frac{1}{2kE_\text{L}E_\text{R}}\Big(\epsilon_{AB}k_A\bar{k}_{B'}+\epsilon_{AB'}\bar{k}_Ak_B\Big) \,,\\
        \int e^{-(k_1+k_2)z}&\langle\Psi_{AA}(-k,z)\Phi_{B,B'}(k,z')\rangle^\text{inh}\,e^{-(k_3+k_4)z'} =\\
        &=\frac{\epsilon_{AB}}{EE_\text{L}E_\text{R}}\big(3_A\bar{3}_{B'}+4_A\bar{4}_{B'}\big)-\frac{1}{2kE_\text{L}E_\text{R}}\Big(\epsilon_{AB}k_A\bar{k}_{B'}+\epsilon_{AB'}\bar{k}_Ak_B\Big)  \,.
    \end{aligned}
\end{align}

The integrals of the double derivative of the propagators become
\begin{align}
    \begin{aligned}
        \int e^{-(k_1+k_2)z}&\langle F_{AA}(-k,z)F_{BB}(k,z')\rangle = -\frac{\epsilon_{AB}\epsilon_{AB}}{E} \,,\\
        \int e^{-(k_1+k_2)z}&\langle F_{AA}(-k,z) F_{BB}(k,z')\rangle_\text{D}\,e^{-(k_3+k_4)z'} =-\frac{\epsilon_{AB}\epsilon_{AB}}{E}+\frac{\bar{k}_A\bar{k}_Ak_{B}k_{B}}{2kE_\text{L}E_\text{R}} \,,\\
        \int e^{-(k_1+k_2)z}&\langle F_{AA}(-k,z)F_{BB}(k,z')\rangle_\text{N}\,e^{-(k_3+k_4)z'} =-\frac{\epsilon_{AB}\epsilon_{AB}}{E}-\frac{\bar{k}_A\bar{k}_Ak_{B}k_{B}}{2kE_\text{L}E_\text{R}} \,.
    \end{aligned}
\end{align}
The residue of the energy pole again coincides with the flat space two-point function. Lastly, for propagators with one leg on the boundary one finds
\begin{align}
    \begin{aligned}
        \int e^{-(k_1+k_2)z}G(k;z,z'=0)&=\frac{1}{E_\text{L}}G(k;0,0) \,,\\
        \int G(k;z=0,z')e^{-(k_3+k_4)z'}&=\frac{1}{E_\text{R}}G(k;0,0) \,.
    \end{aligned}
\end{align}
Here, $G$ contains $e^{-kz}$ or $e^{-kz'}$, which leads to $E_{\text{L,R}}$. On the r.h.s. this exponent is eliminated by having $z=z'=0$.

%%%%%%%%%%%%%%%%%%%%%%%%%%%%%%%%%%%%%%%%%%%%%%%%%%%%%%%%%%%%%
\section{Boundary terms, mixed boundary conditions}
\label{app:boundaryterms}
%%%%%%%%%%%%%%%%%%%%%%%%%%%%%%%%%%%%%%%%%%%%%%%%%%%%%%%%%%%%%
A toy model is the scalar field $\mathcal{L}=\tfrac12 (\pl_z\phi)^2+...$ that generates the following variation
\begin{align}
    \delta S&= \text{bulk eom}+ \int _{\pl M} \delta \phi \,\pl_z \phi \,.
\end{align}
This variation is suitable for Dirichlet boundary conditions and no additional terms are needed. As it was pointed out in e.g. \cite{Arutyunov:1998ve, Henneaux:1998ch,Papadimitriou:2004ap}, it is instructive to use the Hamiltonian language within the AdS/CFT correspondence, i.e. to write the bulk action in the form $S=\int p \dot{q} -H(q,p)$, where $\dot{q}=\dot{\phi}=\pl_z \phi$. Such an action is again designed for Dirichlet boundary conditions as it produces $p\delta q$ on the boundary. Within the AdS/CFT dictionary, the coordinate $q$ is fixed on the boundary to be the source $q|_{z=0}=q_0$ and the momentum, when the action is evaluated on-shell, will be equal to the one-point function $p=\langle O \rangle_{q_0}$ in the presence of $q_0$ according to $\delta S= \int_{\pl M} p\, \delta q$. This also gives two closely related recipes to compute tree-level correlators: either as the on-shell action as a functional of $q_0$ or as $p$ at the boundary as a functional of $q_0$. Now, suppose that we would like to impose different/mixed (Robin) boundary conditions, e.g. to keep $Q=aq+bp$ fixed at the boundary instead of $q$. Its variation implies $a\delta q+ b\delta p =0$, with $a\neq 0$, $b=0$ corresponding to the Dirichlet problem and $a=0$, $b\neq 0$ to the Neumann one.

One can always find a canonically conjugate variable, $P=cq+dp$, so that changing $(q,p)$ to $(Q,P)$ is a canonical transformation 
\begin{align}
    \begin{pmatrix}
        Q \\P
    \end{pmatrix} &= \begin{pmatrix}
        a & b \\c & d 
    \end{pmatrix} \begin{pmatrix}
        q \\ p
    \end{pmatrix}\,, && A=\begin{pmatrix}
        a & b \\c & d 
    \end{pmatrix}\in Sp(2) \,.
\end{align}
Therefore, any other (mixed) boundary condition is a Dirichlet one for another pair of canonical variables. In the new variables, the variation is (assuming $\delta Q=0$)
\begin{align}
    \delta S&= \text{bulk eom}+ \int _{\pl M}  (c Q-a P) b \delta P \,.
\end{align}
It can be compensated by adding a boundary term
\begin{align}
    S_{\text{bndry}}&= \int _{\pl M} \tfrac{\mu}{2} Q^2 -bc QP +\tfrac12 ab P^2 \,.
\end{align}
The coefficient of $Q^2$ is not fixed, but it is a local counter-term on the boundary. The on-shell variation of the action is $S=\int_{\pl M} P\delta Q$, as desired. The on-shell value of the action is $S=\tfrac12\int_{\pl M} P Q$ (always the same in the canonical variables and dropping the contact term). 

In the main text we choose to solve the equations of motion in terms of the initial variables $p$ and $q$ while imposing the mixed boundary conditions. It is also instructive to see what happens in terms of the new variables. The free equations have the form $\pl_z\xi= M \xi$, where $\xi=(q,p)$. The new equations are $\pl_z \eta=M'\eta$, $\eta=(Q,P)$ with $M'=A^{-1}MA$. For a generic $M'$ we find the following second order equation
\begin{align}
 \pl_z^2Q&= \pl_z Q\, \Tr M' - Q \det M' \,,
 && M'=\begin{pmatrix}
        \alpha & \beta \\\gamma & \delta 
    \end{pmatrix} \,,&&
    M=\begin{pmatrix}
        0 & 1 \\ k^2 & 0 
    \end{pmatrix}\,.
\end{align}
In our case, $M$ is as above and the equations turn out to have the same form $(\pl_z^2-k^2)Q=0$ (note that $\Tr M'=\Tr M$, $ \det M'= \det M$). The Dirichlet propagator is $e^{-kz}$ and the on-shell action leads to the standard kernel $k/2$. Therefore, there is absolutely no noticeable effect in the new canonical variables if the initial equations are $(\pl_z^2-k^2)q=0$. Interactions will make a difference, unless the theory is $Sp(2)$-invariant.  

With mixed boundary conditions for the scalar field it may still be convenient to stick to the initial field variables.  Therefore, we can use $a \delta q +b \delta p=0$ to find 
\begin{align}\label{boundarynonNeumann}
    S_{\text{bndry}}&= \int _{\pl M}\frac{b}{2a} p^2 +\frac{\mu'}2 Q^2 \,,
\end{align}
where $\mu'$ is again a free parameter and $Q=aq+bp$ (not meant as a new canonical variable). The on-shell action in the old variables is
\begin{align}\label{nonNeu}
    S_{\text{on-shell}}&= \int _{\pl M}\frac12 pq+S_{\text{bndry}}= \int _{\pl M}\frac12 pq +\frac{b}{2a} p^2 +\frac{\mu'}2 Q^2=\int _{\pl M}\frac{1}{2a} Qp +\frac{\mu'}2 Q^2 \,,
\end{align}
which can be massaged into
\begin{align}
    S_{\text{on-shell}}&= \int _{\pl M}\frac12 PQ+\frac{1}2(\mu' -\tfrac{c}{a}) Q^2 \,.
\end{align}
Therefore, $\mu=\mu' -\tfrac{c}{a}$ and the two on-shell actions agree, of course. An important point here is that the non-contact part of the on-shell action, which is due to $Qp/(2a)$, does not even depend on what $P$ is. Indeed, generically, $P\sim p+\bullet Q$, i.e. the difference between old $p$ and new $P$ is a contact term. Therefore, the only thing to decide is what the boundary condition is (equivalently, what the source to be kept fixed is or what the ratio $a/b$ is). There is no physical information in the choice of $P$ (ratio $c/d$), unless the coefficient of the contact term has a meaning. The overall scale factors of $P$ and $Q$ also do not matter. The simplest choice for \eqref{nonNeu} would be to have $P=p/a$.

In practice, we also face a situation where a contact term is already present (e.g. a Chern--Simons term)
\begin{align}
    \delta S= \text{bulk eom} + \int_{\pl M} (p+\xi q)\delta q \,.
\end{align}
In this case we just need to subtract the last piece to reduce the problem to the one we have already solved
\begin{align}
    S_{\text{bndry}}&= \int _{\pl M}-\frac{\xi}2 q^2+\frac{b}{2a} p^2 +\frac{\mu'}2 Q^2 \,.
\end{align}
The on-shell action does not change. In the Yang--Mills theory, $q\sim A_i$, $p\sim \pl_z A_i$ and the contact term is the Chern--Simons term, $AdA$, on the boundary (an additional power of the momentum as compared to just $Q^2$ does not change anything). Another special case is when we would like to approach the Neumann boundary condition as $a=0$. As it is seen from \eqref{boundarynonNeumann} this limit appears subtle. With a small rearrangement between the main and the contact term we can find another form, which is smooth for $a\rightarrow0$,
\begin{align}\label{boundarynonDirichlet}
    S_{\text{bndry}}&= \int _{\pl M}-pq -\frac{a}{2b} q^2 +\frac{\mu''}2 Q^2 \,.
\end{align}
With this choice the on-shell action is 
\begin{align}
    S_{\text{on-shell}}&= \int _{\pl M}\frac12 pq+S_{\text{bndry}}= \int _{\pl M}-\frac12 pq -\frac{a}{2b} q^2 +\frac{\mu''}2 Q^2=\int _{\pl M}-\frac{1}{2b} qQ +\frac{\mu''}2 Q^2 \,.
\end{align}
Lastly, let us note that there is an ambiguity in what $P$ is. The simplest option is to keep $P=p$, i.e. $c=0$, $ad=1$, which works for all cases, but the Neumann, where $Q=p$. In fact, from the AdS/CFT perspective it is the Dirichlet condition that is special (the leading falloff vanishes), while all mixed ones, Neumann's including, can be considered together since the leading falloff never vanishes. Therefore, for non-Dirichlet boundary conditions ($\gamma\neq0$ in the main text), we can adopt $P=c q$, $d=0$, $bc=-1$. This means that we can read off the correlation function by extracting the leading fall-off of the bulk field $\phi$ (modulo the $-1/b$-factor per leg).

\paragraph{Perturbation theory with mixed boundary conditions.} Suppose we have a mixed (Robin) boundary problem:
\begin{align}
   \Delta q&=\rho\,, &  aq + bq'\Big|_{\pl M}&= f\,, & aG+bG'\Big|_{\pl M}=0 \,.
\end{align}
Green's theorem implies ($s$ is a point on the boundary and $'$ is the normal derivative)
\begin{align}
    q(x)&= \int_M G(x,y)\,\rho(y) +\int_{\pl M} q(s) G'(s,x)-q'(s) G(s,x)   \,.
\end{align}
With the help of the boundary conditions one can rewrite the latter in two different ways\footnote{The Green's identity is by no means a statement that one can specify two boundary data independently.}
\begin{align}
    q(x)&= \int_M G(x,y)\,\rho(y) -\int_{\pl M} \frac{f(s)}{b} G(s,x)=   \int_M G(x,y)\,\rho(y) +\int_{\pl M} \frac{f(s)}{a} G'(s,x) \,.
\end{align}
From the AdS/CFT viewpoint, $\rho$ encodes nonlinear corrections, which will be taken into account order by order, while $f$ is the source and the combination of $q$ and $p=q'$ to be kept fixed at the boundary. The Dirichlet condition can be approached either as $b=0$ or $a=\infty$ (and the Neumann as $a=0$ or $b=\infty$). In the former case, $b=0$, we need to use the second expression above. In the latter case, $a=\infty$, we have to use the second and to rescale the source by $a$, as otherwise it will be suppressed. The first expression is suitable for the Neumann or generic mixed boundary conditions. The set of boundary conditions is optimistically $\mathbb{C(R)P}^1$, i.e. it is about the ratio $a/b$ or $b/a$. At generic values of $a$, $b$ one can think of $a$, $b$ as 'renormalizing' the source. When some of them becomes infinite one also has to rescale the source to approach the Dirichlet (Neumann) limits as $a=\infty$ ($b=\infty$). Since we are computing correlation functions, a finite rescaling of the source does not modify the properly normalized OPE coefficients, etc. Therefore, we will drop $a$ or $b$ from the denominator of the formula. 

\paragraph{Scalar field example.} As an example, let us impose mixed boundary condition on the scalar field, i.e. to keep
\begin{align}
    Q&=a q+ bp = a \phi +b \pl_z \phi
\end{align}
fixed at the boundary. The Green's function should satisfy $a G(0,z') +b \pl_z G(0,z')=0$. The solution is
\begin{align}
    G&=-\frac{1}{2k}\left(e^{-k|z-z'|}-\frac{a+bk}{a-bk}e^{-k(z+z')}\right) \,.
\end{align}
$b=0$ or $a\rightarrow\infty$ (Dirichlet) and $a=0$ or $b\rightarrow\infty$ (Neumann) are correctly recovered 
\begin{align}
    G&=-\frac{1}{2k}\left(e^{-k|z-z'|}\mp e^{-k(z+z')}\right) \,.
\end{align}
In particular, the boundary-to-bulk propagator is $\tfrac{b}{a-bk} e^{-kz'}$ for $b\neq0$. It is not conformally invariant since the mixed boundary conditions for the scalar field are not, except for the pure Neumann or Dirichlet. Dirichlet is a somewhat a special case since the leading falloff changes and one has $G \sim z K$ instead of $G\sim z^0 K$ for all the other boundary conditions. The boundary-to-bulk propagator is $e^{-k z'}$. Its first term in the small $z'$ expansion is a contact term, $\delta^3 (x-y)$, and it is the second term that determines the two-point function $\langle O(-k) O(k)\rangle\sim k$, which is of a dimension-$2$ operator. In the Neumann case the boundary-to-bulk propagator is $-\frac{1}{k}e^{-kz}$ and it is the $\mathcal{O}(z)$-coefficient that is a contact term. The two-point function is $\langle O(-k) O(k)\rangle\sim k^{-1}$, which is of a dimension-$1$ operator. As discussed in the main text, the difference between the Dirichlet and Neumann propagators
\begin{align}
    G_\text{D}-G_\text{N}&= \frac{1}{k}e^{-k(z+z')}= \frac{1}{k} e^{-kz} \langle O_2 O_2 \rangle \frac{1}{k} e^{-kz'} 
\end{align}
is the product of two $\Delta=1$ boundary-to-bulk propagators times the two-point function of the Hubbard-Stratonovich field.  

As one can see, the mixed boundary conditions are not conformally invariant for the scalar field. Nevertheless, one can partially replicate the Yang--Mills case by imposing different mixed boundary conditions, keeping $Q=a k q+ b p$ fixed. Here, multiplying by $k$ mimics the magnetic field, $B^i \sim \epsilon^{ijk} \pl_j A_k$, while $p\sim \pl_z \phi$, corresponding to the subleading falloff, mimics the electric field $E_i\sim \pl_z A_i$. While this is fine for the Maxwell theory, taking $k \phi$ is a nonlocal operation in position space for the scalar field, but it makes sense and it is just the shadow transform that maps an operator of dimension $\Delta$ to the one of dimension $d-\Delta$. Indeed, $\phi$ at the boundary has conformal dimension $1$, its shadow has dimension $2$ (note that $\tilde O = (k^2)^{d/2-\Delta} O$ up to a numerical factor for a dimension-$\Delta$ scalar operator $O$). The new (conformally-invariant) Green's function is
\begin{align}
    G&=-\frac{1}{2k}\left(e^{-k|z-z'|}-\frac{a+b}{a-b}e^{-k(z+z')}\right) \,.
\end{align}
The scalar field example is, perhaps, the simplest one where a one-parameter family of boundary conditions can be imposed and it is a toy model for the case of a gauge field.\footnote{In principle, one can consider one-parameter family of boundary conditions for a massive scalar field with an arbitrary mass since the dimensions of the dual operators are $\Delta$ and $d-\Delta$, i.e. the shadow transform still works!} We will see that this is also the propagator for the helicity states in Yang--Mills theory with mixed boundary conditions.  

The Hamiltonian form of the action has certain advantages for AdS/CFT applications. Some theories are naturally in the first order form (e.g. SDYM or fermions), the rest can be put into such a form. For instance, the scalar field toy model can be recast in the first order form as $p \pl_z \phi-\epsilon p^2/2+...$, where $\epsilon p=\pl_z \phi$ is the equation of motion for $p$ (here, $\epsilon$ is reminiscent of the one in the Chalmers--Siegel action). When written in the first order form, there is also a $\langle p \phi \rangle$ two-point function, which has the inhomogeneous term $\sign(z-z') \exp[-k|z-z'|]$, similar to the SDYM propagator. Introducing $p$ can even be helpful for the Dirichlet boundary conditions as $\pl_z \phi$ picks the subleading falloff's coefficient. 

\paragraph{Toy model of SDYM and Chalmers--Siegel action.} Let us consider a toy model with the action
\begin{align}
    S_\epsilon&=\int \psi D_+ \phi- \tfrac{\epsilon}2 \psi^2 +\text{interactions}\,,
\end{align}
where $D_\pm=\pl_z \pm |k|$. For the interactions one can take $\psi \phi^2/2$ to mimic the main story. This model may not look Lorentz invariant (from the bulk point of view), but it is. Integrating out $\psi$ we get $\psi=\epsilon^{-1} D_+ \phi$ and
\begin{align}
    S_\epsilon&=\epsilon^{-1}\tfrac12\int D_+\phi D_+ \phi+...=-\epsilon^{-1}\tfrac12\int \phi D_-D_+ \phi+... \,.
\end{align}
Note that $D_-D_+=\pl^2_z-k^2$, which is the usual kinetic term in the flat space.\footnote{The trick here is that $\vec{k}^2$ come originally as $\vec{k}\cdot \vec{k}$, but one can reinterpret it as $|k|^2$, which leads to $D_\pm$. } As is shown in the main text, the theories of $s=1/2$ and $s=1$ can be represented via a pair of actions $S_\epsilon$. For $\epsilon\neq0$ we can safely get rid of $\psi$ and the discussion above about the Dirichlet/Neumann/mixed boundary conditions applies. 

Let us consider the $\epsilon=0$ limit as a toy-model of SDYM vs. Chalmers--Siegel relation. Firstly, the equations loose one derivative and become first order, which drastically changes the options for boundary conditions: Dirichlet only. The amount of boundary data is preserved: one can fix $\phi$, $\psi$ in ``SDYM'' vs. $\phi$, $\pl_z \phi= D_+\phi-k \phi=\epsilon\psi-k \phi$ in ``Chalmers--Siegel''. The variation and the free equations of motion are (the minus comes from the fact that $z=0$ is the lower limit)
\begin{align}
    \delta S& =-\int _{\pl M}\psi \delta \phi \,,&&D_+\phi=0 \,,&& D_-\psi=0 \,.
\end{align}
The boundary data are $\psi_0$ and $\phi_0$. However, only solutions of $D_+\phi=0$ are admissible since we have $\phi=e^{-kz}\phi_0$ and $\psi=e^{+kz}\psi_0$, the solutions for $\psi$ being irregular in the bulk. Therefore, only $\phi_0$ is free whereas $\psi=0$. This is consistent with the variation $\delta S$ and no boundary terms are needed. The on-shell action vanishes, but a two-point function can be generated via a boundary term $k\phi^2$.\footnote{In this simple model, $\phi$, $\pl_z \phi$ have dimensions $\Delta=1,2$ in the original second order theory, hence, $\phi^2$ is not conformally invariant.} The bulk-to-bulk propagator is 
\begin{align}
    \langle \psi(z)\phi(z')\rangle &= -\frac{1}2 e^{-k|z-z'|}[\sign(z-z')-1]=e^{-k|z-z'|} \theta(z'-z) \,.
\end{align}
It obeys $D_+^{z'}\langle \psi(z)\phi(z')\rangle=\delta(z-z')$, $D_-^{z}\langle \psi(z)\phi(z')\rangle=-\delta(z-z')$. Note that there is no regular homogeneous solution that can be added since $D_-\psi=0$ implies $e^{+kz}$, which is consistent with us not having much choice in the boundary conditions. The boundary limits of the two-point function are 
\begin{align}
    \langle \psi(0)\phi(z')\rangle &=e^{-kz'} \,, &
    \langle \psi(z)\phi(0)\rangle &= 0\,,
\end{align}
which is consistent with the boundary problem and ``Fefferman-Graham'' expansion $\phi=e^{-kz} \phi_0$. 

Let us try to relate ``SDYM'' to ``Chalmers--Siegel'' and to the initial $\phi$-theory. One can introduce $F_\pm=\pm D_\pm \phi$. The mixed boundary condition is of type $a k q + bp=Q$, $\delta Q=0$, where $a=\cos(\gamma)$, $b=i\sin(\gamma)$. The boundary condition can be rewritten as $2Q=F_+e^{+i\gamma}+F_- e^{-i\gamma}$. Here, we impose the conformally-invariant boundary condition by replacing $q$ with $k q$ (see the comment above about the shadow transform), which also mimics $B=\slashed k \Phi$ for YM. In the initial $\phi$-theory we are free to impose mixed boundary conditions. This requires a boundary term, which we adjust to the neighborhood of the Neumann boundary condition:
\begin{align}
    \epsilon\delta S&= \int_{\pl M} \delta \phi \pl_z \phi\equiv \int_{\pl M} p\, \delta q && \Longrightarrow && \epsilon S_{\text{bndry}}= \int _{\pl M}-pq -\frac{a}{2b} q^2\,.
\end{align}
In order to make the ``Chalmers--Siegel'' action stationary, one has to add a different boundary term (various forms of this term are possible and we have chosen the simplest one)
\begin{align}\label{toyBoundaryterm}
    \delta S_\epsilon&= \int_{\pl M}  \psi \delta \phi \equiv \epsilon^{-1} \int_{\pl M}F_+ \delta \phi&& \Longrightarrow && S_{\text{bndry}}= -\tfrac14\epsilon (1-e^{2i\gamma})\int_{\pl M} \psi \tfrac{1}{k}\psi  \,.
\end{align}
In ``SDYM'' we cannot impose mixed boundary conditions. Therefore, it is interesting to see how we can reproduce ``Chalmers--Siegel'' as a perturbation of ``SDYM''. Let us try to interpret $\tfrac{\epsilon}2\psi^2$ as a deformation of ``SDYM''. In the ``Chalmers--Siegel'' theory the $\psi-\phi$ propagator remains the same, but there is also $\phi-\phi$:
\begin{align}
    \langle \phi(z) \phi(z')\rangle = -\frac{\epsilon}{2k}[e^{-k|z-z'|}  - e^{2i\gamma} e^{-k(z+z')}]\,.
\end{align}
From the ``SDYM'' vantage point the $\phi-\phi$ propagator corresponds to the diagram with $\tfrac{\epsilon}2\psi^2$-insertion. If we integrate two $\psi-\phi$ propagators we get
\begin{align}
    \epsilon \int dz'' \,\langle \psi(z'')\phi(z)\rangle\,\langle \psi(z'')\phi(z')\rangle&= -\langle \phi(z) \phi(z')\rangle\big|_{\gamma=0}\,,
\end{align}
where we recall that $\gamma=0$ is the Dirichlet boundary condition. Indeed, the Dirichlet boundary condition is the only possibility for $\phi$ in ``SDYM''. The additional term comes from the boundary vertex $S_{\text{bndry}}$. Therefore, if we treat all $\epsilon$-terms as a perturbation, we get the correct $\phi-\phi$ propagator with mixed boundary conditions out of the Dirichlet ``SDYM'' propagator. 

One more comment is that $S_{\text{bndry}}$ \eqref{toyBoundaryterm} looks nonlocal as it contains $\frac{1}{k}$. This would definitely be true in the ``SDYM'' picture. However, in the ``Chalmers--Siegel'' picture $\psi=\epsilon^{-1}F_+$ and the coefficient of $1/k$ is a total derivative $\pl_z \phi \pl_z \phi$ on the boundary. One can adjust this boundary term to make it explicitly local. The most general ansatz (of the right weight and including the 'non-local' piece from ``SDYM'') reads
\begin{align}
    S_{\text{bndry}}= \int_{\pl M} \tfrac{\alpha}2 q kq + \beta q F_+ +\tfrac{\kappa}{2} F_+ \tfrac1{k}F_+\,.
\end{align}
We can fine tune $\kappa=0$ and still satisfy the mixed boundary conditions with 
\begin{align}
    S_{\text{bndry}}= \epsilon^{-1}\int_{\pl M} -\tfrac{ (a-b)}{2 b } qk q-{F_+ q} = \epsilon^{-1}\int_{\pl M}-F_+ q+\tfrac{1}{1-e^{2 i \gamma }} qk q\,.
\end{align}
The variation of the complete action gives on-shell
\begin{align}
    \delta S&= \epsilon^{-1}\int _{\pl M} -\frac{\delta Q q}{b }=\epsilon^{-1}\int _{\pl M} \frac{i \delta Q q}{\sin(\gamma) }\,.
\end{align}
Note that the boundary terms are chosen to be well-behaved in the Neumann limit, as usual. 

One can try to add the $\psi\phi^2/2$ interaction and see what ``YM'', ``SDYM'' and ``Chalmers--Siegel'' give. Since the $\langle \psi(z)\phi(0)\rangle $-propagator vanishes in SDYM the $3$-, $4$- and all tree-level correlators vanish as well (one-loop diagrams do not have to vanish). In the genuine SDYM there are two $\phi$ and two $\psi$ that have opposite $D_\pm$ operators acting on them, which allows one to have one dynamical $\phi$ and one dynamical $\psi$, making the model much more nontrivial, see the main text. Therefore, in ``SDYM'' tree-level $=0$. By contrast, the ``Chalmers--Siegel'' and ``YM'' theories do seem to have tree-level amplitudes, but very simple. Let us define $\mathcal{F}_+=D_+\phi+\phi^2/2$.\footnote{One can go further and define $\mathcal{F}_-=-D_-\phi+\phi^2/2$ and covariant derivatives $\mathcal{D}_+=D_+ +\phi$ and $\mathcal{D}_-=-D_- +\phi$. Given this, we also have $\mathcal{D}_+ \mathcal{F}_--\mathcal{D}_-\mathcal{F}_+\equiv0$.} With the Lagrangian $(\mathcal{F}_+)^2/2$ the cubic vertex is $\phi^2 D_+ \phi$, which vanishes on-shell since $D_+$ annihilates the boundary-to-bulk propagator. One contribution to the quartic amplitude is through $\phi^4$, which gives the $1/E$ energy pole. Another contribution comes from two $D_+$ hitting the $\phi-\phi$ bulk-to-bulk propagator. This yields the $\psi-\psi$ propagator, which is $-\delta(z-z')$, the net result being the same $1/E$ (there is a cancellation between this pseudo-exchange and the quartic as in YM). Had we changed the relative coefficient between the cubic and the quartic, so that the Lagrangian is not a square, the amplitude would have been nonzero. In ``Chalmers--Siegel'' theory, $\psi(z)=0$ as in ``SDYM''. At this point, the toy model misses the key aspects of SDYM and of the Chalmers--Siegel theories where the correlators do not vanish.

\paragraph{Chiral symmetry breaking, a toy-model of self-duality.} As we have already seen, the spin-half theory's action, when expressed in terms of helicity components is very close to the toy-model above. Let us work out the propagators as well. Let us consider a massless bulk Dirac fermion $\chi^\alpha=(\Psi^A, \Phi^{A'})$ and its conjugate $\bar\chi_\alpha=(\bar\Psi_{A'}, \bar\Phi_{A})$, which are related by complex conjugation. Simultaneously, we consider the case of Weyl fermions where $\Psi^{A}$ and $\Phi^{A'}$ are related via a reality condition. There are $4$ blocks in the two-point function $\langle \chi \bar \chi\rangle$:
\begin{align}
    \langle \chi^\alpha (k,z) \bar \chi^\beta(-k,z')\rangle&= \begin{pmatrix}
        \langle \Psi^A \bar\Phi^B\rangle & \langle \Psi^A \bar\Psi^{B'}\rangle \\
        \langle \Phi^{A'} \bar\Phi^B\rangle & \langle \Phi^{A'} \bar\Psi^{B'}\rangle
    \end{pmatrix} \,.
\end{align}
The diagonal blocks satisfy the homogeneous Weyl equation and, in fact, have to vanish unless there is a chiral symmetry breaking. The off-diagonal blocks are easy to find by Fourier transforming $\slashed p/p^2$, see Appendix \ref{app:fourier} for more detail:
\besubeqs\label{spinhalfpropag}
\begin{align}
    \langle \Psi^A \bar\Psi^{B'}\rangle&= -\tfrac{1}{2k} e^{-k|z-z'|}[k^{AB'}+k \epsilon^{AB'}\sign(z-z')] \,,\\
     \langle \Phi^{A'} \bar\Phi^B\rangle&=-\tfrac{1}{2k} e^{-k|z-z'|}[k^{A'B}-k \epsilon^{A'B}\sign(z-z')] \,.
\end{align}
\esubeqs
As reviewed above, the Dirichlet/Neumann boundary conditions pick (anti)-chiral components of $\chi$ (or $\bar\chi$). Note that at $m=0$ the two falloffs have the same $\Delta=3/2$. In our convention $\Gamma_Z$ is off-diagonal and the chirality projection $1\pm i \Gamma_z$ can be imposed as $\Psi^A=\pm i \Phi^A$. A more general one-parameter family of conformally-invariant boundary conditions is possible \cite{Sezgin:2003pt}
\begin{align}
    e^{i\alpha} \Psi^A=-e^{-i\alpha} \Phi^A \,.
\end{align}
In order to impose this boundary condition we have to activate the diagonal blocks:
\begin{align}
    \langle \Psi^A \bar\Phi^{B}\rangle&= \tfrac{1}{2k} e^{-k(z+z')} k^A \brk^{B} e^{-2i\alpha} \,,&
     \langle \Phi^{A'} \bar\Psi^{B'}\rangle&=\tfrac{1}{2k} e^{-k(z+z')} \brk^{A'} k^{B'}  e^{+2i\alpha} \,,
\end{align}
where we have already imposed the boundary condition, which is manifested by the $\alpha$-dependence. One sees that the $AB$ and $A'B'$ two-point functions never vanish for the usual boundary conditions, i.e. Dirichlet, Neumann and even mixed ones. In principle, one can consider the ``(anti) self-dual limit'' where $\alpha \rightarrow \pm i \infty$.  

Massless fields of spin-half admit a one-parameter family of conformally-invariant boundary conditions, which illustrates some of the features we later find for gauge fields. Generic $\alpha$ can be associated with mixed boundary conditions. In $x$-space the inhomogeneous solution reads
\begin{align}
    \langle \Psi^A(x_1,z_1) \bar\Psi^{B'}(x_2,z_2)\rangle&= \frac{x_{12}^{AB'}+ i\epsilon^{AB'} z_{12} }{[x^2_{12}+z^2_{12}+i0]^2} \,.
\end{align}
The homogeneous solutions are just the images of the above under $z'\rightarrow -z'$. The above seems to have an unambiguous limit for the two-point function. However, this is not true. The $\epsilon^{AB}$ term is a delta-sequence $\delta_z(x_{12})$ and cannot be just dropped at $z=0$. Indeed, in the Fourier space we see that the double-boundary limit
\begin{align}
        \langle \Psi^A \bar\Psi^{B'}\rangle&= -\tfrac{1}{2k} e^{-k|z-z'|}[k^{AB'}+k \epsilon^{AB'}\sign(z-z')]\rightarrow -\tfrac{1}{2k}[k^{AB'} \pm k \epsilon^{AB'}]
\end{align}
depends on whether we approach it from $z<z'$ or from $z>z'$. The difference is a contact term. Another way so see there has to be a jump across $z=0$ is to recall the equation $(k_{AC'} -\pl_z\epsilon_{AC'}) S^{AB'}=\delta(z)\epsilon\fdu{C'}{B'}$ for the propagator. Integrating over a small interval around $z=0$ one confirms $S$ has to jump. The two-point function of a $\Delta=3/2$ operator should be regularized, e.g. one can use the differential regularization to find 
\begin{align}
    \frac{x_{AA}}{(x^2)^2} \sim \pl_{AA}\square \log (x^2 M^2) \,.
\end{align}
The same $\sign(z-z')$ will be present in the gauge boson propagator in Chalmers--Siegel formulation and SDYM. Since $\Delta=3/2$ the spin-half case is also extremal in the sense that the shadow dimension $d-\Delta$ coincides with $\Delta$. Therefore, one can also consider $\delta^3(x-y)$ as a legit contribution to the two-point function.

%%%%%%%%%%%%%%%%%%%%%%%%%%%%%%%%%%%%%%%%%%%%%%%%%%%%%%%%%%%%%
\section{Color ordering}
\label{app:color}
%%%%%%%%%%%%%%%%%%%%%%%%%%%%%%%%%%%%%%%%%%%%%%%%%%%%%%%%%%%%%

In Appendix \ref{app:fourFlat} and Section \ref{sec:fourAdS}, we derived the color-stripped four-point flat-space amplitudes and AdS correlation functions, respectively. In this section we discuss the appropriate color ordering for the exchange diagrams and the four-point contact diagram. In principle, color ordering is a standard material, but we make sure everything works the same for SDYM since the action/vertices are different from YM.

The exchange diagrams contain two vertices, each of which is accompanied by
\begin{align}
    f^{abc} = \Tr\Big([T^a,T^b]T^c\Big)\,,
\end{align}
where the generators $T^a$ are in the fundamental representation. Consequently, every exchange diagram comes with a double color trace
\begin{align}\label{doubleTrace}
    f^{abe}f^{cde}=\Tr\Big([T^a,T^b]T^e\Big)\Tr\Big([T^c,T^d]T^e\Big)\,.
\end{align}
At tree level, $SU(N)$ and $U(N)$ gauge theories are equivalent due to a well-known phenomenon called photon-decoupling. For $U(N)$ we have the identity
\begin{align}
    \Tr(T^aA)\Tr(T^aB)=\frac{1}{2}\Tr(AB)\,.
\end{align}
Hence, the double trace in \eqref{doubleTrace} can be expressed in terms of a single trace,
\begin{align}
    \Tr\Big([T^a,T^b]T^e\Big)\Tr\Big([T^c,T^d]T^e\Big)=\frac{1}{2}\Tr\Big([T^a,T^b][T^c,T^d]\Big)\,.
\end{align}
We simplify notation by replacing the Lie algebra generator $T^a$ by $i=1,2,3,4$, i.e. the same labels with which the momenta of the corresponding fields are dressed. Expanding out the commutators, we now find
\begin{align}
    f^{12e}f^{34e} = \frac{1}{2}\Big(\Tr[1234]-\Tr[2134]-\Tr[1243]+\Tr[2143]\Big)\,.
\end{align}
The full $s$-channel contribution to the amplitude is given by
\begin{align}
    \mathcal{M}_s(1234)=\frac{1}{2}\mathcal{A}_s(1234)\Big(\Tr[1234]-\Tr[2134]-\Tr[1243]+\Tr[2143]\Big)\,.
\end{align}
Here, $\mathcal{A}_s(1234)$ is the color-stripped amplitude/correlation function, c.f. \eqref{As}; we added the ordering $1234$ for convenience. The $t$-channel is obtained by swapping $1 \leftrightarrow 3$ in the $s$-channel, while for the $u$-channel one should swap $2 \leftrightarrow 3$. In other words, the channels are related by
\begin{align} \label{stu}
    \begin{aligned}
        \mathcal{M}_t(1234)&=\mathcal{M}_s(3214)\,, & \mathcal{M}_u(1234) = \mathcal{M}_s(1324)\,.
    \end{aligned}
\end{align}
Thus, we find
\begin{align}
    \begin{aligned}
        \mathcal{M}_t(1234)&=\frac{1}{2}\mathcal{A}_t(1234)\Big(\Tr[3214]-\Tr[2314]-\Tr[3241]+\Tr[2341]\Big)\,,\\
        \mathcal{M}_u(1234)&=\frac{1}{2}\mathcal{A}_u(1234)\Big(\Tr[1324]-\Tr[3124]-\Tr[1342]+\Tr[3142]\Big)\,.
    \end{aligned}
\end{align}
In total, this amounts to 12 different terms. Due to the cylicity condition of the trace, the full amplitude, $\mathcal{M}(1234)\equiv \mathcal{M}_s(1234)+\mathcal{M}_t(1234)+\mathcal{M}_u(1234)$, can be written as 6 independent terms of acyclic permutations of $\text{Tr[1234]}$. Collecting similar terms, one obtains
\begin{align} \label{fullAmp}
    \begin{aligned}
        \mathcal{M}(1234) &= \frac{1}{2}\Big(\Tr[1234]\big(\mathcal{A}_s(1234)+\mathcal{A}_t(1234)\big)-\Tr[1342]\big(\mathcal{A}_s(1234)+\mathcal{A}_u(1234)\big)+\\
        &-\Tr[1243]\big(\mathcal{A}_s(1234)+\mathcal{A}_u(1234)\big)+\Tr[1432]\big(\mathcal{A}_s(1234)+\mathcal{A}_t(1234)\big)+\\
        &+\Tr[1423]\big(-\mathcal{A}_t(1234)+\mathcal{A}_u(1234)\big)+\Tr[1324]\big(-\mathcal{A}_t(1234)+\mathcal{A}_u(1234)\big)\Big)\,.
    \end{aligned}
\end{align}
Note that each term is obtained from the first by a non-cyclic permutation of the legs. As a result, to know the full amplitude/correlator it suffices to compute only the color-ordered partial amplitude/correlator
\begin{align}
    \mathcal{M}_{\text{partial}}(1234) = \frac{1}{2}\big(\mathcal{A}_s(1234)+\mathcal{A}_t(1234)\big) \,.
\end{align}

As is discussed in Section \ref{subsec:cYM} and Appendix \ref{app:flat}, the YM exchange diagram (partially) cancels against the four-point contact diagram. This necessarily means that the four-point function contains the color-structure of the $s$-, $t$- and $u$-channel. Indeed, the four-point vertex reads
\begin{align}
    \begin{aligned}
        V^{\text{quartic}}_{AA',BB',CC',DD'}&= 2g^2\Big[f^{abe}f^{cde}\epsilon_{A'B'}\epsilon_{C'D'}(\epsilon_{AD}\epsilon_{BC}+\epsilon_{AC}\epsilon_{BD})+\\
        &-f^{ade}f^{bce}\epsilon_{A'D'}\epsilon_{B'C'}(\epsilon_{AB}\epsilon_{CD}+\epsilon_{AC}\epsilon_{BD})+f^{ace}f^{bde}\epsilon_{A'C'}\epsilon_{B'D'}(\epsilon_{AB}\epsilon_{CD}-\epsilon_{AD}\epsilon_{BC})\Big] \,.
    \end{aligned}
\end{align}
For simplicity, the four-point vertex used in Section \ref{subsec:cYM} and Appendix \ref{app:flat} contains only the $s$-channel component.

%%%%%%%%%%%%%%%%%%%%%%%%%%%%%%%%%%%%%%%%%%%%%%%%%%%%%%%%%%%%%
\section{Flat Space}
\label{app:flat}
%%%%%%%%%%%%%%%%%%%%%%%%%%%%%%%%%%%%%%%%%%%%%%%%%%%%%%%%%%%%%
All amplitudes in flat space are textbook. Nevertheless, there is a reason to recompute them first: the leading term in the anti-de Sitter amplitude in the limit $E\rightarrow0$ is exactly the flat-space amplitude. Therefore, it is useful to understand how to compute it in the optimal way and how to relate results in different gauges to each other. Of course, the amplitude is gauge invariant, but different gauges can often-times be related to each other channel by channel, which simplifies the analysis. 

%%%%%%%%%%%%%%%%%%%%%%%%%%%%%%%%%%%%%%%%%%%%%%%%%%%%%%%%%%%%%
\subsection{Three point}
%%%%%%%%%%%%%%%%%%%%%%%%%%%%%%%%%%%%%%%%%%%%%%%%%%%%%%%%%%%%%
\subsubsection{SDYM}
\label{sec:flatSDYM3pt}
%%%%%%%%%%%%%%%%%%%%%%%%%%%%%%%%%%%%%%%%%%%%%%%%%%%%%%%%%%%%%
Let us recall that the SDYM action reads
\begin{align} \label{SDYMaction}
    S=\Tr\int \Psi^{AA}F_{AA}\,.
\end{align}
The self-dual field strength is defined by $F_{AA}=\partial_{AA'}\Phi\fdu{A,}{A'}+\Phi_{A,A'}\Phi\fdu{A,}{A'}$. The gauge field $\Phi_{A,A'}$ carries positive helicity, while the field $\Psi^{AA}$ carries negative helicity. Expanding the field strength in the action gives the cubic interaction
\begin{align}
    S_3= \Tr\int \Phi_{A,A'}\Phi\fdu{A,}{A'}\Psi^{AA}=\epsilon^{AC}\epsilon^{BC}\epsilon^{A'B'}\Tr\int\Phi_{A,A'}\Phi_{B,B'}\Psi_{CC}\,.
\end{align}
From this we can read off the Feynman rule for the cubic scattering amplitude, which is displayed in Figure \eqref{fig:feynmanRulesSDYM}. We will omit color factors from here on. For more details, see Appendix \ref{app:color}. The polarization vectors in Lorenz gauge are given by
\begin{align}\label{sdympol}
    (\epsilon^{+}_i)^{AA'} &= \frac{q_i^{A} \bar{i}^{A'}}{\langle q_i i \rangle}  \,, & (\epsilon^-_i)^{AA} &= i^{A}i^{A}\,.
\end{align}
Here, $q_i^A$ are reference spinors and $i^A$, $\bar{i}^{A'}$ are the spinors that factorize the $i$-th (null) momentum, $p_i^{AA'}=i^A\bar{i}^{A'}$. We also introduce the notation\footnote{While in most (flat-space) literature angle brackets and square brackets are used for spinors in different representations of the Lorentz group, i.e. the barred and unbarred spinors, we adopt the convention to use only angle brackets. This allows for a smooth transition to AdS, where both barred and unbarred spinors are in the same representation, even allowing contractions of the type $\langle i\bar{j}\rangle$.}
\begin{align}
    \langle i j \rangle \equiv -\epsilon_{AA'} i^A j^{A'} = i^Aj_A \,.
\end{align}

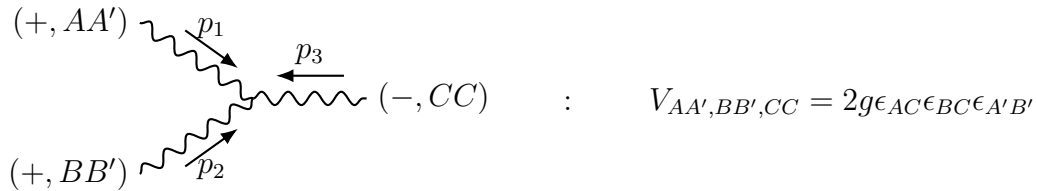
\begin{figure}[h]
\centering
\begin{tabular}{c c c}
    % Vertex
    \begin{tikzpicture}
        \coordinate (v) at (0,0);
        
        % Define external legs
        \coordinate (a) at (-1.5,1);
        \coordinate (b) at (-1.5,-1);
        \coordinate (c) at (1.5,0);
        
        % Draw gauge boson lines 
        \draw[decorate,decoration={snake},thick] (a) -- (v);
        \draw[decorate,decoration={snake},thick] (b) -- (v);
        \draw[decorate,decoration={snake},thick] (v) -- (c);
        
        % Draw parallel momentum arrows 
        \draw[-{Latex},thick] (-0.9,0.9) -- (-0.2,0.4) node[midway,above] {\( p_1 \)};
        \draw[-{Latex},thick] (-0.9,-0.9) -- (-0.2,-0.4) node[midway,below] {\( p_2 \)};
        \draw[{Latex}-,thick] (0.3,0.3) --  (1.2,0.3) node[midway,above] {\( p_3 \)};
        
        % Add labels for fields next to endpoints
        \node[left] at (a) {\( (+,AA') \)};
        \node[left] at (b) {\( (+,BB') \)};
        \node[right] at (c) {\( (-,CC) \)};
    \end{tikzpicture}
    & \raisebox{12mm}{\quad : \quad} &
    \raisebox{12mm}{\( V_{AA',BB',CC}=2g\epsilon_{AC}\epsilon_{BC}\epsilon_{A'B'}\)}
\end{tabular}
\caption{SDYM Lorenz gauge Feynman rules in flat space.}
\label{fig:feynmanRulesSDYM}
\end{figure}%
\noindent The SDYM color-stripped amplitude is depicted in Figure \ref{fig:cubicSDYMFlat} and reads
\begin{align} \label{SDYM3ptRefSpin}
    \mathcal{A}_3 = V_{AA',BB',CC}(\epsilon_1^+)^{AA'}(\epsilon_2^+)^{BB'}(\epsilon_3^-)^{CC}=-2g\frac{\langle q_13 \rangle \langle q_23 \rangle}{\langle q_11 \rangle \langle q_22 \rangle} \langle \bar{1}\bar{2} \rangle \,.
\end{align}
\begin{figure}[h]
\centering
\begin{tabular}{c c c}
    % Vertex
    \begin{tikzpicture}
        \coordinate (v) at (0,0);
        
        % Define external legs
        \coordinate (a) at (-1.5,1);
        \coordinate (b) at (-1.5,-1);
        \coordinate (c) at (1.5,0);
        
        % Draw gauge boson lines 
        \draw[decorate,decoration={snake},thick] (a) -- (v);
        \draw[decorate,decoration={snake},thick] (b) -- (v);
        \draw[decorate,decoration={snake},thick] (v) -- (c);
        
        % Draw parallel momentum arrows 
        \draw[-{Latex},thick] (-0.9,0.9) -- (-0.2,0.4) node[midway,above] {\( p_1 \)};
        \draw[-{Latex},thick] (-0.9,-0.9) -- (-0.2,-0.4) node[midway,below] {\( p_2 \)};
        \draw[{Latex}-,thick] (0.3,0.3) --  (1.2,0.3) node[midway,above] {\( p_3 \)};
        
        % Add labels for fields next to endpoints
        \node[left] at (a) {\( (+) \)};
        \node[left] at (b) {\( (+) \)};
        \node[right] at (c) {\( (-) \)};
    \end{tikzpicture}
    & \raisebox{12mm}{\quad : \quad} &
    \raisebox{12mm}{\( \mathcal{A}_3=V_{AA',BB',CC}(\epsilon_1^+)^{AA'}(\epsilon_2^+)^{BB'}(\epsilon_3^-)^{CC}\)}
\end{tabular}
\caption{SDYM three-point function.}
\label{fig:cubicSDYMFlat}
\end{figure}
\noindent The amplitude is gauge-invariant and should not depend on the reference spinors. To show that their dependence disappears, we use momentum conservation
\begin{align}
    1^A\bar{1}^{A'}+2^A\bar{2}^{A'}+3^A\bar{3}^{A'}=0
\end{align}
and contract this with $q_{1A}\bar{2}_{A'}$ and $q_{2A}\bar{1}_{A'}$ to obtain
\begin{align} \label{refSpinors}
    \begin{aligned}
        \langle q_1 3 \rangle &=\frac{\langle q_1 1 \rangle \langle \bar{1}\bar{2} \rangle}{\langle \bar{2}\bar{3} \rangle}  \,,& \langle q_2 3 \rangle &= -\frac{\langle q_22 \rangle \langle \bar{1}\bar{2} \rangle}{\langle \bar{1}\bar{3} \rangle} \,,
    \end{aligned}
\end{align}
respectively. The amplitude can now be written in the well-known form
\begin{align} \label{3ptSDYM}
    \mathcal{A}_3 = 2g\frac{\langle \bar{1}\bar{2} \rangle^3}{\langle \bar{1}\bar{3}\rangle \langle \bar{2}\bar{3} \rangle}\,.
\end{align}

%%%%%%%%%%%%%%%%%%%%%%%%%%%%%%%%%%%%%%%%%%%%%%%%%%%%%%%%%%%%%
\subsubsection{YM}
\label{sec:YMFlat}
%%%%%%%%%%%%%%%%%%%%%%%%%%%%%%%%%%%%%%%%%%%%%%%%%%%%%%%%%%%%%
SDYM is a closed subsector of the full YM theory. Effectively, this means that SDYM has only access to scattering amplitudes with a specific helicity configuration, i.e. all but one $+$ for tree-level diagrams. While the amplitudes for these diagrams coincide for SDYM and YM, the formulations of both theories are vastly different. For example, the YM action contains a second-order differential operator, as opposed to the first-order differential operator in SDYM. Moreover, the Yang--Mills three-point vertex contains a derivative, while SDYM's does not. Here we will provide techniques that show how the amplitudes in SDYM and YM are related and we prove their equivalence in flat space.

\paragraph{Flat space.} To facilitate the comparison between YM and SDYM, we derive the Feynman rules for YM in a way that resembles the SDYM Feynman rules. To this end, we recall 
\begin{align} \label{FFbar}
    \begin{aligned}
        F_{AA} &= \partial_{AA'}\Phi\fdu{A,}{A'}+\Phi_{A,A'}\Phi\fdu{A,}{A'}\,, &
        \bar{F}_{A'A'}&=\partial_{AA'}\Phi\fud{A}{,A'}+\Phi_{A,A'}\Phi\fud{A}{,A'}
    \end{aligned}
\end{align}
and the YM action
\begin{align} \label{cYMAction}
    S_{\text{YM}} = \frac{1}{4g^2}\Tr\int \big(F^{AA}F_{AA}+\bar{F}^{A'A'}\bar{F}_{A'A'}\big) \,,
\end{align}
i.e. we set $a=\frac{1}{4}$ in \eqref{YMaction}. The topological term
\begin{align}
    S_{\text{top}} = \frac{1}{4g^2}\Tr\int \big(F^{AA}F_{AA}-\bar{F}^{A'A'}\bar{F}_{A'A'}\big)
\end{align}
can be added to the action to obtain
\begin{align}
    S_{\text{cYM}} = \frac{1}{2g^2}\Tr\int F^{AA}F_{AA}\,.
\end{align}
For completeness, let us decompose the cubic YM action into the self-dual and anti-self-dual parts, $S_3=S_3^{\text{SD}}+S_3^{\text{ASD}}$, coming from the $F^{AA}F_{AA}$ -- and $\bar{F}^{A'A'}\bar{F}_{A'A'}$ -- term, respectively. Using \eqref{FFbar}, one finds these actions to be
\begin{align}
    \begin{aligned}
        S_3^{\text{SD}} &= \frac{1}{2g^2}\Tr\int \partial_{AA'}\Phi\fdu{A,}{A'}\Phi_\fud{A,}{B'}\Phi^{A,B'} = \frac{1}{4g^2}\Tr\int\epsilon^{B'C'}(\epsilon^{AC}\partial^{BA'}+\epsilon^{AB}\partial^{CA'})\Phi_{A,A'}\Phi_{B,B'}\Phi_{C,C'} \,,\\
        S_3^{\text{ASD}} &= \frac{1}{2g^2}\Tr\int \partial_{AA'}\Phi\fud{A,}{A'}\Phi\fdu{B,}{A'}\Phi^{B,A'} = \frac{1}{4g^2}\Tr\int\epsilon^{BC}(\epsilon^{A'C'}\partial^{AB'}+\epsilon^{A'B'}\partial^{AC'})\Phi_{A,A'}\Phi_{B,B'}\Phi_{C,C'}\,,
    \end{aligned}
\end{align}
where the derivative is acting only on the first field.

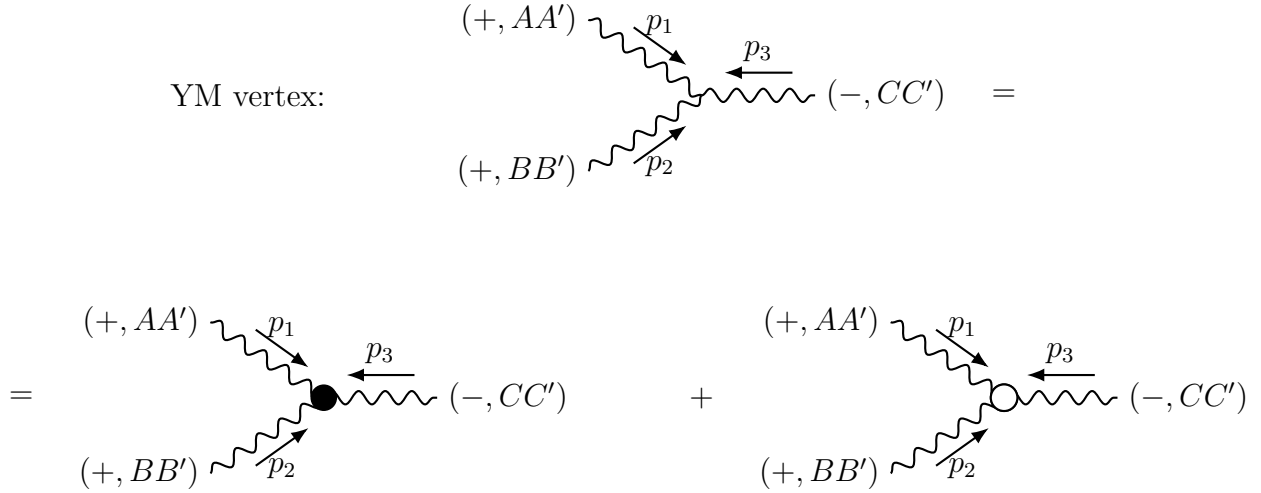
\begin{figure}[h]
\centering
    % Vertex
    \begin{tikzpicture}
        \coordinate (v) at (0,0);
        
        % Define external legs
        \coordinate (a) at (-1.5,1);
        \coordinate (b) at (-1.5,-1);
        \coordinate (c) at (1.5,0);
        
        % Draw gauge boson lines 
        \draw[decorate,decoration={snake},thick] (a) -- (v);
        \draw[decorate,decoration={snake},thick] (b) -- (v);
        \draw[decorate,decoration={snake},thick] (v) -- (c);
        
        % Draw parallel momentum arrows 
        \draw[-{Latex},thick] (-0.9,0.9) -- (-0.2,0.4) node[midway,above] {\( p_1 \)};
        \draw[-{Latex},thick] (-0.9,-0.9) -- (-0.2,-0.4) node[midway,below] {\( p_2 \)};
        \draw[{Latex}-,thick] (0.3,0.3) --  (1.2,0.3) node[midway,above] {\( p_3 \)};
        
        % Add labels for fields next to endpoints
        \node[left] at (a) {\( (+,AA') \)};
        \node[left] at (b) {\( (+,BB') \)};
        \node[right] at (c) {\( (-,CC') \)};
        \node at (4,0) {$=$};
        
        \begin{scope}[xshift=4cm,yshift=-4cm]
        \node at (-4,0) {$+$};
        
        \coordinate (v) at (0,0);
        
        % Define external legs
        \coordinate (a) at (-1.5,1);
        \coordinate (b) at (-1.5,-1);
        \coordinate (c) at (1.5,0);
        
        % Draw gauge boson lines 
        \draw[decorate,decoration={snake},thick] (a) -- (v);
        \draw[decorate,decoration={snake},thick] (b) -- (v);
        \draw[decorate,decoration={snake},thick] (v) -- (c);
        
        % Draw parallel momentum arrows 
        \draw[-{Latex},thick] (-0.9,0.9) -- (-0.2,0.4) node[midway,above] {\( p_1 \)};
        \draw[-{Latex},thick] (-0.9,-0.9) -- (-0.2,-0.4) node[midway,below] {\( p_2 \)};
        \draw[{Latex}-,thick] (0.3,0.3) --  (1.2,0.3) node[midway,above] {\( p_3 \)};

        \node at (-10,4) {\text{YM vertex:}} ;
        
        % Add labels for fields next to endpoints
        \node[left] at (a) {\( (+,AA') \)};
        \node[left] at (b) {\( (+,BB') \)};
        \node[right] at (c) {\( (-,CC') \)};

        % Draw vertex as a filled dot
        \draw[thick] (v) circle (5pt);
        \fill[white] (v) circle (5pt);
        \draw[black, thick]
        (v) node[draw, circle, minimum size=10pt, inner sep=0pt,
              pattern=north east lines, pattern color=gray,
              fill=white] {} ;
        \end{scope}

        \begin{scope}[xshift=-5cm,yshift=-4cm]
        \node at (-4,0) {$=$};
        
        \coordinate (v) at (0,0);
        
        % Define external legs
        \coordinate (a) at (-1.5,1);
        \coordinate (b) at (-1.5,-1);
        \coordinate (c) at (1.5,0);
        
        % Draw gauge boson lines 
        \draw[decorate,decoration={snake},thick] (a) -- (v);
        \draw[decorate,decoration={snake},thick] (b) -- (v);
        \draw[decorate,decoration={snake},thick] (v) -- (c);
        
        % Draw parallel momentum arrows 
        \draw[-{Latex},thick] (-0.9,0.9) -- (-0.2,0.4) node[midway,above] {\( p_1 \)};
        \draw[-{Latex},thick] (-0.9,-0.9) -- (-0.2,-0.4) node[midway,below] {\( p_2 \)};
        \draw[{Latex}-,thick] (0.3,0.3) --  (1.2,0.3) node[midway,above] {\( p_3 \)};
        
        % Add labels for fields next to endpoints
        \node[left] at (a) {\( (+,AA') \)};
        \node[left] at (b) {\( (+,BB') \)};
        \node[right] at (c) {\( (-,CC') \)};

        % Draw vertex as a filled dot
        \fill (v) circle (5pt);
        \end{scope}
    \end{tikzpicture}
\caption{The YM vertex is the sum of the SD and ASD vertex.}
\label{fig:YM&top1}
\end{figure}

An immediate difference from SDYM is that the YM propagator and vertex do not discriminate between different helicity structures. Especially, the SD and ASD vertex take on a much more complicated form,\footnote{In order to evaluate the field theory using perturbative methods, we rescale $\Phi\rightarrow g\Phi$.}
\begin{align} \label{vertex1}
    \begin{aligned}
        V^{\text{SD}}_{AA',BB',CC'}&= \frac{g}{2}\Big[\epsilon_{B'C'}(\epsilon_{AC}p^1_{BA'}+\epsilon_{AB}p^1_{CA'})+\epsilon_{C'A'}(\epsilon_{BA}p^2_{CB'}+\epsilon_{BC}p^2_{AB'})+\\
        &+\epsilon_{A'B'}(\epsilon_{CB}p^3_{AC'}+\epsilon_{CA}p^3_{BC'})\Big] \,,\\
        V^{\text{ASD}}_{AA',BB',CC'}&= \frac{g}{2} \Big[\epsilon_{BC}(\epsilon_{A'C'}p^1_{AB'}+\epsilon_{A'B'}p^1_{AC'})+\epsilon_{CA}(\epsilon_{B'A'}p^2_{BC'}+\epsilon_{B'C'}p^2_{BA'})+\\
        &+\epsilon_{AB}(\epsilon_{C'B'}p^3_{CA'}+\epsilon_{C'A'}p^3_{CB'})\Big] \,,
    \end{aligned}
\end{align}
respectively. The YM vertex and the topological vertex read
\begin{align} \label{vertex2}
    \begin{aligned}
        V_{AA',BB',CC'} &= V_{AA',BB',CC'}^{\text{SD}}+V_{AA',BB',CC'}^{\text{ASD}}\,,\\
        V_{AA',BB',CC'}^{\text{top}} &= V_{AA',BB',CC'}^{\text{SD}}-V_{AA',BB',CC'}^{\text{ASD}}\,,
    \end{aligned}
\end{align}
respectively. This is also depicted in Figure \ref{fig:YM&top1} and \ref{fig:YM&top2}, where the SD and ASD vertices are represented by a black and white dot, respectively.

\begin{figure}[h]
\centering
    % Vertex
    \begin{tikzpicture}
        \begin{scope}[yshift=-8cm]
            \coordinate (v) at (0,0);
        
        % Define external legs
        \coordinate (a) at (-1.5,1);
        \coordinate (b) at (-1.5,-1);
        \coordinate (c) at (1.5,0);
        
        % Draw gauge boson lines 
        \draw[decorate,decoration={snake},thick] (a) -- (v);
        \draw[decorate,decoration={snake},thick] (b) -- (v);
        \draw[decorate,decoration={snake},thick] (v) -- (c);
        
        % Draw parallel momentum arrows 
        \draw[-{Latex},thick] (-0.9,0.9) -- (-0.2,0.4) node[midway,above] {\( p_1 \)};
        \draw[-{Latex},thick] (-0.9,-0.9) -- (-0.2,-0.4) node[midway,below] {\( p_2 \)};
        \draw[{Latex}-,thick] (0.3,0.3) --  (1.2,0.3) node[midway,above] {\( p_3 \)};

        \node at (-6,0) {\text{Topological vertex:}} ;
        
        % Add labels for fields next to endpoints
        \node[left] at (a) {\( (+,AA') \)};
        \node[left] at (b) {\( (+,BB') \)};
        \node[right] at (c) {\( (-,CC') \)};
        \node at (4,0) {$=$};

        \fill[white] (0,0) circle (5pt);
        \draw[black, thick]
        (v) node[draw, circle, minimum size=10pt, inner sep=0pt,
              pattern=north east lines, pattern color=gray] {} ;
        
        \begin{scope}[xshift=4cm,yshift=-4cm]
        \node at (-4,0) {$-$};
        
        \coordinate (v) at (0,0);
        
        % Define external legs
        \coordinate (a) at (-1.5,1);
        \coordinate (b) at (-1.5,-1);
        \coordinate (c) at (1.5,0);
        
        % Draw gauge boson lines 
        \draw[decorate,decoration={snake},thick] (a) -- (v);
        \draw[decorate,decoration={snake},thick] (b) -- (v);
        \draw[decorate,decoration={snake},thick] (v) -- (c);
        
        % Draw parallel momentum arrows 
        \draw[-{Latex},thick] (-0.9,0.9) -- (-0.2,0.4) node[midway,above] {\( p_1 \)};
        \draw[-{Latex},thick] (-0.9,-0.9) -- (-0.2,-0.4) node[midway,below] {\( p_2 \)};
        \draw[{Latex}-,thick] (0.3,0.3) --  (1.2,0.3) node[midway,above] {\( p_3 \)};
        
        % Add labels for fields next to endpoints
        \node[left] at (a) {\( (+,AA') \)};
        \node[left] at (b) {\( (+,BB') \)};
        \node[right] at (c) {\( (-,CC') \)};

        % Draw vertex as a filled dot
        \draw[thick] (v) circle (5pt);
        \fill[white] (v) circle (5pt);

        \end{scope}

        \begin{scope}[xshift=-5cm,yshift=-4cm]
        \node at (-4,0) {$=$};
        
        \coordinate (v) at (0,0);
        
        % Define external legs
        \coordinate (a) at (-1.5,1);
        \coordinate (b) at (-1.5,-1);
        \coordinate (c) at (1.5,0);
        
        % Draw gauge boson lines 
        \draw[decorate,decoration={snake},thick] (a) -- (v);
        \draw[decorate,decoration={snake},thick] (b) -- (v);
        \draw[decorate,decoration={snake},thick] (v) -- (c);
        
        % Draw parallel momentum arrows 
        \draw[-{Latex},thick] (-0.9,0.9) -- (-0.2,0.4) node[midway,above] {\( p_1 \)};
        \draw[-{Latex},thick] (-0.9,-0.9) -- (-0.2,-0.4) node[midway,below] {\( p_2 \)};
        \draw[{Latex}-,thick] (0.3,0.3) --  (1.2,0.3) node[midway,above] {\( p_3 \)};
        
        % Add labels for fields next to endpoints
        \node[left] at (a) {\( (+,AA') \)};
        \node[left] at (b) {\( (+,BB') \)};
        \node[right] at (c) {\( (-,CC') \)};

        % Draw vertex as a filled dot
        \fill (v) circle (5pt);
        \end{scope}
        \end{scope}
    \end{tikzpicture}
\caption{The topological vertex is the difference between the SD and ASD vertex.}
\label{fig:YM&top2}
\end{figure}

Any topological contributions are expected to vanish, as flat space has no boundary. This can also be seen from the vertices directly, as we have
\begin{align}
    \begin{aligned}
    V^{\text{top}}_{AA',BB',CC'} &\equiv V^{\text{SD}}_{AA',BB',CC'}-V^{\text{ASD}}_{AA',BB',CC'} =\\
    &= g\Big[\epsilon_{AB}\epsilon_{A'C'}(p_1+p_2+p_3)_{CB'}-\epsilon
    _{AC}\epsilon_{A'B'}(p_1+p_2+p_3)_{BC'}\Big]=0\,,
    \end{aligned}
\end{align}
which vanishes because of the four-momentum conservation. To obtain this expression, we used the identity
\begin{align}
    X^{A}Y^B=X^BY^A+\epsilon^{AB}X_CY^C\,.
\end{align}
For future use, we relate\footnote{In AdS, $V^{\text{top}}_{AA',BB',CC'}$ does not vanish and it should be taken into account that cYM differs from YM by exactly this vertex. This will be discussed in Section \ref{sec:cYMAdS}.}
\begin{align}
    V_{AA',BB',CC'} = 2V^{\text{SD}}_{AA',BB',CC'} \,,
\end{align}
which is visualized in Figure \ref{fig:2SDVertex}.

\begin{figure}[h!]
\centering
\begin{tikzpicture}
    \begin{scope}[xshift=-4cm]
        \coordinate (v) at (0,0);
        
        % Define external legs
        \coordinate (a) at (-1.5,1);
        \coordinate (b) at (-1.5,-1);
        \coordinate (c) at (1.5,0);
        
        % Draw gauge boson lines 
        \draw[decorate,decoration={snake},thick] (a) -- (v);
        \draw[decorate,decoration={snake},thick] (b) -- (v);
        \draw[decorate,decoration={snake},thick] (v) -- (c);
        
        % Draw parallel momentum arrows 
        \draw[-{Latex},thick] (-0.9,0.9) -- (-0.2,0.4) node[midway,above] {\( p_1 \)};
        \draw[-{Latex},thick] (-0.9,-0.9) -- (-0.2,-0.4) node[midway,below] {\( p_2 \)};
        \draw[-{Latex},thick] (0.3,0.3) --  (1.2,0.3) node[midway,above] {\( p_3 \)};
        
        % Add labels for fields next to endpoints
        \node[left] at (a) {\( (+,AA') \)};
        \node[left] at (b) {\( (+,BB') \)};
        \node[right] at (c) {\( (-,CC') \)};
              
    \end{scope}
    % Equal sign between top diagrams
    \node at (0,0) {$=2\times$};
    \begin{scope}[xshift=4cm]
        \coordinate (v) at (0,0);
        
        % Define external legs
        \coordinate (a) at (-1.5,1);
        \coordinate (b) at (-1.5,-1);
        \coordinate (c) at (1.5,0);
        
        % Draw gauge boson lines 
        \draw[decorate,decoration={snake},thick] (a) -- (v);
        \draw[decorate,decoration={snake},thick] (b) -- (v);
        \draw[decorate,decoration={snake},thick] (v) -- (c);
        
        % Draw parallel momentum arrows 
        \draw[-{Latex},thick] (-0.9,0.9) -- (-0.2,0.4) node[midway,above] {\( p_1 \)};
        \draw[-{Latex},thick] (-0.9,-0.9) -- (-0.2,-0.4) node[midway,below] {\( p_2 \)};
        \draw[-{Latex},thick] (0.3,0.3) --  (1.2,0.3) node[midway,above] {\( p_3 \)};
        
        % Add labels for fields next to endpoints
        \node[left] at (a) {\( (+,AA') \)};
        \node[left] at (b) {\( (+,BB') \)};
        \node[right] at (c) {\( (-,CC') \)};
        
        \fill (v) circle (5pt);
    \end{scope}
\end{tikzpicture}
\caption{The full YM vertex is equivalent to two times the SD vertex, due to the vanishing topological term in $S_{\text{cYM}}=S_{\text{YM}}+S_{\text{top}}$.}
\label{fig:2SDVertex}
\end{figure}

\begin{figure}[h!]
\centering
\begin{tikzpicture}
    \begin{scope}[xshift=-4cm]
        \coordinate (v) at (0,0);
        
        % Define external legs
        \coordinate (a) at (-1.5,1);
        \coordinate (b) at (-1.5,-1);
        \coordinate (c) at (1.5,0);
        
        % Draw gauge boson lines 
        \draw[decorate,decoration={snake},thick] (a) -- (v);
        \draw[decorate,decoration={snake},thick] (b) -- (v);
        \draw[decorate,decoration={snake},thick] (v) -- (c);
        
        % Draw parallel momentum arrows 
        \draw[-{Latex},thick] (-0.9,0.9) -- (-0.2,0.4) node[midway,above] {\( p_1 \)};
        \draw[-{Latex},thick] (-0.9,-0.9) -- (-0.2,-0.4) node[midway,below] {\( p_2 \)};
        \draw[-{Latex},thick] (0.3,0.3) --  (1.2,0.3) node[midway,above] {\( p_3 \)};
        
        % Add labels for fields next to endpoints
        \node[left] at (a) {\( (+,AA') \)};
        \node[left] at (b) {\( (+,BB') \)};
        \node[right] at (c) {\( (-,CC') \)};
        
        % Draw vertex as a filled dot
        \fill (v) circle (5pt);        
    \end{scope}
    % Equal sign between top diagrams
    \node at (-0.5,0) {=};
    \begin{scope}[xshift=3cm]
        \coordinate (v) at (0,0);
        
        % Define external legs
        \coordinate (a) at (-1.5,1);
        \coordinate (b) at (-1.5,-1);
        \coordinate (c) at (1.5,0);
        
        % Draw gauge boson lines 
        \draw[decorate,decoration={snake},thick] (a) -- (v);
        \draw[decorate,decoration={snake},thick] (b) -- (v);
        \draw[decorate,decoration={snake},thick] (v) -- (c);
        
        % Draw parallel momentum arrows 
        \draw[-{Latex},thick] (-0.9,0.9) -- (-0.2,0.4) node[midway,above] {\( p_1 \)};
        \draw[-{Latex},thick] (-0.9,-0.9) -- (-0.2,-0.4) node[midway,below] {\( p_2 \)};
        \draw[-{Latex},thick] (0.3,0.3) --  (1.2,0.3) node[midway,above] {\( p_3 \)};
        
        % Add labels for fields next to endpoints
        \node[left] at (a) {\( (+,AA') \)};
        \node[left] at (b) {\( (+,BB') \)};
        \node[right] at (c) {\( (-,CC') \)};
        
        \fill[blue] (-0.5,0.3) circle (5pt);
    \end{scope}
        % Plus sign after top diagrams
    \node at (7,0) {+};
    \begin{scope}[xshift=-4cm,yshift=-4cm]
        % Plus sign in front of bottom diagrams
        \node at (-4.5,0) {+};
        \coordinate (v) at (0,0);
        
        % Define external legs
        \coordinate (a) at (-1.5,1);
        \coordinate (b) at (-1.5,-1);
        \coordinate (c) at (1.5,0);
        
        % Draw gauge boson lines 
        \draw[decorate,decoration={snake},thick] (a) -- (v);
        \draw[decorate,decoration={snake},thick] (b) -- (v);
        \draw[decorate,decoration={snake},thick] (v) -- (c);
        
        % Draw parallel momentum arrows 
        \draw[-{Latex},thick] (-0.9,0.9) -- (-0.2,0.4) node[midway,above] {\( p_1 \)};
        \draw[-{Latex},thick] (-0.9,-0.9) -- (-0.2,-0.4) node[midway,below] {\( p_2 \)};
        \draw[-{Latex},thick] (0.3,0.3) --  (1.2,0.3) node[midway,above] {\( p_3 \)};
        
        % Add labels for fields next to endpoints
        \node[left] at (a) {\( (+,AA') \)};
        \node[left] at (b) {\( (+,BB') \)};
        \node[right] at (c) {\( (-,CC') \)};
        \fill[blue] (-0.5,-0.3) circle (5pt);
            \begin{scope}[xshift=7cm]
            % Plus sign between bottom diagrams
            \node at (-3.5,0) {+};
            \coordinate (v) at (0,0);
        
            % Define external legs
            \coordinate (a) at (-1.5,1);
            \coordinate (b) at (-1.5,-1);
            \coordinate (c) at (1.5,0);
        
            % Draw gauge boson lines 
            \draw[decorate,decoration={snake},thick] (a) -- (v);
            \draw[decorate,decoration={snake},thick] (b) -- (v);
            \draw[decorate,decoration={snake},thick] (v) -- (c);
        
            % Draw parallel momentum arrows 
            \draw[-{Latex},thick] (-0.9,0.9) -- (-0.2,0.4) node[midway,above] {\( p_1 \)};
            \draw[-{Latex},thick] (-0.9,-0.9) -- (-0.2,-0.4) node[midway,below] {\( p_2 \)};
            \draw[-{Latex},thick] (0.3,0.3) --  (1.2,0.3) node[midway,above] {\( p_3 \)};
        
            % Add labels for fields next to endpoints
            \node[left] at (a) {\( (+,AA') \)};
            \node[left] at (b) {\( (+,BB') \)};
            \node[right] at (c) {\( (-,CC') \)};
            \fill[blue] (0.5,0) circle (5pt);
        
        \end{scope}
    \end{scope}
\end{tikzpicture}
\caption{The SD vertex distributes the derivative evenly over all legs. The derivative is denoted by the blue dot.}
\label{fig:cYMSD}
\end{figure}

\begin{figure}[h!]
\centering
\begin{tikzpicture}
    \begin{scope}[xshift=-4cm]
        \coordinate (v) at (0,0);
        
        % Define external legs
        \coordinate (a) at (-1.5,1);
        \coordinate (b) at (-1.5,-1);
        \coordinate (c) at (1.5,0);
        
        % Draw gauge boson lines 
        \draw[decorate,decoration={snake},thick] (a) -- (v);
        \draw[decorate,decoration={snake},thick] (b) -- (v);
        \draw[decorate,decoration={snake},thick] (v) -- (c);
        
        % Draw parallel momentum arrows 
        \draw[-{Latex},thick] (-0.9,0.9) -- (-0.2,0.4) node[midway,above] {\( p_1 \)};
        \draw[-{Latex},thick] (-0.9,-0.9) -- (-0.2,-0.4) node[midway,below] {\( p_2 \)};
        \draw[-{Latex},thick] (0.3,0.3) --  (1.2,0.3) node[midway,above] {\( p_3 \)};
        
        % Add labels for fields next to endpoints
        \node[left] at (a) {\( (+,AA') \)};
        \node[left] at (b) {\( (+,BB') \)};
        \node[right] at (c) {\( (-,CC') \)};
        
        % Draw vertex as a filled dot
        \draw[thick] (v) circle (5pt);
        \fill[white] (v) circle (5pt);        
    \end{scope}
    % Equal sign between top diagrams
    \node at (-0.5,0) {=};
    \begin{scope}[xshift=3cm]
        \coordinate (v) at (0,0);
        
        % Define external legs
        \coordinate (a) at (-1.5,1);
        \coordinate (b) at (-1.5,-1);
        \coordinate (c) at (1.5,0);
        
        % Draw gauge boson lines 
        \draw[decorate,decoration={snake},thick] (a) -- (v);
        \draw[decorate,decoration={snake},thick] (b) -- (v);
        \draw[decorate,decoration={snake},thick] (v) -- (c);
        
        % Draw parallel momentum arrows 
        \draw[-{Latex},thick] (-0.9,0.9) -- (-0.2,0.4) node[midway,above] {\( p_1 \)};
        \draw[-{Latex},thick] (-0.9,-0.9) -- (-0.2,-0.4) node[midway,below] {\( p_2 \)};
        \draw[-{Latex},thick] (0.3,0.3) --  (1.2,0.3) node[midway,above] {\( p_3 \)};
        
        % Add labels for fields next to endpoints
        \node[left] at (a) {\( (+,AA') \)};
        \node[left] at (b) {\( (+,BB') \)};
        \node[right] at (c) {\( (-,CC') \)};
        
        \fill[yellow] (-0.5,0.3) circle (5pt);
    \end{scope}
        % Plus sign after top diagrams
    \node at (7,0) {+};
    \begin{scope}[xshift=-4cm,yshift=-4cm]
        % Plus sign in front of bottom diagrams
        \node at (-4.5,0) {+};
        \coordinate (v) at (0,0);
        
        % Define external legs
        \coordinate (a) at (-1.5,1);
        \coordinate (b) at (-1.5,-1);
        \coordinate (c) at (1.5,0);
        
        % Draw gauge boson lines 
        \draw[decorate,decoration={snake},thick] (a) -- (v);
        \draw[decorate,decoration={snake},thick] (b) -- (v);
        \draw[decorate,decoration={snake},thick] (v) -- (c);
        
        % Draw parallel momentum arrows 
        \draw[-{Latex},thick] (-0.9,0.9) -- (-0.2,0.4) node[midway,above] {\( p_1 \)};
        \draw[-{Latex},thick] (-0.9,-0.9) -- (-0.2,-0.4) node[midway,below] {\( p_2 \)};
        \draw[-{Latex},thick] (0.3,0.3) --  (1.2,0.3) node[midway,above] {\( p_3 \)};
        
        % Add labels for fields next to endpoints
        \node[left] at (a) {\( (+,AA') \)};
        \node[left] at (b) {\( (+,BB') \)};
        \node[right] at (c) {\( (-,CC') \)};
        \fill[yellow] (-0.5,-0.3) circle (5pt);
            \begin{scope}[xshift=7cm]
            % Plus sign between bottom diagrams
            \node at (-3.5,0) {+};
            \coordinate (v) at (0,0);
        
            % Define external legs
            \coordinate (a) at (-1.5,1);
            \coordinate (b) at (-1.5,-1);
            \coordinate (c) at (1.5,0);
        
            % Draw gauge boson lines 
            \draw[decorate,decoration={snake},thick] (a) -- (v);
            \draw[decorate,decoration={snake},thick] (b) -- (v);
            \draw[decorate,decoration={snake},thick] (v) -- (c);
        
            % Draw parallel momentum arrows 
            \draw[-{Latex},thick] (-0.9,0.9) -- (-0.2,0.4) node[midway,above] {\( p_1 \)};
            \draw[-{Latex},thick] (-0.9,-0.9) -- (-0.2,-0.4) node[midway,below] {\( p_2 \)};
            \draw[-{Latex},thick] (0.3,0.3) --  (1.2,0.3) node[midway,above] {\( p_3 \)};
        
            % Add labels for fields next to endpoints
            \node[left] at (a) {\( (+,AA') \)};
            \node[left] at (b) {\( (+,BB') \)};
            \node[right] at (c) {\( (-,CC') \)};
            \fill[yellow] (0.5,0) circle (5pt);
        
        \end{scope}
    \end{scope}
\end{tikzpicture}
\caption{The ASD vertex distributes the derivative evenly over all legs. The derivative is denoted by the yellow dot.}
\label{fig:cYMASD}
\end{figure}

\begin{figure}[h]
\centering
\begin{tabular}{c c c}
    % Vertex
    \begin{tikzpicture}
        \begin{scope}[xshift=-6cm]
        \coordinate (v) at (0,0);
        
        % Define external legs
        \coordinate (a) at (-1.5,1);
        \coordinate (b) at (-1.5,-1);
        \coordinate (c) at (1.5,0);
        
        % Draw gauge boson lines 
        \draw[decorate,decoration={snake},thick] (a) -- (v);
        \draw[decorate,decoration={snake},thick] (b) -- (v);
        \draw[decorate,decoration={snake},thick] (v) -- (c);
        
        % Draw parallel momentum arrows 
        \draw[-{Latex},thick] (-0.9,0.9) -- (-0.2,0.4) node[midway,above] {\( p_1 \)};
        \draw[-{Latex},thick] (-0.9,-0.9) -- (-0.2,-0.4) node[midway,below] {\( p_2 \)};
        \draw[{Latex}-,thick] (0.3,0.3) --  (1.2,0.3) node[midway,above] {\( p_3 \)};
        
        % Add labels for fields next to endpoints
        \node[left] at (a) {\( \epsilon^+_1 \)};
        \node[left] at (b) {\( \epsilon^+_2 \)};
        \node[right] at (c) {\( \epsilon_3^- \)};
        \end{scope}

        \node at (-3,0) {$=$};
        \node at (-2.5,0) {$\,\,2\,\times$};
        
        \coordinate (v) at (0,0);
        
        % Define external legs
        \coordinate (a) at (-1.5,1);
        \coordinate (b) at (-1.5,-1);
        \coordinate (c) at (1.5,0);
        
        % Draw gauge boson lines 
        \draw[decorate,decoration={snake},thick] (a) -- (v);
        \draw[decorate,decoration={snake},thick] (b) -- (v);
        \draw[decorate,decoration={snake},thick] (v) -- (c);
        
        % Draw parallel momentum arrows 
        \draw[-{Latex},thick] (-0.9,0.9) -- (-0.2,0.4) node[midway,above] {\( p_1 \)};
        \draw[-{Latex},thick] (-0.9,-0.9) -- (-0.2,-0.4) node[midway,below] {\( p_2 \)};
        \draw[{Latex}-,thick] (0.3,0.3) --  (1.2,0.3) node[midway,above] {\( p_3 \)};
        
        % Add labels for fields next to endpoints
        \node[left] at (a) {\( \epsilon_1^+ \)};
        \node[left] at (b) {\( \epsilon^+_2 \)};
        \node[right] at (c) {\( \epsilon^-_3 \)};

        % Draw vertex as a filled dot
        \fill (v) circle (5pt);
        \node at (3,0) {$=$};

        \begin{scope}[xshift=-6cm,yshift=-4cm]
            % Plus sign between bottom diagrams
            \node at (-3,0) {=};
            \node at (-2.5,0) {$\,\,2\,\times$};
            
            \coordinate (v) at (0,0);
        
            % Define external legs
            \coordinate (a) at (-1.5,1);
            \coordinate (b) at (-1.5,-1);
            \coordinate (c) at (1.5,0);
        
            % Draw gauge boson lines 
            \draw[decorate,decoration={snake},thick] (a) -- (v);
            \draw[decorate,decoration={snake},thick] (b) -- (v);
            \draw[decorate,decoration={snake},thick] (v) -- (c);
        
            % Draw parallel momentum arrows 
            \draw[-{Latex},thick] (-0.9,0.9) -- (-0.2,0.4) node[midway,above] {\( p_1 \)};
            \draw[-{Latex},thick] (-0.9,-0.9) -- (-0.2,-0.4) node[midway,below] {\( p_2 \)};
            \draw[-{Latex},thick] (0.3,0.3) --  (1.2,0.3) node[midway,above] {\( p_3 \)};
        
            % Add labels for fields next to endpoints
            \node[left] at (a) {\( \epsilon^+_1 \)};
            \node[left] at (b) {\( \epsilon^+_2 \)};
            \node[right] at (c) {\( \epsilon^-_3 \)};
            \fill[blue] (0.5,0) circle (5pt);
        \end{scope}
    \end{tikzpicture}
\end{tabular}
\caption{The YM three-point function is twice the SD three-point function, which only picks up a nonzero value when the derivative acts on the negative helicity external state.}
\label{fig:2SD}
\end{figure}
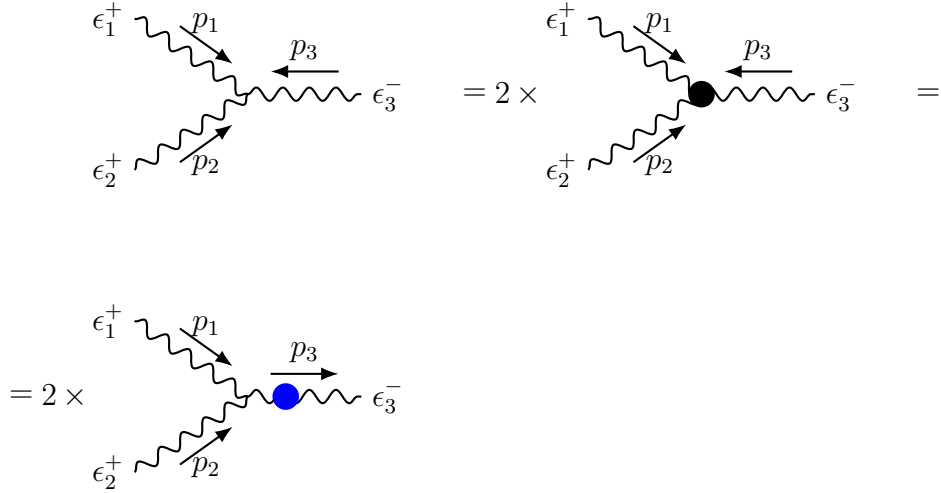

In Figure \ref{fig:cYMSD} and \ref{fig:cYMASD} it is shown diagrammatically how the derivative is distributed over the legs with the blue and yellow dot representing the derivative inside the SD and ASD vertex, respectively. Although this leads to a wealth of diagrams for higher order correlators, the vertex simplifies greatly if positive helicity polarization vectors are attached. The polarization vectors for YM read\footnote{Note that the sign in $(\epsilon^-)^{AA'}$ may differ from some sources in the literature. Our sign convention is chosen such that the YM and SDYM negative helicity polarization vectors are related through \eqref{Feps1}.}
\begin{align}
    (\epsilon^+_i)^{AA'} &= \frac{q_i^A\bar{i}^{A'}}{\langle q_i i\rangle} \,, &  (\epsilon^-_i)^{AA'} &= \frac{i^A\bar{q}_i^{A'}}{\langle \bar{q}_i\bar{i}\rangle}\,.
\end{align} 
Namely, the SD vertex simplifies using
\begin{align} \label{Feps1}
    p\fud{A}{A'}(\epsilon^+_i)^{AA'}&=0\,, & p\fud{A}{A'}(\epsilon_i^-)^{AA'} &= i^Ai^A \,.  
\end{align}
We observe that when momenta act on positive helicity states it vanishes and it gives back the polarization vector for the $\Psi$ field when acting on negative helicity states. For the ASD vertex we have
\begin{align} \label{Feps2}
    p^{AA'}(\epsilon^-_i)\fdu{A}{A'} &= 0 \,, & p^{AA'}(\epsilon_i^+)\fdu{A}{A'}&= \bar{i}^{A'}\bar{i}^{A'}\,,
\end{align}
i.e. when the momentum acts on negative helicity states it gives zero and when it acts on the positive helicity states it gives the ``anti-SDYM'' polarization vectors. The three-point scattering amplitude obtained from the SD vertex is
\begin{align}
    \mathcal{A}^{\text{SD}}_3=V^{\text{SD}}_{AA',BB',CC'}(\epsilon_1^+)^{AA'}(\epsilon_2^+)^{BB'}(\epsilon_3^-)^{CC'}=-g\frac{\langle q_13 \rangle \langle q_2 3 \rangle}{\langle q_11 \rangle \langle q_22 \rangle }\langle \bar{1}\bar{2} \rangle=g\frac{\langle \bar{1}\bar{2} \rangle^3}{\langle \bar{1}\bar{3} \rangle \langle \bar{2} \bar{3} \rangle}\,,
\end{align}
where \eqref{refSpinors} was used, and the ASD vertex yields\footnote{Although it also seems possible to eliminate $\bar{q}_3$ using similar relations as \eqref{refSpinors}, these relation would contain contractions between unbarred momentum spinors, e.g. $\langle 12 \rangle$. For three-particle kinematics one must choose either $1^A \sim 2^A \sim 3^A$ or $\bar{1}^A \sim \bar{2}^A\sim\bar{3}^A$. The contraction $\langle \bar{1}\bar{2}\rangle$ already appears in the amplitude before applying any identities, so it is natural to choose the former. This makes the identities associated to $\bar{q}_3$ ill-defined.}
\begin{align}
    \begin{aligned}
        \mathcal{A}^{\text{ASD}}_3=V^{\text{ASD}}_{AA',BB',CC'}&(\epsilon_1^+)^{AA'}(\epsilon_2^+)^{BB'}(\epsilon_3^-)^{CC'}=\\
        &=g\frac{\langle \bar{1}\bar{2} \rangle}{\langle q_11\rangle \langle q_22\rangle \langle \bar{q}_3\bar{3}\rangle}\Big(\langle q_11 \rangle \langle q_23 \rangle \langle \bar{q}_3\bar{1} \rangle + \langle q_13 \rangle \langle q_22 \rangle \langle \bar{q}_3 \bar{2} \rangle\Big)=\\
        &= g\frac{\langle \bar{1}\bar{2} \rangle^2}{\langle \bar{q}_3\bar{3}\rangle} \frac{\langle \bar{q}_3\bar{2}\rangle \langle \bar{1}\bar{3} \rangle - \langle \bar{q}_3 \bar{1} \rangle \langle \bar{2}\bar{3} \rangle}{\langle \bar{1}\bar{3} \rangle \langle \bar{2}\bar{3} \rangle} = g\frac{\langle \bar{1}\bar{2} \rangle^3}{\langle \bar{1}\bar{3} \rangle \langle \bar{2}\bar{3} \rangle}\,,
    \end{aligned}
\end{align}
where \eqref{refSpinors} was used, together with the Fierz identity
\begin{align}
    \langle ij \rangle \langle km \rangle + \langle ki \rangle \langle jm \rangle + \langle jk \rangle \langle im \rangle = 0\,.
\end{align} 
It is then straightforward to see that the YM amplitude is
\begin{align}
    \mathcal{A}_3 = \mathcal{A}_3^{\text{SD}} + \mathcal{A}_3^{\text{ASD}} = 2g\frac{\langle \bar{1}\bar{2} \rangle^3}{\langle \bar{1}\bar{3} \rangle \langle \bar{2}\bar{3} \rangle} \,,
\end{align}
which matches the SDYM result \eqref{3ptSDYM}. Furthermore, the topological term vanishes,
\begin{align}
    \mathcal{A}_3^{\text{top}}=\mathcal{A}_3^{\text{SD}} - \mathcal{A}_3^{\text{ASD}} = 0\,.
\end{align}
The latter implies that $\mathcal{A}_3=2\mathcal{A}_{3}^{\text{SD}}$, which is visualized in Figure \ref{fig:2SD}.

%%%%%%%%%%%%%%%%%%%%%%%%%%%%%%%%%%%%%%%%%%%%%%%%%%%%%%%%%%%%%
\subsection{Four point}
\label{app:fourFlat}
%%%%%%%%%%%%%%%%%%%%%%%%%%%%%%%%%%%%%%%%%%%%%%%%%%%%%%%%%%%%%

%%%%%%%%%%%%%%%%%%%%%%%%%%%%%%%%%%%%%%%%%%%%%%%%%%%%%%%%%%%%%
\subsubsection{SDYM: Feynman/Lorenz gauge}
\label{sec:}
%%%%%%%%%%%%%%%%%%%%%%%%%%%%%%%%%%%%%%%%%%%%%%%%%%%%%%%%%%%%%

We start by computing the flat-space scattering amplitude for SDYM as a warm-up. This will also facilitate the comparison with the $\text{AdS}_4$ correlator in the flat limit. To this end we first compute the propagator. The final answer was already discussed in Section \ref{sec:props}; here we provide more detail. The kinetic term in the action can be written as
\begin{align}
    S_{\text{kin}} = 
    -\Tr\int \Psi^{A,B}\epsilon_{BC}p_{AC'}\Phi^{C,C'} \,.
\end{align}
Here, the Feynman gauge is used where the Nakanishi-Lautrup field is absorbed into $\Psi^{A,B}$. The kinetic operator reads
\begin{align}
    K_{A,B;C,C'}(p) = -\epsilon_{BC}p_{AC'}
\end{align}
and the propagator $G_{A,B';C,C'}(p)$ should satisfy
\begin{align} \label{propEq}
    K(p)\fdu{A,B;}{E,E'}G_{C,D;E,E'}(p)=-\epsilon_{AC}\epsilon_{BD}\,.
\end{align}
It is useful to write the Feynman gauge propagator as
\begin{align}
    \langle \Psi_{A,B}(-p)\Phi_{C,C'}(p)\rangle_\text{F} \equiv G_{A,B;C,C'}(p)\,,
\end{align}
which allows one to obtain $\langle \Phi_{A,A'}(p)\Psi_{B,C}(p)\rangle_\text{F}$ by replacing $p\rightarrow -p$ and renaming the indices accordingly. It is then clear that on the other leg one imposes
\begin{align}\label{propEq2}
    K(-p)\fud{M,N}{;A,A'}G_{M,N;BB'}=-\epsilon_{AB}\epsilon_{A'B'} \,.
\end{align}
Now, \eqref{propEq} and \eqref{propEq2} become
\begin{align}
    \begin{aligned}
        -p\fdu{D}{M'}\langle \Psi_{A,B}(-p)\Phi_{C,M'}(p)\rangle_\text{F}&=-\epsilon_{AD}\epsilon_{BC} \,, &
        p\fud{M}{A'}\langle \Psi_{M,A}(-p)\Phi_{B,B'}(p)\rangle_\text{F}&=-\epsilon_{AB}\epsilon_{A'B'} \,.
    \end{aligned}
\end{align}
The left-hand sides of these relations come with a different sign, because the kinetic operator acts on the leg with momentum $-p$.
As a check, the propagator should also satisfy the equations of motion for the $\langle\Phi\Psi\rangle$ propagator,
\begin{align}
    \begin{aligned}
        p\fdu{B}{M'}\langle \Phi_{A,M'}(-p)\Psi_{C,D}(p)\rangle_\text{F}&=-\epsilon_{AD}\epsilon_{BC} \,, &
        -p\fud{M}{B'}\langle \Phi_{A,A'}(-p)\Psi_{M,B}(p)\rangle_\text{F}&=-\epsilon_{AB}\epsilon_{A'B'} \,.        
    \end{aligned}
\end{align}
Inverting the kinetic term now gives
\begin{align}\label{flatProp}
    \begin{aligned}
        \langle \Psi_{A,B}(-p) \Phi_{C,C'}(p)\rangle_\text{F}&=  -\frac{1}{p^2}\epsilon_{BC} p_{AC'} \,, &
        \langle \Phi_{A,A'}(-p)\Psi_{B,C}(p)\rangle_\text{F}&= -\frac{1}{p^2}\epsilon_{AC}p_{BA'}\,.
    \end{aligned}
\end{align}
Note that they are indeed related by swapping the indices and reversing the sign of the momentum, $p\leftrightarrow -p$. The interaction term is given by
\begin{align}
    S_{\text{int}}=\Tr\int\Psi^{AA}\Phi_{A,A'}\Phi\fdu{A,}{A'}=\frac{1}{2}\Tr\int (\epsilon^{AC}\epsilon^{BD}+\epsilon^{AD}\epsilon^{BC})\epsilon^{A'B'}\Phi_{A,A'}\Phi_{B,B'}\Psi_{C,D}\,.
\end{align}
Please note that the indices on $\Psi$ are symmetrized in the interaction term. This allowed us to add the Nakanishi-Lautrup field introduced in Section \ref{sec:SDYMprops}, i.e. replace $\Psi^{AA}\rightarrow \Psi^{A,B}$; due to the explicit symmetrization, the Nakanishi-Lautrup field actually does not enter the interaction. This has the advantage that the Feynman rules can now be written completely in terms of $\Psi^{A,B}$. However, for simplicity we stick to the symmetrized $\Psi^{AA}$, which effectively leads to choosing the Lorenz gauge. The propagator and vertex in this gauge is given in Figure \ref{fig:feynmanRules}.
\begin{figure}[h]
\centering
\begin{tabular}{c c c}
    % Propagator
    \begin{tikzpicture}
        \coordinate (a) at (-1.5,0);
        \coordinate (b) at (1.5,0);
        
        % Draw wavy propagator line
        \draw[decorate,decoration={snake},thick] (a) -- (b);
        
        % Draw momentum arrow
        \draw[-{Latex},thick] (-0.6,0.3) -- (0.6,0.3) node[midway,above] {\( p \)};
        
        % Labels for fields
        \node[left] at (a) {\( (-,AA) \)};
        \node[right] at (b) {\( (+,BB') \)};
    \end{tikzpicture}
    & \raisebox{3mm}{\quad : \quad} &
    \raisebox{3mm}{\( \langle\Psi_{AA}(-p)\Phi_{B,B'}(p)\rangle_\text{F}=-\frac{1}{p^2}\epsilon_{AB}p_{AB'} \)}
    \\[1.5cm]
    % Vertex
    \begin{tikzpicture}
        \coordinate (v) at (0,0);
        
        % Define external legs
        \coordinate (a) at (-1.5,1);
        \coordinate (b) at (-1.5,-1);
        \coordinate (c) at (1.5,0);
        
        % Draw gauge boson lines 
        \draw[decorate,decoration={snake},thick] (a) -- (v);
        \draw[decorate,decoration={snake},thick] (b) -- (v);
        \draw[decorate,decoration={snake},thick] (v) -- (c);
        
        % Draw parallel momentum arrows 
        \draw[-{Latex},thick] (-0.9,0.9) -- (-0.2,0.4) node[midway,above] {\( p_1 \)};
        \draw[-{Latex},thick] (-0.9,-0.9) -- (-0.2,-0.4) node[midway,below] {\( p_2 \)};
        \draw[-{Latex},thick] (0.3,0.3) --  (1.2,0.3) node[midway,above] {\( p_3 \)};
        
        % Add labels for fields next to endpoints
        \node[left] at (a) {\( (+,AA') \)};
        \node[left] at (b) {\( (+,BB') \)};
        \node[right] at (c) {\( (-,CC) \)};
    \end{tikzpicture}
    & \raisebox{12mm}{\quad : \quad} &
    \raisebox{12mm}{\( V_{AA',BB',CC}= 2g\epsilon_{AC}\epsilon_{BC}\epsilon_{A'B'}\)}
\end{tabular}
\caption{SDYM Lorenz gauge Feynman rules in flat space.}
\label{fig:feynmanRules}
\end{figure}

The $s$-channel diagram is given in Figure \ref{fig:schannel}. Explicitly, using $p^2=(p_1+p_2)^2=-\langle 12 \rangle \langle \bar{1}\bar{2} \rangle$, the $s$-channel amplitude becomes
\begin{figure}[h!]
\centering
\begin{tabular}{c c}
    % S-channel diagram
    \begin{tikzpicture}
        % Define the vertices
        \coordinate (v1) at (-1,0);
        \coordinate (v2) at (1,0);
        
        % Define the external legs
        \coordinate (a) at (-2,1);
        \coordinate (b) at (-2,-1);
        \coordinate (c) at (2,1);
        \coordinate (d) at (2,-1);
        
        % Draw the gauge boson lines
        \draw[decorate,decoration={snake},thick] (a) -- (v1);
        \draw[decorate,decoration={snake},thick] (b) -- (v1);
        \draw[decorate,decoration={snake},thick] (c) -- (v2);
        \draw[decorate,decoration={snake},thick] (d) -- (v2);
        
        % Draw the internal propagator 
        \draw[decorate,decoration={snake},thick] (v1) -- (v2);
        
        % Draw parallel momentum arrows
        \draw[-{Latex},thick] (-1.5,0.8) -- (-1.1,0.4) node[midway,above right] {\( p_1 \)};
        \draw[-{Latex},thick] (-1.5,-0.8) -- (-1.1,-0.4) node[midway,below right] {\( p_2 \)};
        \draw[-{Latex},thick] (1.1,0.4) -- (1.5,0.8) node[midway,above left] {\( p_4 \)};
        \draw[-{Latex},thick] (1.1,-0.4) -- (1.5,-0.8) node[midway,below left] {\( p_3 \)};
        \draw[-{Latex},thick] (-0.3,0.2) -- (0.3,0.2) node[midway,above] {\( p_1+p_2 \)};
        
        % Add labels for fields next to endpoints
        \node[left] at (a) {\( \epsilon^+_1 \)};
        \node[left] at (b) {\( \epsilon^+_2 \)};
        \node[right] at (c) {\( \epsilon^-_4 \)};
        \node[right] at (d) {\( \epsilon^+_3 \)};
    \end{tikzpicture}
    & \raisebox{12mm}{\(\quad : \quad\)}
\end{tabular}

\vspace{0.5cm}

\[
\mathcal{A}_s = (\epsilon^+_1)^{AA'}(\epsilon^+_2)^{BB'}V\fdu{AA',BB',}{EE} \, \langle \Psi_{EE}(-p)\Phi_{F,F'}(p)\rangle_\text{F} \, V\fud{FF'}{,CC',DD}(\epsilon^+_3)^{CC'}(\epsilon^-_4)^{DD} 
\]

\caption{SDYM Lorenz gauge $s$-channel diagram in flat space.}
\label{fig:schannel}
\end{figure}
\begin{align}\label{As}
    \mathcal{A}_s = 4g^2\frac{\langle q_14 \rangle \langle q_24 \rangle \langle q_34 \rangle}{\langle q_11 \rangle\langle q_22 \rangle\langle q_33 \rangle} \frac{\langle \bar{3}\bar{4} \rangle}{\langle 12 \rangle}  \,.
\end{align}
The $t$-channel amplitude is simply obtained by swapping $1 \leftrightarrow 3$, $\bar{1}\leftrightarrow\bar{3}$ and $q_1 \leftrightarrow q_3$ and reads
\begin{align}
    \mathcal{A}_t = -4g^2\frac{\langle q_14 \rangle \langle q_24 \rangle \langle q_34 \rangle}{\langle q_11 \rangle\langle q_22 \rangle\langle q_33 \rangle} \frac{\langle \bar{1}\bar{4} \rangle}{\langle 23 \rangle} \,.
\end{align}
The color-ordered partial amplitude is obtained by summing the $s$-channel and $t$-channel amplitude,
\begin{align} \label{As+At}
    \mathcal{A}&=\mathcal{A}_s+\mathcal{A}_t=
    4g^2\frac{\langle q_14 \rangle \langle q_24 \rangle \langle q_34 \rangle}{\langle q_11 \rangle\langle q_22 \rangle\langle q_33 \rangle}\left(\frac{\langle 23 \rangle\langle \bar{3}\bar{4} \rangle - \langle 12 \rangle \langle \bar{1}\bar{4} \rangle}{\langle 12 \rangle \langle 23 \rangle}\right)=0 \,.
\end{align}
In the last equality we used momentum conservation,
\begin{align} \label{pcons}
    \frac{\langle 23 \rangle \langle \bar{3}\bar{4} \rangle-\langle 12 \rangle \langle \bar{1}\bar{4} \rangle}{\langle 12 \rangle \langle 23 \rangle} = -\frac{(1^A\bar{1}^{A'}+2^A\bar{2}^{A'}+3^A\bar{3}^{A'}+4^A\bar{4}^{A'})2_{A}\bar{4}_{A'}}{\langle 12 \rangle \langle 23 \rangle}=0 \,.
\end{align}

%%%%%%%%%%%%%%%%%%%%%%%%%%%%%%%%%%%%%%%%%%%%%%%%%%%%%%%%%%%%%
\subsubsection{SDYM: Axial gauge} \label{sec:SDYMAxial}
%%%%%%%%%%%%%%%%%%%%%%%%%%%%%%%%%%%%%%%%%%%%%%%%%%%%%%%%%%%%%
Scattering amplitudes are gauge-invariant, but the Feynman rules are not. It is therefore instructive to see why a gauge transformation does not affect the amplitudes. Here we will investigate axial gauge, but the reasoning extends to other gauges. The boundary of anti-de Sitter space will lead to important corrections to the arguments below.

In axial gauge, the gauge field $\Phi_{A,A'}$ is orthogonal to a spacelike vector $n^{AA'}$. The Feynman rules are summarized in Figure \ref{fig:feynmanRulesAxial}. Note that the vertex used is just the same as in Lorenz gauge, while it should actually be projected onto its $n$-orthogonal component in its first two pairs of indices,
\begin{align}
    V_{AA',BB',CC} = \epsilon_{AC}\epsilon_{BC}\epsilon_{A'B'}-\frac{1}{2n^2}\big(\epsilon_{AC}n_{CA'}n_{BB'}-\epsilon_{BC}n_{CB'}n_{AA'}\big)\,.
\end{align}
However, since the legs on the vertices will be contracted with $n$-orthogonal states, this is realized automatically.

\begin{figure}[h]
\centering
\begin{tabular}{c c c}
    % Propagator
    \begin{tikzpicture}
        \coordinate (a) at (-1.5,0);
        \coordinate (b) at (1.5,0);
        
        % Draw wavy propagator line
        \draw[decorate,decoration={snake},thick] (a) -- (b);
        
        % Draw momentum arrow
        \draw[-{Latex},thick] (-0.6,0.3) -- (0.6,0.3) node[midway,above] {\( p \)};
        
        % Labels for fields
        \node[left] at (a) {\( (-,AA) \)};
        \node[right] at (b) {\( (+,BB') \)};
    \end{tikzpicture}
    & \raisebox{3mm}{\quad : \quad} &
    \raisebox{3mm}{\( \begin{aligned} &\langle \Psi^{AA}(-p)\Phi^{BB'}(p) \rangle_\text{A}=\\
    &=-\frac{1}{p^2} \left(\epsilon^{AB} p^{AB'}+\frac{1}{2(p\cdot n)} p\fud{A}{M'}n^{AM'} p^{BB'}\right) \end{aligned} \)}
    \\[1.5cm]
    % Vertex
    \begin{tikzpicture}
        \coordinate (v) at (0,0);
        
        % Define external legs
        \coordinate (a) at (-1.5,1);
        \coordinate (b) at (-1.5,-1);
        \coordinate (c) at (1.5,0);
        
        % Draw gauge boson lines 
        \draw[decorate,decoration={snake},thick] (a) -- (v);
        \draw[decorate,decoration={snake},thick] (b) -- (v);
        \draw[decorate,decoration={snake},thick] (v) -- (c);
        
        % Draw parallel momentum arrows 
        \draw[-{Latex},thick] (-0.9,0.9) -- (-0.2,0.4) node[midway,above] {\( p_1 \)};
        \draw[-{Latex},thick] (-0.9,-0.9) -- (-0.2,-0.4) node[midway,below] {\( p_2 \)};
        \draw[-{Latex},thick] (0.3,0.3) --  (1.2,0.3) node[midway,above] {\( p_3 \)};
        
        % Add labels for fields next to endpoints
        \node[left] at (a) {\( (+,AA') \)};
        \node[left] at (b) {\( (+,BB') \)};
        \node[right] at (c) {\( (-,CC) \)};
    \end{tikzpicture}
    & \raisebox{12mm}{\quad : \quad} &
    \raisebox{12mm}{\( V_{AA',BB',CC}= 2g\,\epsilon_{AC}\epsilon_{BC}\epsilon_{A'B'}\)}
\end{tabular}
\caption{SDYM axial gauge Feynman rules in flat space.}
\label{fig:feynmanRulesAxial}
\end{figure}
The polarization vectors are again given by \eqref{sdympol}. In axial gauge, the reference spinors are restricted by the condition that the polarization vectors need to be $n$-transverse, i.e. $n^{AA'}(\epsilon_i^+)_{AA'}=0$.\footnote{In AdS we will choose $n_{AA'}=\epsilon_{AA'}$ and the polarization vectors become symmetric in axial gauge, $(\epsilon_i^+)^{AA'}=\frac{\bar{i}^A\bar{i}^{A'}}{2k_i}$ and $(\epsilon_i^-)^{AA}=i^Ai^A$.} For our purposes it is not necessary to solve this explicitly; we only need to keep in mind that these polarization vectors are also admissible in the Lorenz gauge. We see that the first term in the propagator is just the Lorenz gauge propagator. In fact, the second term can be written as a gauge transformation of the gauge field,
\begin{align}
    \langle \Psi^{AA}(-p)\Phi^{BB'}(p)\rangle_{\text{A}} = \langle \Psi^{AA}(-p)\Phi^{BB'}(p)\rangle_{\text{F}} + p^{BB'}\xi^{AA}\,,
\end{align}
with $\xi_{AA}=-\frac{1}{p^2}\frac{p\fud{A}{M'}n^{AM'}}{2(p\cdot n)}$. It is then not hard to see that
\begin{align}
    p^{AA'}V_{AA',BB',CC}\Big|_{\text{two legs on-shell}}=0
\end{align}
and the gauge variation disappears from the amplitude. We thus conclude that the scattering amplitude in axial gauge agrees with the amplitude in the Lorenz gauge, but with specialized reference spinors. This result holds channel by channel.

%%%%%%%%%%%%%%%%%%%%%%%%%%%%%%%%%%%%%%%%%%%%%%%%%%%%%%%%%%%%%
\subsubsection{YM: Feynman gauge} \label{sec:cYMFlat}
%%%%%%%%%%%%%%%%%%%%%%%%%%%%%%%%%%%%%%%%%%%%%%%%%%%%%%%%%%%%%
In Section \ref{sec:YMFlat}, we encountered a significant simplification of the three-point amplitude for YM: instead of two different diagrams involving the SD and ASD vertex, the three-point amplitude was expressed using only the SD vertex with the appropriate coefficient. Moreover, this diagram turned out to be easier to evaluate than the diagram containing the ASD vertex, due to the helicities configuration $(++-)$, where most external states have positive helicity.

\begin{figure}[h!]
\centering
\begin{tikzpicture}
    % Top diagram (original)
    \begin{scope}[yshift=4cm]
        \begin{scope}[xshift=-4cm]
        \coordinate (v1) at (-1,0);
        \coordinate (v2) at (1,0);
        \coordinate (a) at (-2,1);
        \coordinate (b) at (-2,-1);
        \coordinate (c) at (2,1);
        \coordinate (d) at (2,-1);
        
        \draw[decorate,decoration={snake},thick] (a) -- (v1);
        \draw[decorate,decoration={snake},thick] (b) -- (v1);
        \draw[decorate,decoration={snake},thick] (c) -- (v2);
        \draw[decorate,decoration={snake},thick] (d) -- (v2);
        \draw[decorate,decoration={snake},thick] (v1) -- (v2);
        
        \draw[-{Latex},thick] (-1.5,0.8) -- (-1.1,0.4) node[midway,above right] {\( p_1 \)};
        \draw[-{Latex},thick] (-1.5,-0.8) -- (-1.1,-0.4) node[midway,below right] {\( p_2 \)};
        \draw[-{Latex},thick] (1.1,0.4) -- (1.5,0.8) node[midway,above left] {\( p_4 \)};
        \draw[-{Latex},thick] (1.1,-0.4) -- (1.5,-0.8) node[midway,below left] {\( p_3 \)};
        \draw[-{Latex},thick] (-0.3,0.2) -- (0.3,0.2) node[midway,above] {\( p_1+p_2 \)};
        
        \node[left] at (a) {\( \epsilon^+_1 \)};
        \node[left] at (b) {\( \epsilon^+_2 \)};
        \node[right] at (c) {\( \epsilon^-_4 \)};
        \node[right] at (d) {\( \epsilon^+_3 \)};
        \end{scope}
        \node at (0,0) {=};
        \node at (0.5,0) {$\,\,4\,\times$};
        \begin{scope}[xshift=4cm]
            \coordinate (v1) at (-1,0);
        \coordinate (v2) at (1,0);
        \coordinate (a) at (-2,1);
        \coordinate (b) at (-2,-1);
        \coordinate (c) at (2,1);
        \coordinate (d) at (2,-1);
        
        \draw[decorate,decoration={snake},thick] (a) -- (v1);
        \draw[decorate,decoration={snake},thick] (b) -- (v1);
        \draw[decorate,decoration={snake},thick] (c) -- (v2);
        \draw[decorate,decoration={snake},thick] (d) -- (v2);
        \draw[decorate,decoration={snake},thick] (v1) -- (v2);
        
        \draw[-{Latex},thick] (-1.5,0.8) -- (-1.1,0.4) node[midway,above right] {\( p_1 \)};
        \draw[-{Latex},thick] (-1.5,-0.8) -- (-1.1,-0.4) node[midway,below right] {\( p_2 \)};
        \draw[-{Latex},thick] (1.1,0.4) -- (1.5,0.8) node[midway,above left] {\( p_4 \)};
        \draw[-{Latex},thick] (1.1,-0.4) -- (1.5,-0.8) node[midway,below left] {\( p_3 \)};
        \draw[-{Latex},thick] (-0.3,0.2) -- (0.3,0.2) node[midway,above] {\( p_1+p_2 \)};
        
        \node[left] at (a) {\( \epsilon^+_1 \)};
        \node[left] at (b) {\( \epsilon^+_2 \)};
        \node[right] at (c) {\( \epsilon^-_4 \)};
        \node[right] at (d) {\( \epsilon^+_3 \)};
        \node at (3,0) {=}; 

        \fill (v1) circle (5pt);
        \fill (v2) circle (5pt);
        \end{scope}
    \end{scope}
    
    % First bottom diagram
    \begin{scope}[xshift=-4cm]
        \node at (-3.5,0) {=};
        \node at (-3,0) {$\,\,4\, \times$};
        \coordinate (v1) at (-1,0);
        \coordinate (v2) at (1,0);
        \coordinate (a) at (-2,1);
        \coordinate (b) at (-2,-1);
        \coordinate (c) at (2,1);
        \coordinate (d) at (2,-1);
        
        \draw[decorate,decoration={snake},thick] (a) -- (v1);
        \draw[decorate,decoration={snake},thick] (b) -- (v1);
        \draw[decorate,decoration={snake},thick] (c) -- (v2);
        \draw[decorate,decoration={snake},thick] (d) -- (v2);
        \draw[decorate,decoration={snake},thick] (v1) -- (v2);
        
        \draw[-{Latex},thick] (-1.5,0.8) -- (-1.1,0.4) node[midway,above right] {\( p_1 \)};
        \draw[-{Latex},thick] (-1.5,-0.8) -- (-1.1,-0.4) node[midway,below right] {\( p_2 \)};
        \draw[-{Latex},thick] (1.1,0.4) -- (1.5,0.8) node[midway,above left] {\( p_4 \)};
        \draw[-{Latex},thick] (1.1,-0.4) -- (1.5,-0.8) node[midway,below left] {\( p_3 \)};
        \draw[-{Latex},thick] (-0.3,0.2) -- (0.3,0.2) node[midway,above] {\( p_1+p_2 \)};
        
        \node[left] at (a) {\( \epsilon^+_1 \)};
        \node[left] at (b) {\( \epsilon^+_2 \)};
        \node[right] at (c) {\( \epsilon^-_4 \)};
        \node[right] at (d) {\( \epsilon^+_3 \)};
        
        \fill[blue] (-0.5,0) circle (5pt);
        \fill[blue] (1.6,0.5) circle (5pt);
    \end{scope}
    
    % Plus sign between bottom diagrams
    \node at (0,0) {+};
    \node at (0.5,0) {$\,\,4\,\times$};

    % Second bottom diagram
    \begin{scope}[xshift=4cm]
        \coordinate (v1) at (-1,0);
        \coordinate (v2) at (1,0);
        \coordinate (a) at (-2,1);
        \coordinate (b) at (-2,-1);
        \coordinate (c) at (2,1);
        \coordinate (d) at (2,-1);
        
        \draw[decorate,decoration={snake},thick] (a) -- (v1);
        \draw[decorate,decoration={snake},thick] (b) -- (v1);
        \draw[decorate,decoration={snake},thick] (c) -- (v2);
        \draw[decorate,decoration={snake},thick] (d) -- (v2);
        \draw[decorate,decoration={snake},thick] (v1) -- (v2);
        
        \draw[-{Latex},thick] (-1.5,0.8) -- (-1.1,0.4) node[midway,above right] {\( p_1 \)};
        \draw[-{Latex},thick] (-1.5,-0.8) -- (-1.1,-0.4) node[midway,below right] {\( p_2 \)};
        \draw[-{Latex},thick] (1.1,0.4) -- (1.5,0.8) node[midway,above left] {\( p_4 \)};
        \draw[-{Latex},thick] (1.1,-0.4) -- (1.5,-0.8) node[midway,below left] {\( p_3 \)};
        \draw[-{Latex},thick] (-0.3,0.2) -- (0.3,0.2) node[midway,above] {\( p_1+p_2 \)};
        
        \node[left] at (a) {\( \epsilon^+_1 \)};
        \node[left] at (b) {\( \epsilon^+_2 \)};
        \node[right] at (c) {\( \epsilon^-_4 \)};
        \node[right] at (d) {\( \epsilon^+_3 \)};
        
        \fill[blue] (-0.5,0) circle (5pt);
        \fill[blue] (0.5,0) circle (5pt);
    \end{scope}
\end{tikzpicture}

\caption{The four-point function for the YM theory can be represented using only the SD vertex. This diagram is then decomposed into a diagram with one derivative on the internal line, similar to the SDYM exchange diagram, and a diagram with two derivatives on the internal line.}
\label{fig:YMflatDer}
\end{figure}
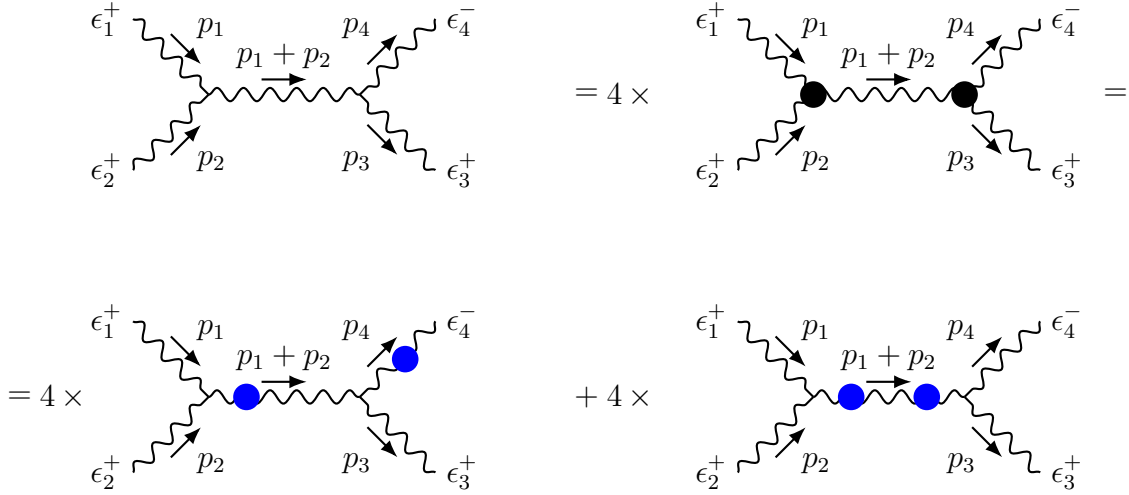

In this section, we aim to extend this simplification to the four-point exchange diagram with helicity configuration $(+++-)$ and compute the four-point amplitude. The simplification of the three-point amplitude arose from an analogous simplifcation at the level of the vertex: the YM vertex can be written entirely in terms of the SD vertex,
\begin{align}
    V_{AA',BB',CC'}=2V^{\text{SD}}_{AA',BB',CC'} \,,
\end{align}
as a consequence of the vanishing boundary term. As a result, the $s$-channel diagram can also be expressed using only the SD vertex, see Figure \ref{fig:YMflatDer}. This diagram contains two vertices, which leads to $9$ distinct diagrams corresponding to different derivative placements, according to the decomposition shown in Figure \ref{fig:cYMSD}. Fortunately, \eqref{Feps1} and \eqref{Feps2} imply that any diagram vanishes when a derivative acts on a positive helicity external state. For the helicity configuration $(+++-)$, only $2$ non-vanishing diagrams remain, as is shown in Figure \ref{fig:YMflatDer}.

The $\langle\Phi\Phi\rangle$ propagator is displayed in Figure \ref{fig:cYMprop}, see Section \ref{sec:Fgauge} for more details. The first diagram in Figure \ref{fig:YMflatDer} is then easily seen to be equivalent to the SDYM $s$-channel diagram through the identities \eqref{Feps1} and
\begin{align}
    p\fdu{A}{A'}\langle \Phi_{A,A'}(-p) \Phi_{B,B'}(p) \rangle_\text{F} &= \langle \Psi_{AA}(-p) \Phi_{B,B'}(p) \rangle_\text{F} \,.
\end{align}
The former relates the negative helicity on-shell states of cYM and SDYM and the latter relates the cYM and SDYM propagator through the derivative that sits inside the cYM vertex. Notice that this diagram is independent of the reference spinor $\bar{q}_4$ due to the first identity.

\begin{figure}[h]
\centering
\begin{tabular}{c c c}
    % Propagator
    \begin{tikzpicture}
        \coordinate (a) at (-1.5,0);
        \coordinate (b) at (1.5,0);
        
        % Draw wavy propagator line
        \draw[decorate,decoration={snake},thick] (a) -- (b);
        
        % Draw momentum arrow
        \draw[-{Latex},thick] (-0.6,0.3) -- (0.6,0.3) node[midway,above] {\( p \)};
        
        % Labels for fields
        \node[left] at (a) {\( (-,AA') \)};
        \node[right] at (b) {\( (+,BB') \)};
    \end{tikzpicture}
    & \raisebox{3mm}{\quad : \quad} &
    \raisebox{3mm}{\( \langle\Phi_{A,A'}(-p)\Phi_{B,B'}(p)\rangle_\text{F}=-\frac{1}{p^2}\epsilon_{AB}\epsilon_{A'B'} \)}
    \\[1.5cm]
\end{tabular}
\caption{cYM Feynman gauge propagator Feynman rule in flat space.}
\label{fig:cYMprop}
\end{figure}

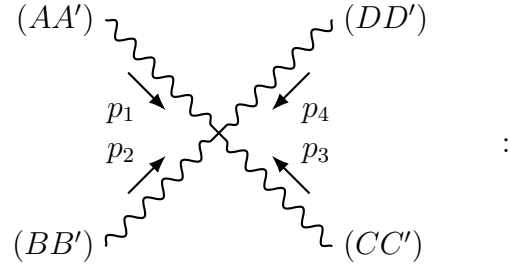
\begin{figure}[h!]
\centering
\begin{tabular}{c c}
    % Quartic vertex diagram
    \begin{tikzpicture}
        % Define the central vertex
        \coordinate (v) at (0,0);
        
        % Define the external legs
        \coordinate (a) at (-1.5,1.5);
        \coordinate (b) at (-1.5,-1.5);
        \coordinate (c) at (1.5,1.5);
        \coordinate (d) at (1.5,-1.5);
        
        % Draw the gauge boson lines
        \draw[decorate,decoration={snake},thick] (a) -- (v);
        \draw[decorate,decoration={snake},thick] (b) -- (v);
        \draw[decorate,decoration={snake},thick] (c) -- (v);
        \draw[decorate,decoration={snake},thick] (d) -- (v);
        
        % Draw momentum arrows
        \draw[-{Latex},thick] (-1.2,0.8) -- (-0.7,0.3) node[midway,below left] {\( p_1 \)};
        \draw[-{Latex},thick] (-1.2,-0.8) -- (-0.7,-0.3) node[midway,above left] {\( p_2 \)};
        \draw[-{Latex},thick] (1.2,0.8) -- (0.7,0.3) node[midway,below right] {\( p_4 \)};
        \draw[-{Latex},thick] (1.2,-0.8) -- (0.7,-0.3) node[midway,above right] {\( p_3 \)};
        
        % Add labels for polarization vectors
        \node[left] at (a) {\( (AA') \)};
        \node[left] at (b) {\( (BB') \)};
        \node[right] at (c) {\( (DD') \)};
        \node[right] at (d) {\( (CC') \)};
    \end{tikzpicture}
    & \raisebox{16mm}{\quad : \quad} 
\end{tabular}
\[
    \begin{aligned} V^{\text{quartic}}_{AA',BB',CC',DD'}&= 2g^2\epsilon_{A'B'}\epsilon_{C'D'}(\epsilon_{AD}\epsilon_{BC}+\epsilon    _{AC}\epsilon_{BD})
    \end{aligned}
\]

\caption{cYM quartic vertex Feynman rule.}
\label{fig:quartic}
\end{figure}

The right diagram in Figure \ref{fig:YMflatDer} is absent in SDYM and it produces a non-vanishing contribution to the amplitude in the YM theory. With the derivatives from both vertices acting on the propagator, we find the identity
\begin{align}
    -p\fdu{A}{A'}p\fdu{B}{B'}\langle \Phi_{A,A'}(-p)\Phi_{B,B'}(p) \rangle_\text{F} = \langle \Psi_{AA}(-p) \Psi_{BB}(p) \rangle_\text{F} = -\epsilon_{AB}\epsilon_{AB}\,,
\end{align}
with the $\langle\Phi\Phi\rangle$ propagator given in \eqref{phiphiProp}. After contracting the three-point vertices from the $s$-channel diagram with this second-derivative propagator, it resembles the Feynman rule for the quartic vertex, which is present in cYM, but absent in SDYM. The quartic Feynman rule is depicted in Figure \ref{fig:quartic}. Using the above identities, the exchange diagram in the $s$-channel in cYM reads
\begin{align}
    \mathcal{A}_s^{\text{cYM}} &= \mathcal{A}_s^{\text{SDYM}} -2g^2(\epsilon_1^+)^{AA'}(\epsilon_2^+)^{BB'}\epsilon_{A'B'}\epsilon_{C'D'}(\epsilon_{AC}\epsilon_{BD}+\epsilon_{AD}\epsilon_{BC})(\epsilon_3^+)^{CC'}(\epsilon_4^-)^{DD'} \,.
\end{align}
The contact diagram gives\footnote{More precisely, the contact diagram gives three different contributions, proportional to $f^{abe}f^{cde}$, $f^{ade}f^{bce}$ and $f^{ace}f^{bde}$. The first corresponds to the color structure found in the $s$-channel diagram, the second is the structure found in the $t$-channel and the third in the $u$-channel. See Appendix \ref{app:color} for more details. Here, we only focus on the color factors of the $s$-channel.}
\begin{align}
    \mathcal{A}_{\text{contact}} = 2g^2(\epsilon_1^+)^{AA'}(\epsilon_2^+)^{BB'}\epsilon_{A'B'}\epsilon_{C'D'}(\epsilon_{AC}\epsilon_{BD}+\epsilon_{AD}\epsilon_{BC})(\epsilon_3^+)^{CC'}(\epsilon_4^-)^{DD'} \,.
\end{align}
Thus, the bottom-right diagram in Figure \ref{fig:YMflatDer} cancels the contact diagram. In other words,
\begin{align}
    \mathcal{A}_s^\text{cYM}+\mathcal{A}_\text{contact}=\mathcal{A}_s^\text{SDYM} \,.
\end{align}

\paragraph{Gauge independence.}
It is now easy to see that the scattering amplitude in cYM in Feynman gauge and axial gauge are identical. The same arguments as for SDYM in Section \ref{sec:SDYMAxial} are used to show that the diagram with one derivative on the internal line is gauge-invariant. As we have seen in this section, only the propagator (with a gauge field on at least one of the legs) changes in axial gauge, while other quantities remain the same. However, the new exchange diagram that appears in cYM contains the $\langle \Psi_{AA}(-p)\Psi_{BB}(p) \rangle$ propagator, which is constructed out of gauge-invariant quantities and the contact diagram contains no propagators. In fact, these two diagrams cancel against each other in any gauge.

%%%%%%%%%%%%%%%%%%%%%%%%%%%%%%%%%%%%%%%%%%%%%%%%%%%%%%%%%%%%%
\section{Gauge (in)dependence day}
\label{app:gaugedependendence}
%%%%%%%%%%%%%%%%%%%%%%%%%%%%%%%%%%%%%%%%%%%%%%%%%%%%%%%%%%%%%
Let us consider the difference between two four-point functions computed in different gauges. Two gauges differ by pure gauge terms for the bulk-to-bulk propagator (we omit the indices). 
\begin{align}
    G'&= G+\Delta G \,,& \Delta G(k,z;k',z')&= \nabla_\text{L} \xi_\text{L} +\nabla_\text{R} \xi_\text{R} +\nabla_\text{L}\nabla_\text{R} \xi_\text{c} \,.
\end{align}
Let us consider the simplest example of an exchange between two conserved currents built from scalar fields, which can occur in scalar electrodynamics. The difference between two gauges is
\begin{align}
    \Delta \langle OOOO\rangle &= \int J^\mu(k_1,k_2,z) \Delta G(k,z;k',z')_{\mu\nu} J^\nu(k_3,k_4,z') \,.
\end{align}
The bulk terms disappear due to $\nabla_\mu J^\mu=0$ when the two legs of the current are on-shell, i.e. on the boundary. The remnants are the boundary terms (note that $\vec{k}_{12}=-\vec{k}_{34}$):\footnote{Note that for the Lorenz/axial propagators the $\xi_\mu$ are such that $\xi_\mu\sim n_\mu$, i.e. only the bulk integral against $J^z$ can contribute, see below.}
\begin{align}
    \begin{aligned}
        \Delta \langle OOOO\rangle &= \int dz'\, J^z(k_1,k_2,0) \xi_\nu^\text{L}(\vec{k}_{12};z=0;z')J^\nu(k_3,k_4,z')+\\
        &+\int dz\, J^\mu(k_1,k_2,z) \xi_\mu^\text{R}(\vec{k}_{12};z;z'=0)J^z(k_3,k_4,0)+\\
        &\qquad +J^z(k_1,k_2,0)\xi_\text{c}(\vec{k}_{12};z=0;z'=0)J^z(k_3,k_4,0) \,.
    \end{aligned}
\end{align}
Let us restrict to the toy model with four scalar fields interacting with $A_\mu$ pairwise (so that there are no additional symmetrizations). The current, its $z$-component we need, is
\begin{align}
    J_z&= \phi_1 \overleftrightarrow{\pl_z}\phi_2= f_1 f_2\exp[-(k_1+k_2)z](k_1-k_2)\,,
\end{align}
where $f_{1,2}$ are $k_i^{-1}$ for Neumann and $f_{1,2}=1$ for Dirichlet. At the boundary the current is just $J_z=f_1f_2(k_1-k_2)$. For the Dirichlet boundary condition $f_i=1$ and 
\begin{align}\notag
    \int dz'\, &J^z(k_1,k_2,0) \xi_\nu^\text{L}(\vec{k}_{12};z=0;z')J^\nu(k_3,k_4,z')=(k_1-k_2)\int dz'\, \xi_\nu(\vec{k}_{12},z=0;z')J^\nu(k_3,k_4,z')\,,
\end{align}
which, when split into two terms, better be analytic in either $k_1$ or $k_2$. A typical pure gauge term would contain $\exp[-k_{12}z']$ at $z=0$ and it is impossible to get rid of $k_{1,2}$ dependence! Therefore, the only possibility seems to be that $\xi$ vanishes at the boundary.

To understand the origin of contact terms, let us consider an example of a four-point correlator
\begin{align}
    \int \prod_{i=1}^{i=4} d^3k_i \, \delta^3(\sum k_i) e^{i \sum k_j x_j} F(k_1,k_2,k_3)&=
    \int \prod_{i=1}^{i=3} d^3k_i \, e^{i \sum_i k_i(x_i-x_4)} F(k_1,k_2,k_3)\,,
\end{align}
where $F$ is assumed not to depend on, say, $k_4$ thanks to the momentum conservation. 
Provided, $F$ depends on, say, $k_2$ analytically, e.g. it does not depend on it at all, $\exp ik_2(x_2-x_4)$ factors out and produces $\delta^3(x_2-x_4)$. This is how one can get a sum of contact terms
\begin{align}
    \Delta\langle OOOO\rangle & = \sum\delta(x_i-x_j) V(...) \,.
\end{align}
Whenever, the external operators are currents themselves, such contact terms mimic the Ward identities
\begin{align}
    \pl_1\cdot \langle JJJJ\rangle & = \sum\delta(x_1-x_i) \langle JJJ\rangle \,.
\end{align}

%%%%%%%%%%%%%%%%%%%%%%%%%%%%%%%%%%%%%%%%%%%%%%%%%%%%%%%%%%%%%
\subsection{Dirichlet Yang--Mills theory}
\label{sec:DirYM}
%%%%%%%%%%%%%%%%%%%%%%%%%%%%%%%%%%%%%%%%%%%%%%%%%%%%%%%%%%%%%
It is important to make sure that the results for gauge-invariant observables do not depend on the gauge. For the Dirichlet boundary conditions the dual operator is a current $J_i$ and via Witten diagrams we are computing the generating functional $W[A^i]$ of its correlators. The difference between the Dirichlet propagator in Feynman gauge and axial gauge reads
\begin{align} 
    \begin{aligned} \label{deltaphiphi2}
        \Delta\langle \Phi_{A,A'}(-k,z)&\Phi_{B,B'}(k,z')\rangle_\text{D} = \frac{\epsilon_{AA'}\epsilon_{BB'}}{4k}\Big(e^{-k|z-z'|} + e^{-k(z+z')}\Big) +\\
        &- \frac{k_{AA'}k_{BB'}}{4k^3}\Big(e^{-k|z-z'|} - e^{-k(z+z')}+k\big(|z-z'|-(z+z')\big)\Big)\,,
    \end{aligned}
\end{align}
which can be written as a total derivative (pure gauge),
\begin{align} \label{deltaphiphi}
    \begin{aligned}
        \Delta\langle \Phi_{A,A'}(-k,z)\Phi_{B,B'}(k,z')\rangle_\text{D} &= \nabla_{AA'}^{-k,z}\xi_{BB'}^\text{R}+\nabla_{BB'}^{k,z'}\xi^\text{L}_{AA'}+\nabla_{AA'}^{-k,z}\nabla_{BB'}^{k,z'}\xi^\text{c}\,,\\
        \xi^\text{L}_{AA'} &= \frac{\epsilon_{AA'}}{4k^2}\Big(\text{sign}(z-z')\big(1-e^{-k|z-z'|}\big)-\big(1-e^{-k(z+z')}\big)\Big) \,,\\
        \xi^\text{R}_{BB'} &= -\frac{\epsilon_{BB'}}{4k^2}\Big(\text{sign}(z-z')\big(1-e^{-k|z-z'|}\big)+\big(1-e^{-k(z+z')}\big)\Big) \,,\\
        \xi^\text{c} &= \frac{1}{4k^3}\Big(e^{-k|z-z'|}-e^{-k(z+z')}+k\big(|z-z'|-(z+z')\big)\Big) \,.
    \end{aligned}
\end{align}
Now, the crucial observation is that the ``gauge parameters'' vanish at the boundary:
\begin{align} \label{xi0}
    \begin{aligned}
        \xi^\text{L}_{AA'}(z'=0) &= 0 \,, & \xi^\text{R}_{BB'}(z=0) &=0 \,, & \xi^\text{c}(z=0)=\xi^\text{c}(z'=0)=0\,.
    \end{aligned}
\end{align}
In other words, the difference between the Feynman and axial gauge computations can be reduced via integration by parts to a boundary term. The latter vanishes thanks to the above (otherwise, it would be impossible to produce even a contact term). The integration by parts lands $\nabla$ onto the two on-shell fields of a vertex, which gives zero by the equations of motion. This proves the gauge independence in Yang--Mills theory for the Dirichlet boundary condition. 

In the Chalmers--Siegel formulation or, whenever the derivative from the Yang--Mills vertex hits the $\langle \Phi\Phi\rangle$ bulk-to-bulk propagator, the $\langle \Psi\Phi\rangle$ two-point function emerges. The difference between the Dirichlet propagator in Feynman gauge and axial gauge reads
\begin{align} \label{deltapsiphi}
    \begin{aligned}
        \Delta \langle \Psi_{AA}(-k,z)&\Phi_{B,B'}(k,z')\rangle_\text{D} = -\frac{k_{AA}\epsilon_{BB'}}{4k}\Big(e^{-k|z-z'|} + e^{-k(z+z')}\Big) +\\
        &+\frac{k_{AA}k_{BB'}}{4k^2}\Big(\text{sign}(z-z')\big(e^{-k|z-z'|}-1\big) - \big(e^{-k(z+z')}-1\big)\Big) \,.
    \end{aligned}
\end{align}
This difference can be written as pure gauge:
\begin{align} \label{deltapsiphietaD}
    \begin{aligned}
        \Delta &\langle \Psi_{AA}(-k,z)\Phi_{B,B'}(k,z')\rangle_\text{D} = \nabla_{BB'}^{k,z'}\eta_{AA} \,,\\
        &\eta_{AA} = -\frac{k_{AA}}{4k^2}\Big(\text{sign}(z-z')\big(1-e^{-k|z-z'|}\big)-\big(1-e^{-k(z+z')}\big)\Big) \,,
    \end{aligned}
\end{align}
where the gauge parameter $\eta_{AA}$ vanishes at $z'=0$. Therefore, the gauge independence is also manifest for $\langle\Psi \Phi\rangle$-propagator.

\paragraph{Neumann comment.} The difference between the Neumann propagator in Feynman gauge and axial gauge reads
\begin{align}
    \begin{aligned}
        \Delta\langle \Phi_{A,A'}(-k,z)\Phi_{B,B'}(k,z')\rangle_\text{N} &= \frac{1}{4k}(\epsilon_{AA'}\epsilon_{BB'}-\frac{k_{AA'}k_{BB'}}{k^2})(e^{-k|z-z'|}+e^{-k(z+z')})+\\
        &-\frac{k_{AA'}k_{BB'}}{4k^2}(|z-z'|-(z+z'))-\frac{\epsilon_{AA'}\epsilon_{BB'}}{2k}e^{-k(z+z')} \,.
    \end{aligned}
\end{align}
The inhomogeneous part is the same as for the Dirichlet case. However, there is a discrete difference in the homogeneous terms. This gauge variation can be expressed as
\begin{align} \label{deltapsiphietaN}
    \begin{aligned}
        \Delta\langle \Phi_{A,A'}(-k,z)\Phi_{B,B'}(k,z')\rangle_\text{N} &= \nabla_{AA'}^{-k,z}\xi_{BB'}^\text{R}+\nabla_{BB'}^{k,z'}\xi^\text{L}_{AA'}+\nabla_{AA'}^{-k,z}\nabla_{BB'}^{k,z'}\xi^\text{c}\,,\\
        \xi^\text{L}_{AA'} &= \frac{\epsilon_{AA'}}{4k^2}\Big(\text{sign}(z-z')\big(1-e^{-k|z-z'|}\big)-\big(1+e^{-k(z+z')}\big)\Big) \,,\\
        \xi^\text{R}_{BB'} &= -\frac{\epsilon_{BB'}}{4k^2}\Big(\text{sign}(z-z')\big(1-e^{-k|z-z'|}\big)+\big(1+e^{-k(z+z')}\big)\Big) \,,\\
        \xi^\text{c} &= \frac{1}{4k^3}\Big(e^{-k|z-z'|}+e^{-k(z+z')}+k\big(|z-z'|-(z+z')\big)\Big) \,.
    \end{aligned}
\end{align}
What has changed as compared to the Dirichlet case is the sign of the $\exp[-k(z+z')]$-terms. The latter forces the gauge parameters not to vanish at the boundary. However, for Neumann boundary conditions the situation is more complicated once the dual field is a gauge field $A_i$ on the boundary. Via AdS/CFT we are computing the correlation functions of this gauge field, which, of course, do depend on the gauge choice (the bulk gauge choice also implies a certain gauge choice on the boundary as is visible from the boundary limit of $\langle \Phi \Phi\rangle$). For $U(1)$ the gauge field $A_i$ can be mapped to the dual current $J=*dA$, which is manifestly conserved, and for this case we would expect the difference between two gauges to suddenly become a contact term. However, we have a nonabelian gauge field in the bulk and, hence, on the boundary. The dual current is $J=*(dA+gAA)$ and is only covariantly conserved $D_A J=0$. There are two complications: (i) to build a current we need higher order corrections $gAA$; (ii) $D_A$ depends on the gauge field. Below, in Section \ref{sec:neumannqed}, we consider the simplest example where the gauge field is still abelian.

To streamline the calculation of the gauge variation between Feynman gauge and axial gauge, it is convenient to consider the difference between the gauge variation for the Neumann and Dirichlet case, which we call $\Delta\Delta\langle \Phi_{A,A'}(-k,z)\Phi_{B,B'}(k,z')\rangle$. This difference between the gauge variations is
\begin{align}\label{deltanablaxi1}
    \begin{aligned}
        &\Delta\Delta \langle \Phi_{A,A'}(-k,z)\Phi_{B,B'}(k,z')\rangle \equiv \Delta \langle \Phi_{A,A'}(-k,z)\Phi_{B,B'}(k,z')\rangle_\text{N}-\Delta\langle \Phi_{AA'}(-k,z)\Phi_{BB'}(k,z')\rangle_\text{D}=\\
        &=\nabla_{AA'}^{-k,z}\Delta\xi^\text{R}_{BB'} + \nabla_{BB'}^{k,z'}\Delta\xi^\text{L}_{AA'}+\nabla_{AA'}^{-k,z}\nabla_{BB'}^{k,z'}\Delta\xi^\text{c}\,,
    \end{aligned}
\end{align}
with
\begin{align}\label{deltanablaxi2}
    \begin{aligned}
        \Delta\xi_{AA'}^\text{L} &= -\frac{\epsilon_{AA'}}{2k^2}e^{-k(z+z')} \,, &
        \Delta\xi_{BB'}^\text{R} &= -\frac{\epsilon_{BB'}}{2k^2}e^{-k(z+z')} \,, &
        \Delta\xi^\text{c} &= \frac{1}{2k^3}e^{-k(z+z')} \,.
    \end{aligned}
\end{align}
After performing the bulk integral, the gauge variation for the Dirichlet case vanishes and we have
\begin{align} \label{ints}
    \begin{aligned}
        &\int e^{-(k_1+k_2)z}\langle \Phi_{AA'}(-k,z)\Phi_{BB'}(k,z')\rangle_{\text{A}}\,e^{-(k_3+k_4)z'} =\\
        &=\int e^{-(k_1+k_2)z}\Big(\langle \Phi_{AA'}(-k,z)\Phi_{BB'}(k,z')\rangle_\text{L} + \Delta\Delta\langle \Phi_{AA'}(-k,z)\Phi_{BB'}(k,z')\rangle\Big)e^{-(k_3+k_4)z'} \,.
    \end{aligned}
\end{align}
Thus, in the exchange diagram $\Delta\Delta\langle \Phi\Phi\rangle$ contains all the relevant information. Taking one derivative, one finds the propagator
\begin{align}\label{deltadeltanablaxi1}
    \begin{aligned}
        &\Delta\Delta \langle \Psi_{AA}(-k,z)\Phi_{B,B'}(k,z')\rangle \equiv \Delta \langle \Psi_{AA}(-k,z)\Phi_{B,B'}(k,z')\rangle_\text{N}-\Delta\langle \Psi_{AA}(-k,z)\Phi_{B,B'}(k,z')\rangle_\text{D}=\\
        &=\nabla_{BB'}^{k,z'}\Delta\eta^\text{L}_{AA}\,,
    \end{aligned}
\end{align}
where
\begin{align} \label{deltadeltanablaxi2}
    \Delta\eta^\text{L}_{AA}=k\fdu{A}{A'}\Delta\xi^\text{L}_{AA'} \,.
\end{align}
The double derivative gives
\begin{align}
    \Delta\Delta \langle \Psi_{AA}(-k,z)\Psi_{BB}(k,z')\rangle = 0 \,.
\end{align}
For the single and double derivative cases a similar relation to \eqref{ints} exists and $\Delta\Delta\langle\Psi\Phi\rangle$ and $\Delta\Delta\langle\Psi\Psi\rangle$ are sufficient to compute the gauge variation of exchange diagrams.

%%%%%%%%%%%%%%%%%%%%%%%%%%%%%%%%%%%%%%%%%%%%%%%%%%%%%%%%%%%%%
\subsection{Neumann scalar QED} 
\label{sec:neumannqed}
%%%%%%%%%%%%%%%%%%%%%%%%%%%%%%%%%%%%%%%%%%%%%%%%%%%%%%%%%%%%%

\begin{figure}[h!]
\centering
\begin{tabular}{c c}
    % S-channel Witten diagram
    \begin{tikzpicture}
        % Define the boundary circle
        \draw[thick] (0,0) circle (2.1cm);
        
        % Define the vertices inside the bulk
        \coordinate (v1) at (-0.7,0);
        \coordinate (v2) at (0.7,0);
        
        % Define the external legs on the boundary
        \coordinate (a) at (-1.7,1.2);
        \coordinate (b) at (-1.7,-1.2);
        \coordinate (c) at (1.35,1.61);
        \coordinate (d) at (1.35,-1.61);
        \coordinate (e) at (2.1,0);
        
        % Draw the gauge boson lines
        \draw[decorate,thick] (a) -- (v1);
            \draw[decorate,thick] (b) -- (v1);
        \draw[decorate,thick] (c) -- (v2);
        \draw[decorate,thick] (d) -- (v2);
        \draw[decorate,decoration={snake},thick] (e) -- (v2);
        
        % Draw the internal propagator 
        \draw[decorate,decoration={snake},thick] (v1) -- (v2);
        
        % Draw parallel momentum arrows
        \draw[-{Latex},thick] (-1.3,1) -- (-0.9,0.52) node[midway,above right] {\( k_1 \)};
        \draw[-{Latex},thick] (-1.3,-1) -- (-0.9,-0.52) node[midway,below right] {\( k_2 \)};
        \draw[{Latex}-,thick] (0.686,0.42) -- (0.92,1) node[midway,above left] {\( k_5 \)};
        \draw[{Latex}-,thick] (0.686,-0.42) -- (0.92,-1) node[midway,below left] {\( k_3 \)};
        \draw[{Latex}-,thick] (1.05,0.2) -- (1.675,0.2) node[midway,above] {\( k_4 \)};
        
    \end{tikzpicture}
\end{tabular}

\caption{The simplest correlator that has both a gauge field on internal line and one external leg with a gauge field.}
\label{fig:scalarQED}
\end{figure}
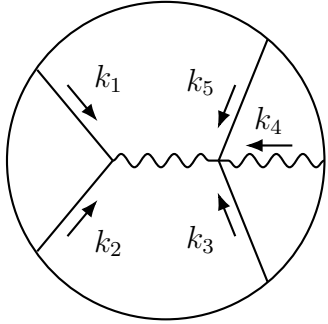

The simplest toy model is scalar electrodynamics in the bulk. It features the cubic vertex $\bar\phi \overleftrightarrow{\pl_\mu}\phi$ and the quartic vertex $A_\mu A^\mu \bar\phi \phi$. Importantly, the current, $J=-\bar \phi \pl \phi+\pl \bar\phi \phi-2 \bar\phi A \phi$ is abelian, i.e. $\pl \cdot J=0$ on-shell. To check the Neumann case we need a diagram with both internal and external $A$-lines, see Fig. \ref{fig:scalarQED}. We evaluate the difference between two gauges in the Neumann case in two steps. Firstly, we already know that the difference between the Feynman and axial gauge vanishes for Dirichlet boundary conditions. Secondly, there is an additional difference to account for the jump to the Neumann case (the boundary-to-bulk propagator for the external $A$ also needs to be changed). The latter difference has a very special structure: all $\xi_{\text{L,R}}$ are proportional to $n_\mu$ and the $\pl_z$-part of $\nabla$ that gives a contribution upon integration by parts is also along $n_\mu$. Now, $\xi_\text{R}$ upon integration by parts leads to $\pl^\mu J_\mu=0$ for free fields represented by the propagators. The boundary term gives
\begin{align}
    J^z(k_1,k_2,0) \int \xi_\text{R}(k_1+k_2,0,z') A_z(k_4,z')\phi(k_3,z')\bar\phi(k_5,z') 
\end{align}
where we note that all $\xi$ are proportional to $n_\mu$. As a result we pick the $A_z$-component on the boundary, while in the Neumann case we need $A_i$ on the boundary. In other words, the contribution to $A_i$ from the boundary terms vanishes. There is also another diagram, with three cubic vertices to which this one 'talks' via gauge invariance when the $\nabla$ that belongs to the quartic vertex is integrated by parts. The bulk contribution vanishes as in flat space, but we do not go as far as to analyze the additional terms for this diagram.    

%%%%%%%%%%%%%%%%%%%%%%%%%%%%%%%%%%%%%%%%%%%%%%%%%%%%%%%%%%%%%
\subsection{Relation to flat space}
\label{sec:reltoflat}
%%%%%%%%%%%%%%%%%%%%%%%%%%%%%%%%%%%%%%%%%%%%%%%%%%%%%%%%%%%%%

The difference between the Dirichlet propagator in Lorenz gauge and axial gauge was given in \eqref{deltaphiphi}. It is instructive to perform the bulk integrals one encounters in the exchange diagram for this gauge variation. This reads
\begin{align}
    \begin{aligned}
        \int e^{-(k_1+k_2)z}\nabla_{AA'}^{-k,z}\xi^\text{R}_{BB'}e^{-(k_3+k_4)z'} &= \frac{1}{4k}\frac{E_\text{L}+E_\text{R}+E}{EE_\text{L}E_\text{R}(E_\text{R}-k)}\epsilon_{BB'}\, (1_A\bar{1}_{A'}+2_A\bar{2}_{A'}) \,,\\
        \int e^{-(k_1+k_2)z}\nabla_{BB'}^{k,z'}\xi^\text{L}_{AA'}e^{-(k_3+k_4)z'} &= \frac{1}{4k}\frac{E_\text{L}+E_\text{R}+E}{EE_\text{L}E_\text{R}(E_\text{L}-k)}\epsilon_{AA'}(3_B\bar{3}_{B'}+4_B\bar{4}_{B'}) \,,\\
        \int e^{-(k_1+k_2)z}\nabla_{AA'}^{-k,z}\nabla_{BB'}^{k,z'}\xi^\text{c}e^{-(k_3+k_4)z'} &=-\frac{1}{4k}\frac{E_\text{L}+E_\text{R}+E}{EE_\text{L}E_\text{R}(E_\text{L}-k)(E_\text{R}-k)}\times\\
        &\times(1_A\bar{1}_{A'}+2_A\bar{2}_{A'})(3_B\bar{3}_{B'}+4_B\bar{4}_{B'}) \,.
    \end{aligned}
\end{align}
Together, the integral of the total gauge variation is
\begin{align}
    \begin{aligned}
    &\int e^{-(k_1+k_2)z}\Delta\langle \Phi_{A,A'}(-k,z)\Phi_{B,B'}(k,z')\rangle e^{-(k_3+k_4)z'} = \frac{1}{4k}\frac{E_\text{L}+E_\text{R}+E}{EE_\text{L}E_\text{R}} \times\\
    &\times\Big(\frac{(1_A\bar{1}_{A'}+2_A\bar{2}_{A
    })\epsilon_{BB'}}{k_1+k_2} +\frac{\epsilon_{AA'} (3_{B}\bar{3}_{B'}+4_B\bar{4}_{B'})}{k_3+k_4} - \frac{(1_A\bar{1}_{A'}+2_A\bar{2}_{A'})(3_B\bar{3}_{B'}+4_B\bar{4}_{B'})}{(k_1+k_2)(k_3+k_4)}\Big) \,,
    \end{aligned}
\end{align}
which is reminiscent of the difference between the flat-space axial gauge propagator and the Feynman one, c.f. \eqref{flatphiphiaxial}. This allows us to recycle the flat space argument found in \ref{sec:SDYMAxial} that explains why the flat-space amplitude is gauge invariant, to argue that the AdS correlator for Dirichlet boundary conditions is gauge invariant too. Note that this is an alternative proof of gauge independence to the one given around \eqref{xi0}. While the latter is a shorter proof, the one laid out here bears resemblance with flat-space results.

It now becomes obvious from the decomposition of the gauge parameters into their inhomogeneous and homogeneous components that the Dirichlet boundary condition is a very special case. Precisely the right homogeneous terms are present to recover the flat-space four-momentum (in the flat limit). As will be proved soon, for other boundary conditions an additional structure, that is unfamiliar to flat space, must be added.

Another important observation is that the exchange diagram integral of the gauge variation recovers the flat-space gauge variation exactly, while the same integral of the gauge-independent part of the propagator only gives the flat-space propagator as the residue of the leading energy pole, while also producing higher orders in $E$. Of course, it is not very surprising that these higher orders do not show up here, as gauge invariance is an exact requirement. Meanwhile, the flat-space amplitude is only expected to be found in the flat-space limit of the AdS correlation function, leaving the possibility for higher order corrections in $E$.

It is again instructive to consider the integral for the difference between the gauge variation for the Neumann and Dirichlet case. We find
\begin{align}
    \begin{aligned}
        \int e^{-(k_1+k_2)z}\nabla_{AA'}^{-k,z}\Delta\xi_{BB'}^\text{R} e^{-(k_3+k_4)z'} &= \frac{\epsilon_{BB'}}{2k^2E_\text{L}E_\text{R}}(k_{AA'}-k\epsilon_{AA'}) \,,\\
        \int e^{-(k_1+k_2)z}\nabla_{BB'}^{k,z}\Delta\xi_{AA'}^\text{L} e^{-(k_3+k_4)z'} &= -\frac{\epsilon_{AA'}}{2k^2E_\text{L}E_\text{R}}(k_{BB'}+k\epsilon_{BB'}) \,,\\
        \int e^{-(k_1+k_2)z}\nabla_{AA'}^{-k,z}\nabla_{BB'}^{k,z'}\Delta\xi^\text{c} e^{-(k_3+k_4)z'} &= -\frac{1}{2k^3E_\text{L}E_\text{R}}(k_{AA'}-k\epsilon_{AA'})(k_{BB'}+k\epsilon_{BB'})
    \end{aligned}
\end{align}
and
\begin{align}
    \begin{aligned}
        &\int e^{-(k_1+k_2)z}\big(\nabla_{AA'}^{-k,z}\Delta\xi_{BB'}^\text{R}+\nabla_{BB'}^{k,z'}\Delta\xi_{AA'}^\text{L}+\nabla_{AA'}^{-k,z}\nabla_{BB'}^{k,z'}\Delta\xi^\text{c}\big)e^{-(k_3+k_4)z'} =\\
        &=\frac{1}{2kE_\text{L}E_\text{R}}\Big(-\frac{(-k_{AA'}+k\epsilon_{AA'})\epsilon_{BB'}}{k}-\frac{\epsilon_{AA'}(k_{BB'}+k\epsilon_{BB'})}{k}+\frac{(-k_{AA'}+k\epsilon_{AA'})(k_{BB'}+k\epsilon_{BB'})}{k^2}\Big) \,.
    \end{aligned}
\end{align}
Here we recognize a similar structure as for Dirichlet, but now the flat-space momentum is replaced by the fake null momenta $-k_{AA'}+k\epsilon_{AA'}$ and $k_{BB'}+k\epsilon_{BB'}$.

To summarize, for Dirichlet boundary conditions the exchange channel integral of the gauge variation takes a form akin to the gauge variation of the propagator in flat space. In the Neumann and mixed case, this is supplemented by a similar structure containing the fake null-momentum instead of the flat-space momentum. Because the fake null-momentum is an AdS object, and gauge invariance does not hold for the Neumann boundary condition, it can be said that the Dirichlet case is closest to the flat-space treatment of Yang--Mills.

The same steps can be repeated for the gauge variation $\Delta\langle \Psi\Phi\rangle$. For Dirichlet boundary conditions this yields
\begin{align} \label{psiphivar}
    \begin{aligned}
        \int e^{-(k_1+k_2)z}&\Delta\langle \Psi_{AA}(-k,z)\Phi_{B,B'}(k,z')\rangle_\text{D}e^{-(k_3+k_4)z'} =\\
        &= -\frac{1}{4k}\frac{E_\text{L}+E_\text{R}+E}{EE_\text{L}E_\text{R}(E_\text{R}-k)}k_{AA}(3_B\bar{3}_{B'}+4_B\bar{4}_{B'}) \,.
    \end{aligned}
\end{align}
This is reminiscent of the gauge variation of the flat-space propagator, c.f. \eqref{axialProp}. The difference between the gauge variation for Neumann and Dirichlet gives a similar expression with the flat-space momentum replaced by the fake null momentum $k_{BB'}+k\epsilon_{BB'}$,
\begin{align}
    \int e^{-(k_1+k_2)z}\Delta\Delta\langle \Psi_{AA}(-k,z)\Phi_{B,B'}(k,z')\rangle e^{-(k_3+k_4)z'} = \frac{1}{2k^2E_\text{L}E_\text{R}}k_{AA}(k_{BB'}+k\epsilon_{BB'}) \,.
\end{align}

%%%%%%%%%%%%%%%%%%%%%%%%%%%%%%%%%%%%%%%%%%%%%%%%%%%%%%%%%%%%%
\section{Gauge-covariant correlator}
\label{app:sdymleftovers}
\begin{figure}[h!]
\centering
\begin{tabular}{c c}
    % S-channel Witten diagram
    \begin{tikzpicture}
        % Define the boundary circle
        \draw[thick] (0,0) circle (2.1cm);
        
        % Define the vertices inside the bulk
        \coordinate (v1) at (-0.7,0);
        
        % Define the external legs on the boundary
        \coordinate (a) at (-1.7,1.2);
        \coordinate (b) at (-1.7,-1.2);
        \coordinate (c) at (1.7,1.2);
        \coordinate (d) at (1.7,-1.2);
        \coordinate (e) at (1.49,1.49);
        
        % Draw the gauge boson lines
        \draw[decorate,decoration={snake},thick] (a) -- (v1);
        \draw[decorate,decoration={snake},thick] (b) -- (v1);
        \draw[decorate,decoration={snake},thick] (e) -- (v1);
        \draw[decorate,decoration={snake},thick] (d) -- (c);
        
        % Add labels for fields next to endpoints
        \node[left] at (a) {\( \epsilon^+_1 \)};
        \node[left] at (b) {\( \epsilon^+_2 \)};
        \node[xshift=5pt, yshift=9pt] at (c) {\( \epsilon^-_4 \)};
        \node[right] at (d) {\( \epsilon^+_3 \)};
        
        % Draw parallel momentum arrows
        \draw[-{Latex},thick] (-1.3,1.0) -- (-0.9,0.6) node[midway,above right] {\( k_1 \)};
        \draw[-{Latex},thick] (-1.3,-1) -- (-0.9,-0.6) node[midway,below right] {\( k_2 \)};
        \draw[{Latex}-,thick] (1.5,0.285) -- (1.5,-0.285) node[midway,left] {\( k_3 \)};
        \draw[-{Latex},thick] (-0.1,0.65) -- (0.52,1.07) node[midway,yshift=12pt] {\( k_1+k_2 \)};

        \fill[blue] (-0.3,0.28) circle (5pt);
    \end{tikzpicture}
\end{tabular}

\vspace{0.5cm}

\caption{Witten diagram that yields the component of the correlator with a composite operator $aa$ on the boundary replacing the right vertex.}
\label{fig:aa}
\end{figure}
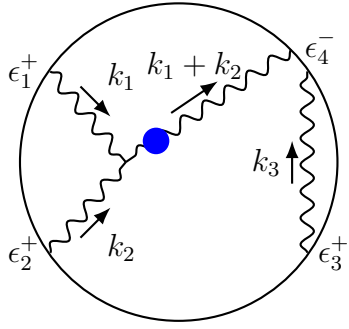

%%%%%%%%%%%%%%%%%%%%%%%%%%%%%%%%%%%%%%%%%%%%%%%%%%%%%%%%%%%%%
In Section \ref{sec:Correlators} it was discussed that the four-point correlation functions we computed are not expected to be gauge-covariant for SDYM and for cYM with mixed/Neumann(-like) boundary conditions. Indeed, we observed a difference between the correlators in Feynman gauge and axial gauge. The reason for this is that the  gauge field in the bulk is dual to a gauge field in the CFT for non-Dirichlet boundary conditions. However, one can construct a gauge-covariant object in the CFT, the magnetic field $B=*(da+g\,a\wedge a)$. The SDYM correlation functions computed in Section \ref{sec:Correlators} have a gauge field on the boundary only for the negative helicity. The first term effectively multiplies the diagram by $-k_4$, but we already did this implicitly by choosing the Dirichlet scaling of the boundary-to-bulk propagators. The second term can be computed by having not one, but two bulk/boundary-to-boundary propagators connected to the boundary point at which the negative helicity CFT data is inserted on the boundary, see Figure \ref{fig:aa}. cYM is less constrained; a bulk $\Phi$ field, for instance, can come from a $\langle\Psi\Phi\rangle$ or a $\langle \Phi\Phi\rangle$ boundary-to-bulk propagator. It is therefore also possible for the three bulk $\Phi$-legs to be coupled to a gauge field on the boundary and the magnetic field must then be constructed on all legs. 

Another observation from Section \ref{subsec:cYM} is that SDYM and Chalmers-Siegel do not seem to coincide in the self-dual limit: there is one topological diagram belonging to Chalmers-Siegel that, due to the presence of the $\Phi\Phi$-propagator, survives in the limit, while it is absent in SDYM. We will show that the composite diagrams are responsible for canceling the contribution of this topological diagram and that SDYM agrees with the self-dual limit of Chalmers-Siegel. As we shall show, only the $\gamma$-dependent component of the correlators associated with the composite diagrams is relevant to this end, so we restrict ourselves to use only the Neumann-like propagators and the same result follows automatically for the Dirichlet-like propagators. 

The $a^2$ term in the magnetic field gives rise to a two-point vertex on the boundary. However, at this boundary point a polarization vector is attached, so it effectively looks like a three-point vertex. It is expressed by
\begin{align}
    V^{aa}_{AA',BB',CC}=2g\,\epsilon_{A'B'}\epsilon_{AC}\epsilon_{BC} \,,
\end{align}
where the pair of indices is contracted with $\epsilon^\pm_{CC}$. Since the SDYM vertex was constructed from the same $\Phi^2$ term in the field strength, it is not surprising to see that this vertex has the same expression; the only difference is that it takes place on the boundary.

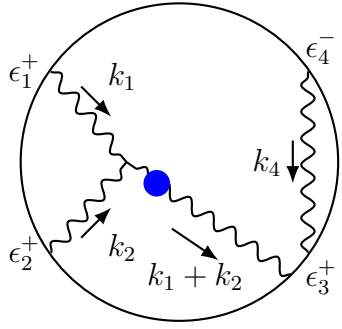
\begin{figure}[h!]
\centering
\begin{tabular}{c c}
    % S-channel Witten diagram
    \begin{tikzpicture}
        % Define the boundary circle
        \draw[thick] (0,0) circle (2.1cm);
        
        % Define the vertices inside the bulk
        \coordinate (v1) at (-0.7,0);
        
        % Define the external legs on the boundary
        \coordinate (a) at (-1.7,1.2);
        \coordinate (b) at (-1.7,-1.2);
        \coordinate (c) at (1.7,1.2);
        \coordinate (d) at (1.7,-1.2);
        \coordinate (e) at (1.49,-1.49);
        
        % Draw the gauge boson lines
        \draw[decorate,decoration={snake},thick] (a) -- (v1);
        \draw[decorate,decoration={snake},thick] (b) -- (v1);
        \draw[decorate,decoration={snake},thick] (e) -- (v1);
        \draw[decorate,decoration={snake},thick] (d) -- (c);
        
        % Add labels for fields next to endpoints
        \node[left] at (a) {\( \epsilon^+_1 \)};
        \node[left] at (b) {\( \epsilon^+_2 \)};
        \node[xshift=5pt, yshift=9pt] at (c) {\( \epsilon^-_4 \)};
        \node[xshift=5pt, yshift=-9pt] at (d) {\( \epsilon^+_3 \)};
        
        % Draw parallel momentum arrows
        \draw[-{Latex},thick] (-1.3,1.0) -- (-0.9,0.6) node[midway,above right] {\( k_1 \)};
        \draw[-{Latex},thick] (-1.3,-1) -- (-0.9,-0.6) node[midway,below right] {\( k_2 \)};
        \draw[-{Latex},thick] (1.5,0.285) -- (1.5,-0.285) node[midway,left] {\( k_4 \)};
        \draw[-{Latex},thick] (-0.1,-0.88) -- (0.52,-1.3) node[midway,yshift=-12pt] {\( k_1+k_2 \)};

        \fill[blue] (-0.3,-0.28) circle (5pt);
    \end{tikzpicture}
\end{tabular}

\vspace{0.5cm}

\caption{Witten diagram that yields the component of the correlator with a composite operator $aa$ on the boundary replacing the right vertex.}
\label{fig:aa2}
\end{figure}

For cYM, the contribution to the correlation function of the diagram in Figure \ref{fig:aa} reads
\begin{align}
    \begin{aligned}
        \mathcal{W}^{1}_{s,\text{L},aa}&=\int \Phi_+^{A,A'}(k_1,z)\Phi_+^{B,B'}(k_2,z)V\fdu{AA',BB'}{EE}\langle \Psi
        _{EE}(-k,z)\Phi_{F,F'}(k,0)\rangle_\text{L} \times\\
        &\times V\fdud{aa}{F,F'}{,CC',DD}\Phi_+^{C,C'}(k_3,0)\Phi_-^{D,D}(k_4,0) =\\
        &=-\frac{g^2}{16k^2E_\text{L}}\frac{\langle \bar{1}\bar{2} \rangle \langle \bar{3}4 \rangle}{k_1k_2k_3k_4}\Big(\langle \bar{1}\bar{2} | kk\bar{k}\bar{k}|\bar{3}4\rangle + e^{-2i\gamma}\langle \bar{1}\bar{2} | \bar{k}\bar{k} kk|\bar{3} 4\rangle-2k^2\big(\langle \bar{1}\bar{3}\rangle \langle \bar{2}4\rangle+\langle \bar{2}\bar{3} \rangle \langle \bar{1}4 \rangle\big)\Big)\,.
    \end{aligned}
\end{align}
The SDYM result is obtained by taking the limit $\gamma\rightarrow -i\infty$ and multiplying the result by $-2k_4$.\footnote{The latter is necessary for passing from cYM to Chalmers--Siegel. It is equivalent to the freedom we have to choose the negative helicity field to come from a $\langle \Psi\Phi\rangle$ propagator or a $\langle \Phi\Phi \rangle$ one.} Another diagram, yielding correlator $\mathcal{W}^{2}_{s,\text{L},aa}$, in which the right bulk vertex is replaced by a composite vertex is shown in Figure \ref{fig:aa2}. Evaluating the correlator naively, we find $\mathcal{W}^{2}_{s,\text{L},aa}=-\mathcal{W}^{1}_{s,\text{L},aa}$ and $\mathcal{W}^{1}_{s,\text{L},aa}$ and $\mathcal{W}^{2}_{s,\text{L},aa}$ cancel each other. However, one should take into account the structure constant $f^{abc}$ that is implicitly present in the vertices. We then find $\mathcal{W}^{2}_{s,\text{L},aa}=\mathcal{W}^{1}_{s,\text{L},aa}$ instead, so they add up.

\begin{figure}[h!]
\centering
\begin{tabular}{c c}
    % S-channel Witten diagram
    \begin{tikzpicture}
        \begin{scope}[xshift=3cm]
        % Define the boundary circle
        \draw[thick] (0,0) circle (2.1cm);
        
        % Define the vertices inside the bulk
        \coordinate (v1) at (-0.7,0);
        \coordinate (v2) at (0.7,0);
        
        % Define the external legs on the boundary
        \coordinate (a) at (-1.7,1.2);
        \coordinate (b) at (-1.7,-1.2);
        \coordinate (c) at (1.7,1.2);
        \coordinate (d) at (1.7,-1.2);
        \coordinate (e) at (-1.49,-1.49);
        
        % Draw the gauge boson lines
        \draw[decorate,decoration={snake},thick] (a) -- (b);
        \draw[decorate,decoration={snake},thick] (e) -- (v2);
        \draw[decorate,decoration={snake},thick] (d) -- (v2);
        \draw[decorate,decoration={snake},thick] (c) -- (v2);
        
        % Add labels for fields next to endpoints
        \node[left] at (a) {\( \epsilon^+_1 \)};
        \node[xshift=-5pt, yshift=-9pt] at (b) {\( \epsilon^+_2 \)};
        \node[xshift=5pt, yshift=9pt] at (c) {\( \epsilon^-_4 \)};
        \node[right] at (d) {\( \epsilon^+_3 \)};
        
        % Draw parallel momentum arrows
        \draw[-{Latex},thick] (1.3,1.0) -- (0.9,0.6) node[midway,above left] {\( k_4 \)};
        \draw[-{Latex},thick] (1.3,-1) -- (0.9,-0.6) node[midway,below left] {\( k_3 \)};
        \draw[-{Latex},thick] (-1.5,0.285) -- (-1.5,-0.285) node[midway,right] {\( k_1 \)};
        \draw[-{Latex},thick] (-0.8,-1.3) -- (-0.18,-0.88) node[midway,yshift=-12pt] {\( k_1+k_2 \)};
        \node at (-3,0) {$+$};
        \fill[blue] (1,0.36) circle (5pt);
        \end{scope}

        \begin{scope}[xshift=-3cm]
        % Define the boundary circle
        \draw[thick] (0,0) circle (2.1cm);
        
        % Define the vertices inside the bulk
        \coordinate (v1) at (-0.7,0);
        \coordinate (v2) at (0.7,0);
        
        % Define the external legs on the boundary
        \coordinate (a) at (-1.7,1.2);
        \coordinate (b) at (-1.7,-1.2);
        \coordinate (c) at (1.7,1.2);
        \coordinate (d) at (1.7,-1.2);
        \coordinate (e) at (-1.49,1.49);
        
        % Draw the gauge boson lines
        \draw[decorate,decoration={snake},thick] (a) -- (b);
        \draw[decorate,decoration={snake},thick] (e) -- (v2);
        \draw[decorate,decoration={snake},thick] (d) -- (v2);
        \draw[decorate,decoration={snake},thick] (c) -- (v2);
        
        % Add labels for fields next to endpoints
        \node[xshift=-5pt, yshift=9pt] at (a) {\( \epsilon^+_1 \)};
        \node[xshift=-6pt,yshift=-6pt] at (b) {\( \epsilon^+_2 \)};
        \node[xshift=5pt, yshift=9pt] at (c) {\( \epsilon^-_4 \)};
        \node[right] at (d) {\( \epsilon^+_3 \)};
        
        % Draw parallel momentum arrows
        \draw[-{Latex},thick] (1.3,1.0) -- (0.9,0.6) node[midway,above left] {\( k_4 \)};
        \draw[-{Latex},thick] (1.3,-1) -- (0.9,-0.6) node[midway,below left] {\( k_3 \)};
        \draw[{Latex}-,thick] (-1.5,0.285) -- (-1.5,-0.285) node[midway,right] {\( k_2 \)};
        \draw[-{Latex},thick] (-0.8,1.3) -- (-0.18,0.88) node[midway,yshift=12pt] {\( k_1+k_2 \)};

        \fill[blue] (1,0.36) circle (5pt);
        \end{scope}        
    \end{tikzpicture}
\end{tabular}

\caption{Witten diagram that yields the component of the correlator with the composite operator $aa$ on the boundary replacing the left vertex.}
\label{fig:12}
\end{figure}
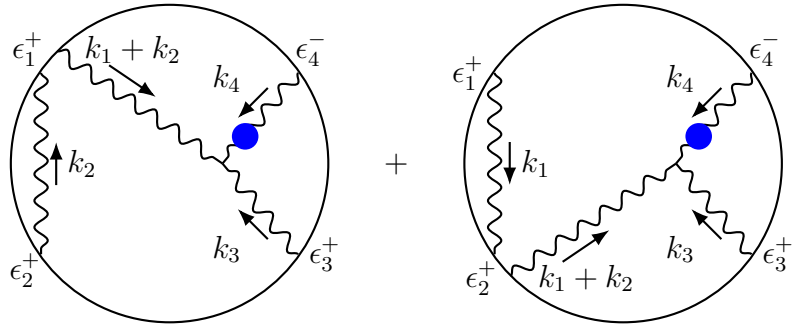

Two other diagrams with the composite operator $a^2$ on the boundary contributing to the $s$-channel are found in Figure \eqref{fig:12}. The left and right diagram amount to correlators $\mathcal{W}^{3}_{s,\text{L},aa}$ and $\mathcal{W}^{4}_{s,\text{L},aa}$, respectively. The left diagram yields
\begin{align}
    \begin{aligned}
        &\mathcal{W}^{3}_{s,\text{L},aa}=\int \Phi_+^{A,A'}(k_1,0)\Phi_+^{B,B'}(k_2,0)V\fudud{aa}{BB',}{EE'}{,AA}\langle \Psi
        _{EE}(-k,0)\Phi_{F,F'}(k,z')\rangle_\gamma \times\\
        &\times V\fud{FF'}{,CC',DD}\Phi_+^{C,C'}(k_3,0)\Psi_-^{DD}(k_4,0) =\\
        &=-\frac{g^2}{16k^3 E_\text{R}}\frac{\langle\bar{1}\bar{2}\rangle\langle\bar{3}4\rangle}{k_1k_2k_3}\Big(8k^2\langle \bar{1}\bar{3} \rangle \langle \bar{2}4 \rangle - \big(1+e^{2i\gamma}\big)\langle \bar{1}\bar{2}|kk\bar{k}\bar{k}|\bar{3}4\rangle - \big(1+e^{-2i\gamma}\big)\langle \bar{1}\bar{2}|\bar{k}\bar{k}kk|\bar{3}4\rangle\Big)\,.
    \end{aligned}
\end{align}
Again, taking into account the color factors, we have $\mathcal{W}^{4}_{s,\text{L},aa}=\mathcal{W}^{3}_{s,\text{L},aa}$.

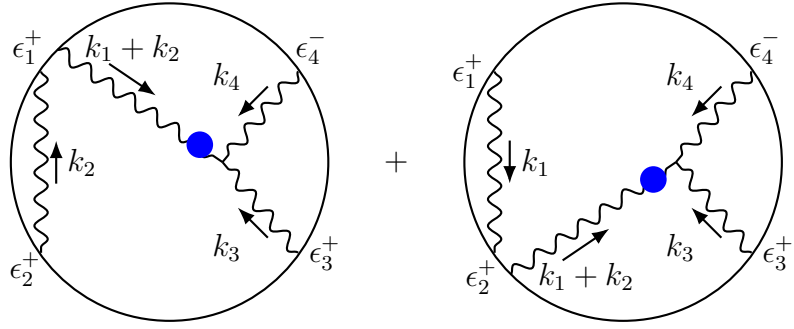
\begin{figure}[h!]
\centering
\begin{tabular}{c c}
    % S-channel Witten diagram
    \begin{tikzpicture}
        \begin{scope}[xshift=3cm]
        % Define the boundary circle
        \draw[thick] (0,0) circle (2.1cm);
        
        % Define the vertices inside the bulk
        \coordinate (v1) at (-0.7,0);
        \coordinate (v2) at (0.7,0);
        
        % Define the external legs on the boundary
        \coordinate (a) at (-1.7,1.2);
        \coordinate (b) at (-1.7,-1.2);
        \coordinate (c) at (1.7,1.2);
        \coordinate (d) at (1.7,-1.2);
        \coordinate (e) at (-1.49,-1.49);
        
        % Draw the gauge boson lines
        \draw[decorate,decoration={snake},thick] (a) -- (b);
        \draw[decorate,decoration={snake},thick] (e) -- (v2);
        \draw[decorate,decoration={snake},thick] (d) -- (v2);
        \draw[decorate,decoration={snake},thick] (c) -- (v2);
        
        % Add labels for fields next to endpoints
        \node[left] at (a) {\( \epsilon^+_1 \)};
        \node[xshift=-5pt, yshift=-9pt] at (b) {\( \epsilon^+_2 \)};
        \node[xshift=5pt, yshift=9pt] at (c) {\( \epsilon^-_4 \)};
        \node[right] at (d) {\( \epsilon^+_3 \)};
        
        % Draw parallel momentum arrows
        \draw[-{Latex},thick] (1.3,1.0) -- (0.9,0.6) node[midway,above left] {\( k_4 \)};
        \draw[-{Latex},thick] (1.3,-1) -- (0.9,-0.6) node[midway,below left] {\( k_3 \)};
        \draw[-{Latex},thick] (-1.5,0.285) -- (-1.5,-0.285) node[midway,right] {\( k_1 \)};
        \draw[-{Latex},thick] (-0.8,-1.3) -- (-0.18,-0.88) node[midway,yshift=-12pt] {\( k_1+k_2 \)};
        \node at (-3,0) {$+$};
        \fill[blue] (0.4,-.23) circle (5pt);
        \end{scope}

        \begin{scope}[xshift=-3cm]
        % Define the boundary circle
        \draw[thick] (0,0) circle (2.1cm);
        
        % Define the vertices inside the bulk
        \coordinate (v1) at (-0.7,0);
        \coordinate (v2) at (0.7,0);
        
        % Define the external legs on the boundary
        \coordinate (a) at (-1.7,1.2);
        \coordinate (b) at (-1.7,-1.2);
        \coordinate (c) at (1.7,1.2);
        \coordinate (d) at (1.7,-1.2);
        \coordinate (e) at (-1.49,1.49);
        
        % Draw the gauge boson lines
        \draw[decorate,decoration={snake},thick] (a) -- (b);
        \draw[decorate,decoration={snake},thick] (e) -- (v2);
        \draw[decorate,decoration={snake},thick] (d) -- (v2);
        \draw[decorate,decoration={snake},thick] (c) -- (v2);
        
        % Add labels for fields next to endpoints
        \node[xshift=-5pt, yshift=9pt] at (a) {\( \epsilon^+_1 \)};
        \node[xshift=-6pt,yshift=-6pt] at (b) {\( \epsilon^+_2 \)};
        \node[xshift=5pt, yshift=9pt] at (c) {\( \epsilon^-_4 \)};
        \node[right] at (d) {\( \epsilon^+_3 \)};
        
        % Draw parallel momentum arrows
        \draw[-{Latex},thick] (1.3,1.0) -- (0.9,0.6) node[midway,above left] {\( k_4 \)};
        \draw[-{Latex},thick] (1.3,-1) -- (0.9,-0.6) node[midway,below left] {\( k_3 \)};
        \draw[{Latex}-,thick] (-1.5,0.285) -- (-1.5,-0.285) node[midway,right] {\( k_2 \)};
        \draw[-{Latex},thick] (-0.8,1.3) -- (-0.18,0.88) node[midway,yshift=12pt] {\( k_1+k_2 \)};

        \fill[blue] (0.4,.23) circle (5pt);
        \end{scope}        
    \end{tikzpicture}
\end{tabular}

\caption{Witten diagrams that yield the component of the correlator with the composite operator $aa$ on the boundary replacing the left vertex.}
\label{fig:aa3}
\end{figure}

Figure \ref{fig:aa3} shows yet another pair of diagrams that should be considered. The left and right diagram yield $\mathcal{W}_{s,\text{L},aa}^5$ and $\mathcal{W}_{s,\text{L},aa}^6$, respectively. Taking into account the color factors, the correlators are
\begin{align}
    \begin{aligned}
        \mathcal{W}_{s,\text{L},aa}^5 &= -\frac{g^2}{16k^2E_\text{R}}\frac{\langle \bar{1}\bar{2} \rangle}{k_1k_2k_3k_4} \Big(k\big(\langle \bar{1}\bar{3} \rangle \langle 4|k\bar{k} |\bar{2} \rangle + \langle \bar{1}4\rangle \langle \bar{3}|k\bar{k}|\bar{2}\rangle + \langle \bar{2}\bar{3}\rangle \langle 4|k\bar{k}|\bar{1} \rangle + \langle \bar{2}4 \rangle \langle \bar{3}|k\bar{k}| \bar{1}\rangle\big) +\\
        &\qquad\qquad\qquad\qquad\qquad- (1+e^{-2i\gamma})\langle \bar{1}\bar{2} |kk\bar{k}\bar{k} |\bar{3} 4\rangle\Big) \,,\\
        \mathcal{W}_{s,\text{L},aa}^6&=\mathcal{W}_{s,\text{L},aa}^5 \,.
    \end{aligned}
\end{align}

\begin{figure}[h!]
\centering
\begin{tabular}{c c}
    % S-channel Witten diagram
    \begin{tikzpicture}
    \begin{scope}[xshift=-12cm]
        % Define the boundary circle
        \draw[thick] (0,0) circle (2.1cm);
        
        % Define the vertices inside the bulk
        \coordinate (v1) at (-1.4,0);
        
        % Define the external legs on the boundary
        \coordinate (a) at (-1.7,1.2);
        \coordinate (b) at (-1.7,-1.2);
        \coordinate (c) at (1.7,1.2);
        \coordinate (d) at (1.7,-1.2);
        \coordinate (e) at (1.49,1.49);

        \filldraw[fill=gray!20, draw=black,rotate=-90] 
                (1.61,-1.35) arc[start angle=40, end angle=140, radius=2.1cm];
            \filldraw[fill=gray!20, draw=black,rotate=-90] 
            (-1.61,-1.35) arc[start angle=220, end angle=320, radius=2.1cm];
        
        % Draw the gauge boson lines
        \draw[decorate,decoration={snake},thick] (a) -- (v1);
        \draw[decorate,decoration={snake},thick] (b) -- (v1);
        \draw[decorate,decoration={snake},thick] (e) -- (v1);
        \draw[decorate,decoration={snake},thick] (d) -- (c);
        
        % Add labels for fields next to endpoints
        \node[left] at (a) {\( \epsilon^+_1 \)};
        \node[left] at (b) {\( \epsilon^+_2 \)};
        \node[xshift=5pt, yshift=9pt] at (c) {\( \epsilon^-_4 \)};
        \node[right] at (d) {\( \epsilon^+_3 \)};
        
        % Draw parallel momentum arrows
        \draw[-{Latex},thick,rotate=-90] (-0.9,-1.4) -- (-0.3,-1.2) node[midway,xshift=0.3 cm,yshift=0.2 cm] {\( k_1 \)};
        \draw[-{Latex},thick,rotate=-90] (0.9,-1.4) -- (0.3,-1.2) node[midway,xshift=0.25 cm,yshift=-0.2 cm] {\( k_2 \)};
        \draw[{Latex}-,thick] (1.5,0.285) -- (1.5,-0.285) node[midway,left] {\( k_3 \)};
        \draw[-{Latex},thick] (-0.1,0.95) -- (0.52,1.27) node[midway,yshift=12pt] {\( k_1+k_2 \)};
    \end{scope}
    \begin{scope}[xshift=-3cm]
    % Define the boundary circle
        \draw[thick] (0,0) circle (2.1cm);
        
        % Define the vertices inside the bulk
        \coordinate (v1) at (-1.4,0);
        
        % Define the external legs on the boundary
        \coordinate (a) at (-1.7,1.2);
        \coordinate (b) at (-1.7,-1.2);
        \coordinate (c) at (1.7,1.2);
        \coordinate (d) at (1.7,-1.2);
        \coordinate (e) at (1.49,-1.49);

        \filldraw[fill=gray!20, draw=black,rotate=-90] 
                (1.61,-1.35) arc[start angle=40, end angle=140, radius=2.1cm];
            \filldraw[fill=gray!20, draw=black,rotate=-90] 
            (-1.61,-1.35) arc[start angle=220, end angle=320, radius=2.1cm];
        
        % Draw the gauge boson lines
        \draw[decorate,decoration={snake},thick] (a) -- (v1);
        \draw[decorate,decoration={snake},thick] (b) -- (v1);
        \draw[decorate,decoration={snake},thick] (e) -- (v1);
        \draw[decorate,decoration={snake},thick] (d) -- (c);
        
        % Add labels for fields next to endpoints
        \node[left] at (a) {\( \epsilon^+_1 \)};
        \node[left] at (b) {\( \epsilon^+_2 \)};
        \node[xshift=5pt, yshift=9pt] at (c) {\( \epsilon^-_4 \)};
        \node[xshift=5pt, yshift=-9pt] at (d) {\( \epsilon^+_3 \)};
        
        % Draw parallel momentum arrows
        \draw[-{Latex},thick,rotate=-90] (-0.9,-1.4) -- (-0.3,-1.2) node[midway,xshift=0.3 cm,yshift=0.2 cm] {\( k_1 \)};
        \draw[-{Latex},thick,rotate=-90] (0.9,-1.4) -- (0.3,-1.2) node[midway,xshift=0.25 cm,yshift=-0.2 cm] {\( k_2 \)};
        \draw[-{Latex},thick] (1.5,0.285) -- (1.5,-0.285) node[midway,left] {\( k_4 \)};
        \draw[-{Latex},thick] (-0.1,-0.95) -- (0.52,-1.27) node[midway,yshift=-12pt] {\( k_1+k_2 \)};
    \end{scope}
    \begin{scope}[xshift=-3cm,yshift=-6cm,rotate=180]
    % Define the boundary circle
        \draw[thick] (0,0) circle (2.1cm);
        
        % Define the vertices inside the bulk
        \coordinate (v1) at (-1.4,0);
        
        % Define the external legs on the boundary
        \coordinate (a) at (-1.7,1.2);
        \coordinate (b) at (-1.7,-1.2);
        \coordinate (c) at (1.7,1.2);
        \coordinate (d) at (1.7,-1.2);
        \coordinate (e) at (1.49,1.49);

        \filldraw[fill=gray!20, draw=black,rotate=-90] 
                    (1.61,-1.35) arc[start angle=40, end angle=140, radius=2.1cm];
                \filldraw[fill=gray!20, draw=black,rotate=-90] 
                (-1.61,-1.35) arc[start angle=220, end angle=320, radius=2.1cm];
        
        % Draw the gauge boson lines
        \draw[decorate,decoration={snake},thick] (a) -- (v1);
        \draw[decorate,decoration={snake},thick] (b) -- (v1);
        \draw[decorate,decoration={snake},thick] (e) -- (v1);
        \draw[decorate,decoration={snake},thick] (d) -- (c);
        
        % Add labels for fields next to endpoints
        \node[xshift=0.25cm,yshift=-0.25cm] at (a) {\( \epsilon^+_3 \)};
        \node[xshift=0.25cm,yshift=0.25cm] at (b) {\( \epsilon^-_4 \)};
        \node[xshift=-0.25cm,yshift=-0.25cm] at (c) {\( \epsilon^+_2 \)};
        \node[xshift=-0.25cm,yshift=0.25cm] at (d) {\( \epsilon^+_1 \)};
        
        % Draw parallel momentum arrows
        \draw[-{Latex},thick,rotate=-90] (-0.9,-1.4) -- (-0.3,-1.2) node[midway,xshift=-0.3 cm,yshift=-0.2 cm] {\( k_3 \)};
        \draw[-{Latex},thick,rotate=-90] (0.9,-1.4) -- (0.3,-1.2) node[midway,xshift=-0.25 cm,yshift=0.2 cm] {\( k_4 \)};
        \draw[{Latex}-,thick] (1.5,0.285) -- (1.5,-0.285) node[midway,right] {\( k_1 \)};
        \draw[{Latex}-,thick] (-0.1,0.95) -- (0.52,1.27) node[midway,yshift=-12pt] {\( k_1+k_2 \)};
    \end{scope}
    \begin{scope}[xshift=-12cm,yshift=-6cm,rotate=180]
    % Define the boundary circle
        \draw[thick] (0,0) circle (2.1cm);
        
        % Define the vertices inside the bulk
        \coordinate (v1) at (-1.4,0);
        
        % Define the external legs on the boundary
        \coordinate (a) at (-1.7,1.2);
        \coordinate (b) at (-1.7,-1.2);
        \coordinate (c) at (1.7,1.2);
        \coordinate (d) at (1.7,-1.2);
        \coordinate (e) at (1.49,-1.49);

        \filldraw[fill=gray!20, draw=black,rotate=-90] 
                    (1.61,-1.35) arc[start angle=40, end angle=140, radius=2.1cm];
                \filldraw[fill=gray!20, draw=black,rotate=-90] 
                (-1.61,-1.35) arc[start angle=220, end angle=320, radius=2.1cm];
        
        % Draw the gauge boson lines
        \draw[decorate,decoration={snake},thick] (a) -- (v1);
        \draw[decorate,decoration={snake},thick] (b) -- (v1);
        \draw[decorate,decoration={snake},thick] (e) -- (v1);
        \draw[decorate,decoration={snake},thick] (d) -- (c);
        
        % Add labels for fields next to endpoints
        \node[xshift=0.25cm,yshift=-0.25cm] at (a) {\( \epsilon^+_3 \)};
        \node[xshift=0.25cm,yshift=0.25cm] at (b) {\( \epsilon^-_4 \)};
        \node[xshift=-0.25cm,yshift=-0.25cm] at (c) {\( \epsilon^+_2 \)};
        \node[xshift=-0.25cm,yshift=0.25cm] at (d) {\( \epsilon^+_1 \)};
        
        % Draw parallel momentum arrows
        \draw[-{Latex},thick,rotate=-90] (-0.9,-1.4) -- (-0.3,-1.2) node[midway,xshift=-0.3 cm,yshift=-0.2 cm] {\( k_3 \)};
        \draw[-{Latex},thick,rotate=-90] (0.9,-1.4) -- (0.3,-1.2) node[midway,xshift=-0.25 cm,yshift=0.2 cm] {\( k_4 \)};
        \draw[-{Latex},thick] (1.5,0.285) -- (1.5,-0.285) node[midway,right] {\( k_2 \)};
        \draw[{Latex}-,thick] (-0.1,-0.95) -- (0.52,-1.27) node[midway,yshift=12pt] {\( k_1+k_2 \)};
    \end{scope}
    \end{tikzpicture}
\end{tabular}

\vspace{0.5cm}

\caption{Witten diagrams that yields the component of the correlator with a composite operators $aa$ and a topological vertex on the boundary.}
\label{fig:aaTop}
\end{figure}
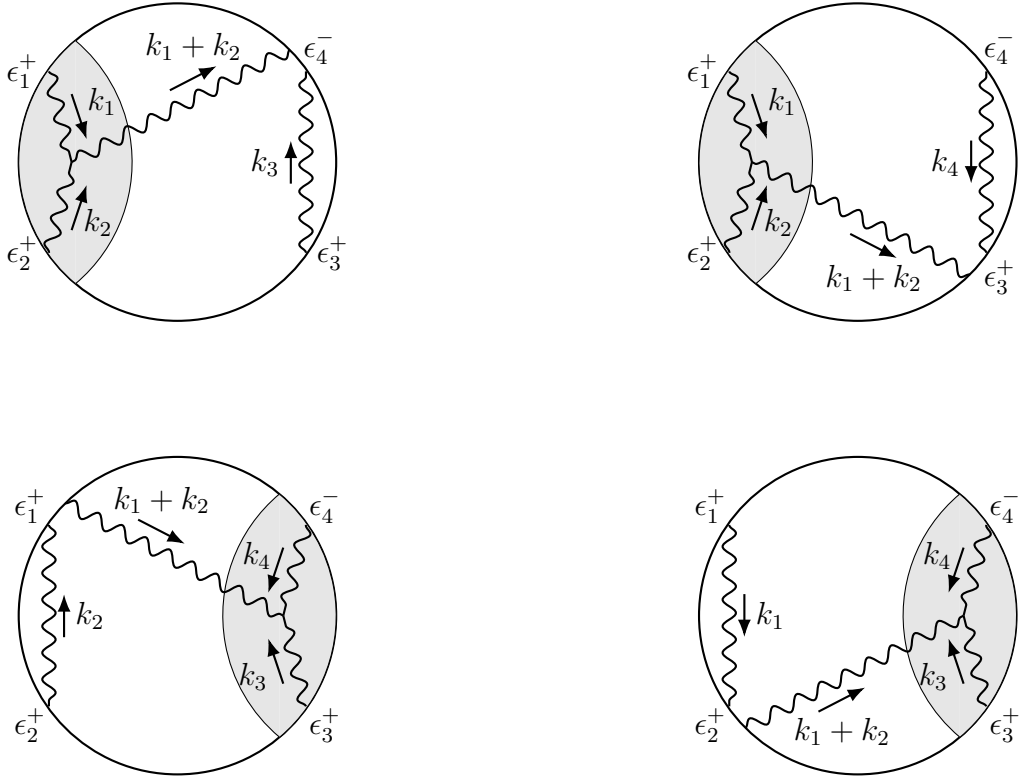

Of course, we have changed only one of the two vertices present in the exchange diagram so far. The other vertex can either be replaced by a topological vertex or by another insertion of a composite operator. The former type of diagrams are depicted in Figure \ref{fig:aaTop}. From the top to the bottom and from left to right we call the corresponding correlators $\mathcal{T}_{s,\text{L},aa}^i$, with $i=1,2,3,4$. They read
\begin{align}
    \begin{aligned}
        \mathcal{T}_{s,\text{L},aa}^1 &= \frac{g^2}{32k^3}\frac{\langle \bar{1}\bar{2} \rangle \langle \bar{3}4 \rangle}{k_1k_2k_3k_4}\Big(2k^2\big(\langle \bar{1}\bar{3} \rangle \langle \bar{2}4 \rangle + \langle \bar{1}4 \rangle \langle \bar{2}\bar{3} \rangle\big) - (1+e^{2i\gamma}) \langle \bar{1}\bar{2}|kk\bar{k}\bar{k}|\bar{3}4 \rangle +\\
        &\qquad\qquad\qquad\qquad- (1+e^{-2i\gamma})\langle \bar{1}\bar{2}| \bar{k}\bar{k}kk|\bar{3}4 \rangle\Big) \,,\\
        \mathcal{T}_{s,\text{L},aa}^2&=\mathcal{T}_{s,\text{L},aa}^1\,,\\
        \mathcal{T}_{s,\text{L},aa}^3 &= -\frac{g^2}{32k^3}\frac{\langle \bar{1}\bar{2} \rangle \langle \bar{3}4 \rangle}{k_1k_2k_3k_4}\Big(2k^2\big(\langle \bar{1}\bar{3} \rangle \langle \bar{2}4 \rangle + \langle \bar{1}4 \rangle \langle \bar{2}\bar{3} \rangle\big) - (1+e^{2i\gamma}) \langle \bar{1}\bar{2}|kk\bar{k}\bar{k}|\bar{3}4 \rangle +\\
        &\qquad\qquad\qquad\qquad- (1+e^{-2i\gamma})\langle \bar{1}\bar{2}| \bar{k}\bar{k}kk|\bar{3}4 \rangle\Big)\,,\\
        \mathcal{T}_{s,\text{L},aa}^4&=\mathcal{T}_{s,\text{L},aa}^3 \,,
    \end{aligned}
\end{align}
where again color factors have been taken into account.

\begin{figure}[h!]
\centering
\begin{tabular}{c c}
    % S-channel Witten diagram
    \begin{tikzpicture}
    \begin{scope}[xshift=-12cm]
        % Define the boundary circle
        \draw[thick] (0,0) circle (2.1cm);
        
        % Define the vertices inside the bulk
        \coordinate (v1) at (-1.4,0);
        
        % Define the external legs on the boundary
        \coordinate (a) at (-1.7,1.2);
        \coordinate (b) at (-1.7,-1.2);
        \coordinate (c) at (1.7,1.2);
        \coordinate (d) at (1.7,-1.2);
        \coordinate (e) at (-1.49,1.49);
        \coordinate (f) at (1.49,-1.49);
        
        % Draw the gauge boson lines
        \draw[decorate,decoration={snake},thick] (a) -- (b);
        \draw[decorate,decoration={snake},thick] (e) -- (f);
        \draw[decorate,decoration={snake},thick] (d) -- (c);
        
        % Add labels for fields next to endpoints
        \node[left] at (a) {\( \epsilon^+_1 \)};
        \node[left] at (b) {\( \epsilon^+_2 \)};
        \node[xshift=5pt, yshift=9pt] at (c) {\( \epsilon^-_4 \)};
        \node[right] at (d) {\( \epsilon^+_3 \)};
        
        % Draw parallel momentum arrows
        \draw[{Latex}-,thick] (-1.5,0.285) -- (-1.5,-0.285) node[midway,xshift=0.3 cm,yshift=0.2 cm] {\( k_2 \)};
        \draw[{Latex}-,thick] (1.5,0.285) -- (1.5,-0.285) node[midway,left] {\( k_3 \)};
        \draw[-{Latex},thick] (0,0.4) -- (0.4,0.0) node[midway,yshift=12pt] {\( k_1+k_2 \)};
    \end{scope}
    \begin{scope}[xshift=-3cm]
    % Define the boundary circle
        \draw[thick] (0,0) circle (2.1cm);
        
        % Define the vertices inside the bulk
        \coordinate (v1) at (-1.4,0);
        
        % Define the external legs on the boundary
        \coordinate (a) at (-1.7,1.2);
        \coordinate (b) at (-1.7,-1.2);
        \coordinate (c) at (1.7,1.2);
        \coordinate (d) at (1.7,-1.2);
        \coordinate (e) at (1.49,1.49);
        \coordinate (f) at (-1.49,1.49);
        
        % Draw the gauge boson lines
        \draw[decorate,decoration={snake},thick] (a) -- (b);
        \draw[decorate,decoration={snake},thick] (f) -- (e);
        \draw[decorate,decoration={snake},thick] (c) -- (d);
        
        % Add labels for fields next to endpoints
        \node[left] at (a) {\( \epsilon^+_1 \)};
        \node[left] at (b) {\( \epsilon^+_2 \)};
        \node[xshift=5pt, yshift=9pt] at (c) {\( \epsilon^-_4 \)};
        \node[xshift=5pt, yshift=-9pt] at (d) {\( \epsilon^+_3 \)};
        
        % Draw parallel momentum arrows
        \draw[{Latex}-,thick] (-1.5,0.285) -- (-1.5,-0.285) node[midway,right] {\( k_2 \)};
        \draw[{Latex}-,thick] (1.5,0.285) -- (1.5,-0.285) node[midway,left] {\( k_3 \)};
        \draw[-{Latex},thick] (-0.285,1.3) -- (0.285,1.3) node[midway,yshift=-12pt] {\( k_1+k_2 \)};
    \end{scope}
    \begin{scope}[xshift=-3cm,yshift=-6cm,rotate=180]
    % Define the boundary circle
        \draw[thick] (0,0) circle (2.1cm);
        
        % Define the vertices inside the bulk
        \coordinate (v1) at (-1.4,0);
        
        % Define the external legs on the boundary
        \coordinate (a) at (-1.7,1.2);
        \coordinate (b) at (-1.7,-1.2);
        \coordinate (c) at (1.7,1.2);
        \coordinate (d) at (1.7,-1.2);
        \coordinate (e) at (-1.49,1.49);
        \coordinate (f) at (1.49,1.49);
        
        % Draw the gauge boson lines
        \draw[decorate,decoration={snake},thick] (a) -- (b);
        \draw[decorate,decoration={snake},thick] (e) -- (f);
        \draw[decorate,decoration={snake},thick] (c) -- (d);
        
        % Add labels for fields next to endpoints
        \node[xshift=0.25cm,yshift=-0.25cm] at (a) {\( \epsilon^+_3 \)};
        \node[xshift=0.25cm,yshift=0.25cm] at (b) {\( \epsilon^-_4 \)};
        \node[xshift=-0.25cm,yshift=-0.25cm] at (c) {\( \epsilon^+_2 \)};
        \node[xshift=-0.25cm,yshift=0.25cm] at (d) {\( \epsilon^+_1 \)};
        
        % Draw parallel momentum arrows
       \draw[{Latex}-,thick] (1.5,0.285) -- (1.5,-0.285) node[midway,right] {\( k_1 \)};
        \draw[{Latex}-,thick] (-1.5,0.285) -- (-1.5,-0.285) node[midway,left] {\( k_4 \)};
        \draw[{Latex}-,thick] (-0.285,1.3) -- (0.285,1.3) node[midway,yshift=12pt] {\( k_1+k_2 \)};
    \end{scope}
    \begin{scope}[xshift=-12cm,yshift=-6cm,rotate=180]
    % Define the boundary circle
        \draw[thick] (0,0) circle (2.1cm);
        
        % Define the vertices inside the bulk
        \coordinate (v1) at (-1.4,0);
        
        % Define the external legs on the boundary
        \coordinate (a) at (-1.7,1.2);
        \coordinate (b) at (-1.7,-1.2);
        \coordinate (c) at (1.7,1.2);
        \coordinate (d) at (1.7,-1.2);
        \coordinate (e) at (1.49,1.49);
        \coordinate (f) at (-1.49,-1.49);

        % Draw the gauge boson lines
        \draw[decorate,decoration={snake},thick] (a) -- (b);
        \draw[decorate,decoration={snake},thick] (c) -- (d);
        \draw[decorate,decoration={snake},thick] (e) -- (f);
        
        % Add labels for fields next to endpoints
        \node[xshift=0.25cm,yshift=-0.25cm] at (a) {\( \epsilon^+_3 \)};
        \node[xshift=0.25cm,yshift=0.25cm] at (b) {\( \epsilon^-_4 \)};
        \node[xshift=-0.25cm,yshift=-0.25cm] at (c) {\( \epsilon^+_2 \)};
        \node[xshift=-0.25cm,yshift=0.25cm] at (d) {\( \epsilon^+_1 \)};
        
        % Draw parallel momentum arrows
        \draw[{Latex}-,thick] (1.5,0.285) -- (1.5,-0.285) node[midway,right] {\( k_1 \)};
        \draw[{Latex}-,thick] (-1.5,0.285) -- (-1.5,-0.285) node[midway,left] {\( k_4 \)};
        \draw[{Latex}-,thick] (0,-0.4) -- (0.4,0) node[midway,yshift=12pt] {\( k_1+k_2 \)};
    \end{scope}
    \end{tikzpicture}
\end{tabular}

\vspace{0.5cm}

\caption{Witten diagrams that yield the correlators with two composite operator $aa$ on the boundary.}
\label{fig:aaDouble}
\end{figure}
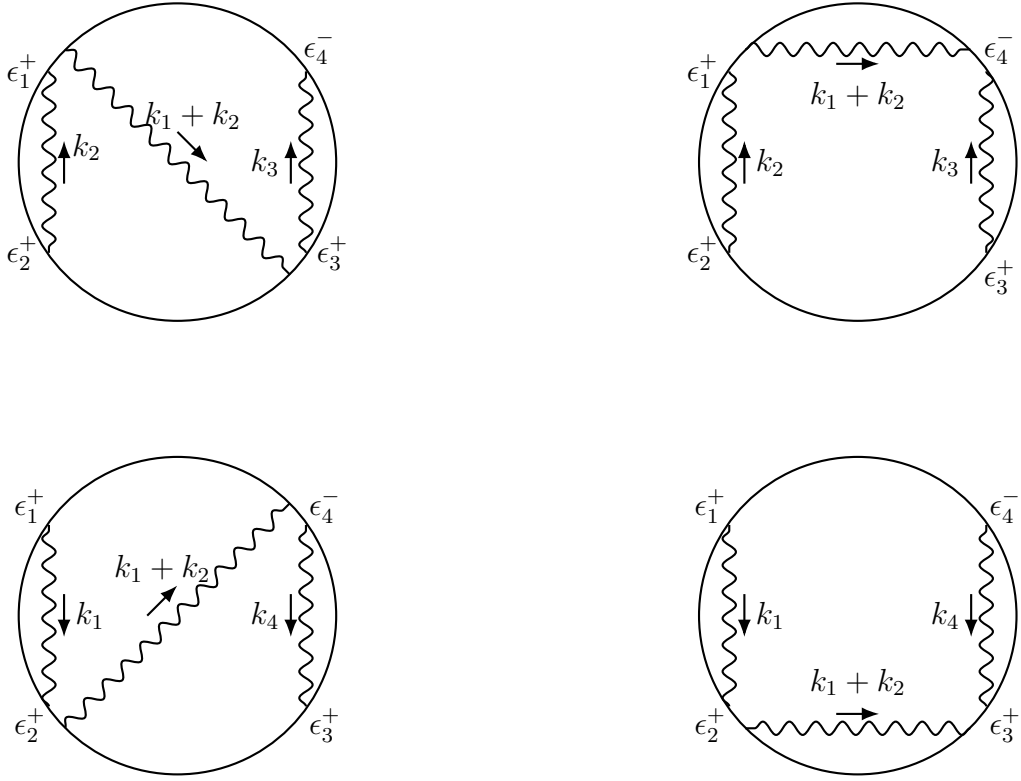

The last type of diagrams that are to be evaluated are depicted in Figure \ref{fig:aaDouble}. From the top to the bottom and from left to right we denote the corresponding correlators $\mathcal{W}_{s,\text{L},(aa)^2}^i$, with $i=1,2,3,4$. Treating the color factors carefully, one obtains
\begin{align}
    \begin{aligned}
        \mathcal{W}_{s,\text{L},(aa)^2}^1 &= -\frac{g^2}{32} \frac{\langle \bar{1}\bar{2} \rangle \langle \bar{3}4\rangle}{k_1k_2k_3k_4} \Big(8k^2\langle \bar{1}\bar{3} \rangle \langle \bar{2}4 \rangle - (1+e^{2i\gamma})\langle \bar{1}\bar{2}|kk\bar{k}\bar{k}|\bar{3}4 \rangle - (1+e^{-2i\gamma}) \langle \bar{1}\bar{2}|\bar{k}\bar{k}kk|\bar{3}4 \rangle\Big) \,,\\
        \mathcal{W}_{s,\text{L},(aa)^2}^2&=\mathcal{W}_{s,\text{L},(aa)^2}^3=\mathcal{W}_{s,\text{L},(aa)^2}^4=\mathcal{W}_{s,\text{L},(aa)^2}^1\,.
    \end{aligned}
\end{align}

\paragraph{Which diagrams survive the self-dual limit?} The above diagrams are accompanied by factors of $e^{in\gamma}$ that are carried by the external lines (and sometimes internal $\langle \Phi\Phi\rangle$ line), and some diagrams come with the coefficient of the Chern--Simons term, which becomes $2(b-a)=4e^{-2i\gamma}$ if we set $a=1$ in the self-dual limit, see Section \ref{sec:boundaryterms} for more details. As an example, the non-topological diagrams considered in the main text come with three positive helicity external lines and no internal $\langle \Phi\Phi \rangle$ ones, so they carry an overall factor of $e^{6i\gamma}$, which will be divided out for all diagrams.

\begin{figure}[h!]
\centering
    \begin{tikzpicture}
        \begin{scope}[xshift=-6cm]
% Define the boundary circle
        \draw[thick] (0,0) circle (2.1cm);
        
        % Define the vertices inside the bulk
        \coordinate (v1) at (-0.7,0);
        \coordinate (v2) at (0.7,0);
        
        % Define the external legs on the boundary
        \coordinate (a) at (-1.7,1.2);
        \coordinate (b) at (-1.7,-1.2);
        \coordinate (c) at (1.7,1.2);
        \coordinate (d) at (1.7,-1.2);
        \coordinate (e) at (-1.49,1.49);
        
        % Draw the gauge boson lines
        \draw[decorate,decoration={snake},thick] (a) -- (b);
        \draw[decorate,decoration={snake},thick] (e) -- (v2);
        \draw[decorate,decoration={snake},thick] (d) -- (v2);
        \draw[decorate,decoration={snake},thick] (c) -- (v2);
        
        % Add labels for fields next to endpoints
        \node[xshift=-5pt, yshift=9pt] at (a) {\( \epsilon^+_1 \)};
        \node[xshift=-6pt,yshift=-6pt] at (b) {\( \epsilon^+_2 \)};
        \node[xshift=5pt, yshift=9pt] at (c) {\( \epsilon^-_4 \)};
        \node[right] at (d) {\( \epsilon^+_3 \)};
        
        % Draw parallel momentum arrows
        \draw[-{Latex},thick] (1.3,1.0) -- (0.9,0.6) node[midway,above left] {\( k_4 \)};
        \draw[-{Latex},thick] (1.3,-1) -- (0.9,-0.6) node[midway,below left] {\( k_3 \)};
        \draw[{Latex}-,thick] (-1.5,0.285) -- (-1.5,-0.285) node[midway,right] {\( k_2 \)};
        \draw[-{Latex},thick] (-0.8,1.3) -- (-0.18,0.88) node[midway,yshift=12pt] {\( k_1+k_2 \)};

        \fill[blue] (1,0.36) circle (5pt);
        
            \node at (3,0) {$+$};
        \end{scope}

            \begin{scope}[xshift=-12cm]
                    % Define the boundary circle
            \draw[thick] (0,0) circle (2.1cm);
            
            % Define the vertices inside the bulk
            \coordinate (v1) at (-1.4,0);
            \coordinate (v2) at (0.7,0);
            
            % Define the external legs on the boundary
            \coordinate (a) at (-1.7,1.23);
            \coordinate (b) at (-1.7,-1.23);
            \coordinate (c) at (1.7,1.2);
            \coordinate (d) at (1.7,-1.2);

            \filldraw[fill=gray!20, draw=black,rotate=-90] 
                (1.61,-1.35) arc[start angle=40, end angle=140, radius=2.1cm];
            \filldraw[fill=gray!20, draw=black,rotate=-90] 
            (-1.61,-1.35) arc[start angle=220, end angle=320, radius=2.1cm];
    
                    % Draw parallel momentum arrows
                \draw[-{Latex},thick,rotate=-90] (-0.9,-1.4) -- (-0.3,-1.2) node[midway,xshift=0.3 cm,yshift=0.2 cm] {\( k_1 \)};
                \draw[-{Latex},thick,rotate=-90] (0.9,-1.4) -- (0.3,-1.2) node[midway,xshift=0.25 cm,yshift=-0.2 cm] {\( k_2 \)};
            
            % Draw the gauge boson lines
            \draw[decorate,decoration={snake},thick] (a) -- (v1);
            \draw[decorate,decoration={snake},thick] (b) -- (v1);
            \draw[decorate,decoration={snake},thick] (c) -- (v2);
            \draw[decorate,decoration={snake},thick] (d) -- (v2);
            
            % Draw the internal propagator 
            \draw[decorate,decoration={snake},thick] (v1) -- (v2);
            
            % Draw parallel momentum arrows
            \draw[{Latex}-,thick] (0.9,0.6) -- (1.3,1) node[midway,above left] {\( k_4 \)};
            \draw[{Latex}-,thick] (0.9,-0.6) -- (1.3,-1) node[midway,below left] {\( k_3 \)};
            \draw[-{Latex},thick] (-0.2,0.2) -- (0.2,0.2) node[midway,above] {\( k_1+k_2 \)};
            
            % Add labels for fields next to endpoints
            \node[left] at (a) {\( \epsilon^+_1 \)};
            \node[left] at (b) {\( \epsilon^+_2 \)};
            \node[right] at (c) {\( \epsilon^-_4 \)};
            \node[right] at (d) {\( \epsilon^+_3 \)};

            \fill[blue] (0.9,0.2) circle (5pt);
                \node at (3,0) {$+$};
            \end{scope}
            \begin{scope}
        % Define the boundary circle
        \draw[thick] (0,0) circle (2.1cm);
        
        % Define the vertices inside the bulk
        \coordinate (v1) at (-0.7,0);
        \coordinate (v2) at (0.7,0);
        
        % Define the external legs on the boundary
        \coordinate (a) at (-1.7,1.2);
        \coordinate (b) at (-1.7,-1.2);
        \coordinate (c) at (1.7,1.2);
        \coordinate (d) at (1.7,-1.2);
        \coordinate (e) at (-1.49,-1.49);
        
        % Draw the gauge boson lines
        \draw[decorate,decoration={snake},thick] (a) -- (b);
        \draw[decorate,decoration={snake},thick] (e) -- (v2);
        \draw[decorate,decoration={snake},thick] (d) -- (v2);
        \draw[decorate,decoration={snake},thick] (c) -- (v2);
        
        % Add labels for fields next to endpoints
        \node[left] at (a) {\( \epsilon^+_1 \)};
        \node[xshift=-5pt, yshift=-9pt] at (b) {\( \epsilon^+_2 \)};
        \node[xshift=5pt, yshift=9pt] at (c) {\( \epsilon^-_4 \)};
        \node[right] at (d) {\( \epsilon^+_3 \)};
        
        % Draw parallel momentum arrows
        \draw[-{Latex},thick] (1.3,1.0) -- (0.9,0.6) node[midway,above left] {\( k_4 \)};
        \draw[-{Latex},thick] (1.3,-1) -- (0.9,-0.6) node[midway,below left] {\( k_3 \)};
        \draw[-{Latex},thick] (-1.5,0.285) -- (-1.5,-0.285) node[midway,right] {\( k_1 \)};
        \draw[-{Latex},thick] (-0.8,-1.3) -- (-0.18,-0.88) node[midway,yshift=-12pt] {\( k_1+k_2 \)};
        \node at (-3,0) {$+$};
        \fill[blue] (1,0.36) circle (5pt);
        \node at (3,0) {$=0$};
            \end{scope}
        
    \end{tikzpicture}
\caption{The sum $(b-a)\mathcal{T}_s^{\text{L},1}+e^{-2i\gamma}\mathcal{W}^3_{s,\text{L},aa}+e^{-2i\gamma}\mathcal{W}^4_{s,\text{L},aa}$ vanishes in the self-dual limit.}
\label{fig:sum1}
\end{figure}
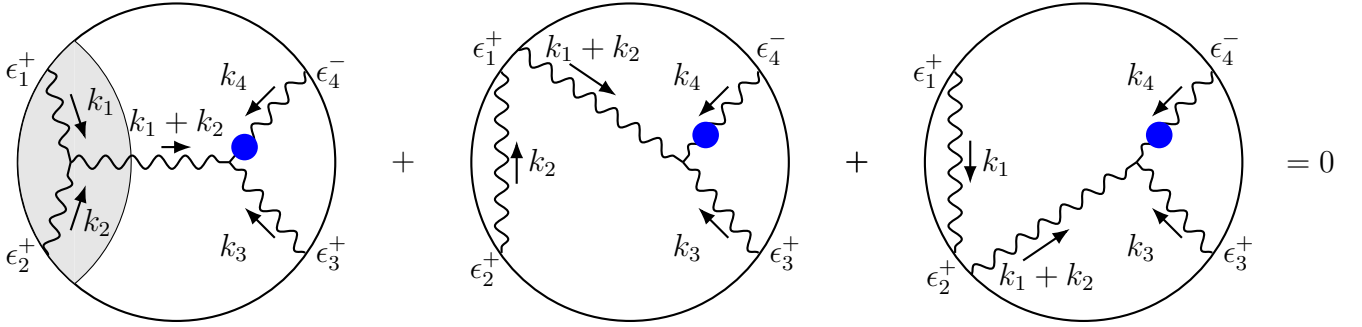

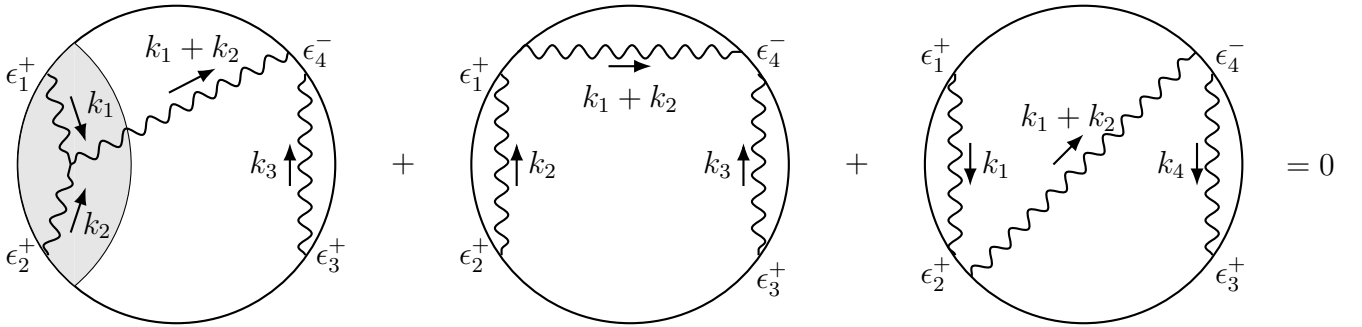
\begin{figure}[h!]
\centering
    \begin{tikzpicture}
        \begin{scope}[xshift=-6cm]
% Define the boundary circle
        \draw[thick] (0,0) circle (2.1cm);
        
        % Define the vertices inside the bulk
        \coordinate (v1) at (-1.4,0);
        
        % Define the external legs on the boundary
        \coordinate (a) at (-1.7,1.2);
        \coordinate (b) at (-1.7,-1.2);
        \coordinate (c) at (1.7,1.2);
        \coordinate (d) at (1.7,-1.2);
        \coordinate (e) at (1.49,1.49);
        \coordinate (f) at (-1.49,1.49);
        
        % Draw the gauge boson lines
        \draw[decorate,decoration={snake},thick] (a) -- (b);
        \draw[decorate,decoration={snake},thick] (f) -- (e);
        \draw[decorate,decoration={snake},thick] (c) -- (d);
        
        % Add labels for fields next to endpoints
        \node[left] at (a) {\( \epsilon^+_1 \)};
        \node[left] at (b) {\( \epsilon^+_2 \)};
        \node[xshift=5pt, yshift=9pt] at (c) {\( \epsilon^-_4 \)};
        \node[xshift=5pt, yshift=-9pt] at (d) {\( \epsilon^+_3 \)};
        
        % Draw parallel momentum arrows
        \draw[{Latex}-,thick] (-1.5,0.285) -- (-1.5,-0.285) node[midway,right] {\( k_2 \)};
        \draw[{Latex}-,thick] (1.5,0.285) -- (1.5,-0.285) node[midway,left] {\( k_3 \)};
        \draw[-{Latex},thick] (-0.285,1.3) -- (0.285,1.3) node[midway,yshift=-12pt] {\( k_1+k_2 \)};
        
            \node at (3,0) {$+$};
        \end{scope}

            \begin{scope}[xshift=-12cm]
                    % Define the boundary circle
        \draw[thick] (0,0) circle (2.1cm);
        
        % Define the vertices inside the bulk
        \coordinate (v1) at (-1.4,0);
        
        % Define the external legs on the boundary
        \coordinate (a) at (-1.7,1.2);
        \coordinate (b) at (-1.7,-1.2);
        \coordinate (c) at (1.7,1.2);
        \coordinate (d) at (1.7,-1.2);
        \coordinate (e) at (1.49,1.49);

        \filldraw[fill=gray!20, draw=black,rotate=-90] 
                (1.61,-1.35) arc[start angle=40, end angle=140, radius=2.1cm];
            \filldraw[fill=gray!20, draw=black,rotate=-90] 
            (-1.61,-1.35) arc[start angle=220, end angle=320, radius=2.1cm];
        
        % Draw the gauge boson lines
        \draw[decorate,decoration={snake},thick] (a) -- (v1);
        \draw[decorate,decoration={snake},thick] (b) -- (v1);
        \draw[decorate,decoration={snake},thick] (e) -- (v1);
        \draw[decorate,decoration={snake},thick] (d) -- (c);
        
        % Add labels for fields next to endpoints
        \node[left] at (a) {\( \epsilon^+_1 \)};
        \node[left] at (b) {\( \epsilon^+_2 \)};
        \node[xshift=5pt, yshift=9pt] at (c) {\( \epsilon^-_4 \)};
        \node[right] at (d) {\( \epsilon^+_3 \)};
        
        % Draw parallel momentum arrows
        \draw[-{Latex},thick,rotate=-90] (-0.9,-1.4) -- (-0.3,-1.2) node[midway,xshift=0.3 cm,yshift=0.2 cm] {\( k_1 \)};
        \draw[-{Latex},thick,rotate=-90] (0.9,-1.4) -- (0.3,-1.2) node[midway,xshift=0.25 cm,yshift=-0.2 cm] {\( k_2 \)};
        \draw[{Latex}-,thick] (1.5,0.285) -- (1.5,-0.285) node[midway,left] {\( k_3 \)};
        \draw[-{Latex},thick] (-0.1,0.95) -- (0.52,1.27) node[midway,yshift=12pt] {\( k_1+k_2 \)};
                \node at (3,0) {$+$};
            \end{scope}
            \begin{scope}[rotate=180]
        % Define the boundary circle
        \draw[thick] (0,0) circle (2.1cm);
        
        % Define the vertices inside the bulk
        \coordinate (v1) at (-1.4,0);
        
        % Define the external legs on the boundary
        \coordinate (a) at (-1.7,1.2);
        \coordinate (b) at (-1.7,-1.2);
        \coordinate (c) at (1.7,1.2);
        \coordinate (d) at (1.7,-1.2);
        \coordinate (e) at (1.49,1.49);
        \coordinate (f) at (-1.49,-1.49);

        % Draw the gauge boson lines
        \draw[decorate,decoration={snake},thick] (a) -- (b);
        \draw[decorate,decoration={snake},thick] (c) -- (d);
        \draw[decorate,decoration={snake},thick] (e) -- (f);
        
        % Add labels for fields next to endpoints
        \node[xshift=0.25cm,yshift=-0.25cm] at (a) {\( \epsilon^+_3 \)};
        \node[xshift=0.25cm,yshift=0.25cm] at (b) {\( \epsilon^-_4 \)};
        \node[xshift=-0.25cm,yshift=-0.25cm] at (c) {\( \epsilon^+_2 \)};
        \node[xshift=-0.25cm,yshift=0.25cm] at (d) {\( \epsilon^+_1 \)};
        
        % Draw parallel momentum arrows
        \draw[{Latex}-,thick] (1.5,0.285) -- (1.5,-0.285) node[midway,right] {\( k_1 \)};
        \draw[{Latex}-,thick] (-1.5,0.285) -- (-1.5,-0.285) node[midway,left] {\( k_4 \)};
        \draw[{Latex}-,thick] (0,-0.4) -- (0.4,0) node[midway,yshift=12pt] {\( k_1+k_2 \)};
        \node at (-3,0) {$=0$};
            \end{scope}
        
    \end{tikzpicture}
\caption{The sum $(b-a)\mathcal{T}_{s,\text{L},aa}^{1}+e^{-2i\gamma}\mathcal{W}^2_{s,\text{L},(aa)^2}+e^{-2i\gamma}\mathcal{W}^3_{s,\text{L},(aa)^2}$ vanishes in the self-dual limit.}
\label{fig:sum2}
\end{figure}

Performing the same counting on the above diagrams, we find the diagrams that survive the self-dual limit to be the non-topological diagram considered in the main text, together with $\mathcal{W}_{s,\text{L},aa}^1$, $\mathcal{W}_{s,\text{L},aa}^3$, $\mathcal{W}_{s,\text{L},aa}^4$,  $\mathcal{T}_{s,\text{L},aa}^1$, $\mathcal{W}_{s,\text{L},(aa)^2}^2$ $\mathcal{W}_{s,\text{L},(aa)^2}^3$ and $\mathcal{T}^{\text{L},1}_{\gamma}$. Only $\mathcal{W}_{s,\text{L},aa}^1$ appears both in YM and in SDYM,\footnote{Differing by a factor $-2k_4$ as mentioned before. For the correct choice of boundary-to-bulk propagator, or choosing Chalmers--Siegel's one, the diagrams are the same. } while all others only exist in YM. Note that the last diagram is a topological diagram, see \eqref{topDiag}. Extracting the factor $e^{6i\gamma}$ and adding the Chern--Simons coefficient to the topological diagram, one finds\footnote{The factor $e^{-2i\gamma}$ that multiplies each diagram justifies the fact that we chose the Neumann-like propagator; the difference between the Neumann-like and Dirichlet-like propagators vanishes in the limit.}
\besubeqs
\begin{align}
        \lim_{\gamma \rightarrow -i\infty}\Big[2(b-a)\mathcal{T}_\gamma^{\text{L},1}+e^{-2i\gamma}\mathcal{W}^3_{s,\text{L},aa}+e^{-2i\gamma}\mathcal{W}^4_{s,\text{L},aa} \Big]&= 0\label{eq1}\,,\\
        \lim_{\gamma\rightarrow -i\infty} \Big[2(b-a)\mathcal{T}_{s,\text{L},aa}^{1}+e^{-2i\gamma}\mathcal{W}^2_{s,\text{L},(aa)^2}+e^{-2i\gamma}\mathcal{W}^3_{s,\text{L},(aa)^2} \Big]&= 0\label{eq2}\,,
\end{align}
\esubeqs
since $\lim_{\gamma\rightarrow -i\infty}2(b-a)=4e^{-2i\gamma}$. This is also shown in Figures \ref{fig:sum1} and \ref{fig:sum2}. We see that \eqref{eq1} implies that correlators in SDYM and the self-dual limit of cYM agree, as the composite diagrams cancel the topological diagram in the self-dual limit. Meanwhile, \eqref{eq2} present the cancellation of various composite diagrams among themselves in the limit. Note that away from the self-dual limit these cancellations do not occur.

To summarize, YM/Chalmers--Siegel theory contains six composite diagrams and one topological diagram that survive in the self-dual limit, while SDYM yields only one composite diagram and no topological ones. In the self-dual limit three composite diagrams in Chalmers--Siegel cancel against each other, while two other composite diagrams cancel with the topological contribution. The end result is that both Chalmers--Siegel and SDYM end up with no topological diagrams and one composite diagram in the self-dual limit, on which they agree. Together with the agreement between the non-topological diagrams of Chalmers--Siegel and SDYM in the self-dual limit from the main text, this proves that the correlators of the self-dual limit Chalmers--Siegel coincide with SDYM.

%%%%%%%%%%%%%%%%%%%%%%%%%%%%%%%%%%%%%%%%%%%%%%%%%%%%%%%%%%%%%
\section{Hamiltonian analysis, Complete gauge}
\label{app:ham}
%%%%%%%%%%%%%%%%%%%%%%%%%%%%%%%%%%%%%%%%%%%%%%%%%%%%%%%%%%%%%
It may be instructive to perform the canonical analysis of cYM and SDYM theories. For one reason, one can reduce the theories down to the physical variables by imposing a complete gauge. We begin with the Chalmers--Siegel action
\begin{align}
    S&= \Tr\int \Psi^{AA}(\nabla_{AB'}\Phi\fdu{A,}{B'}+\Phi_{A,B'}\Phi\fdu{A,}{B} -\tfrac{\epsilon}2 \Psi_{AA}) \,.
\end{align}
We plug in the $3+1$ decomposition $\Phi^{A,A'}=\phi^{AA'}+\epsilon^{AA'} \rho$ and $\pl_{AA'}=-\pl_z \epsilon_{AA'}+ \pll_{AA'}$;  $\pll_{AB}=\pll_{BA}$ is the partial derivative with respect to the spatial coordinates (we replace it by $k_{AB}$ at the end omitting the Fourier $i$). The result is
\begin{align}
    S&= \Tr\int \Psi^{AA}\left(\dot{\phi}_{AA} +\pll_{AB}\phi\fdu{A}{B} + \pll_{AA}\rho +[\phi_{AA},\rho]+\phi_{AB}\phi\fdu{A}{B} -\frac{\epsilon}2 \Psi_{AA}\right)\,.
\end{align}
Since the action is already written in the Hamiltonian-friendly form with $\Psi^{AA}$ playing the role of momentum for $\phi_{AA}$,\footnote{Constraint analysis of an action $S=\int p\dot{q} -H(q,p)$ will give it back. The primary constraints here are $\pi_q-p$ and $\pi_p$, which are second class. Reducing the system onto the constraint surface replaces $\pi_q$ with $p$ and eliminates $\pi_p$. } one can cut the long story short and perform the ``constraint analysis'' for $\rho$. The primary constraint is $\tau^1=\pi_\rho$ and the secondary is
\begin{align}
    \tau^2=\pll_{AA}\Psi^{AA}+[\phi_{AA},\Psi^{AA}]\equiv \pll_i\Psi^i+[\phi_i,\Psi^i]\equiv D_i \Psi^i\,,
\end{align}
which are first class. This is as expected since $\delta \rho=-\dot{\xi}$, $\delta \phi_{AA}=\pll_{AA}\xi$ at the linearized level. We can gauge-fix the theory by choosing two conditions that form together with $\tau^{1,2}$ second class constraints. It is convenient to choose $\tau^3=\rho$ and $\tau^4=\pll_{AA}\phi^{AA}$. 

At the free level, we have $\pll_{AA}\phi^{AA}=0$ and $\pll_{AA}\Psi^{AA}=0$, i.e. $k_i\phi^i=k_i \Psi^i=0$ are two transverse vectors (as it would happen in the Coulomb gauge $\pll_i A^i=0$). The two-dimensional space orthogonal to $k^i$ is spanned by $k^Ak^A$ and $\brk^A \brk^A$. In the flat space (without boundary) the equation of motion $p_{AB'} \phi\fdu{A}{B'}=0$, $p^2=0$ would select $\phi_{AA}\sim \brk_A \brk_A$ and two solutions correspond to positive/negative energy waves $\exp \pm i \brk_A x^{AA'}\brk_{A'}$. In $\text{AdS}_4$, i.e. in the half-flat space, we still have two solution $\phi_{AA}\sim \brk_A\brk_A e^{-kz}$, $\phi_{AA}\sim k_Ak_A e^{+kz}$, but the latter is not regular in the deep interior. 

Eliminating $\rho$ is harmless, it is a Lagrange multiplier for Gauss' law $\tau^2$. We can impose $\pl_i \phi^i=0$ as a gauge condition. Now, the physical variables are the two transverse vectors, we denote $\varphi_{AA}$ and $\pi_{AA}$, that correspond to the transverse components of $\phi_{AA}$ and $\Psi_{AA}$. There is a small complication with $\Psi_{AA}$ since one needs to eliminate its longitudinal component from $\tau^2=0$. We split 
\begin{align}
    \Psi^i&= \pi^i +{\pll^i} \sigma & \text{or}& &\Psi^{AA}=\pi^{AA}+  {k^{AA}}\sigma \,,
\end{align}
to find 
\begin{align}
    D_i {\pll^i} \sigma +[\varphi_i,\pi^i]&=0 \,, && \sigma=-\left(D_i {\pll^i}\right)^{-1} [\varphi_k,\pi^k] \,.
\end{align}
 The final form of the gauge-fixed Hamiltonian action is
\begin{align}
    S&= \Tr\int \pi^{AA}(\dot{\varphi}_{AA} +\pll_{AB}\varphi\fdu{A}{B} + \varphi_{AB}\varphi\fdu{A}{B} -\tfrac{\epsilon}2 \pi_{AA}) +S' \,,
\end{align}
where the additional (instantaneous) interactions $S'$ are
\begin{align}
    S'&= \Tr\int\frac{\epsilon}{2} \sigma {\pll^2} \sigma -\sigma \pll_{AA}(\phi\fud{A}{C}\phi^{AC})  \,.
\end{align}
One can massage these expressions slightly by introducing the nonabelian magnetic field $B_{AA}=\pll_{AB}\varphi\fdu{A}{B} + \varphi_{AB}\varphi\fdu{A}{B}$ and noticing that $\pll^{AA}(\pll_{AB}\varphi\fdu{A}{B})\equiv0$ :
\begin{align}
    S&= \Tr\int \pi^{AA}(\dot{\varphi}_{AA} +B_{AA} -\tfrac{\epsilon}2 \pi_{AA}) +S' \,,& S'&= \Tr\int\frac{\epsilon}{2} \sigma {\pll^2} \sigma -\sigma \pll_{AA}B^{AA} \,.
\end{align}
Note that $\pll_{AA}B^{AA} \neq0$ beyond the free level. Instead, the Bianchi identities imply $D_{AA} B^{AA}\equiv0$. Now, let us set $\epsilon=0$, i.e. SDYM. In the approximation sufficient to compute the quartic amplitude we find $\sigma=-({\pll^2})^{-1}[\varphi_i,\pi^i]$ and 
\begin{align}
    S'&\approx\Tr \int \frac{\epsilon}{2} [\phi_i,\pi^i]\frac{1}{D^2}[\phi_k,\pi^k]+[\phi_k,\pi^k]\frac{\pll^{AA}}{\pll^2}(\phi_{AB}\phi\fdu{A}{B})+... \,.
\end{align}
Only the last term is relevant for SDYM. These terms contain the usual $J_0 \tfrac{1}{\Delta}J_0$-type interactions. In fact, the complete physical gauge is effectively identical to the Coulomb gauge, i.e. only the physical polarizations propagate, while the $J_0$-part of the current (or $\pl_i J^i$) is accounted for by an instantaneous interaction. Terms of this type are well-known in Yang--Mills theory. 

\paragraph{Propagators.} Let us work out the bulk-to-bulk propagators in the complete gauge for the free theory (i.e. the instantaneous interaction will be dropped since there is no $J_0$). The inhomogeneous part of the propagator for the Yang--Mills theory is\footnote{As different from the axial gauge, this expression is already regular in the bulk (there are no $|z-z'|$-terms).}
\begin{align}
    \langle \Phi_{AA}(-k,z) \Phi_{BB}(k,z') \rangle^{\text{inh}}&=-\frac{1}{2k}e^{-k|z-z'|} \Pi^{\text{e}}_{AA,BB} \,,
\end{align}
from which the inhomogeneous part of SDYM's propagator follows
\begin{align}
    \langle \Psi_{AA}(-k,z) \Phi_{BB}(k,z') \rangle^{\text{inh}}&=\frac{1}{2}e^{-k|z-z'|} (\Pi^{\text{e}}_{AA,BB}\sign(z-z')-\Pi^\text{o}_{AA,BB}) \,.
\end{align}
The Yang--Mills propagator with mixed boundary conditions reads
\begin{align}\notag
    \langle \Phi_{AA}(-k,z) \Phi_{BB}(k,z') \rangle&=-\frac{1}{2k}e^{-k|z-z'|} \Pi^{\text{e}}_{AA,BB} +G^{\text{hom},\gamma}_{AA,BB} \,,
\end{align}
where the homogeneous term is the same in all gauges. There are no pure gauge terms of $\exp[-k(z+z')]$-type allowed here. The latter leads to the already discussed contribution to SDYM (to be precise to the $\Psi-\Phi$ propagator in the Chalmers--Siegel formulation):
\begin{align}
    \langle \Psi_{AA}(-k,z) \Phi_{BB}(k,z') \rangle&=\langle \Psi_{AA}(-k,z) \Phi_{BB}(k,z') \rangle^{\text{inh}}-e^{-k(z+z')} e^{-2i\gamma} \frac{\brk_{A}\brk_{A} k_Bk_{B}}{4k^2} \,.
\end{align}
The equation of motion for the $\Phi$-leg is satisfied only in the self-dual limit $\gamma\rightarrow - i \infty$. This just shows that there  exist homogeneous terms that are valid for the Yang--Mills theory but are forbidden for SDYM, thereby also indicating a lesser freedom to impose boundary conditions in SDYM as compared to the Yang--Mills theory. The final result is that the SDYM propagator is just the inhomogeneous part that is borrowed from  flat space!

The boundary-to-bulk propagator for $\gamma\neq0$ is identical to that in the axial gauge \eqref{btobAxial}
\begin{align}
    \langle \Phi_{A,A'}(-k,z)\Phi_{B,B'}(k,0) \rangle&=\frac{1}{2k}\Big[ \Pi^\gamma_{AA',BB'} -\Pi^\text{e}_{AA',BB'}\Big]e^{-kz} \,.
\end{align}
For SDYM we find
\begin{align}
    \begin{aligned}\notag
        \langle \Psi_{AA}(-k,0)\Phi_{BB}(k,z')\rangle &= -\frac{k_Ak_A\bar{k}_B\bar{k}_B}{4k^2}e^{-kz'} \,,&
    \langle\Psi_{AA}(-k,z)\Phi_{BB}(k,0)\rangle &=+  \frac{\bar{k}_A\bar{k}_Ak_Bk_B}{4k^2}e^{-kz}\,.
    \end{aligned}
\end{align}
On contracting the external legs with the physical polarizations we find
\begin{align}
    \begin{aligned}
        \epsilon^-_{AA}(-k)\langle \Psi_{AA}(-k,0)\Phi_{BB}(k,z')\rangle &= -\frac{\bar{k}_B\bar{k}_B}{2k}e^{-kz'} \,,\\
    \langle\Psi_{AA}(-k,z)\Phi_{BB}(k,0)\rangle\epsilon^+_{BB}(k) &=+  \frac{\bar{k}_A\bar{k}_A}{2k}e^{-kz}\,.
    \end{aligned}
\end{align}

\paragraph{Amplitude.} Lastly, let us compute the $4$-point amplitude in SDYM. There are two contributions: the exchange diagram(s) and a contact term. For the $s$-channel exchange the main observation is that the bulk-to-bulk propagator multiplied by the exponents of the boundary-to-bulk propagators integrates to 
\begin{align}\notag
    \int e^{-(k_1+k_2) z} \langle \Psi_{AA}(z,-k) \Phi_{BB}(z',k) \rangle e^{-(k_3+k_4) z'}&=\frac{1}{2E} \left(\frac{1}{E_\text{L}}(\Pi^\text{e}-\Pi^\text{o})-\frac{1}{E_\text{R}}(\Pi^\text{e}+\Pi^\text{o})\right)_{AA,BB} \,.
\end{align}
The quartic vertex can be massaged into\footnote{Once the double commutator is expanded, one sees the same color structures as for $s$-channel, $1234$, $1243$, $2134$, $2143$, see Appendix \ref{app:color}. } 
\begin{align}
    \Tr\,  \pi^{BB}\left[ \phi_{BB},\frac{k^{AA}}{k^2}(\phi_{AC}\phi\fdu{A}{C})\right] \,.
\end{align}
Note that in the flat space our $\Phi-\Phi$ propagator corresponds to 
\begin{align}
    G_{00}&= G_{0i}=0 \,,&& G_{ij}=\frac{1}{p^2} \left(\delta_{ij}-\frac{k_ik_j}{k^2}\right)=\frac{1}{p^2} \Pi^\text{e}_{AA,BB} \,.
\end{align}
The SDYM propagator is then\footnote{One needs to use the vectorial form of $\Pi^\text{e}$ since $p$ is not on-shell, $p^{AA'}=k^{AA'}+\omega \epsilon^{AA'}$.}
\begin{align} \label{PsiPhicomplete}
    \langle \Psi^{AA}\Phi^{BB}\rangle &=-\frac{1}{p^2} p\fud{A}{A'}\Pi_\text{e}^{AA',BB}=\frac{k}{p^2}(k^{AB}\epsilon^{AB}+ \omega\Pi_\text{e}^{AA,BB}) \,.
\end{align}
In the flat space, one can easily see that all gauges give the same results. For the quartic amplitude it is not important where interactions are coming from and one can always think of the usual current interaction $J^\mu A_\mu$. In the Feynman gauge the amplitude is just $J_\mu J^\mu /p^2$. One should remember about the conservation: $p_0J^0 +p_i J^i =0$ and we use $J^i=J^i_\perp+p^i J_\parallel$. The amplitude is 
\begin{align}
    \mathcal{A}&= \frac{1}{p^2} J_\mu J^\mu =\frac{1}{p^2} (J_\perp\cdot J_\perp) +\frac{1}{k^2} (J_0)^2\,.
\end{align}
In the Coulomb gauge the last piece is the instantaneous interaction. In the complete physical gauge $(J_0)^2/k^2$ is directly present in the action. In the temporal gauge the $J_0$-contribution is obtained with the help of $J_\parallel$ via a different $G_{ij}$. 
\paragraph{Flat space.}
To illustrate explicitly how the complete gauge produces the same flat space amplitude as the other gauges, we first observe that the $\langle\Psi\Phi\rangle$ propagator \eqref{PsiPhicomplete} is transverse on both legs. Hence, this is just the $\langle\pi\phi\rangle$ propagator and next we write it as
\begin{align} \label{piphi}
    \langle\pi_{AA}(-p)\phi_{BB'}(p)\rangle = \frac{k}{p^2}\Big(\epsilon_{AB}p_{AB'}+\frac{k_{AA}}{2k^2}\big(\omega p_{BB'}+p^2\epsilon_{BB'}\big)\Big) \,,
\end{align}
where $p^2=k^2-\omega^2$ is used. Since the three-point vertex is the same as in SDYM in axial gauge, i.e. $V_{AA',BB',CC}=2g\,\epsilon_{A'B'}\epsilon_{AC}\epsilon_{BC}$, the first term in the propagator provides the amplitude \eqref{As} when plugged into the $s$-channel diagram, while the second term vanishes due to the Ward identity, as in axial gauge. The novelty of the complete gauge is the third term and it is expected to cancel the contribution from the quartic interaction. We denote the last term in \eqref{piphi} by
\begin{align}\label{DeltaPiPhi}
    \Delta\langle\pi_{AA}(-p)\phi_{BB'}(p)\rangle = \frac{k_{AA}\epsilon_{BB'}}{2k^2}\,.
\end{align}
In the exchange channel it is attached to two vertices, which gives
\begin{align}
    V\fdu{AA',BB'}{EE}\Delta\langle\pi_{EE}(-p)\phi_{FF'}(p)\rangle V\fud{FF'}{,CC',DD} = \frac{2}{k^2}k_{AB}\epsilon_{A'B'}\epsilon_{CD}\epsilon_{DC'}\,.
\end{align}
Meanwhile, the quartic term in the action can be written as
\begin{align}
    S_4=\frac{1}{k^2}\Tr\int k^{AA}(\phi_{AB}\phi\fdu{A}{B})[\phi_{CC},\pi^{CC}]
\end{align}
and the $s$-channel contribution of the four-point vertex reads
\begin{align}
    V_{AA',BB',CC',DD}=-\frac{2}{k^2}k_{AB}\epsilon_{A'B'}\epsilon_{CD}\epsilon_{DC'}\,.
\end{align}
It follows that the instantaneous interaction is canceled by the correction of the exchange diagram.
\paragraph{AdS -- Dirichlet.}
In AdS, the correlator is only expected to be gauge-invariant for Dirichlet boundary conditions, so we consider this first before jumping to SDYM. Since our aim is to compute the correlator in complete gauge for SDYM, we will only consider non-topological diagrams. In particular, we only evaluate the left diagram in Figure \ref{fig:cYM1}, i.e. the diagram that has an analogue in SDYM. The reason is that the other diagram in the same figure contains a $\langle\Psi\Psi\rangle$ propagator, which is gauge-invariant, and the contact diagram contains no bulk-to-bulk propagator.

In AdS the $\langle\pi\phi\rangle$ propagator reads
\begin{align}
    \begin{aligned}
        \langle \pi_{AA}(-k,z)\phi_{BB'}(k,z')\rangle_\text{D} &= \frac{1}{2}\Pi^\text{e}_{AA,BB'}\Big(\text{sign}(z-z')e^{-k|z-z'|}-e^{-k(z+z')}\Big)+\\
        &-\frac{1}{2}\Pi^\text{o}_{AA,BB'}\Big(e^{-k|z-z'|}-e^{-k(z+z')}\Big) \,.
    \end{aligned}
\end{align}
The difference between the $\langle \pi\phi\rangle$ propagator and the $\langle\Psi\Phi\rangle$ propagator in Feynman gauge\footnote{Strictly speaking, both propagators have different dimensions, since $\phi$ has no $z$-component. However, by not automatically symmetrizing indices on $\phi$ we implicitly add a $z$-component to $\phi$ that is zero.} will be denoted by $\Delta\langle \pi_{AA}(-k,z)\phi_{BB'}(k,z')\rangle \equiv \langle \pi_{AA}(-k,z)\phi_{BB'}(k,z')\rangle -\langle \Psi_{AA}(-k,z)\Phi_{BB'}(k,z')\rangle$ and reads
\begin{align}
    \begin{aligned}
        \Delta\langle \pi_{AA}(-k,z)\phi_{BB'}(k,z')\rangle &= \frac{k_{AA}}{4k^2}\Big(k_{BB'}\big(\text{sign}(z-z')e^{-k|z-z'|}-e^{-k(z+z')}\big)+\\
        &-k\epsilon_{BB'}\big(e^{-k|z-z'|}-e^{-k(z+z')}\big)\Big) \,,
    \end{aligned}
\end{align}
which is just the gauge variation between the Feynman gauge and axial gauge propagator \eqref{deltaphiphi2} without the terms linear in $z$ and $z'$. These linear terms were responsible for cancelling a delta function that arise when trying to write the gauge variation as $\nabla \xi$. Thus, due to the absence of these linear terms in the case at hand, it is not possible to write the gauge variation of the propagator as $\nabla\xi$, but instead we find
\begin{align}\label{propDelta}
    \Delta\langle \pi_{AA}(-k,z)\phi_{BB'}(k,z')\rangle = \nabla^{k,z'}_{BB'}\xi_{AA}+\frac{k_{AA}\epsilon_{BB'}}{2k^2}\delta(z-z') \,,
\end{align}
i.e. we need to deal with the delta function by manually cancelling it. Here, the gauge parameter reads
\begin{align}
    \xi_{AA} = \frac{k_{AA}}{4k^2}\Big(\text{sign}(z-z')e^{-k|z-z'|}-e^{-k(z+z')}\Big) \,,
\end{align}
which again resembles the gauge parameter between Feynman and axial gauge, \eqref{deltapsiphietaD}. In the exchange diagram, integration by parts evaluates the gauge parameter on the boundary $z'=0$, where it vanishes. Still, the delta function remains in the gauge variation of the propagator \eqref{propDelta}. Evaluating the bulk integrals, its connection to flat space gauge variation \eqref{DeltaPiPhi} becomes clear:
\begin{align}
    \int e^{-(k_1+k_2)z}\frac{k_{AA}\epsilon_{BB'}}{2k^2}\delta(z-z')e^{-(k_3+k_4)z'}=\frac{k_{AA}\epsilon_{BB'}}{2k^2E}=\frac{1}{E}\Delta\langle \pi_{AA}(-p)\phi_{BB'}(p)\rangle\,.
\end{align}
After contracting the vertices, one obtains
\begin{align}
    \begin{aligned}
        \int e^{-(k_1+k_2)z}V\fdu{AA',BB'}{EE}&\Delta\langle\pi_{EE}(-k,z)\phi_{FF'}(k,z)\rangle V\fud{FF'}{,CC',DD} e^{-(k_3+k_4)z'}\Big|_{\text{on-shell}} =\\
        &= \frac{2}{k^2E}k_{AB}\epsilon_{A'B'}\epsilon_{CD}\epsilon_{DC'}\,.
    \end{aligned}
\end{align}
Meanwhile, the vertex for the instantaneous interaction in AdS and flat space are the same, and the bulk integrals yields
\begin{align}
    \int e^{-(k_1+k_2)z}V_{AA',BB',CC',DD}e^{-(k_3+k_4)z'}=-\frac{2}{k^2E}k_{AB}\epsilon_{A'B'}\epsilon_{CD}\epsilon_{DC'}
\end{align}
and cancels the gauge variation of the exchange diagram. This implies that the linear terms in the propagator in axial gauge take the role of the instantaneous interaction in complete gauge at the level of diagrams. With this we have established the agreement of the four-point correlator between Feynman gauge, axial gauge and complete gauge for Dirichlet boundary conditions in AdS.

\paragraph{AdS -- SDYM.}
It is now easy to evaluate the four-point correlator for SDYM in AdS. The gauge variation
\begin{align}
    \Delta\langle \pi_{AA}(-k,z)\phi_{BB'}(k,z')\rangle_{\text{SDYM}} \equiv\langle \pi_{AA}(-k,z)\phi_{BB'}(k,z')\rangle_{\text{SDYM}}- \langle \Psi_{AA}(-k,z)\Phi_{BB'}(k,z')\rangle_{\text{SDYM}}
\end{align}
reads\footnote{Here we took the Neumann-like propagator. The Dirichlet-like one again leads to gauge independence.}
\begin{align}
    \begin{aligned}
        \Delta\langle \pi_{AA}(-k,z)\phi_{BB'}(k,z')\rangle_{\text{SDYM}}&=\nabla_{BB'}^{k,z'}\xi_{AA}+\frac{k_{AA}\epsilon_{BB'}}{2k^2}\delta(z-z')\,,\\
        \xi_{AA} &= \frac{k_{AA}}{4k^2}\Big(\text{sign}(z-z')e^{-k|z-z'|}+e^{-k(z+z')}\Big)
    \end{aligned}
\end{align}
and the gauge parameter becomes $\xi_{AA}(z'=0)=\frac{k_{AA}}{2k^2}e^{-kz}$ on the boundary. Like before, the delta function cancels with the instantaneous interaction. Similarly to show how the gauge variation between Feynman gauge and axial gauge in SDYM was computed, c.f. \eqref{intbyparts}, we use integration by parts to obtain the gauge variation of the correlator,
\begin{align}
    \Delta\mathcal{W}_s^{\text{SDYM}}=-\frac{g^2}{4 E_\text{L}}\frac{\langle \bar{1}\bar{2}\rangle^2 \langle\bar{3}4\rangle^2}{k_1k_2k_3}\frac{k_1-k_2}{k^2} \,,
\end{align}
which matches the gauge variation between Feynman gauge and axial gauge \eqref{SDYMaxialcorr}. We conclude that the four-point SDYM correlator in axial gauge and complete gauge are equal.

%%%%%%%%%%%%%%%%%%%%%%%%%%%%%%%%%%%%%%%%%%%%%%%%%%%%%%%%%%%%%
\section{Collinear correlators}
\label{app:collinear}
%%%%%%%%%%%%%%%%%%%%%%%%%%%%%%%%%%%%%%%%%%%%%%%%%%%%%%%%%%%%%

In this section, we generalize the recent observation that there are nontrivial $(+\dots+-)$ amplitudes \cite{Guevara:2026qzd} to AdS correlators (there are more details than in \cite{Skvortsov:2026gtq}).

The three-point function is accompanied by the three-momentum conserving delta function $\delta^4(\sum_{i=1}^{3} i_A\bar{i}_{A'}-E\epsilon_{AA'})$, which also takes into account the non-conservation in the radial direction. Since the spinors live in a two-dimensional vector space, we can choose basis vectors $r_A$ and $3_A$. The former is an arbitrary reference spinor $r_A \neq 3_A$. Using this basis, we find
\begin{align}
    \begin{aligned}
        \sum_{i=1}^{3} i_A\bar{i}_{A'}-E\epsilon_{AA'}&=\frac{r_A}{\langle r3\rangle}\big(\sum_{i=1}^{3}\langle i3\rangle\bar{i}_{A'}+E3_{A'}\big) +\frac{3_A}{\langle r3 \rangle}\big(\sum_{i=1}^{3}\langle ri\rangle \bar{i}_{A'}-Er_{A'}\big) \,,\\
        \sum_{i=1}^{3}\langle i3\rangle\bar{i}_{A'}+E3_{A'} &= \bar{1}_{A'}\big(\langle 13 \rangle -E\frac{\langle\bar{2}3\rangle}{\langle \bar{1}\bar{2}\rangle}\big) + \bar{2}_{A'}\big(\langle 23 \rangle - E\frac{\langle 3\bar{2}\rangle}{\langle \bar{1}\bar{2} \rangle}\big) \,,
    \end{aligned}
\end{align}
where in the last line we chose the basis elements $\bar{1}_{A'}$ and $\bar{2}_{A'}$. Using $\delta^2(a\lambda_A+b\mu_A)=\frac{1}{|\langle \lambda\mu\rangle|}\delta(a)\delta(b)$, the delta function can be written as
\begin{align}
    \delta^4(\sum_{i=1}^{3} i_A\bar{i}_{A'}-E\epsilon_{AA'}) = \langle r3 \rangle^2 \frac{\text{sign}\langle\bar{1}\bar{2}\rangle}{\langle \bar{1}\bar{2}\rangle}\delta\big(\langle 13\rangle-E\frac{\langle\bar{2}3\rangle}{\langle \bar{1}\bar{2}\rangle}\big)\delta\big(\langle 23\rangle -E\frac{\langle3\bar{1}\rangle}{\langle \bar{1}\bar{2}\rangle}\big)\delta^2\big(\sum_{i=1}^{3}\langle ri\rangle \bar{i}_{A'}-r_{A'}E \big)\,.
\end{align}
Multiplying this measure with the correlator in \eqref{SDYM3pt} gives the AdS extension of the collinear limit of the three-point function. Note that the flat space case sits in the residue at $E=0$.

The four-point flat space amplitude contains the usual four-momentum conserving delta function $\delta^4(\sum\limits_{i=1}^4 i_A\bar{i}_{A'})$. Another delta function, $\delta(\langle 12 \rangle \langle \bar{1}\bar{2} \rangle) = \frac{\delta(\langle 12 \rangle)}{|\langle \bar{1}\bar{2}\rangle|}$ arises from applying the collinear limit $p_1^{AA'}p^2_{AA'}=0$ on the internal propagator, $\frac{1}{p_{12}^2+i\epsilon}=\text{PV}\frac{1}{p_{12}^2}-i\pi\delta(p_{12}^2)$. In AdS, the first delta function is replaced by $\delta^4(\sum\limits_{i=0}^4 i_A\bar{i}_{A'}-E\epsilon_{AA'})$ to account for the non-conservation in the radial direction. The second delta-function can be recovered by recognizing the leading energy pole of the AdS propagator as the flat space propagator after integration,
\begin{align}
    \int e^{-(k_1+k_2)z}\langle \Phi_{AA'}(-k,z)\Phi_{BB'}(k,z') \rangle e^{-(k_3+k_4)z'} = \frac{1}{E} \big(-\frac{\epsilon_{AB}\epsilon_{A'B'}}{E_\text{L}E_\text{R}}\big) + \dots\,,
\end{align}
where the ellipsis denotes corrections in higher orders of $E$, which we ignore for now. Note that this relation holds  for all boundary conditions in any gauge. With $E_\text{L}E_\text{R}=-\langle 12 \rangle \langle \bar{1}\bar{2} \rangle +EE_\text{L}$, we see that the flat space propagator is recovered, where $i\epsilon$ can be reinstated.\footnote{This will be justified below.} In order to have a smooth limit, we conclude that away from the flat limit, we should write
\begin{align} \label{ieps}
    \int e^{-(k_1+k_2)z}\langle \Phi_{AA'}(-k,z)\Phi_{BB'}(k,z') \rangle e^{-(k_3+k_4)z'} = \frac{1}{E} \big(-\frac{\epsilon_{AB}\epsilon_{A'B'}}{E_\text{L}E_\text{R}-i\epsilon}\big) + \dots\,.
\end{align}
This results in the delta function 
\begin{align} \label{delta}
    \delta(E_\text{L}E_\text{R})=\delta(\langle 12 \rangle \langle \bar{1}\bar{2} \rangle - EE_\text{L})=\delta(\langle 34 \rangle \langle \bar{3}\bar{4} \rangle-EE_\text{R})=\frac{1}{|\langle \bar{3}\bar{4} \rangle|}\delta(\langle 34 \rangle - E\frac{E_\text{R}}{\langle \bar{3}\bar{4} \rangle})
\end{align}    
in the collinear limit.

Again we choose two linearly independent spinors, $4_A$ and a reference spinor $r_A \not\propto 4_A$, to write
\begin{align}
    \sum\limits_{i=0}^4 i_A\bar{i}_{A'} = \frac{1}{\langle r4 \rangle}\Big(-\big(\sum\limits_{i=0}^4 \langle 4i \rangle\bar{i}_{A'}-4_{A'}E\big)r_A+\big(\sum\limits_{i=0}^4 \langle ri \rangle \bar{i}_{A'} - r_{A'}E\big)4_A\Big) \,.
\end{align}
This gives
\begin{align}
    \delta^4(\sum\limits_{i=0}^4 i_A\bar{i}_{A'}) = \delta^2(\sum\limits_{i=0}^4 \langle ri \rangle \bar{i}_{A'}-r_{A'}E)\delta^2(\sum\limits_{i=0}^4\langle4i\rangle \bar{i}_{A'}-4_{A'}E)\langle r4\rangle^2 \,.
\end{align}
The delta function without reference spinor is further reduced to
\begin{align}
    \delta^2(\sum\limits_{i=0}^4\langle4i\rangle \bar{i}_{A'}-4_{A'}E)=\delta^2(\langle 14 \rangle \bar{1}_{A'}+\langle 24\rangle \bar{2}_{A'}+\frac{EE_\text{R}}{\langle \bar{3}\bar{4} \rangle}\bar{3}_{A'}+4_{A'}E) \,,
\end{align}
where the constraint in the delta function \eqref{delta} was used. We now choose a basis spanned by $\bar{1}_{A'}$ and $\bar{2}_{A'}$ to get
\begin{align}
    \begin{aligned}
        \langle 14 \rangle \bar{1}_{A'}+\langle 24\rangle \bar{2}_{A'}+\frac{EE_\text{R}}{\langle \bar{3}\bar{4} \rangle}\bar{3}_{A'}+4_{A'}E &= \Big(\langle 14\rangle - \frac{E}{\langle \bar{1}\bar{2}\rangle}\big(E_\text{R}\frac{\langle \bar{2}\bar{3} \rangle}{\langle \bar{3}\bar{4} \rangle}+\langle \bar{2}4\rangle\big)\Big)\bar{1}_{A'} +\\
        &+ \Big( \langle 24 \rangle + \frac{E}{\langle \bar{1}\bar{2} \rangle} \big( E_\text{R}\frac{\langle \bar{1}\bar{3} \rangle}{\langle \bar{3}\bar{4} \rangle} +\langle \bar{1}4 \rangle\big) \Big)\bar{2}_{A'} \,.
    \end{aligned}
\end{align}
This allows one to factorize the delta function into
\begin{align}
    \delta^2(\sum\limits_{i=0}^4\langle4i\rangle \bar{i}_{A'}-4_{A'}E) = \delta\Big(\langle 14\rangle - \frac{E}{\langle \bar{1}\bar{2}\rangle}\big(E_\text{R}\frac{\langle \bar{2}\bar{3} \rangle}{\langle \bar{3}\bar{4} \rangle}+\langle \bar{2}4\rangle\big)\Big)\delta\Big(\langle 24 \rangle + \frac{E}{\langle \bar{1}\bar{2} \rangle} \big( E_\text{R}\frac{\langle \bar{1}\bar{3} \rangle}{\langle \bar{3}\bar{4} \rangle} +\langle \bar{1}4 \rangle\Big) \,.
\end{align}
The total measure now becomes
\begin{align} \label{measure}
    \begin{aligned}
        &\delta(E_\text{L}E_\text{R})\delta^4(\sum\limits_{i=0}^4i_A\bar{i}_{A'}-\epsilon_{AA'}E)=\frac{\langle r4 \rangle^2}{|\langle \bar{1}\bar{2}\rangle||\langle \bar{3}\bar{4} \rangle|}\delta^2(\sum\limits_{i=0}^4 \langle ri \rangle \bar{i}_{A'}-r_{A'}E)\times\\
        &\times\delta\Big(\langle 14\rangle - \frac{E}{\langle \bar{1}\bar{2}\rangle}\big(E_\text{R}\frac{\langle \bar{2}\bar{3} \rangle}{\langle \bar{3}\bar{4} \rangle}+\langle \bar{2}4\rangle\big)\Big)\times\delta\Big(\langle 24 \rangle + \frac{E}{\langle \bar{1}\bar{2} \rangle} \big( E_\text{R}\frac{\langle \bar{1}\bar{3} \rangle}{\langle \bar{3}\bar{4} \rangle} +\langle \bar{1}4 \rangle\Big)  \delta(\langle 34 \rangle - E\frac{E_\text{R}}{\langle \bar{3}\bar{4} \rangle})\,,
    \end{aligned}
\end{align}
which recovers the flat space measure in the limit $E\rightarrow 0$.

\paragraph{Propagators.} In the above, we assumed, but not derived, that $i\epsilon$ appears in the propagator after integration. Moreover, we did not have any means to know how and if $i\epsilon$ enters the energy correction terms of this integral. Here, we attempt to resolve both issues and derive some useful properties of propagators in the process. We start by relating the homogeneous terms in the propagator to the inhomogeneous ones, after which we solidify the statement that $i\epsilon$ appears in the inhomogeneous term the way we expected. Using the derived relation between the homogeneous and inhomogeneous terms, we argue that $i\epsilon$ enters the homogeneous terms in a similar way.

In Section \ref{sec:Fgauge}, it was discussed that the homogeneous terms in the Feynman gauge propagator for Dirichlet and Neumann boundary conditions can be obtained from the inhomogeneous solution by applying the method of images. It turned out that there are two operations one can apply to the inhomogeneous term of a propagator, $G_\text{inh}^{\mu\nu}(-k,z;k,z')$: reflecting the radial coordinate $z'$ along the boundary, i.e. $z'\rightarrow -z'$, and reflecting the spacetime index $\nu$ along the vector $n_\nu$. Let us denote the operators that implement these transformations by $R(z')$ and $R_{\mu\nu}=\eta_{\mu\nu}-2\frac{n_\mu n_\nu}{n^2}$, respectively. Indeed, if $G_\text{inh}^{\mu\nu}(-k,z;k,z')$ solves the equation of motion for some differential operator $D_{\mu\nu}^{-k,z}$ that acts on the $z$ coordinate and contains three-momentum $-k^{AA'}$,
\begin{align}
    D_{\mu\nu}^{-k,z} G_\text{inh}^{\nu\rho}(-k,z;k,z')=\delta_\mu^\rho \delta(z-z') \,,
\end{align}
then $G_{\text{hom},1}^{\mu\nu}(-k,z;k,z')\equiv R(z')G_\text{inh}^{\mu\nu}(-k,z;k,z')=G_\text{inh}^{\mu\nu}(-k,z;k,-z')$ is a homogeneous solution,
\begin{align}
    D^{-k,z}_{\mu\nu}G_{\text{hom},1}^{\nu\rho}(-k,z;k,z') = 0\,.
\end{align}
Also, if $D^{-k,z}_{\mu\nu}R\fud{\nu}{\rho}=R\fud{\nu}{\rho}D^{-k,z}_{\mu\nu}$, then $G_{\text{hom},2}^{\mu\nu}\equiv R\fud{\nu}{\rho}G_{\text{hom},1}^{\mu\rho}$ is a homogeneous solution.
We note that this is true for Feynman gauge, because $D^{-k,z}_{\mu\nu}=\eta_{\mu\nu}\Box$. In axial gauge, the kinetic operator is $n$-transverse (in both indices), $n^\nu D^{-k,z}_{\mu\nu}=0$, and so $R_{\mu\nu}$ can be used to derive a new homogeneous solution. Since $G^{\mu\nu}_{\text{hom},1}$ itself is also $n$-transverse in axial gauge, $G^{\mu\nu}_{\text{hom},1}=G^{\mu\nu}_{\text{hom},2}$ and reflection of the index does not yield a new solution.

Next, in any gauge the propagator for mixed boundary conditions contains $G_{\text{hom},\gamma}^{\mu\nu}$, which consists of the $\Pi_\text{e}^{\mu\nu}$ and $\Pi_\text{o}^{\mu\nu}$. They can be used to construct $G_{\text{hom},3}^{\mu\nu}\equiv \Pi_\text{o}\fud{\nu}{\rho}G_{\text{hom},1}^{\mu\rho}$ and $G_{\text{hom},4}^{\mu\nu}\equiv \Pi_\text{e}\fud{\nu}{\rho}G_{\text{hom},1}^{\mu\rho}$. If $D^{-k,z}_{\mu\nu}\Pi_{e}^{\nu\rho}=\Pi_{e}^{\nu\rho}D^{-k,z}_{\mu\nu}$, they are homogeneous solutions too. Again, in Feynman gauge this holds because $D_{\mu\nu}^{-k,z}=\eta_{\mu\nu}\Box$. In axial gauge, $D^{-k,z}_{AA,BB}=(\partial_z^2-k^2)\epsilon_{AB}\epsilon_{AB}-\frac{1}{2}k_{AA}k_{BB}$ and both $\Pi_\text{e}^{\mu\nu}$ and $\Pi_{\text{o}}^{\mu\nu}$ are $k$-transverse in both indices. Hence, $G_{\text{hom},3}^{\mu\nu}$ and $G_{\text{hom},4}^{\mu\nu}$ are also homogeneous solutions. Note that in axial gauge $G_{\text{hom},1}^{\mu\nu}\sim \Pi^{\mu\nu}_\text{e}$, so $G_{\text{hom},1}^{\mu\nu}=G_{\text{hom},3}^{\mu\nu}$. This can be generalized to any gauge, as the differential operator before gauge fixing reads $D^{-k,z}_{AA',BB'}=(\partial_z^2-k^2)\epsilon_{AB}\epsilon_{A'B'}-\frac{1}{2}\nabla^{-k,z}_{AA'}\nabla^{-k,z}_{BB'}$, where $\nabla^{-k,z}_{BB'}=-(k_{BB'}+\epsilon_{BB'}\partial_z)$. The fact that the operators $\Pi_\text{e/o}^{\mu\nu}$ are $k$- and $\epsilon$-transverse in both indices allows one to treat $G_{\text{hom},3}^{\mu\nu}$ and $G_{\text{hom},4}^{\mu\nu}$ as homogeneous solutions in any gauge.

To summarize, acting on an inhomogeneous solution of Yang--Mills with the operator $R(z')$, potentially followed by $\Pi_\text{e/o}^{\mu\nu}$, gives a homogeneous solution in any gauge, albeit not always a new solution. It is only the operator $R_{\mu\nu}$ that does not always yield a solution, so this needs to be checked case-by-case. The homogeneous terms of the descendant propagators, e.g. the $\langle \Psi\Phi \rangle$ propagator, can be obtained in the usual way by acting on $\langle \Phi\Phi \rangle$ with derivatives. It can also be shown that the same projectors map the inhomogeneous solutions of the Chalmers--Siegel theory to homogeneous ones.

At this stage, one would like to use the argument that applying the above-mentioned projectors, or derivatives to pass to Chalmer-Siegel, does not affect the $i\epsilon$ in \eqref{ieps}. However, there are two issues: i) the projector $R(z')$ might not commute with the integrals and ii) the integral of the inhomogeneous term already produces higher energy corrections, for which it is still unknown where the $i\epsilon$ appears. To resolve both problems, we consider the $\langle \Phi\Phi \rangle$ propagator suggested in \cite{raju2011recursion}, which explicitly contains $i\epsilon$. It is the propagator in axial gauge for Dirichlet boundary conditions and it reads
\begin{align}
    \langle \Phi_{AA}(-k,z)\Phi_{BB}(k,z')\rangle_\text{D}^{\text{Axial}} = -\frac{2}{\pi}\int_0^\infty dp \frac{\sin(pz)\sin(pz')}{k^2+p^2-i\epsilon}\mathcal{T}_{AA,BB} \,,
\end{align}
with $\mathcal{T}_{AA,BB}=\epsilon_{AB}\epsilon_{A'B'}-\frac{k_{AA}k_{BB}}{2p^2}$. The integrand has poles at $p=\pm i\sqrt{k^2-i\epsilon}$. It is then straightforward to find the inhomogeneous term to be
\begin{align}
    \langle \Phi_{AA}(-k,z)\Phi_{BB}(k,z')\rangle_\text{inh}^{\text{Axial}} = \frac{1}{2\pi}\int_0^\infty dp \frac{e^{ip(z-z')}+e^{-ip(z-z')}}{k^2+p^2-i\epsilon}\mathcal{T}_{AA,BB} \,.
\end{align}
From here one obtains the homogeneous terms for mixed boundary conditions by applying the operators discussed above. We also have an ansatz for the propagator containing $i\epsilon$ in Feynman gauge:
\begin{align}
    \langle \Phi_{A,A'}(-k,z)\Phi_{B,B'}(k,z')\rangle_\text{inh}^{\text{F}} = \frac{1}{2\pi}\int_0^\infty dp \frac{e^{ip(z-z')}+e^{-ip(z-z')}}{k^2+p^2-i\epsilon}\epsilon_{AB}\epsilon_{A'B'} \,.
\end{align}
Again, the procedure discussed above yields the homogeneous terms one may add.

These integral expressions give the propagators from the main text after evaluating the integrals and setting $\epsilon=0$. Now we keep $\epsilon\ll 1$ finite to see how this changes the propagators. It shifts the pole to $p\approx\pm i (k -i\epsilon)$, where $\epsilon$ was rescaled. Now, the propagator schematically looks like\footnote{More precisely, $\epsilon$ also appears in $f$ and $g$, but only in such a way that is not relevant for the collinear limit.}
\begin{align}
    \langle \Phi_{A,A'}(-k,z)\Phi_{B,B'}(k,z')\rangle= f(k;z,z')e^{-(k-i\epsilon)|z-z'|}+g(k;z,z')e^{-(k-i\epsilon)(z+z')} + \text{linear terms}
\end{align}
and we observe that the projectors do not affect the $i\epsilon$. The bulk integral becomes
\begin{align}
    \int e^{-(k_1+k_2)z}\langle \Phi_{A,A'}(-k,z)\Phi_{B,B'}(k,z')\rangle e^{-(k_3+k_4)z'} = \frac{1}{E_\text{L}E_\text{R}-i\epsilon}\Big(-\frac{\epsilon_{AB}\epsilon_{A'B'}}{E}+\dots\Big) \,.
\end{align}
This confirms the assumption from \eqref{ieps} that $i\epsilon$ appears in the denominator of the term containing the energy pole after integration, and it extends this to the higher order terms in $E$. Moreover, derivatives that might hit the propagator in a diagram do not affect the $i\epsilon$ either. Thus, the measure \eqref{measure} can be used for the entire correlator, not just the energy pole.

\footnotesize
\providecommand{\href}[2]{#2}\begingroup\raggedright\endgroup

\end{document}